\documentclass[traditabstract]{aa}
\usepackage{graphicx}
\usepackage[varg]{txfonts}
\usepackage{longtable}
\usepackage{lscape}
\usepackage{natbib}
\usepackage{url}
\usepackage{color}
\usepackage{multirow,bigdelim}
\usepackage{cases}
\bibpunct{(}{)}{;}{a}{}{,} % to follow the A&A style
\newcommand\kms{{\rm\,km\,s^{-1}}}

\newcommand\teff{T_{\rm eff}}

\begin{document}

%=======================================================================
\title{
Chemical evolution of the Galactic bulge as traced by
microlensed dwarf and subgiant stars\thanks{Based on 
data obtained with the European Southern Observatory
telescopes (Proposal ID:s 87.B-0600, 88.B-0349, 89.B-0047, and 90.B-0204),
the Magellan Clay telescope at the Las Campanas observatory,
and the Keck I telescope at the W.M. Keck 
Observatory, which is operated as a scientific partnership among 
the California Institute of Technology, the University of California 
and the National Aeronautics and Space Administration.}
\fnmsep
\thanks{Tables 2, 3, 4, and 5 are available in electronic form at the CDS via 
anonymous ftp to {\tt cdsarc.u-strasbg.fr (130.79.128.5)} or via 
{\tt http://cdsweb.u-strasbg.fr/cgi-bin/qcat?J/A+A/XXX/AXX}.
} 
}
\subtitle{
V. Evidence for a wide age distribution and a complex MDF
}
\titlerunning{Chemical evolution of the Galactic bulge as traced by
microlensed dwarf and subgiant stars. V.}

\author{
T.~Bensby\inst{1}
\and
J.C.~Yee\inst{2}
\and
S.~Feltzing\inst{1}
\and
J.A.~Johnson\inst{2}
\and
A.~Gould\inst{2}
\and
J.G.~Cohen\inst{3}
\and
M.~Asplund\inst{4}
\and
J.~Mel\'endez\inst{5}
\and
S.~Lucatello\inst{6}
\and\\
C.~Han\inst{7}
\and
I.~Thompson\inst{8}
\and
A.~Gal-Yam\inst{9}
\and
A.~Udalski\inst{10}
%\and
%P.~Baumann\inst{5}
%\and
%J.~Chanam\'e\inst{11}
\and
D.P.~Bennett\inst{11}
\and
I.A.~Bond\inst{12}
\and
W.~Kohei\inst{13}
\and
T.~Sumi\inst{13}
\and\\
D.~Suzuki\inst{13}
\and
K.~Suzuki\inst{14}
\and
S.~Takino\inst{14}
\and
P.~Tristram\inst{15}
\and 
N.~Yamai\inst{16}
\and
A.~Yonehara\inst{16}
 }

\institute{Lund Observatory, Department of Astronomy and 
Theoretical physics, 
Box 43, SE-221\,00 Lund, Sweden
%\email{sofia, daniela, jennifer@astro.lu.se}
\and
Department of Astronomy, Ohio State University, 140 W. 18th Avenue, 
Columbus, OH 43210, USA
%\email{jaj,\,gould@astronomy.ohio-state.edu}
\and
Palomar Observatory, Mail Stop 249-17, California Institute of Technology, 
Pasadena, CA 91125, USA
\and
Research School of Astronomy \& Astrophysics, Mount Stromlo Observatory, Cotter Road,
Weston Creek, ACT 2611, Australia
%\email{asplund@mpa-garching.mpg.de}
\and
Departamento de Astronomia do IAG/USP, Universidade de S\~ao Paulo,
Rua do Mat\~ao 1226, S\~ao Paulo, 05508-900, SP, Brasil
%\email{jorge@astro.up.pt}
\and
INAF-Astronomical Observatory of Padova, Vicolo dell'Osservatorio 5, 
35122 Padova, Italy
%\email{sara.lucatello@oapd.inaf.it}
\and
Department of Physics, Chungbuk National University, Cheongju 361-763, Republic of Korea
\and
Carnegie Observatories, 813 Santa Barbara Street, Pasadena, CA 91101, USA
\and
Department of Particle Physics and Astrophysics, Weizmann Institute 
of Science, 76100 Rehovot, Israel
%\email{avishay.gal-yam@weizmann.ac.il}
\and
Warsaw University Observatory, A1. Ujazdowskie 4, 00-478, Warszawa, Poland
%\and
%Carnegie Institution of Washington, 5241 Broad Branch Road, 
%NW Washington, DC 20015-1305, USA
\and
University of Notre Dame, Notre Dame, IN 46556, USA
\and
Institute for Information and Mathematical Sciences, Massey University, 
Auckland 1330, New Zealand
\and
Department of Earth and Space Science, Osaka University, Osaka 560-0043, Japan
\and
Solar-Terrestrial Environment Laboratory, Nagoya University, Nagoya 464-8601, Japan
\and
Mt. John Observatory, P.O. Box 56, Lake Tekapo 8770, New Zealand
\and
Department of Physics, Faculty of Science, Kyoto Sangyo University, Kyoto, 603-8555, Japan
}

%=======================================================================

\date{Received 31 October 2012 / Accepted 28 November 2012}
%=======================================================================
\offprints{T. Bensby, \email{tbensby@astro.lu.se}}

%=======================================================================
 \abstract{
 Based on high-resolution spectra obtained during gravitational 
 microlensing events we present a detailed elemental abundance analysis
 of 32 dwarf and subgiant stars in the Galactic bulge. 
 Combined with the sample of 26 stars from the 
 previous papers in this series, we now have 58 microlensed bulge dwarfs 
 and subgiants that have been homogeneously analysed. The main characteristics 
 of the sample and the findings that can be drawn are: 
 (i) The metallicity distribution (MDF) is wide and spans all 
 metallicities between $\rm [Fe/H]=-1.9$ to $+0.6$;
 (ii) The dip in the MDF around solar metallicity that was apparent in 
 our previous analysis of a smaller sample (26 microlensed stars) is no 
 longer evident; instead it has a complex structure and indications of multiple 
 components are starting to emerge. A tentative interpretation is that there 
 could be different stellar populations at interplay,
 each with a different scale height: the thin disk, the thick disk,
 and a bar population;
 (iii) The stars with $\rm [Fe/H]\lesssim-0.1$ are old with ages 
 between 10 and 12\,Gyr;
 (iv) The metal-rich stars with $\rm [Fe/H]\gtrsim-0.1$ show a wide variety of
 ages, ranging from 2 to 12\,Gyr with a distribution that has a dominant peak 
 around 4-5\,Gyr and a tail towards higher ages;
 (v) There are indications in the  $\rm [\alpha/Fe] - [Fe/H]$ that the
 ``knee'' occurs around $\rm [Fe/H] = -0.3$ to $-0.2$, which is a slightly
 higher metallicity as compared to the ``knee'' for the local thick disk.  
 This suggests that the chemical enrichment of the metal-poor
 bulge has been somewhat faster than what is observed for the local thick disk.
 The results from the microlensed bulge dwarf stars in combination with other 
 findings in the literature, in particular the evidence that the 
 bulge has cylindrical rotation, indicate that the Milky Way could
 be an almost pure disk galaxy. The
 bulge would then just be a conglomerate of the other Galactic stellar populations 
 (thin disk, thick disk, halo, and ...?), residing together in 
 the central parts of the Galaxy, influenced by the Galactic bar.
 }
   \maketitle

%=======================================================================
\section{Introduction}

Galactic bulges are emerging as inherently complex features in 
spiral galaxies and are a defining component of many spiral and disk
galaxies \citep{kormendy2004,athanassoula2005}. Numerous studies have 
shown them to have several spatial structures overlaying each other. 
The Milky Way is no different and over the last decades studies of 
the Galactic bulge have revealed an increasingly complex, and not 
always consistent, picture of the stellar population present in the 
innermost few kpc of our Galaxy. The formation of this central 
concentration of stars is not fully understood and if the observed 
spatial structures are uniquely related to dynamical and chemical 
features is a very actively studied field.
New surveys such as the VISTA Variables in the Via Lactea (VVV) public 
survey \citep{saito2010} are looking into this 
intriguing Galactic component. Several outstanding questions remain 
un-resolved including the true shape of the metallicity distribution 
function (MDF) and how the MDF is connected to different spatial and 
kinematical structures. A recent discussion of these issues can be 
found in, e.g., \cite{babusiaux2010}.

A number of formation scenarios for the Galactic bulge have been 
proposed and were first summarised by \citet{wyse1992} and set out 
in greater detail in later works \citep[e.g.,][]{kormendy2004,athanassoula2005}. 
The first major scenario is in situ star formation from primordial
gas in the gravitational collapse \citep[e.g.,][]{eggen1962,matteucci1990},
or merging of clumps within the disk at high redshift 
\citep[e.g.,][]{noguchi1999,bournaud2009}. Historically, this has 
been thought to be the major component of the Galactic bulge,
and bulges in general, and is referred to as a classical bulge. 
However, in the Milky Way it now seems that the spherical
component that should form from an initial collapse is after all a
rather small component. The mass of the classical bulge might be less
than 8\,\% of the mass of the disk \citep{shen2010}.

Instead, the scenario that is currently emerging as the most compelling
is secular evolution in which a bulge and bar forms from buckling instabilities 
in the disk \citep[e.g.,][]{combes1990,kormendy2004,athanassoula2005,ness2012}.
Bars are common in other galaxies and the existence of a bar in
the Milky Way in addition to its classical, spherical bulge has been
debated for some time. Star counts have shown that there
definitely is a bar \citep{stanek1994,mcwilliam2010,saito2011,saito2012}.
Studies of the stellar kinematics show that a large fraction of the 
stars exhibit cylindrical rotation  \citep[e.g.,][]{shen2010,howard2009,sumi2003}
which is indicative of a boxy bulge, i.e., one with a bar \citep{athanassoula2005}. 
Thus both star counts
as well as the stellar kinematics unambiguously show the presence of a
bar. Since the bar forms from buckling instabilities in a disk, there may be
a strong connection between the Galactic stellar disk and the bar population.  

A realistic approach could be a mixed scenario encompassing two or more 
formation scenarios.  Mixed formation has been observed 
in other galaxies \citep[e.g.,][]{prugniel2001} and a mixed origin of
the Galactic bulge is explored by a number of theoretical studies
predicting both the shape of the MDF as well as the elemental
abundance trends that should be expected
\citep[e.g.,][]{tsujimoto2012,grieco2012}.  The presence of the
first stars in the Galactic bulge as predicted in, e.g., \cite{wyse1992}
who discuss the possibility of merging of dwarf satellite galaxies that
sink into the Galactic bulge, have been further developed
within the context of cosmological simulations \citep{tumlinson2010}.

Elemental abundances in stars are able to provide vital clues to the
formation time scales involved in the stellar population(s) present in
the Galactic bulge as well as to constrain the shape of the MDF and 
slope of the IMF \citep[e.g.,][]{ballero2007b,tsujimoto2012}. 
Initial studies of red giants
revealed the Galactic bulge to be relatively metal-rich
but with a wide metallicity distribution
\citep[e.g.,][]{mcwilliam1994,zoccali2008}.
It also appeared enhanced in
the $\alpha$-elements, indicative of a short formation time
\citep{mcwilliam1994,fulbright2006,fulbright2007}. The elevated levels
of $\alpha$-elements continued to super-solar metallicities, thus
setting the bulge stellar population apart from what was found in the
solar neighbourhood, i.e., with solar levels of most elements at these
metallicities \citep[e.g.,][]{edvardsson1993}.  
As the bulge is far away the stars are faint and hence these
studies were and, essentially are, confined to the most luminous
giants.  Results based on giant spectra are not trivial to interpret.
Physical processes within giant stars also erase
some original abundance signatures.  This is the case for C, N, and Li
in all giants, and for O, Na, Mg, and Al in some cases 
\citep[e.g.,][]{kraft1992}.  Additionally, although recent differential
analysis of giant stars in the bulge and disk have achieved high
precision \citep{alvesbrito2010}, the situation at high metallicities
is still unclear due to heavy blending, as the cool atmospheres of
giants, rich in molecules, are difficult to analyse
\citep[e.g.,][]{lebzelter2012}.  

The spectra of dwarfs, even
metal-rich ones, are fairly straightforward to analyse and are the
best tracers of galactic chemical evolution \citep{edvardsson1993}.
The dwarf stars are also unique in that they give star-by-star 
age information and not just the bulk age-information available through 
analysis of colour magnitude diagrams. Normally dwarf stars in the bulge
are too faint to be observed with high-resolution spectrographs.
However, during microlensing events they can brighten by several magnitudes
and during the last few years there have been several studies
that have utilised this phenomenon and obtained spectra that allowed
detailed elemental abundance analysis of bulge dwarf stars.
In total there have been 26 bulge dwarfs observed
\citep{cavallo2003,johnson2007,johnson2008,cohen2008,cohen2009,bensby2009,
epstein2010,bensby2010,bensby2011} all homogeneously analysed in 
\cite{bensby2011}. These studies have revitalised the study
of chemical evolution and structure of the bulge.

%---------------------------------------------------------------------
\begin{table*}
\centering
\caption{
Summary of the 32 microlensed bulge dwarf stars new in this study (sorted by observation date)$^\dagger$. 
\label{tab:events}
}
\setlength{\tabcolsep}{1.4mm}
\tiny
\begin{tabular}{lccrrrrrrrrrr}
\hline\hline
\noalign{\smallskip}
\multicolumn{1}{c}{Object}     &
RAJ2000                        &
DEJ2000                        &
 \multicolumn{1}{c}{$l$}       &
 \multicolumn{1}{c}{$b$}                           &
 \multicolumn{1}{c}{$T_{E}$}   &
 \multicolumn{1}{c}{$T_{max}$}                     &  
 \multicolumn{1}{c}{$A_{max}$}                     &
 \multicolumn{1}{c}{$T_{obs}$} &  
 Exp.                          &
 $S/N$                         &
 Spec.                         &
 \multicolumn{1}{c}{$R$}       \\
                               &
[hh:mm:ss]                     &
[dd:mm:ss]                     &
 [deg]                         &    
 [deg]                         &  
 [days]                        &
 \multicolumn{1}{c}{[HJD]}                         &
                               &   
 \multicolumn{1}{c}{[MJD]}     &  
 \multicolumn{1}{c}{[s]}       &
                               &
                               &
                               \\  
\noalign{\smallskip}
\hline
\noalign{\smallskip}
MOA-2011-BLG-104S              & 17:54:22.48 & $-$29:50:01.67 & $ 0.21$ & $-2.10$ &  41 &  5670.31  &    44 & 5670.244 & 7200 &  25 & UVES & 42\,000  \\
MOA-2011-BLG-090S              & 18:10:29.85 & $-$26:38:43.03 & $ 4.74$ & $-3.62$ & 187 &  5692.26  &    38 & 5681.149 & 7200 &  50 & UVES & 42\,000  \\
MOA-2011-BLG-174S              & 17:57:20.62 & $-$30:22:47.57 & $ 0.06$ & $-2.93$ &  79 &  5716.13  &   160 & 5715.295 & 7200 &  40 & UVES & 42\,000  \\
MOA-2011-BLG-234S              & 18:10:56.78 & $-$26:33:48.59 & $ 4.86$ & $-3.67$ &  49 &  5736.73  &  1270 & 5736.354 & 4650 &  60 & MIKE & 55\,000  \\
MOA-2011-BLG-278S              & 17:54:11.32 & $-$30:05:21.56 & $-0.03$ & $-2.19$ &  18 &  5744.74  &   320 & 5744.080 & 7200 &  18 & MIKE & 64\,000  \\
\multirow{4}{*}{MOA-2011-BLG-191S}   & \multirow{4}{*}{17:51:40.15} & \multirow{4}{*}{$-$29:53:26.15} & \multirow{4}{*}{$-0.14$} & \multirow{4}{*}{$-1.62$} & \multirow{4}{*}{192} & \multirow{4}{*}{5763.32}   &   \multirow{4}{*}{375} & \ldelim\{{4}{1.0em} 5762.300 & 2400 &  60 & HIRES& 46\,000  \\
                               &             &                &         &         &     &           &       & 5762.991 & 7200 &  25 & UVES & 90\,000  \\
                               &             &                &         &         &     &           &       & 5762.964 & 5400 &  55 & MIKE & 55\,000  \\
                               &             &                &         &         &     &           &       & 5763.155 & 3600 &  50 & MIKE & 55\,000  \\
OGLE-2011-BLG-1072S            & 17:56:53.94 & $-$28:50:53.31 & $ 1.34$ & $-2.08$ &  17 &  5775.40  &   365 & 5774.977 & 7200 &  55 & UVES & 60\,000  \\
OGLE-2011-BLG-1105S            & 17:54:48.38 & $-$31:25:57.86 & $-1.13$ & $-2.98$ &  36 &  5780.77  &    76 & 5780.281 & 3600 &  20 & HIRES& 38\,000  \\
OGLE-2011-BLG-0950S            & 17:57:16.62 & $-$32:39:57.15 & $-1.93$ & $-4.05$ &  73 &  5786.40  &   115 & 5785.202 & 5850 &  25 & UVES & 42\,000  \\
OGLE-2011-BLG-0969S            & 18:09:41.13 & $-$31:11:04.70 & $ 0.65$ & $-5.63$ &  33 &  5790.40  &    37 & 5789.042 & 7071 &  45 & UVES & 42\,000  \\
OGLE-2011-BLG-1410S            & 17:32:49.61 & $-$29:23:10.60 & $-1.87$ & $+2.12$ &  19 &  5835.10  &    26 & 5834.988 & 7200 &  15 & UVES & 42\,000  \\
\multirow{2}{*}{MOA-2011-BLG-445S}   & \multirow{2}{*}{18:04:45.63} & \multirow{2}{*}{$-$28:35:44.06} & \multirow{2}{*}{$ 2.41$} & \multirow{2}{*}{$-3.45$} & \multirow{2}{*}{124}    & \multirow{2}{*}{5873.09}          &    \multirow{2}{*}{33} & \ldelim\{{2}{1.0em}5870.993 & 3600 & \rdelim\}{2}{3mm}\multirow{2}{*}{50}    & UVES & 42\,000  \\
                               &             &                &         &         &     &           &       & 5871.990 & 3600 &     & UVES & 42\,000  \\
OGLE-2012-BLG-0026S            & 17:34:18.70 & $-$27:08:33.90 & $ 0.19$ & $+3.07$ &  94 &  5991.52  &    80 & 5990.314 & 7200 &  45 & UVES & 42\,000  \\
OGLE-2012-BLG-0211S            & 18:10:10.96 & $-$25:01:40.20 & $ 6.12$ & $-2.78$ &  25 &  6013.12  &  2000 & 6013.311 & 7200 &  30 & UVES & 70\,000  \\
OGLE-2012-BLG-0270S            & 17:14:42.46 & $-$29:35:50.40 & $-4.26$ & $+5.27$ &  52 &  6034.67  &    46 & 6033.289 & 7200 &  30 & UVES & 70\,000  \\
MOA-2012-BLG-202S              & 18:12:34.84 & $-$25:02:59.79 & $ 6.36$ & $-3.27$ &  44 &  6039.63  &   200 & 6039.219 & 7200 &  30 & UVES & 42\,000  \\
MOA-2012-BLG-187S              & 18:08:02.16 & $-$29:28:11.45 & $ 1.99$ & $-4.50$ &  19 &  6040.67  &    33 & 6040.211 & 7200 &  50 & UVES & 42\,000  \\
MOA-2012-BLG-022S              & 17:57:40.97 & $-$27:29:56.49 & $ 2.59$ & $-1.55$ & 592 &  6021.40  &   210 & 6055.209 & 7200 &  25 & UVES & 42\,000  \\
OGLE-2012-BLG-0521S            & 18:05:36.74 & $-$25:45:47.42 & $ 4.98$ & $-2.23$ &  29 &  6058.60  &   165 & 6058.103 & 6600 &  55 & UVES & 42\,000  \\
OGLE-2012-BLG-0563S            & 18:05:57.72 & $-$27:42:43.20 & $ 3.31$ & $-3.25$ &  53 &  6069.06  &   440 & 6067.334 & 7200 &  60 & UVES & 42\,000  \\
OGLE-2012-BLG-0617S            & 17:54:53.87 & $-$31:08:19.10 & $-0.86$ & $-2.85$ &  20 &  6068.60  &   500 & 6068.144 & 7200 &  95 & UVES & 42\,000  \\
MOA-2012-BLG-291S              & 18:02:43.08 & $-$28:23:00.86 & $ 2.38$ & $-2.96$ &  21 &  6070.06  &    34 & 6070.241 & 7200 &  30 & UVES & 42\,000  \\
MOA-2012-BLG-391S              & 17:58:56.67 & $-$31:26:30.82 & $-0.69$ & $-3.75$ &  11 &  6101.59  &    48 & 6101.106 & 7200 &  35 & UVES & 42\,000  \\
MOA-2012-BLG-410S              & 18:10:22.62 & $-$25:10:16.83 & $ 6.02$ & $-2.89$ &  77 &  6103.22  &   240 & 6103.005 & 7200 &  25 & UVES & 42\,000  \\
OGLE-2012-BLG-1156S            & 18:00:35.95 & $-$28:11:15.20 & $ 2.32$ & $-2.45$ &  16 &  6134.23  &    21 & 6134.110 & 7200 &  30 & UVES & 42\,000  \\
OGLE-2012-BLG-1217S            & 18:10:16.85 & $-$27:37:59.00 & $ 3.84$ & $-4.05$ &  16 &  6147.34  &    26 & 6146.979 & 7200 &  45 & UVES & 42\,000  \\
MOA-2012-BLG-532S              & 17:58:41.13 & $-$30:02:11.80 & $ 0.50$ & $-3.01$ &  10 &  6151.63  &   151 & 6151.145 & 7200 &  17 & UVES & 42\,000  \\
OGLE-2012-BLG-1274S            & 17:45:00.65 & $-$34:32:49.50 & $-4.86$ & $-2.81$ &  29 &  6165.25  &   960 & 6164.969 & 7200 &  90 & UVES & 42\,000  \\
OGLE-2012-BLG-0816S            & 17:58:15.98 & $-$29:57:21.45 & $ 0.53$ & $-2.89$ &  81 &  6175.52  &    31 & 6175.261 & 1800 &  15 &HIRES & 48\,000  \\
\multirow{2}{*}{OGLE-2012-BLG-1279S}   & \multirow{2}{*}{17:56:52.40} & \multirow{2}{*}{$-$31:46:54.12} & \multirow{2}{*}{$-1.21$} & \multirow{2}{*}{$-3.54$} & \multirow{2}{*}{93}    & \multirow{2}{*}{6184.20}          &    \multirow{2}{*}{1535} & \ldelim\{{2}{1.0em}6183.040 & 2400 & \rdelim\}{2}{3mm}\multirow{2}{*}{115}   & MIKE & 42\,000  \\
                               &             &                &         &         &     &           &       & 6183.972 & 3600 &     & MIKE & 42\,000  \\
OGLE-2012-BLG-1534S            & 18:00:46.35 & $-$28:01:00.30 & $ 2.48$ & $-2.40$ &  27 &  6206.96  &    83 & 6205.980 & 7200 &  80 & UVES & 42\,000  \\
OGLE-2012-BLG-1526S            & 18:09:43.13 & $-$28:48:45.20 & $ 2.75$ & $-4.51$ &  36 &  6214.91  &    13 & 6216.991 & 7200 &  35 & UVES & 42\,000  \\
\noalign{\smallskip}
\hline
\end{tabular}
\flushleft
{\tiny      
{\bf Notes.} $^\dagger$ Given for each microlensing event is: RA and DE coordinates (J2000) read from the
fits headers of the spectra (the direction where the telescope pointed during observation); 
galactic coordinates ($l$ and $b$); 
duration of the event in days ($T_{E}$); time when maximum magnification occured ($T_{max}$); 
maximum magnification ($A_{max}$); time when the event was observed
with high-resolution spectrograph ($T_{obs}$);
the exposure time (Exp.),
the measured signal-to-noise ratio per pixel at $\sim$6400\,{\AA} ($S/N$); the
spectrograph that was used (Spec); the spectral resolution ($R$).
}
\end{table*}
%---------------------------------------------------------------------

For instance, the microlensed
dwarf stars \citep{bensby2010,bensby2011} as well as of red giant branch
(RGB) stars \citep{hill2011,uttenthaler2012} have revealed 
a dual-component bulge MDF - one metal-poor component
around $\rm [Fe/H]\approx -0.6$ and one metal-rich component
around $\rm [Fe/H]\approx +0.3$.  There has also been considerable 
evidence that secular evolution of the thick disk may have played a large 
role in the origin and evolution of the bulge. This is based on
the apparent similarity of the $\rm [\alpha/Fe]-[Fe/H]$ abundance 
trends in the bulge at sub-solar metallicities 
\citep[e.g.,][]{melendez2008,bensby2010,alvesbrito2010,gonzalez2011} and those 
of the nearby thick disk \citep[e.g,][]{bensby2007letter,fuhrmann2008,reddy2006}.
Also, the metal-poor peak in the bi-modal bulge MDF appears to coincide
with the MDF for the thick disk and considering that the
age structure of the metal-poor bulge dwarfs is the same as for the
thick disk, i.e., mainly old, it might well be that the bulge and the 
thick disk have had similar or even shared chemical histories
\citep{melendez2008,alvesbrito2010,bensby2010,gonzalez2011,bensby2011}.
Of course, the comparisons between the bulge and the thick disk have
mainly used nearby thick disk samples, and if the bulge partly 
originated through secular evolution of the thick disk it would be 
desirable to compare with thick disk stellar samples located in situ
in the inner Galactic disk. The only such study is the
\cite{bensby2010letter} study of 44 red giants, located between 
galactocentric radii 4-7\,kpc and up to 3-4\,kpc from the Galactic plane. 
Those results show that the inner disk region has a similar 
abundance structure as seen in the solar neighbourhood,
i.e., a thin disk-thick disk dichotomy, and that the similarities between 
the bulge and the thick disk persist when comparing to an inner disk
sample.

A recent controversy is the overall age distribution in the bulge. 
Photometric observations of different bulge fields point to an 
exclusively old stellar population with ages greater than 10\,Gyr 
\citep{zoccali2003,clarkson2008,brown2010,clarkson2011}.
In addition to the very old stars, the observations of 
microlensed bulge dwarf stars have revealed a significant fraction of 
stars with ages between 3-7\,Gyr \citep{bensby2010,bensby2011}. 
\cite{nataf2012} tried to reconcile the dilemma
of the young bulge dwarfs by showing that a factor of 2 discrepancy 
between spectroscopic and photometric age determinations of the 
Galactic bulge main-sequence turnoff possibly can be explained
if the Galactic bulge is helium-enhanced relative to that assumed 
by standard isochrones. An intriguing recent result, 
based on spectra of 575 bulges from the Sloan 
Digital Sky Survey, is that barred galaxies show a stellar age 
distribution with a young stellar component not present in unbarred 
galaxies \citep{coelho2011}.

Another intriguing result
is the multi-component ($>4$) MDF proposed by Ness et al.~(in prep.)
based on low/intermediate-resolution spectra of 28\,000 RGB stars at 
different longitudes and latitudes in the bulge. Whether these proposed
multiple bulge components are real and where they come from 
needs to be investigated.

In this work we present a detailed elemental abundance analysis 
of 32 new microlensing events towards the Galactic bulge.
Together with the previous 26 events homogeneously analysed in
\cite{bensby2011}, the sample now contains
58 microlensed dwarf and subgiant stars in the bulge.

%-----------------------------------------------------------------------
\begin{figure*}
\resizebox{\hsize}{!}{
\includegraphics[bb=10 40 515 540,clip]{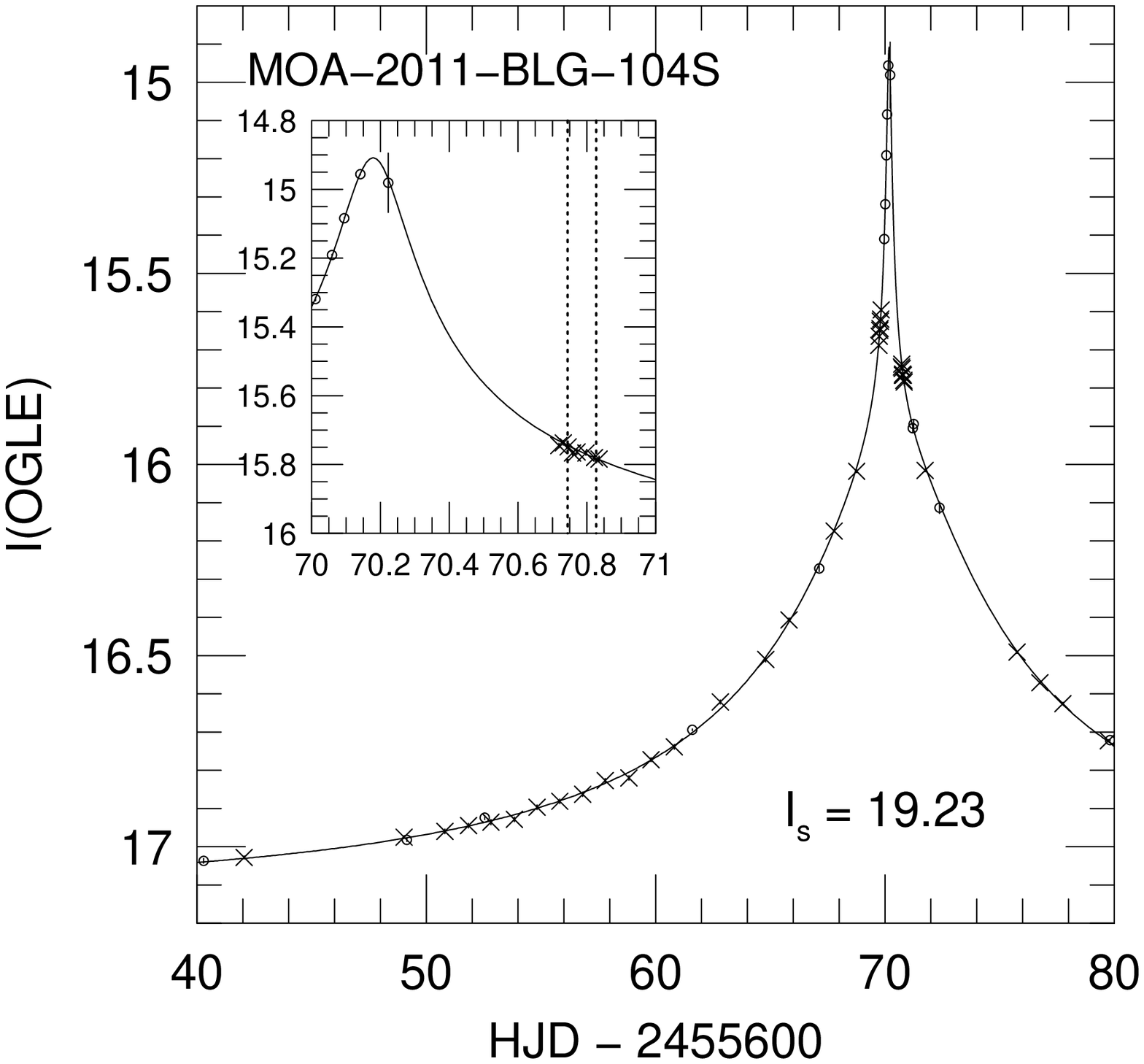}
\includegraphics[bb=67 40 515 540,clip]{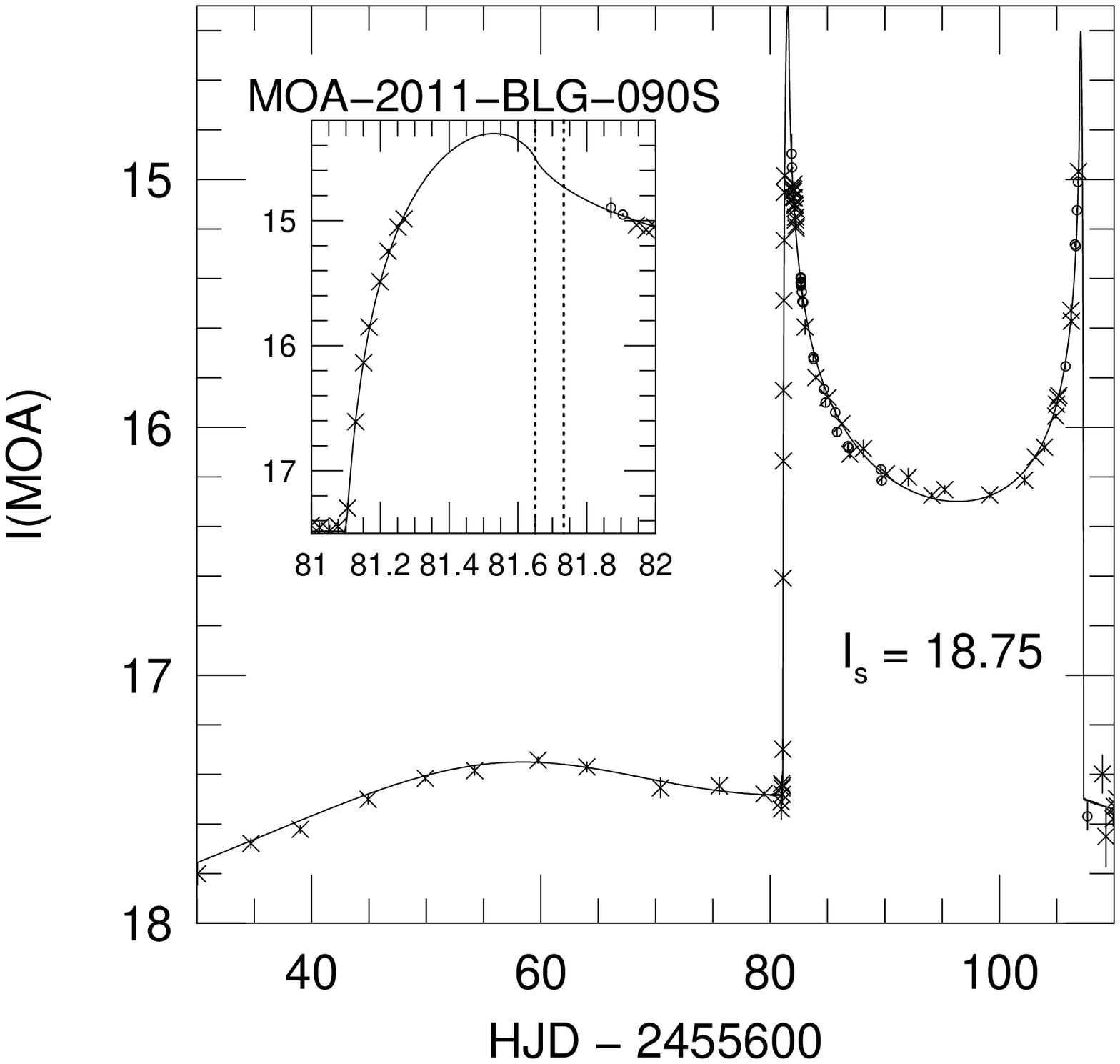}
\includegraphics[bb=67 40 515 540,clip]{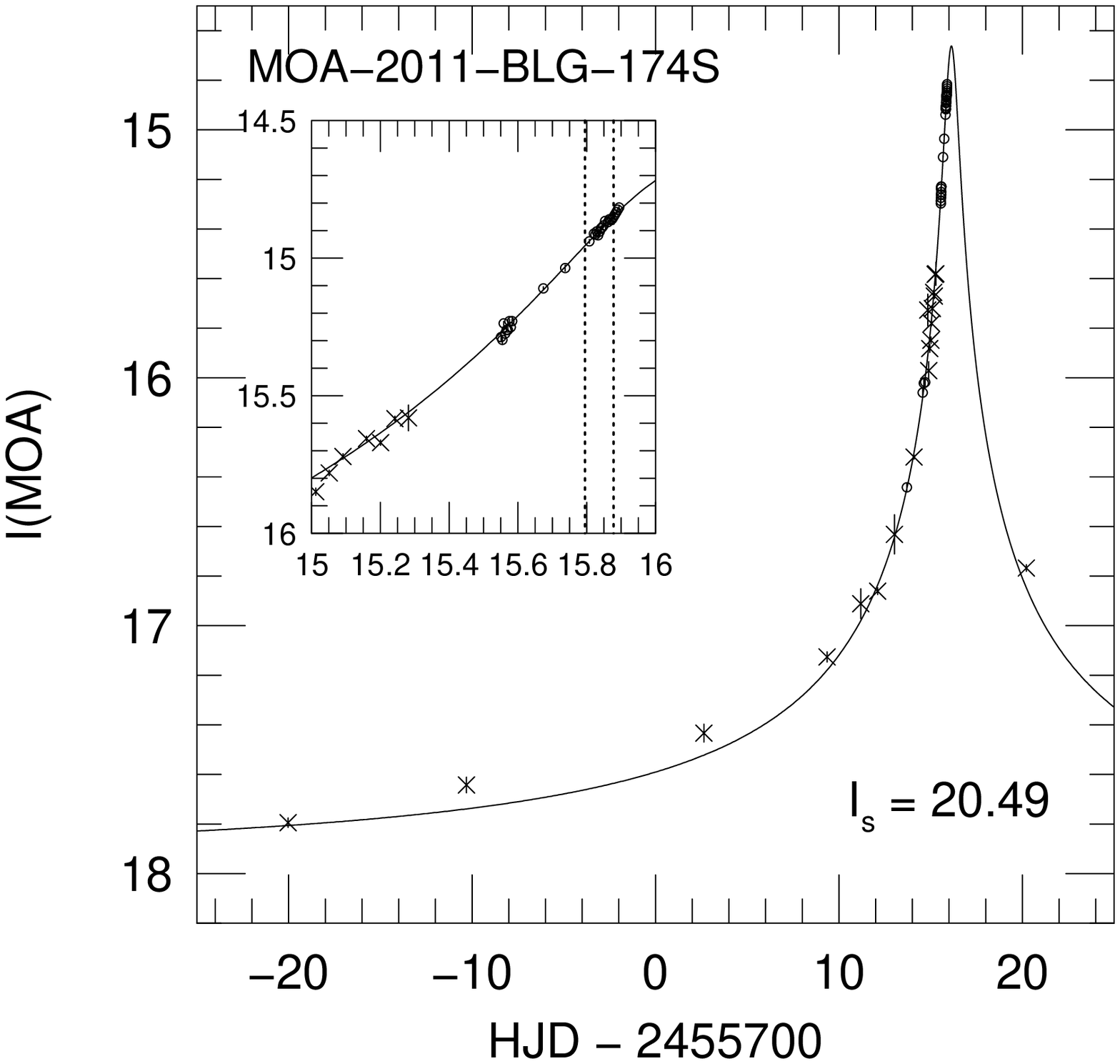}
\includegraphics[bb=67 40 515 540,clip]{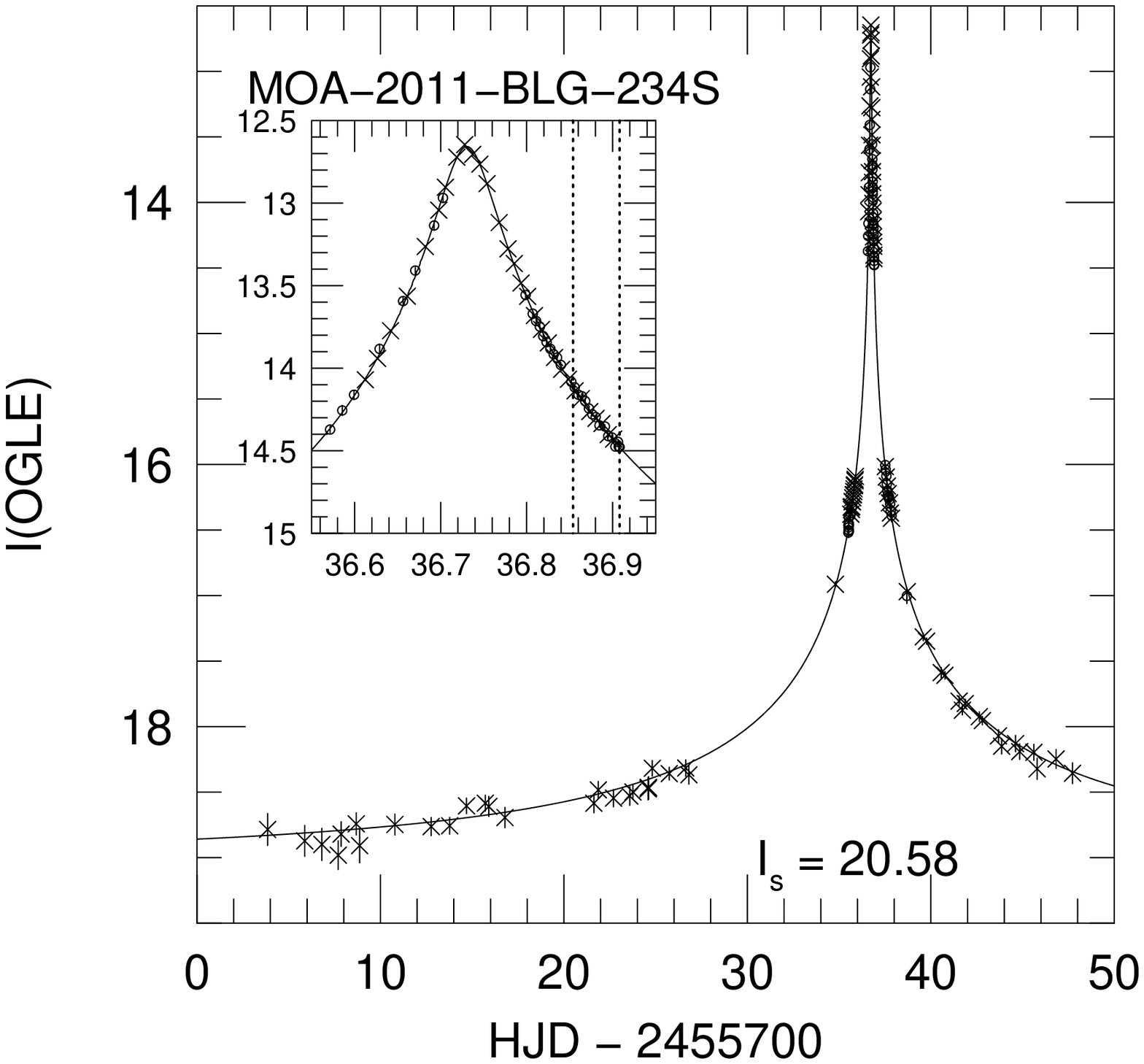}
}
\resizebox{\hsize}{!}{
\includegraphics[bb=10 40 515 525,clip]{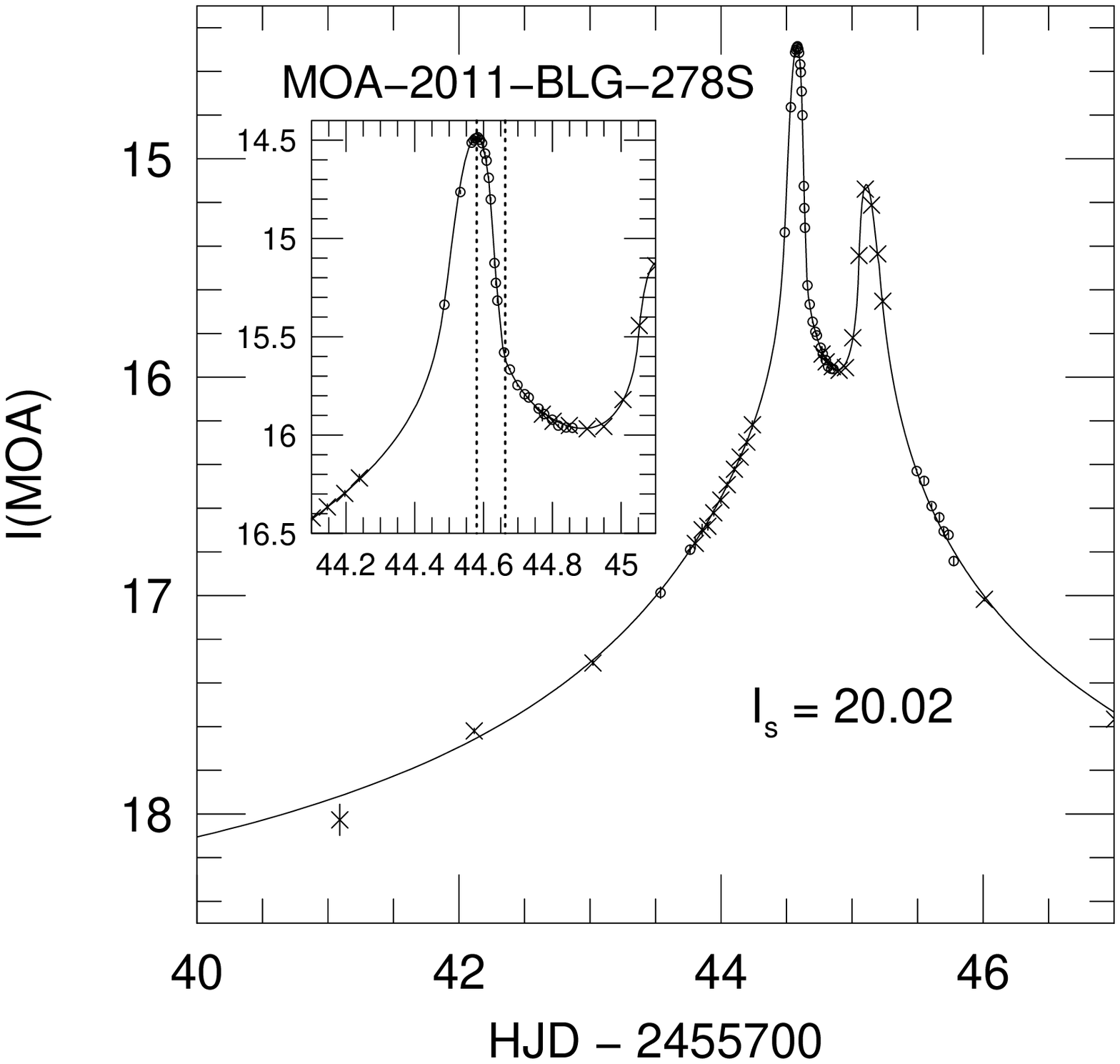}
\includegraphics[bb=67 40 515 525,clip]{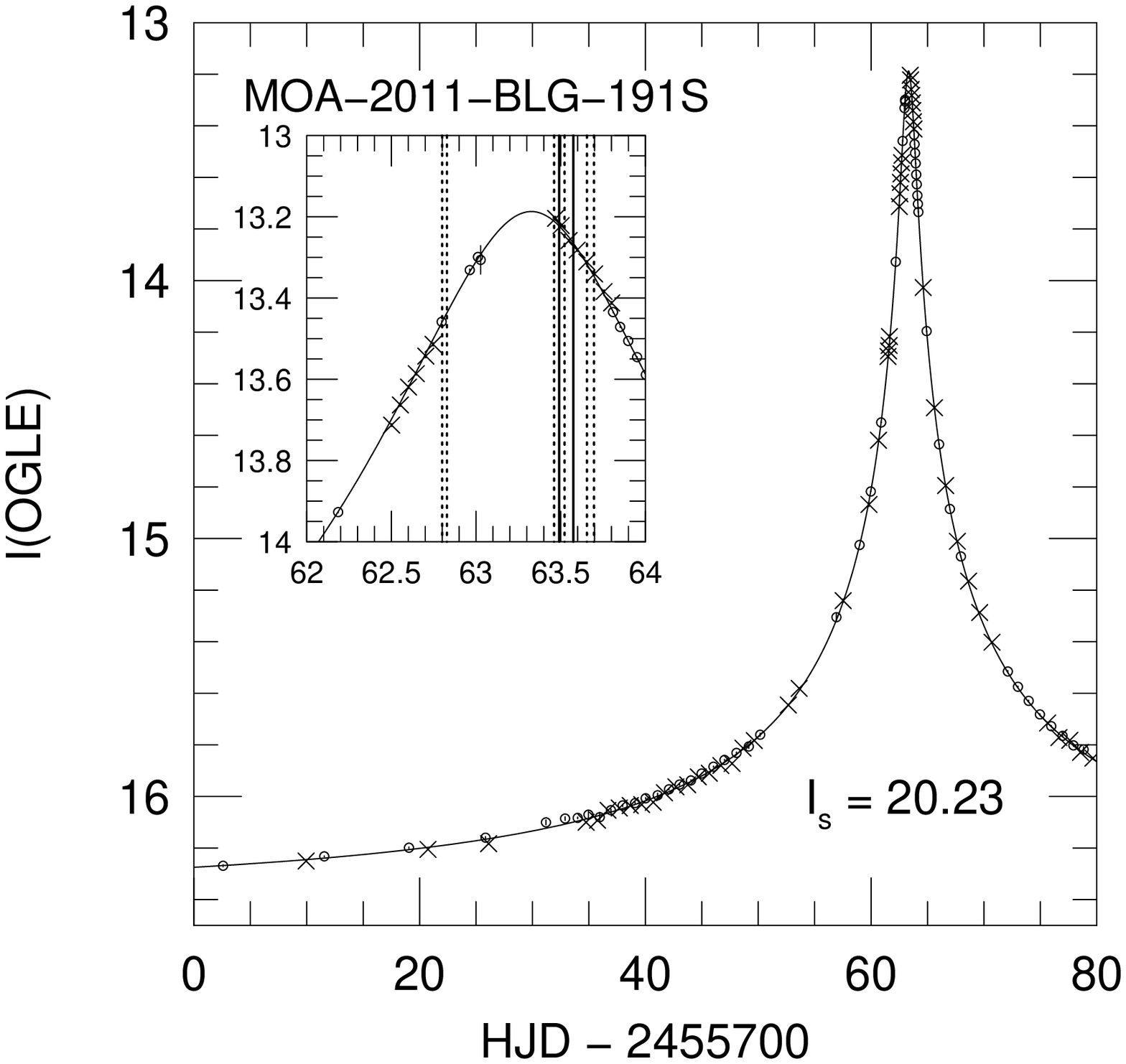}
\includegraphics[bb=67 40 515 525,clip]{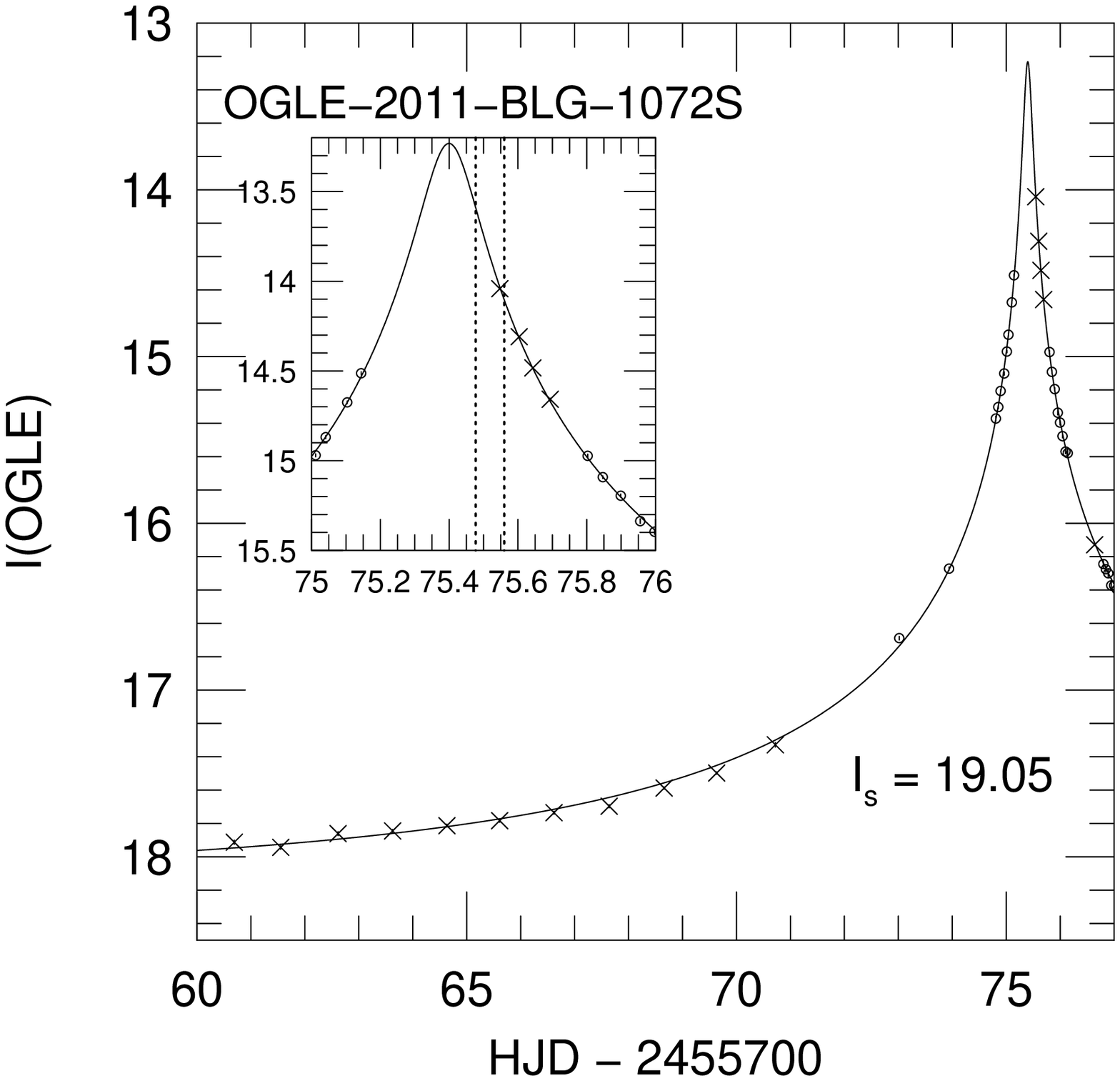}
\includegraphics[bb=67 40 515 525,clip]{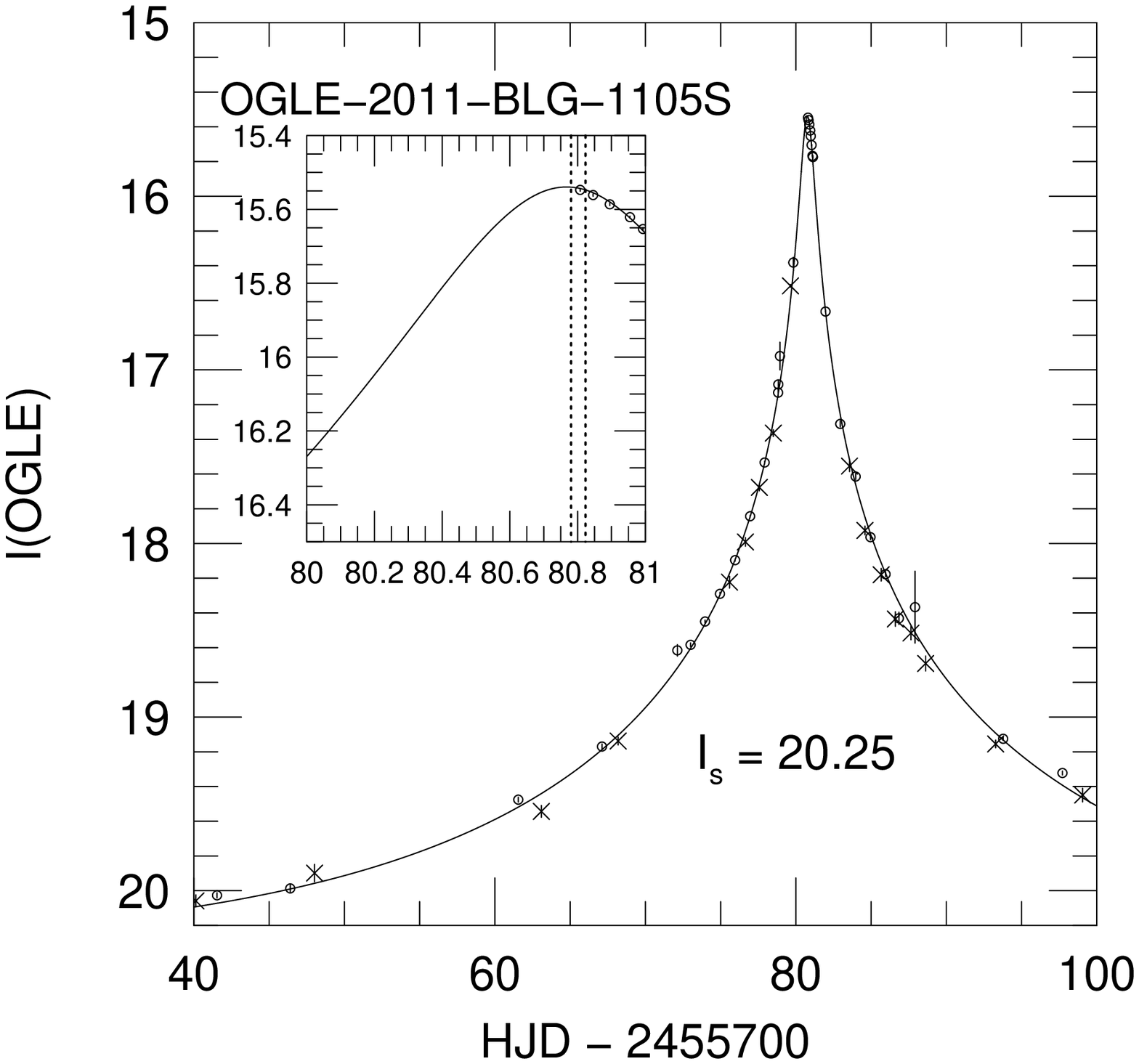}
}
\resizebox{\hsize}{!}{
\includegraphics[bb=10 40 515 525,clip]{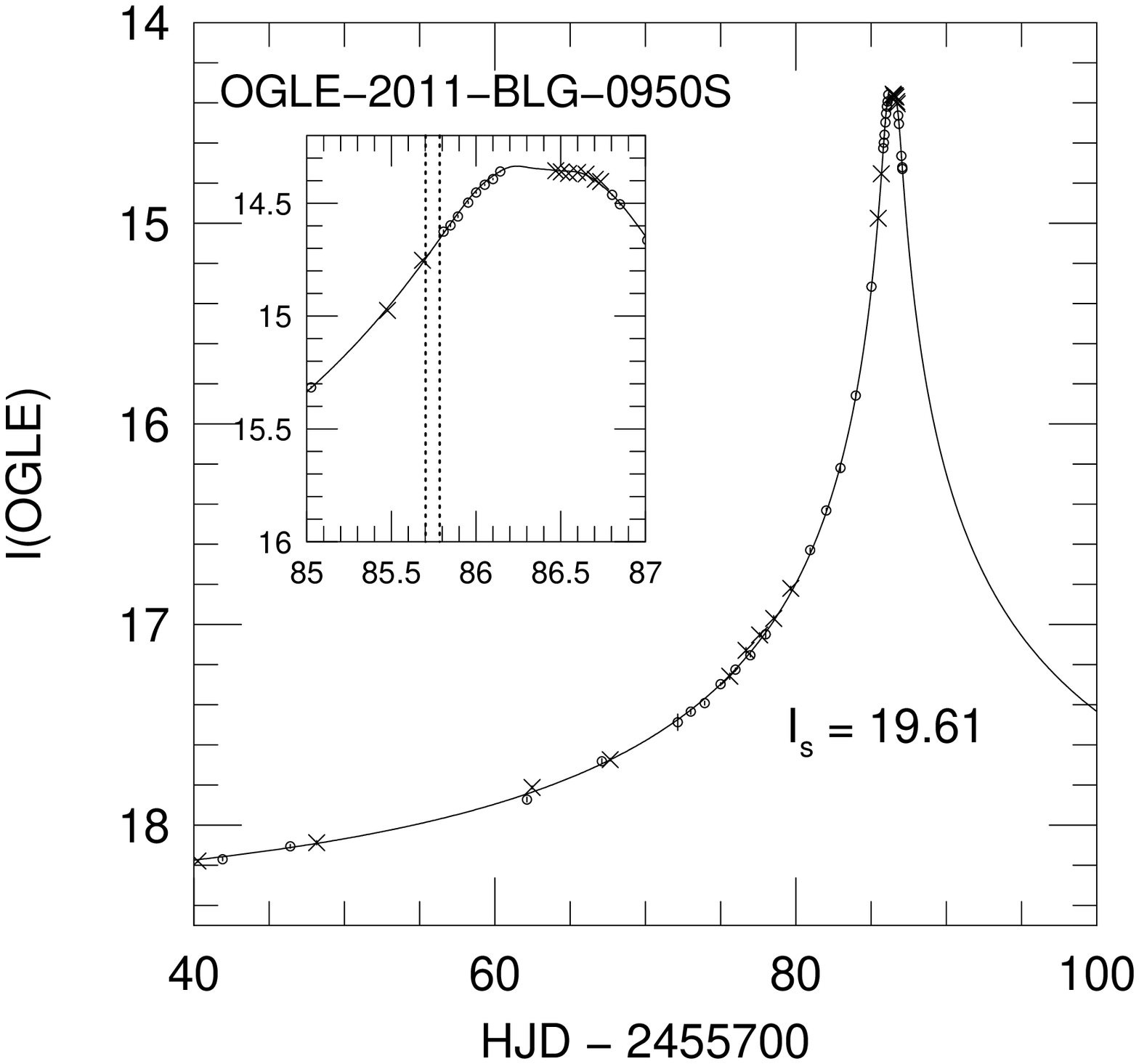}
\includegraphics[bb=67 40 515 525,clip]{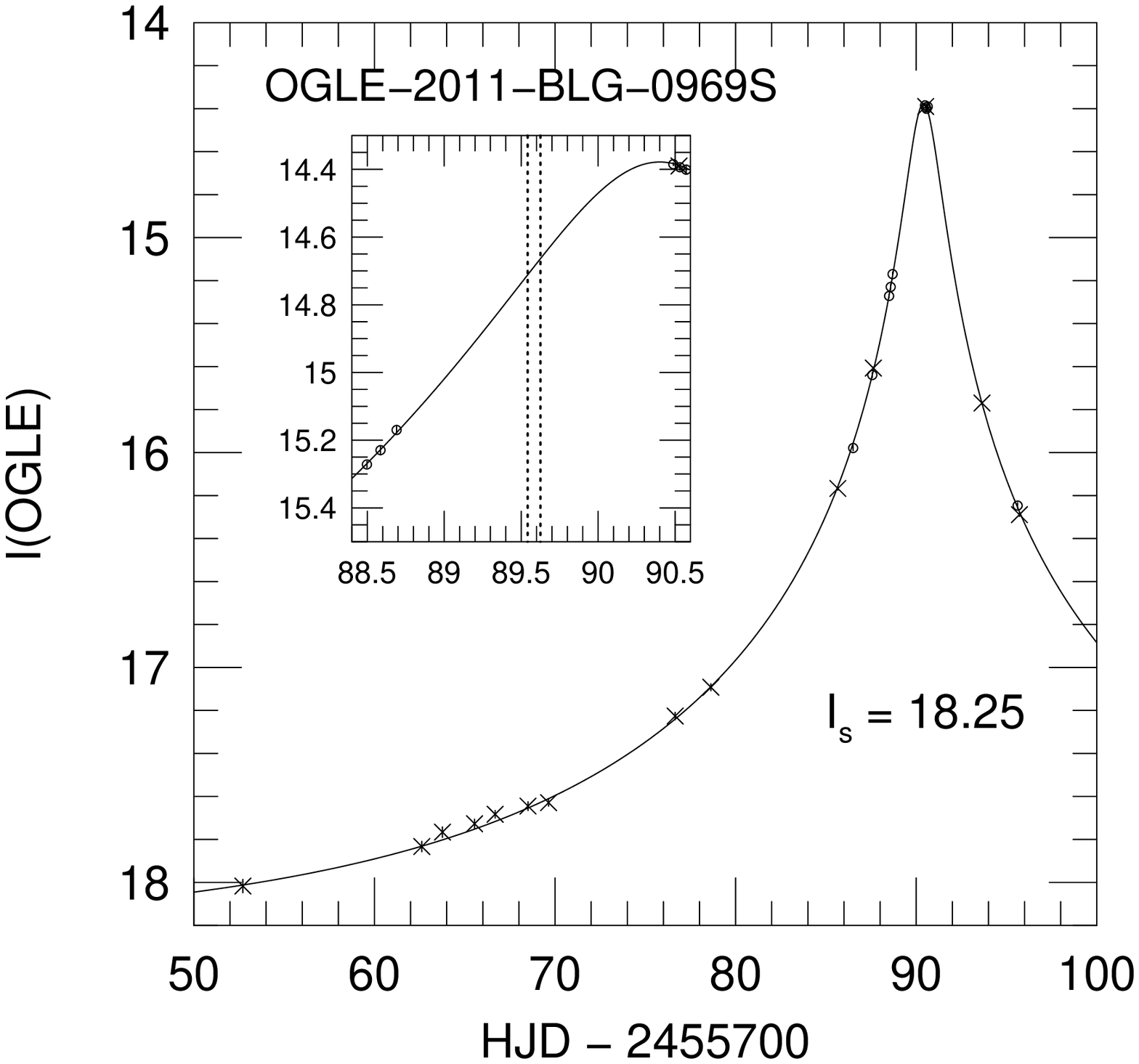}
\includegraphics[bb=67 40 515 525,clip]{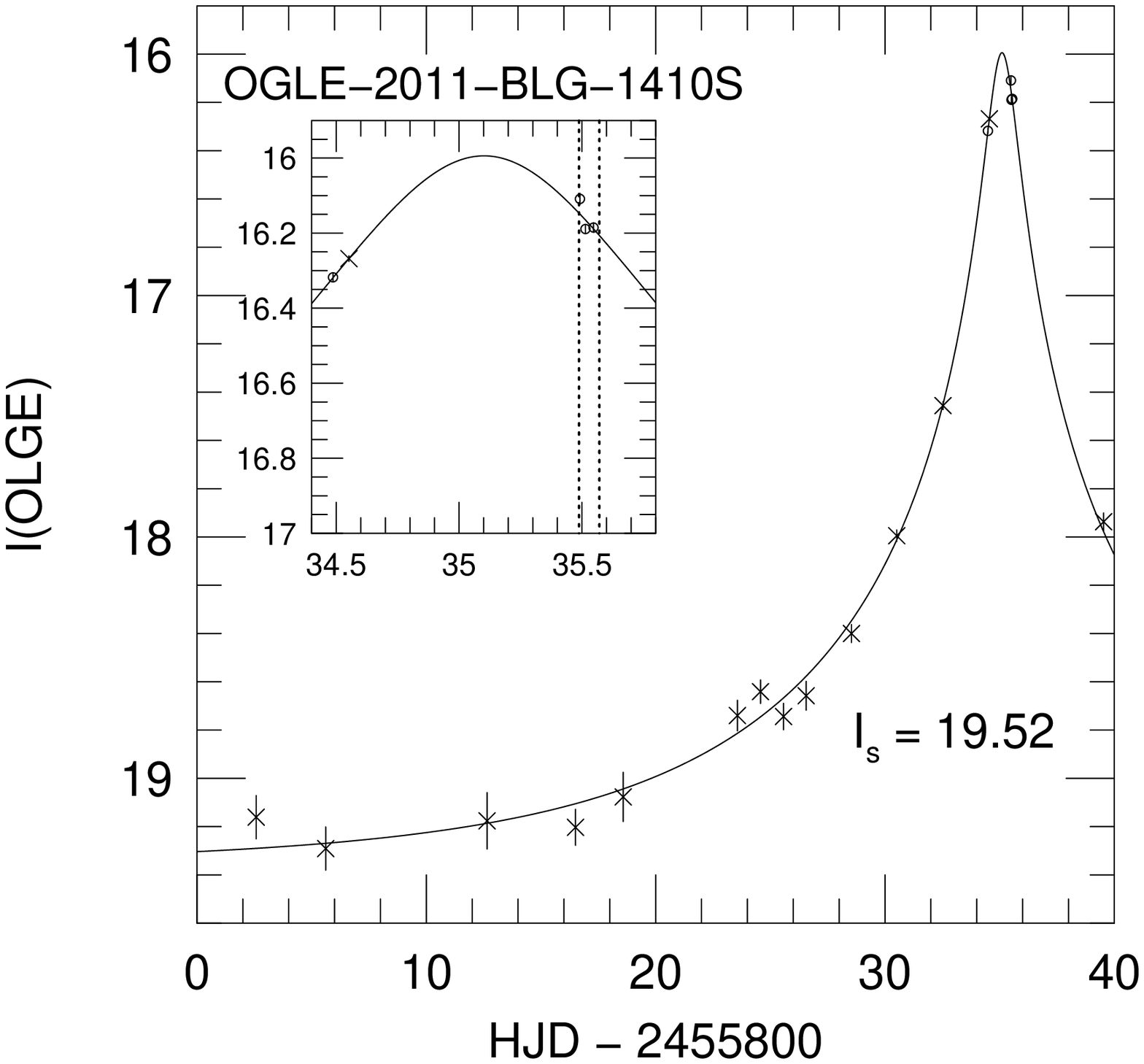}
\includegraphics[bb=67 40 515 525,clip]{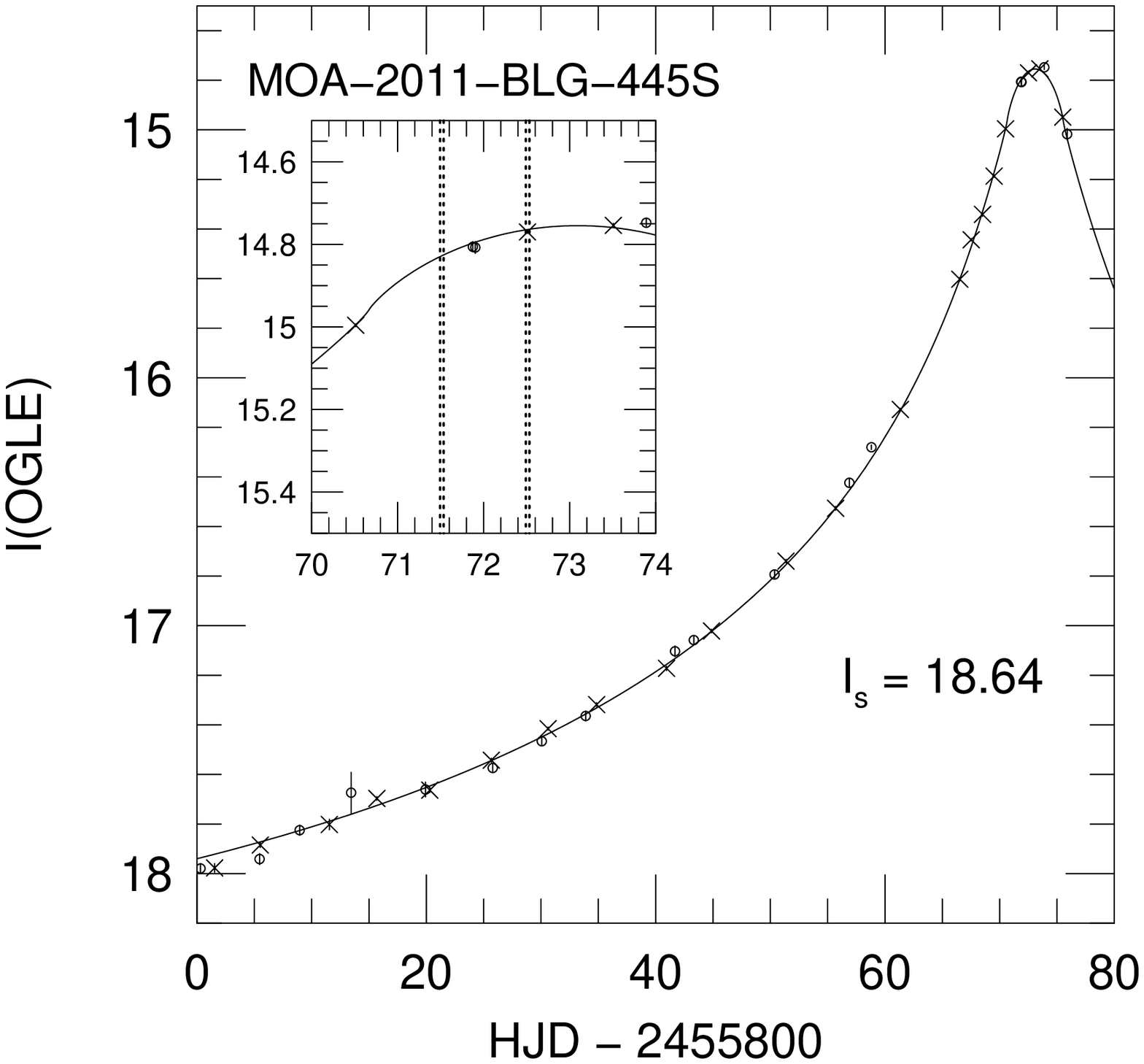}
}
\resizebox{\hsize}{!}{
\includegraphics[bb=10 40 515 525,clip]{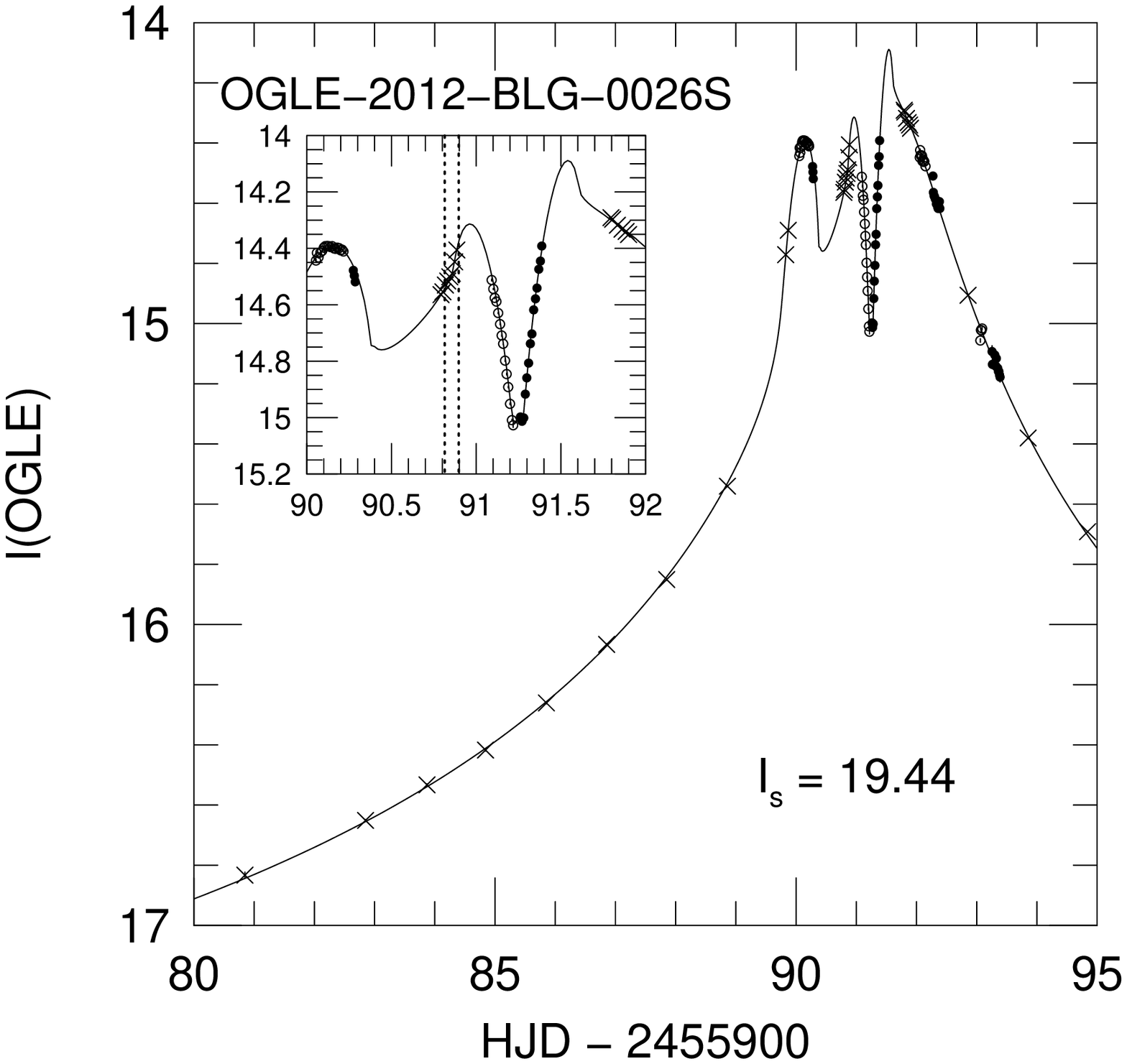}
\includegraphics[bb=67 40 515 525,clip]{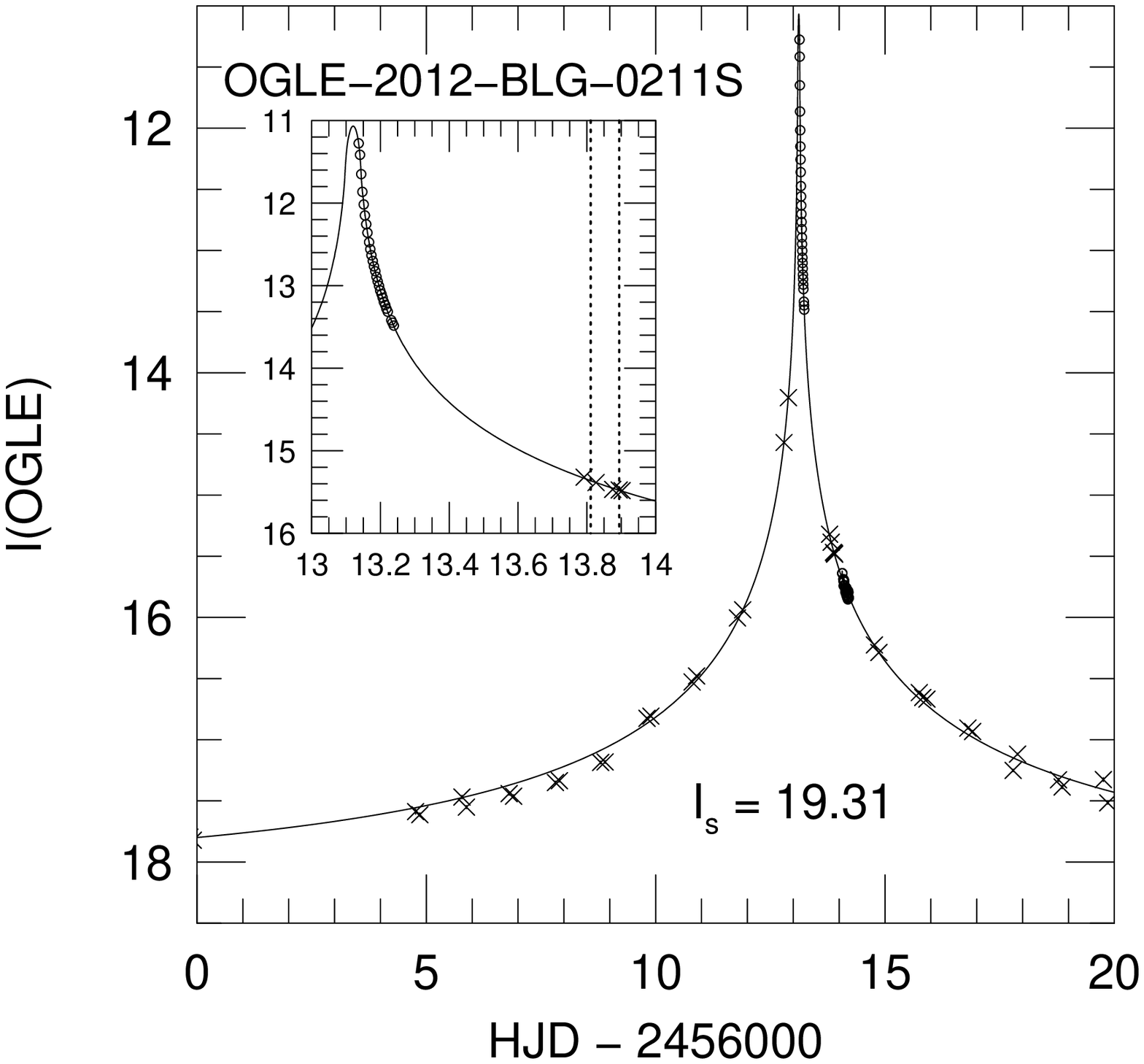}
\includegraphics[bb=67 40 515 525,clip]{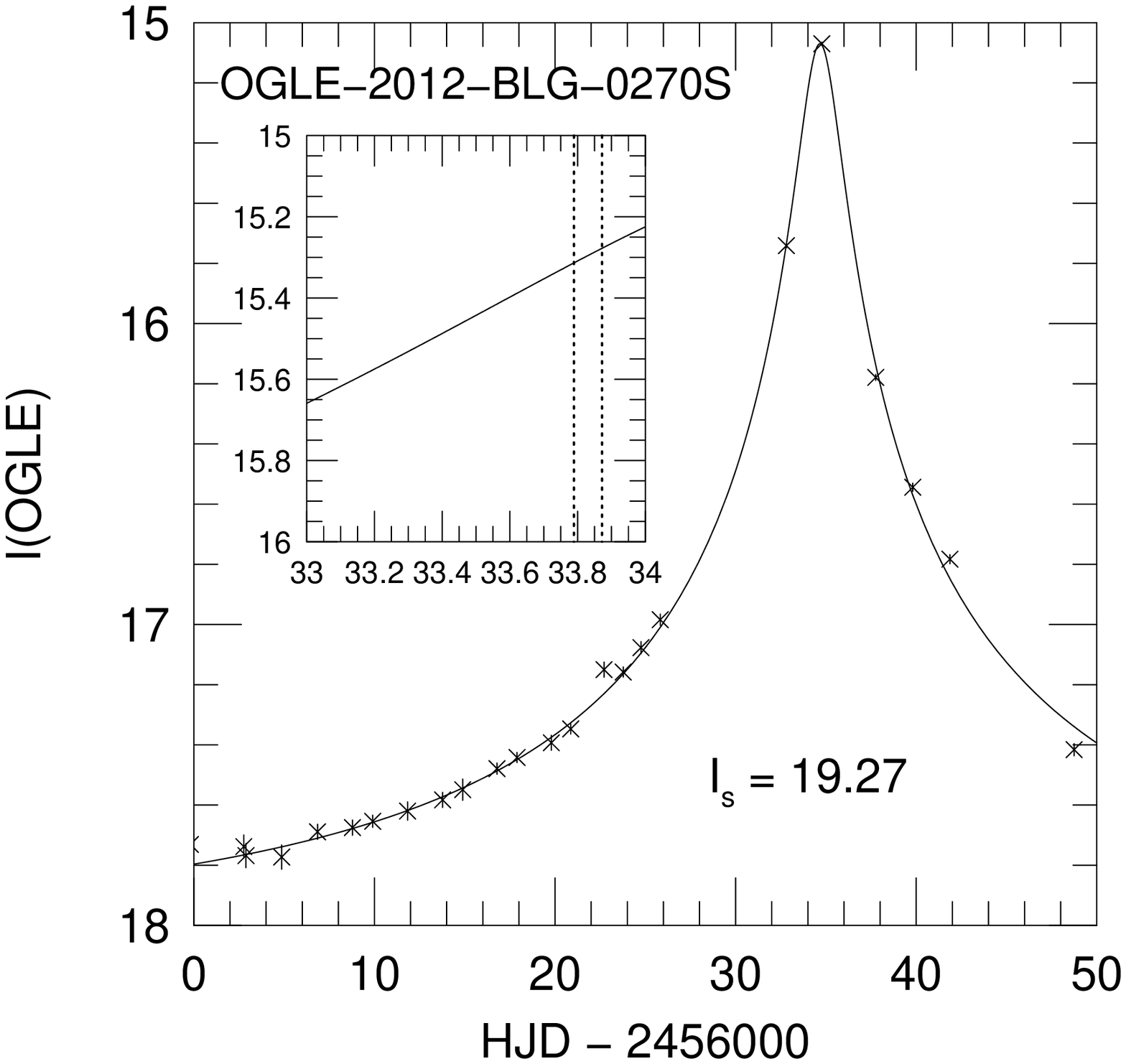}
\includegraphics[bb=67 40 515 525,clip]{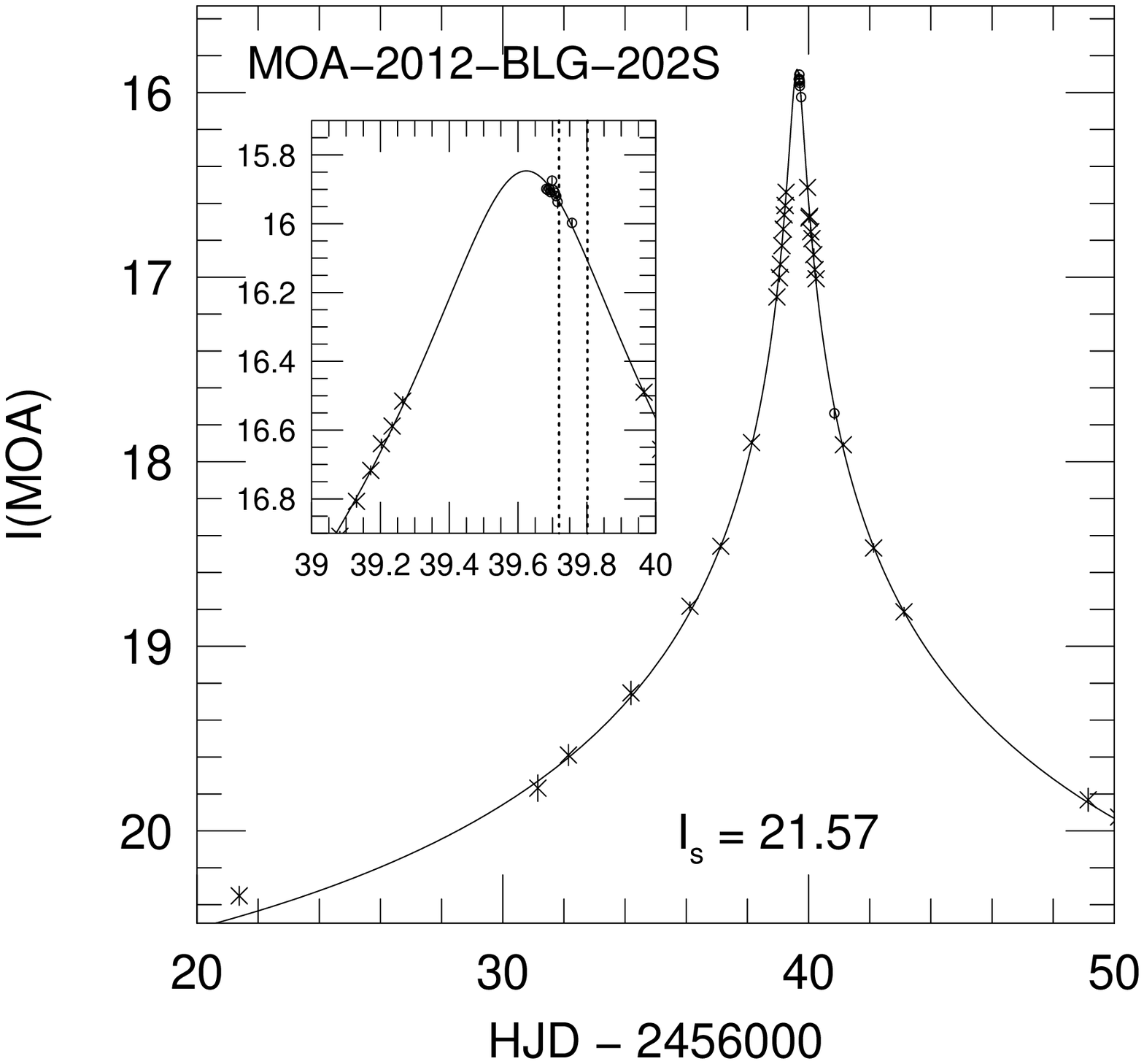}
}
\caption{Light curves for the first 16 of the 32 new microlensing 
events. Each plot has a zoom window, showing the time intervals when 
the source stars were observed with high-resolution spectrographs. 
In each plot the un-lensed magnitude of the source star is also given 
($I_{\rm S}$).
\label{fig:events}
}
\end{figure*}
%-----------------------------------------------------------------------
%-----------------------------------------------------------------------
\begin{figure*}
\resizebox{\hsize}{!}{
\includegraphics[bb=10 40 515 540,clip]{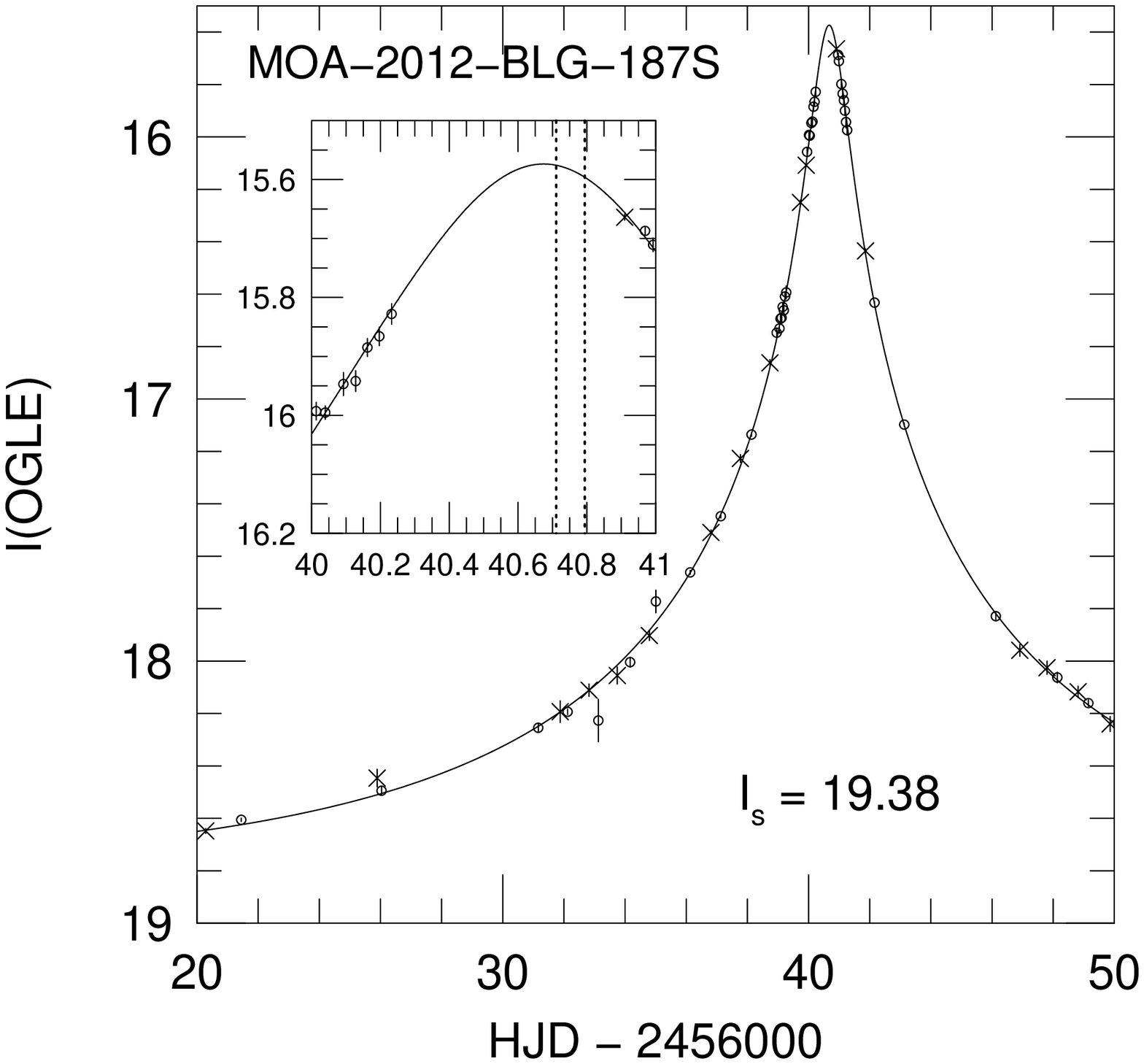}
\includegraphics[bb=67 40 515 540,clip]{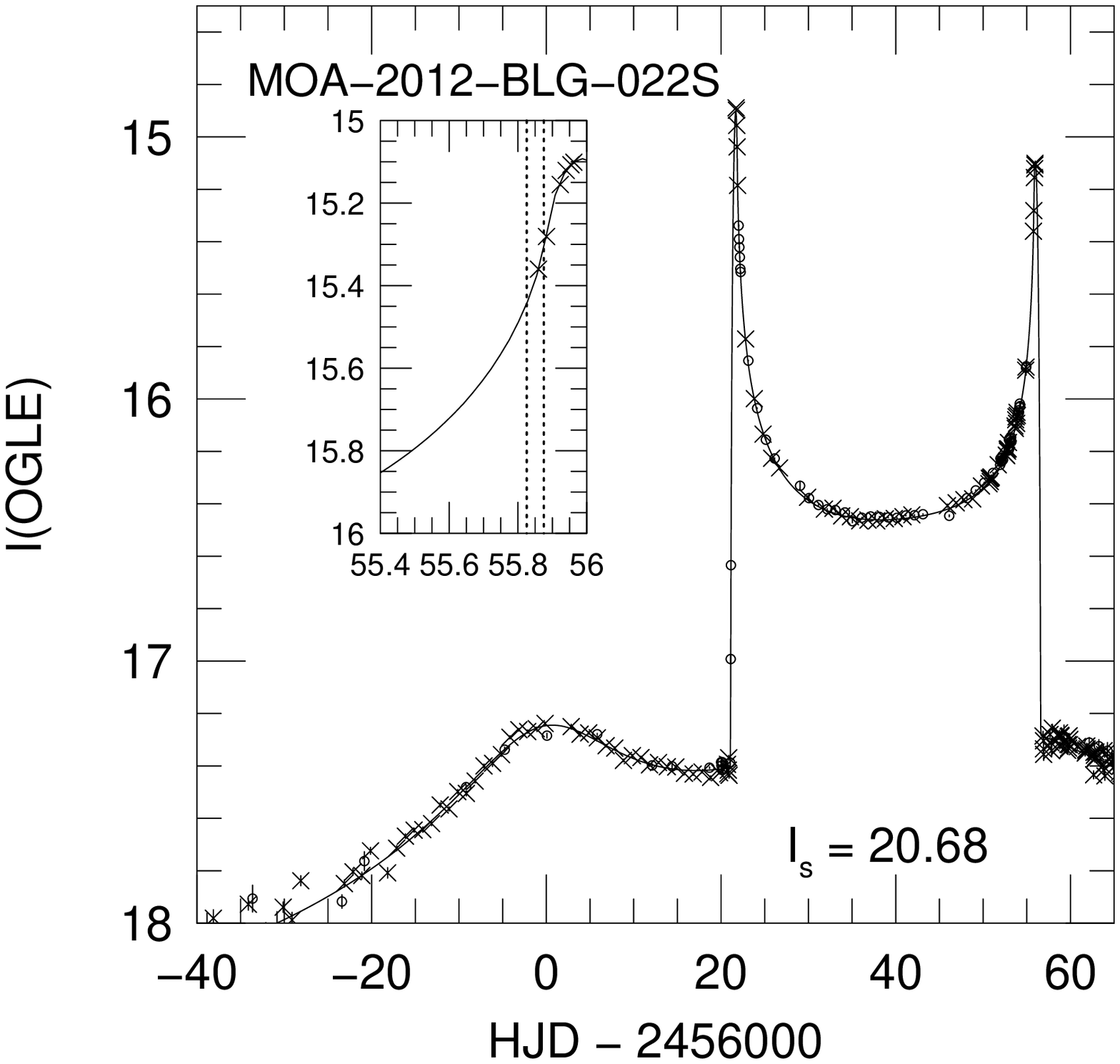}
\includegraphics[bb=67 40 515 540,clip]{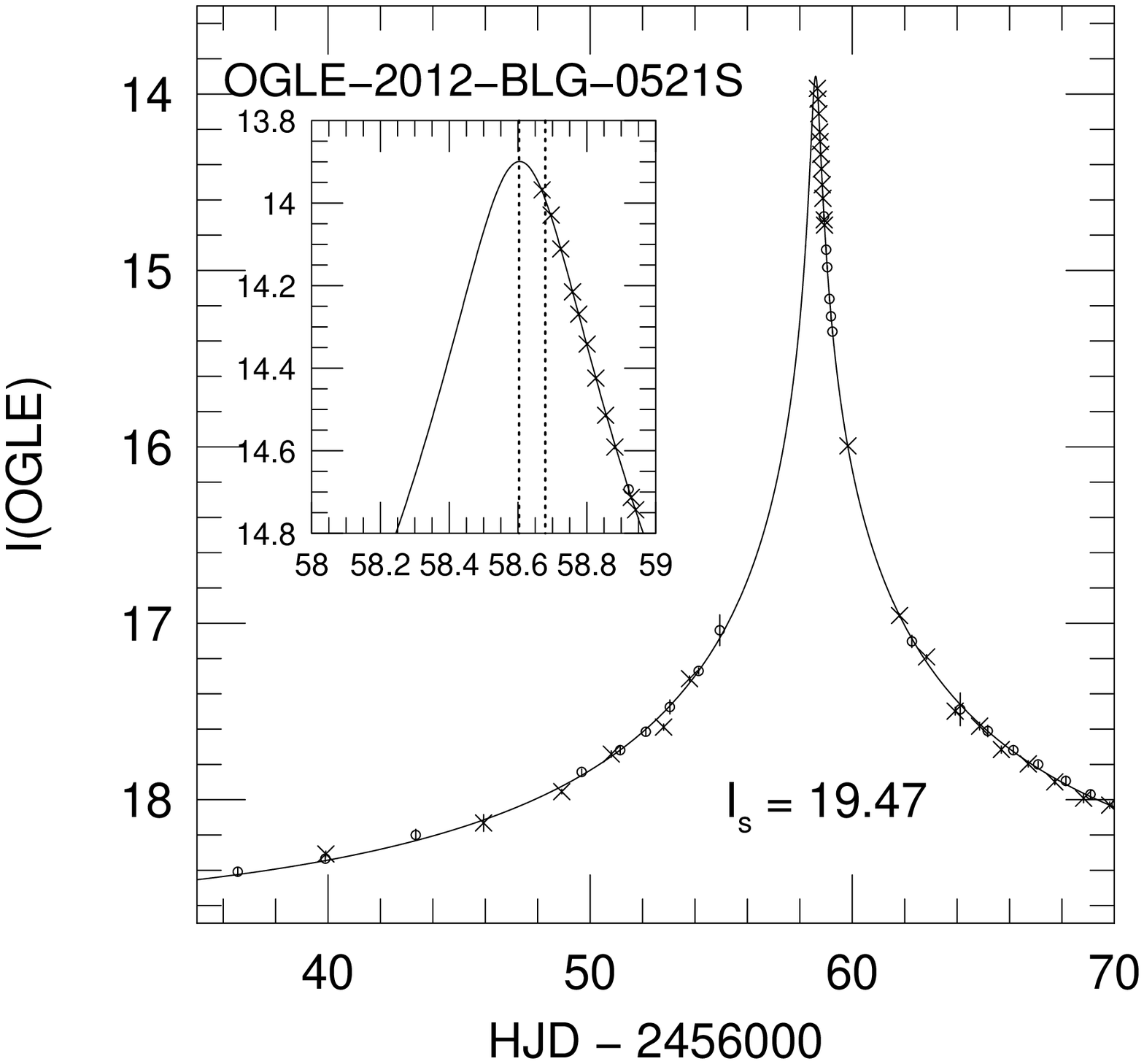}
\includegraphics[bb=67 40 515 540,clip]{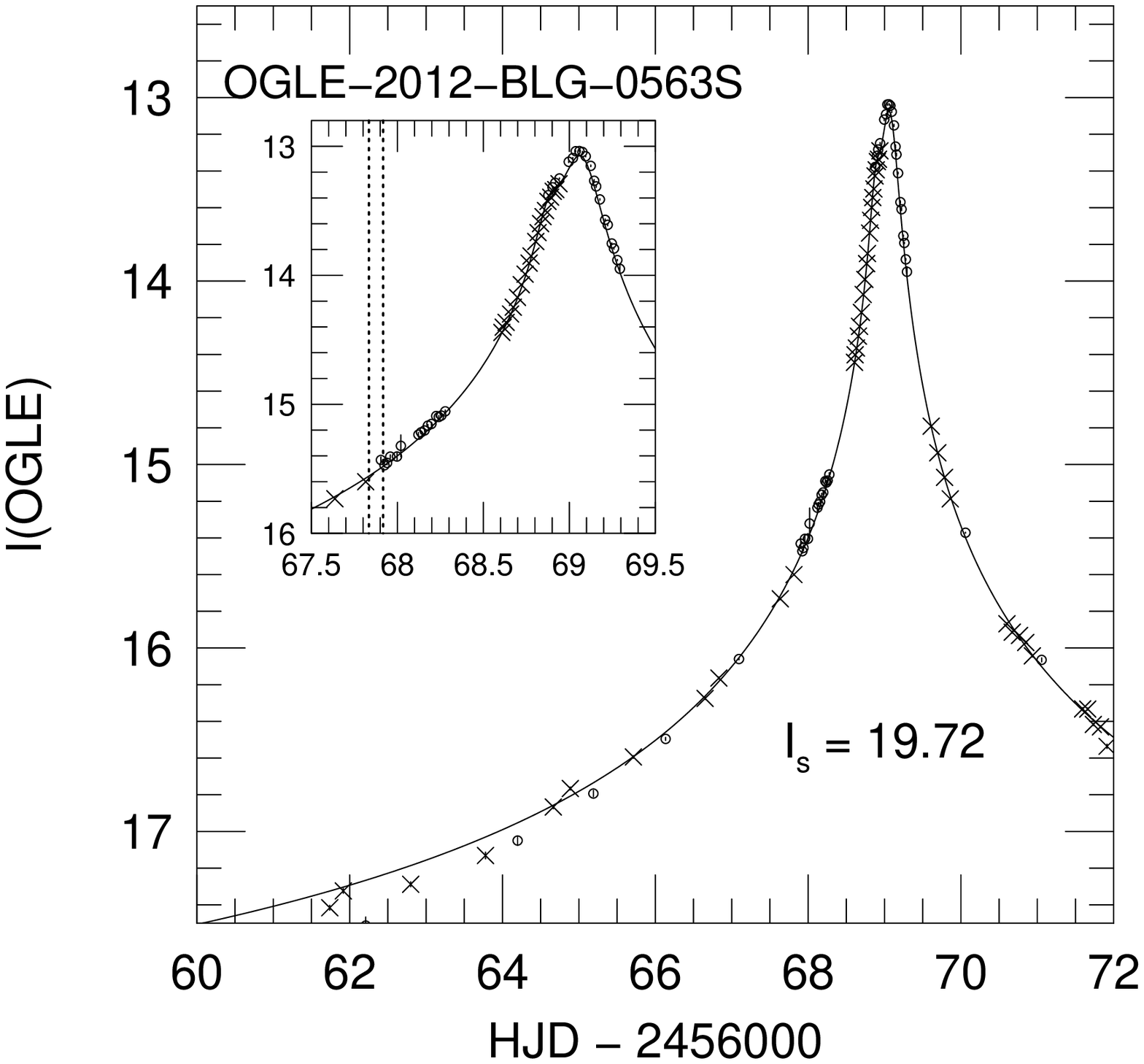}
}
\resizebox{\hsize}{!}{
\includegraphics[bb=10 40 515 525,clip]{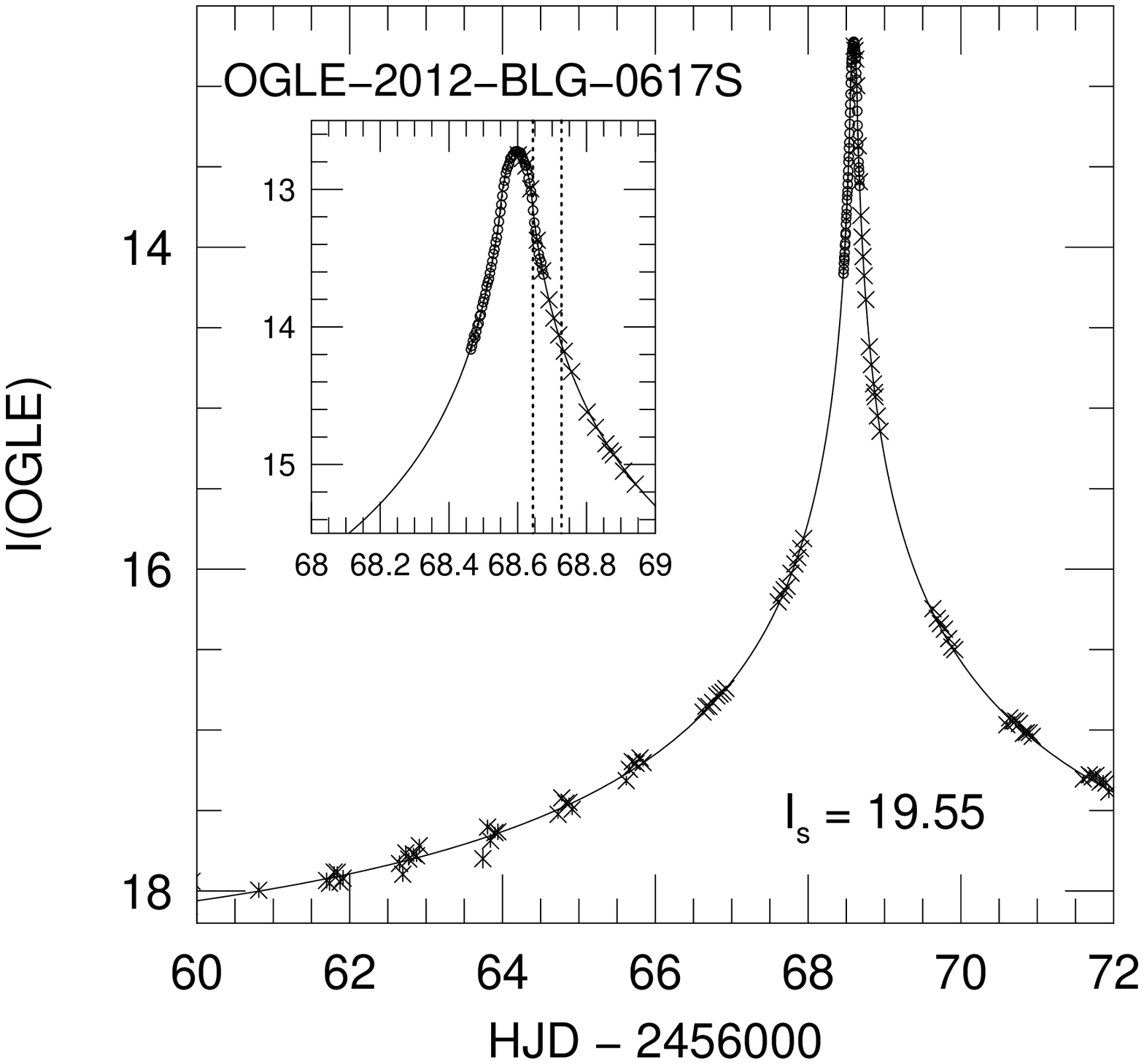}
\includegraphics[bb=67 40 515 525,clip]{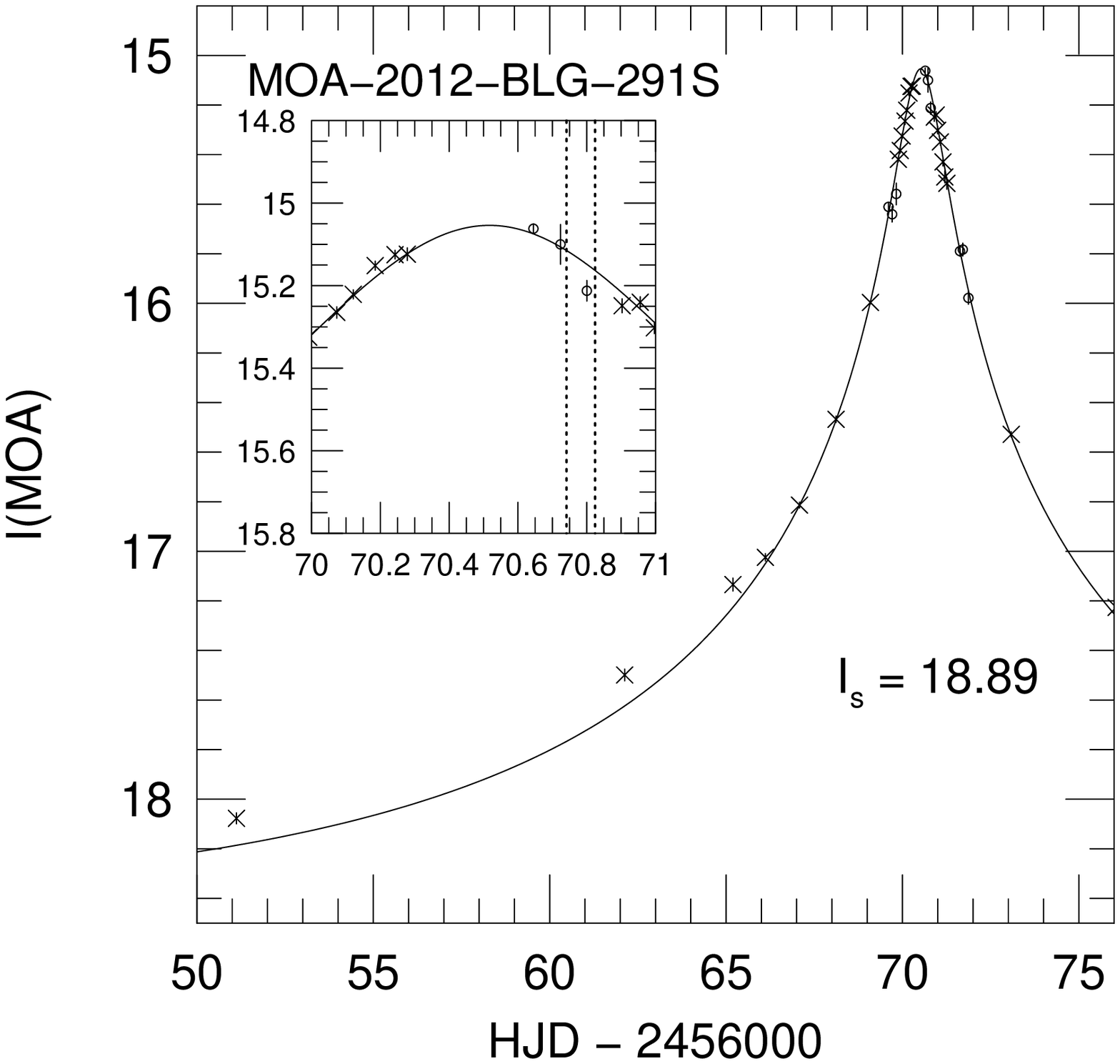}
\includegraphics[bb=67 40 515 525,clip]{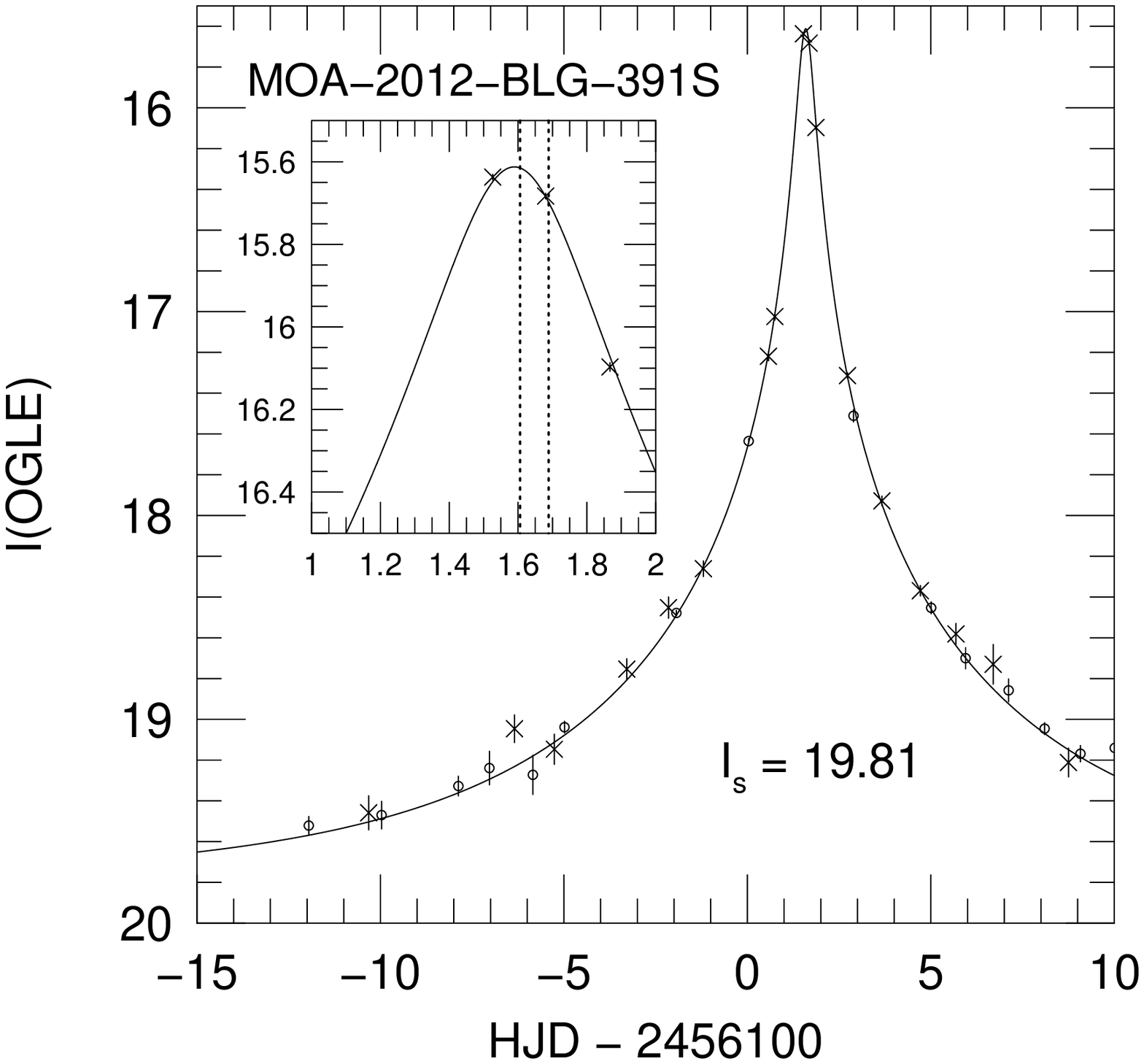}
\includegraphics[bb=67 40 515 525,clip]{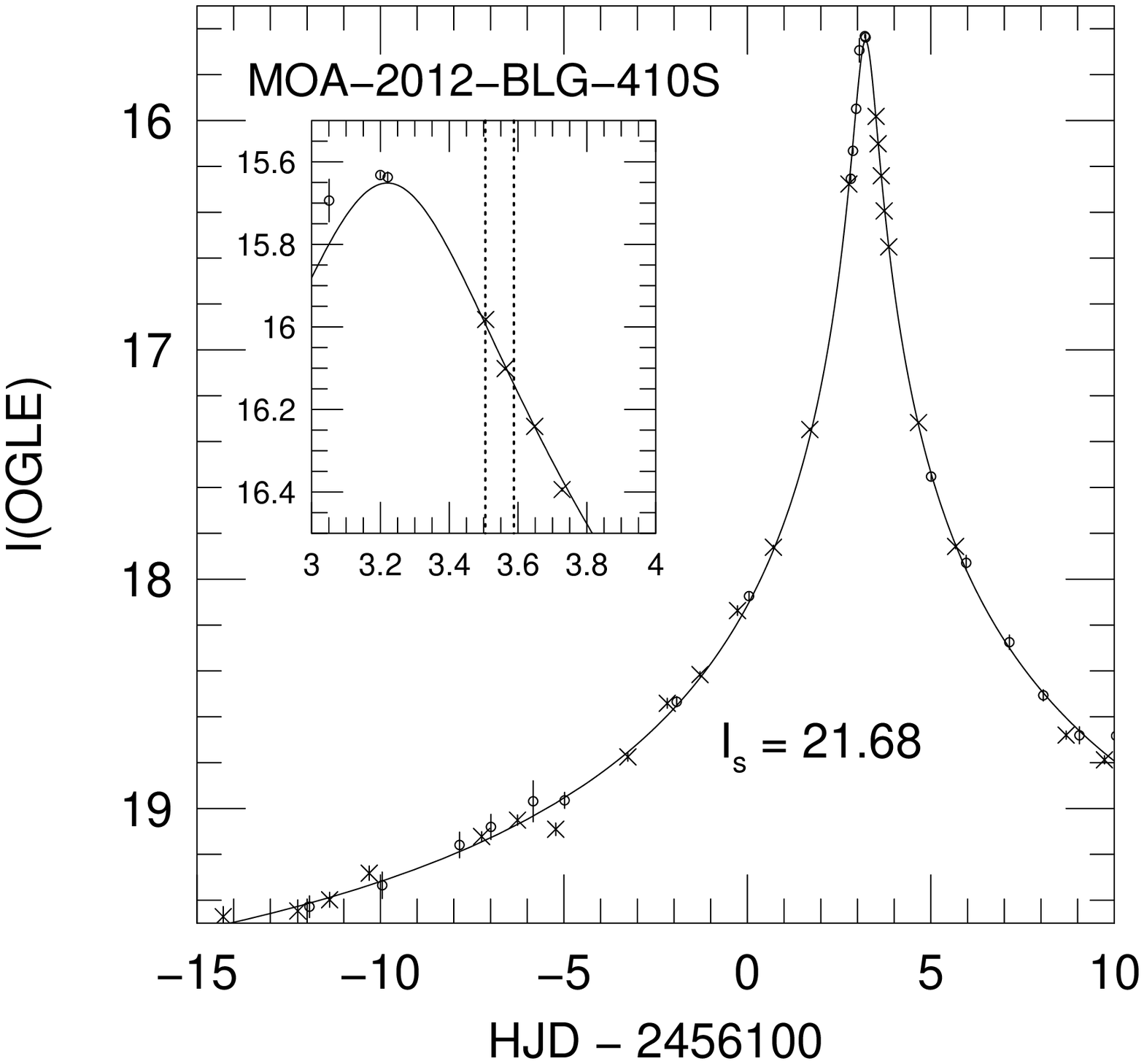}
}
\resizebox{\hsize}{!}{
\includegraphics[bb=10 40 515 525,clip]{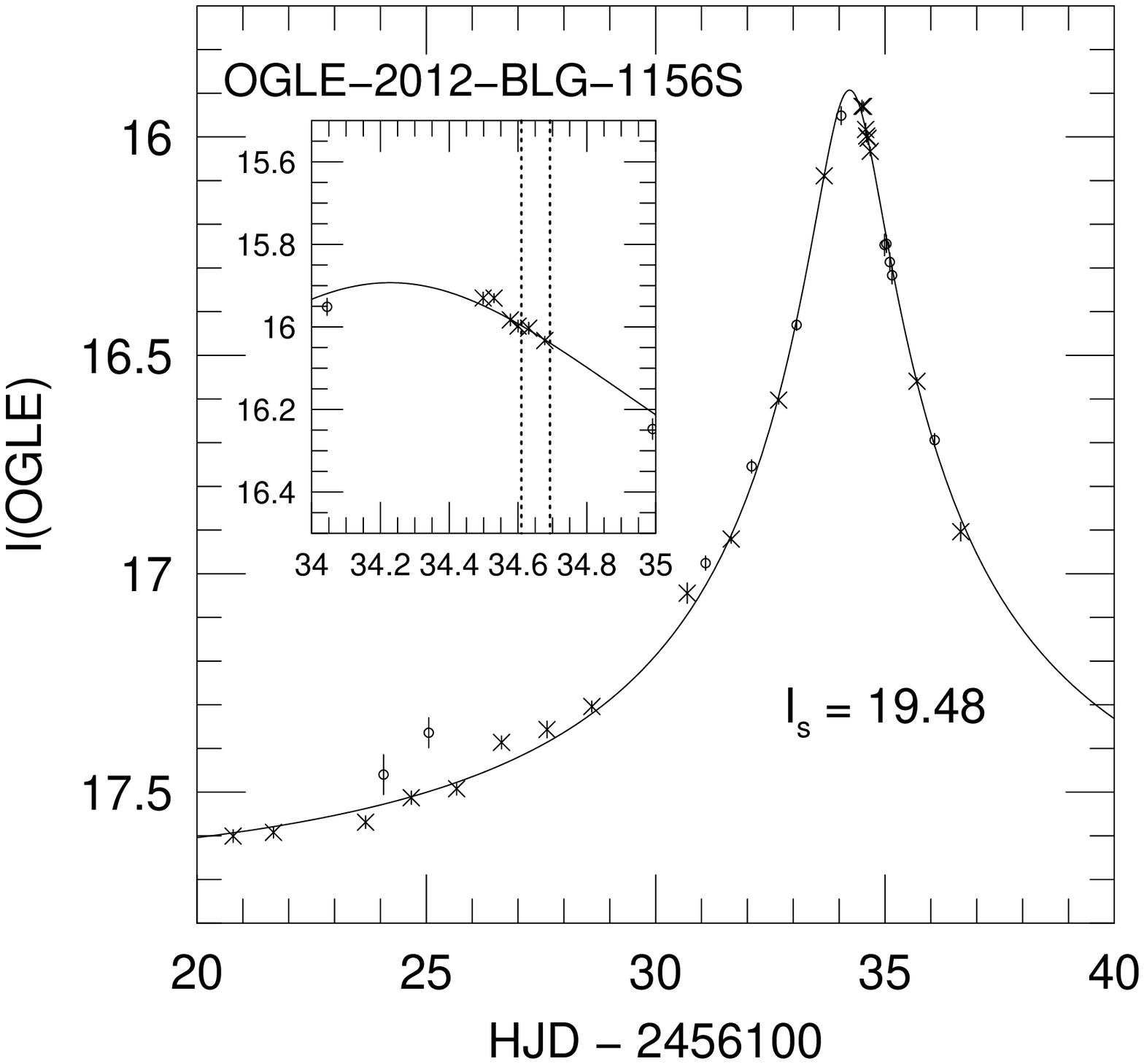}
\includegraphics[bb=67 40 515 525,clip]{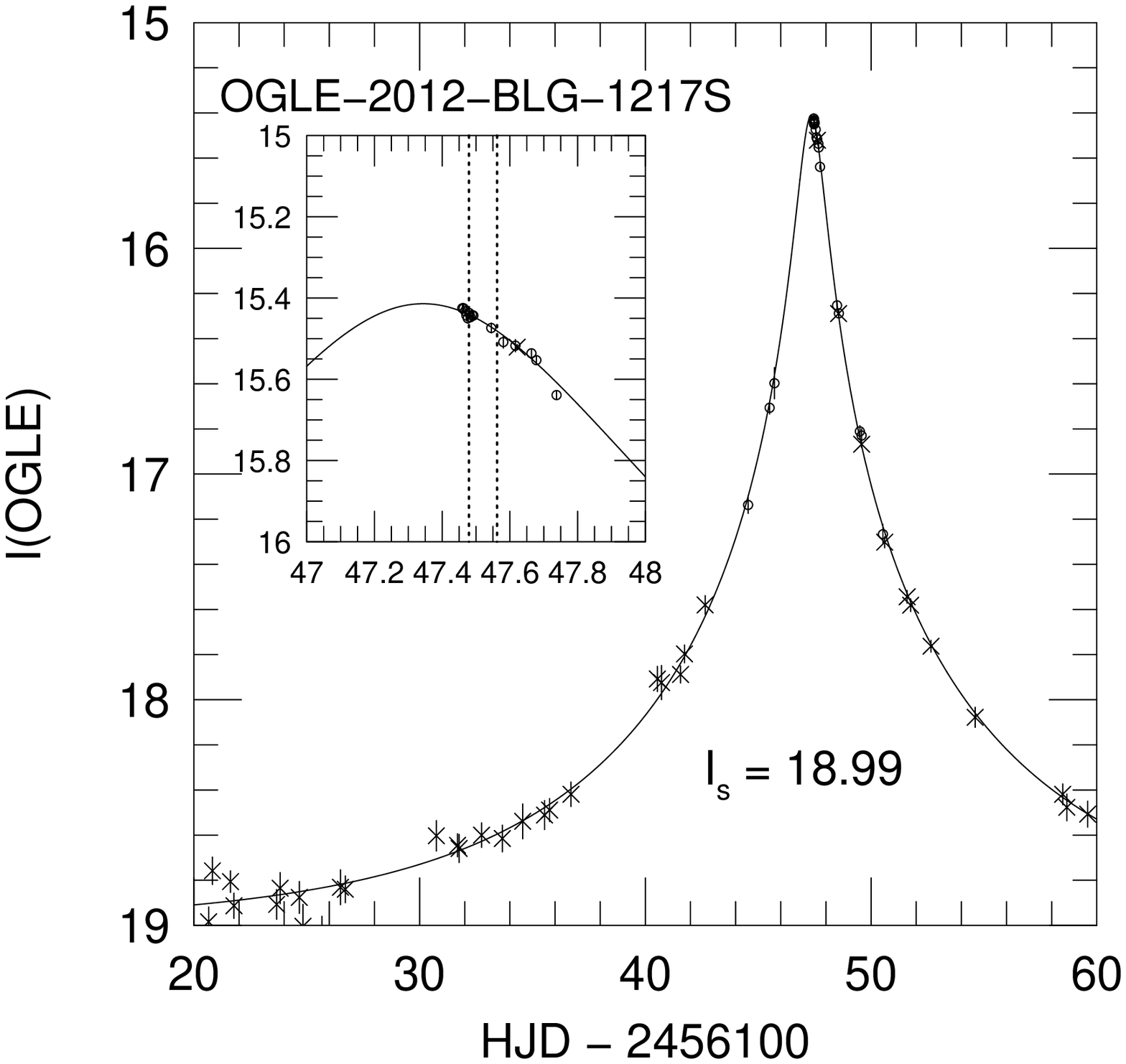}
\includegraphics[bb=67 40 515 525,clip]{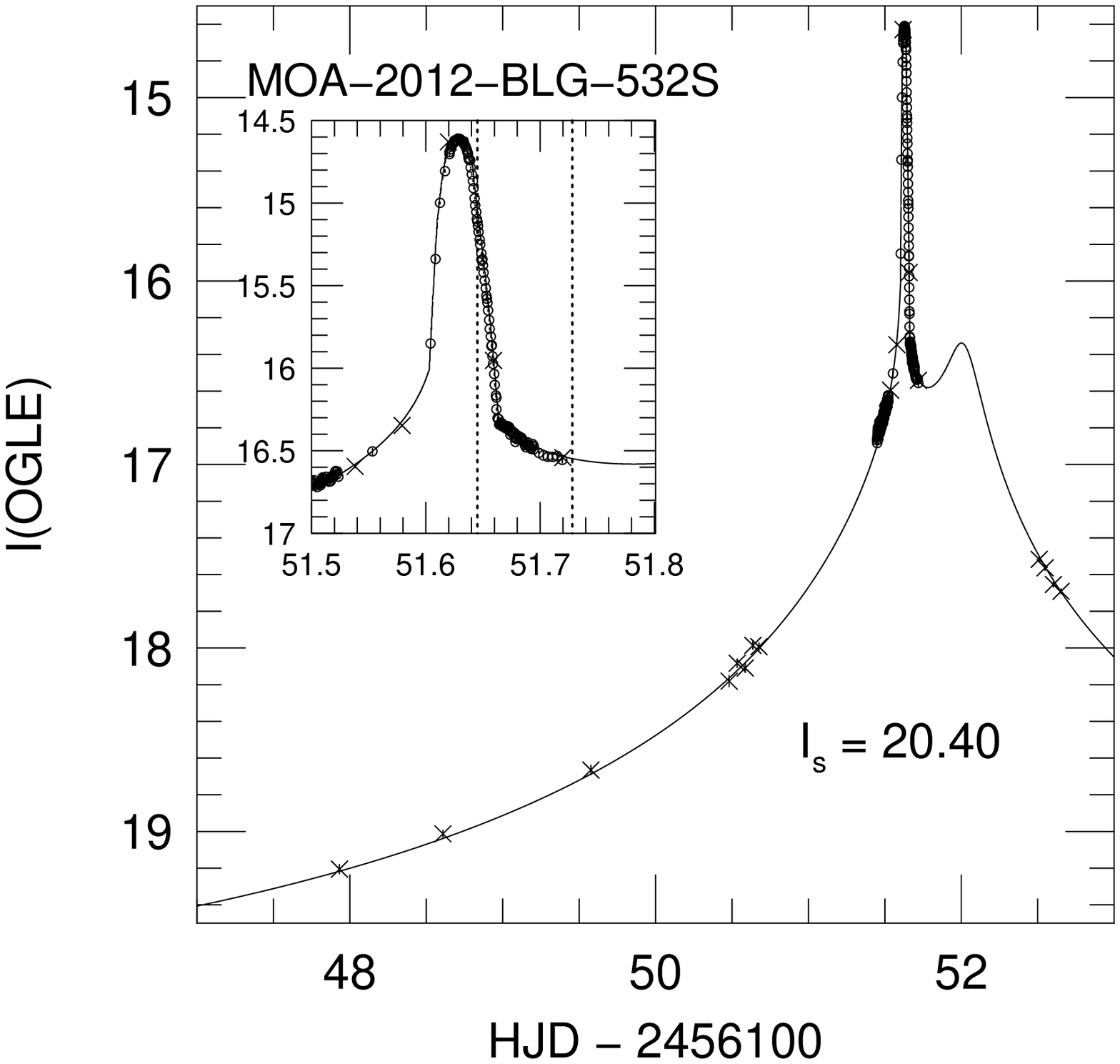}
\includegraphics[bb=67 40 515 525,clip]{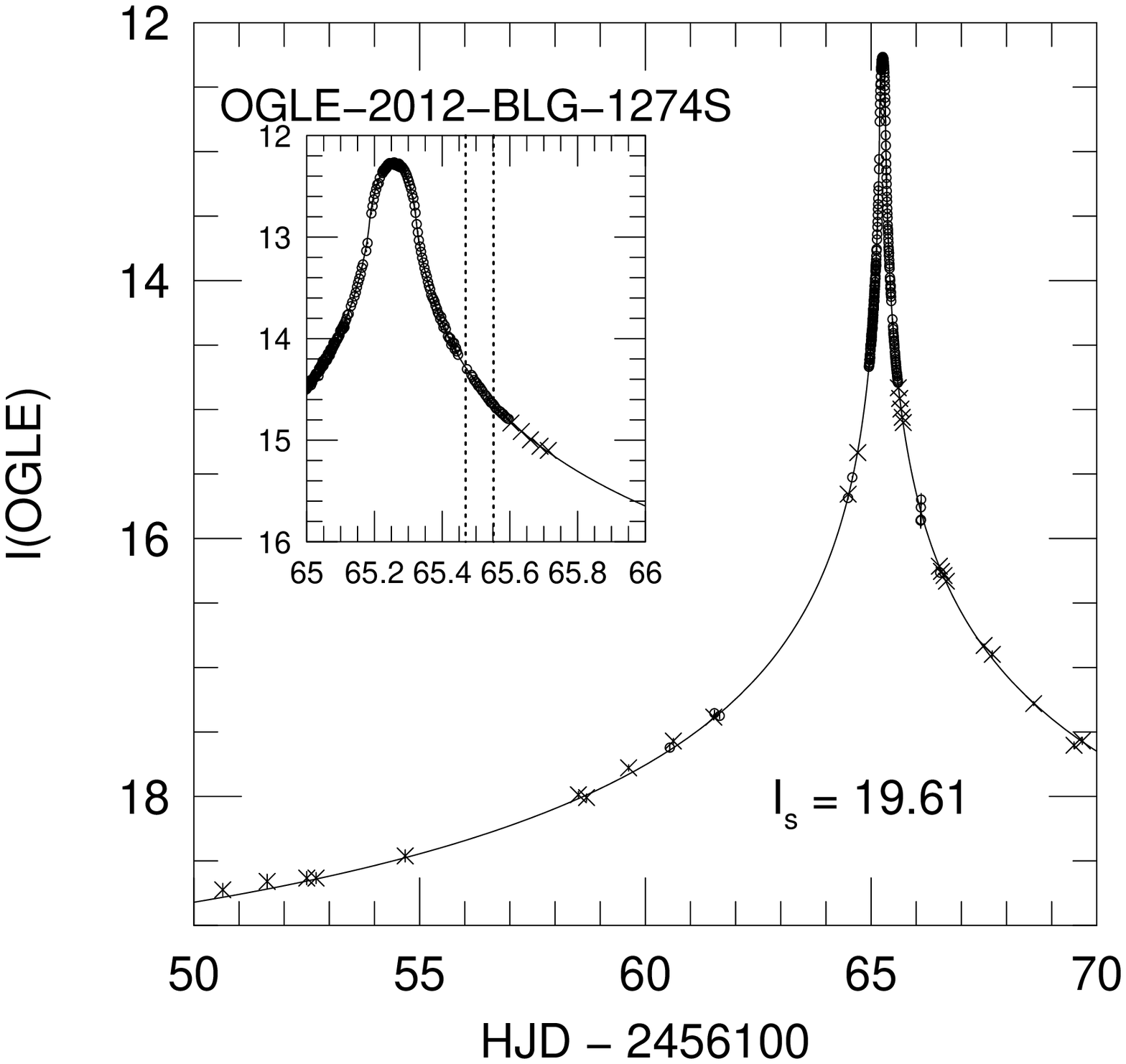}
}
\resizebox{\hsize}{!}{
\includegraphics[bb=10 40 515 525,clip]{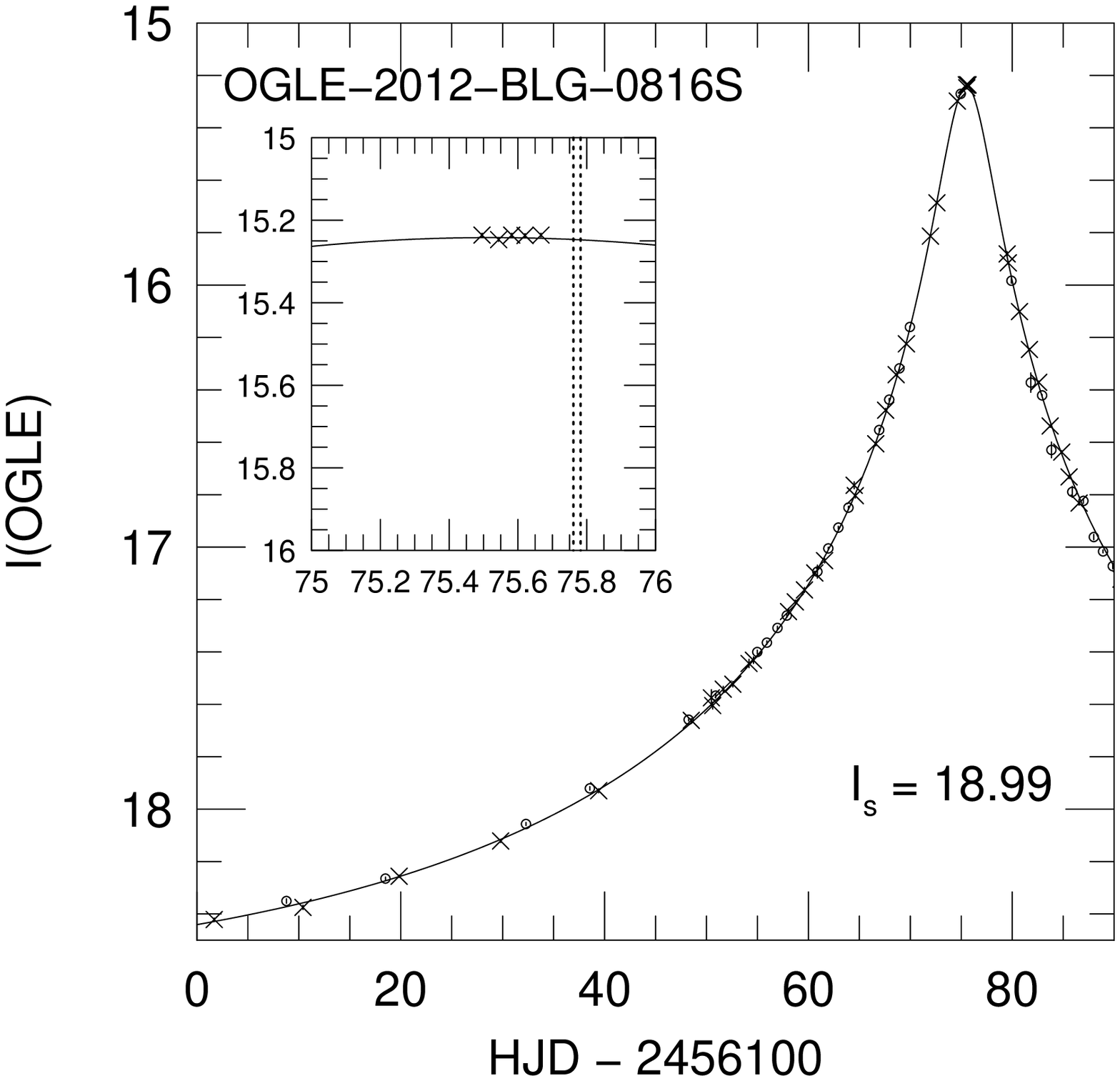}
\includegraphics[bb=67 40 515 525,clip]{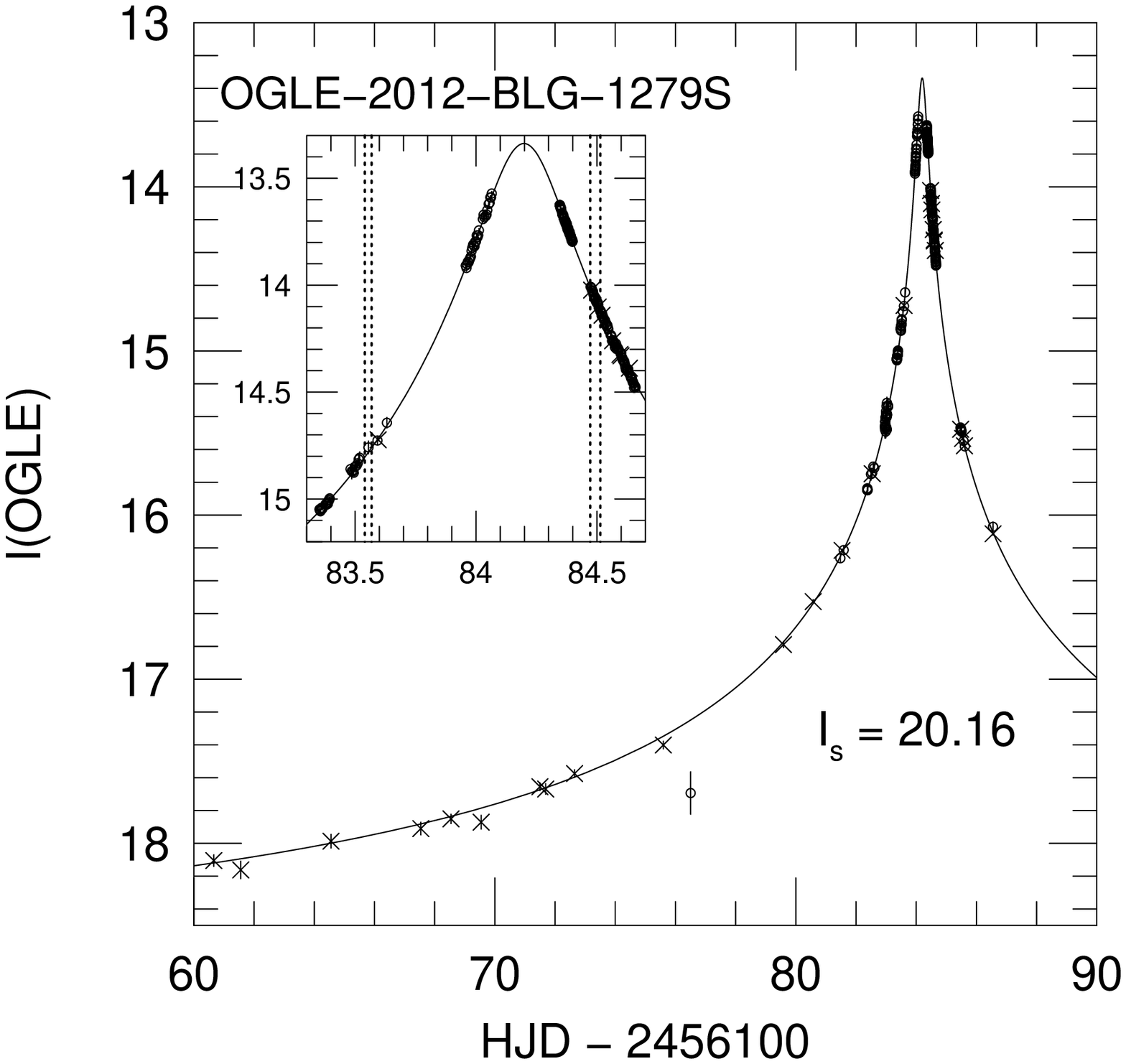}
\includegraphics[bb=67 40 515 525,clip]{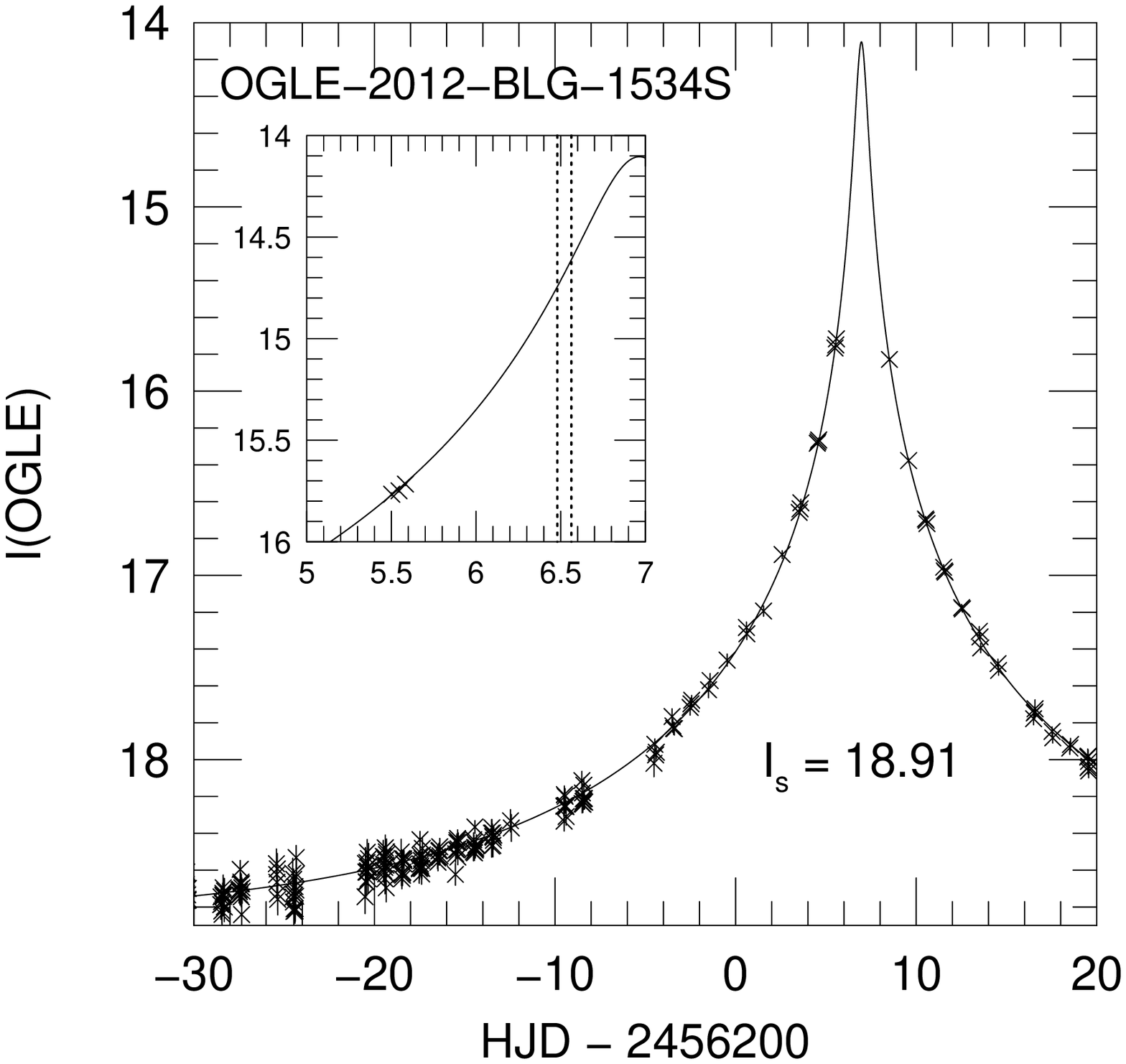}
\includegraphics[bb=67 40 515 525,clip]{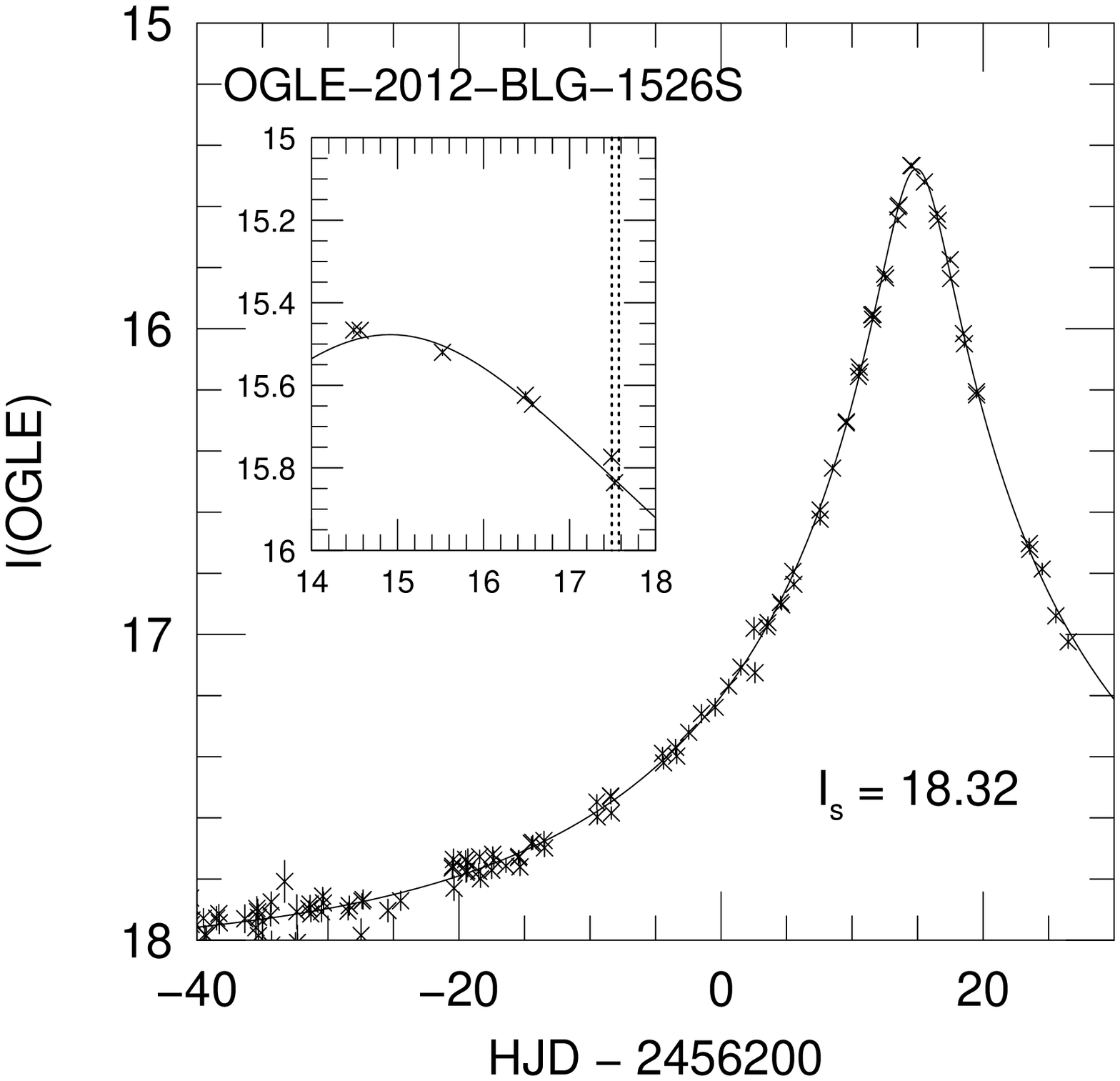}
}
\caption{Light curves for the remaining 16 of the 32 new microlensing events.
Each plot has a zoom window, showing the time intervals when the source 
stars
were observed with high-resolution spectrographs. In each plot the
un-lensed magnitude of the source star is also given ($I_{\rm S}$).
\label{fig:events2}
}
\end{figure*}
%-----------------------------------------------------------------------

%=======================================================================
\section{Observations and data reduction}

Dwarf and main sequence turn-off stars in the Galactic bulge 
have $V$ magnitudes around 19-20 \citep[e.g.,][]{feltzing2000b},
which is far too faint to observe under 
normal circumstances even with today's 8-10\,m class telescopes. To 
obtain a decent high-resolution spectrum ($S/N>50$ and $R>40\,000$)
of $V=19-20$ star would require more than 50 hours of exposure time 
with an instrument such as UVES on VLT. To the rescue, gravitational 
microlensing has surfaced 
as nature's golden magnifying glass to study dwarf stars in the bulge. 
Gravitational microlensing occurs whenever a compact object, by chance, 
passes the line of sight between an observer and a distant object.
The brightness of the distant object can then be magnified by factors of 
several hundred. However, as microlensing events are completely
random it is impossible to predict when and where they will happen. 
The search for microlensing events in the bulge is today mainly done
by the MOA and OGLE surveys, which monitor the bulge every night. 
The OGLE survey is carried out primarily in the $I$ band using
1.3\,m Warsaw Telescope at Las Campanas, Chile and detects roughly
1700 events per year \citep{udalski1994,udalski2003}.
The MOA survey is carried out primarily in a broad $R/I$ filter 
using the 1.8\,m telescope at Mt.\ John, New Zealand, and detects 
roughly 650 events per year \citep{bond2001,sumi2011}.
For a small subset of these events the background source stars are both 
(1) dwarfs (or subgiants), and (2) magnified highly enough to 
allow us to obtain a high-resolution high-S/N spectrum in a few hours.  
To find these spectroscopy targets, we perform real-time modelling of 
OGLE and MOA data as they are updated, and also acquire additional 
data on promising targets using the ANDICAM camera on the 1.3\,m 
SMARTS telescope at Cerro Tololo Interamerican Observatory (CTIO) in 
Chile. We also use the SMARTS telescope to measure the $(V-I)$ colours
of most events when they are near peak brightness.

Due to the unpredictability of microlensing events it is impossible to have 
a standard observing program, which is allocated specific 
nights in visitor mode or put in queue service mode.
Hence, starting in the ESO observing period P82 we have been running
Target-of-Opportunity (ToO) programs using UVES \citep{dekker2000} on the VLT
allowing us to trigger observations with only a few hours notice. 
Up until now, including the second of five triggers in P90, we have submitted
52 triggers with the result of 47 successfully acquired spectra of dwarf stars
in the bulge. Of the five ``failures'' there were two events for which the predicted
magnification was much higher than the actual one, resulting in too low $S/N$ and 
spectra useless for abundance analysis, one event turned out to be a red giant
and will be published in another paper, one spectrum was obtained  when only the
UVES blue arm was available resulting in very low $S/N$ and short wavelength 
coverage, and deemed useless, and one event turned out to be a nova 
towards the bulge (see \citealt{wagner2012}). The ToO observations with VLT
have been augmented with a few observations obtained with
the MIKE spectrograph \citep{bernstein2003} on Magellan or the HIRES 
spectrograph \citep{vogt1994} on Keck. The spectra have 
resolutions between $R\approx 40\,000-90\,000$ (see Table~\ref{tab:events}).

Our goal has been to observe the targets as close to peak brightness
as possible, and to observe targets that reach at least $I\approx 15.0$
during the observations. With a $1"$ wide slit this means that the spectra
should have a minimum signal-to-noise ratio of around $S/N\approx 50-60$ and a 
resolution of $R\approx 42\,000$ in the
UVES 760\,nm setting (according to the UVES exposure time calculator).
However, sometimes the observed microlensing light curve does not follow the 
predicted one, meaning that the obtained spectra could be better or worse
than predicted. Also, observing conditions (seeing and clouds)
have severely affected the quality of some of the observations.
Hence, the signal-to-noise ratios vary between $S/N\approx 30-100$
(see Table~\ref{tab:events}) and for a few events the
$S/N$ is even lower than 20. Figure~\ref{fig:events} shows 
the light curves for the 32 new events (light curves for the first 26 events are
published  in \citealt{bensby2009,bensby2010,bensby2011}). 
In each plot the time interval when the spectroscopic observations were 
carried out has been marked  (see insets in each panel). 
Generally, we were able to catch many  objects very close to peak 
brightness, but in a few cases we were less lucky. 
Especially interesting is the observation of MOA-2012-BLG-532S (Fig.~\ref{fig:events2}).
This event suddenly entered a caustic crossing resulting in a strong a sudden
increase in brightness. We managed to submit a trigger and start observations.
However, its brightness started to decrease again sooner than expected,
resulting in most of the two hour exposure being executed when it had
dropped 2 magnitudes in brightness. Hence, the spectrum for MOA-2012-BLG-532S
is of lower quality. 

Data reductions of the UVES spectra were carried out with the 
UVES pipeline \citep{ballester2000} versions 4.9.0 through 5.0.7,
depending on when the event was observed. 
The MIKE spectra were reduced with the Carnegie Observatories
python pipeline\footnote{Available at 
{\tt http://obs.carnegiescience.edu/Code/mike}}, and the reduction of the 
HIRES spectrum follows the procedure outlined in \cite{cohen2008}.

%=======================================================================
\section{Analysis} 

%=======================================================================
\subsection{Stellar parameters -- with Fe\,I NLTE corrections} 
\label{sec:analysis}

%-----------------------------------------------------------------------
\begin{figure}
\resizebox{\hsize}{!}{
\includegraphics[bb=18 144 592 718,clip]{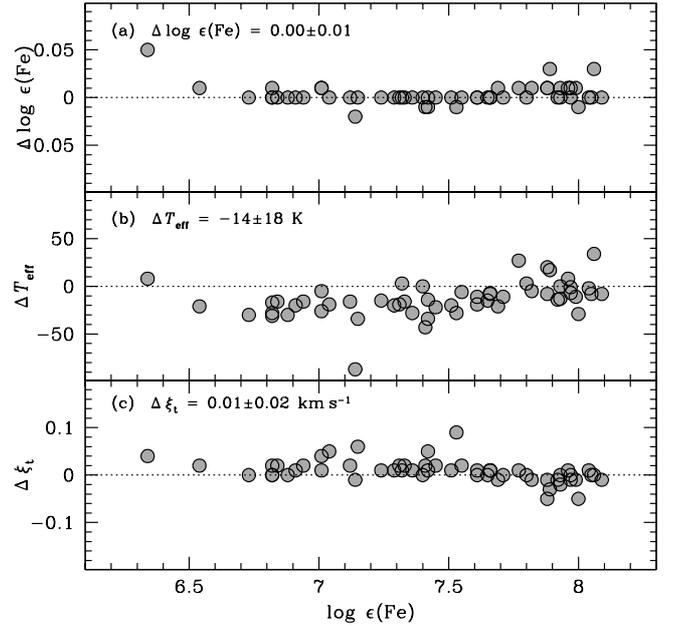}}
\caption{Comparison of stellar parameters before and after implementation
of NLTE corrections for \ion{Fe}{i} lines. The differences are NLTE values
minus LTE values and mean differences and 
$\sigma$ are indicated in each panel.
                    }
         \label{fig:nltecheck}
   \end{figure}
%-----------------------------------------------------------------------

%-----------------------------------------------------------------------
\begin{figure*}
\resizebox{\hsize}{!}{
\includegraphics[bb=18 178 580 718,clip]{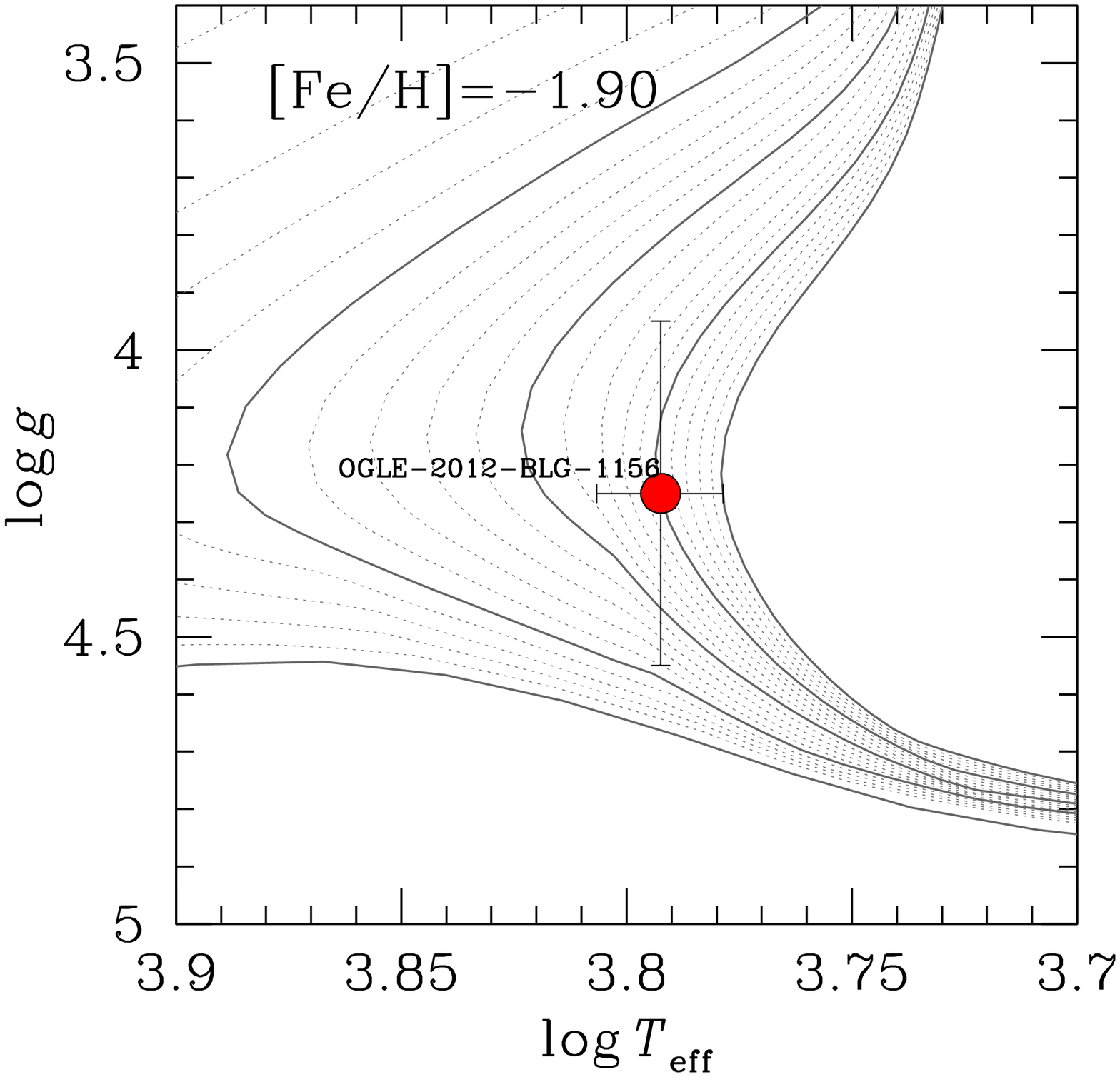}
\includegraphics[bb=47 178 580 718,clip]{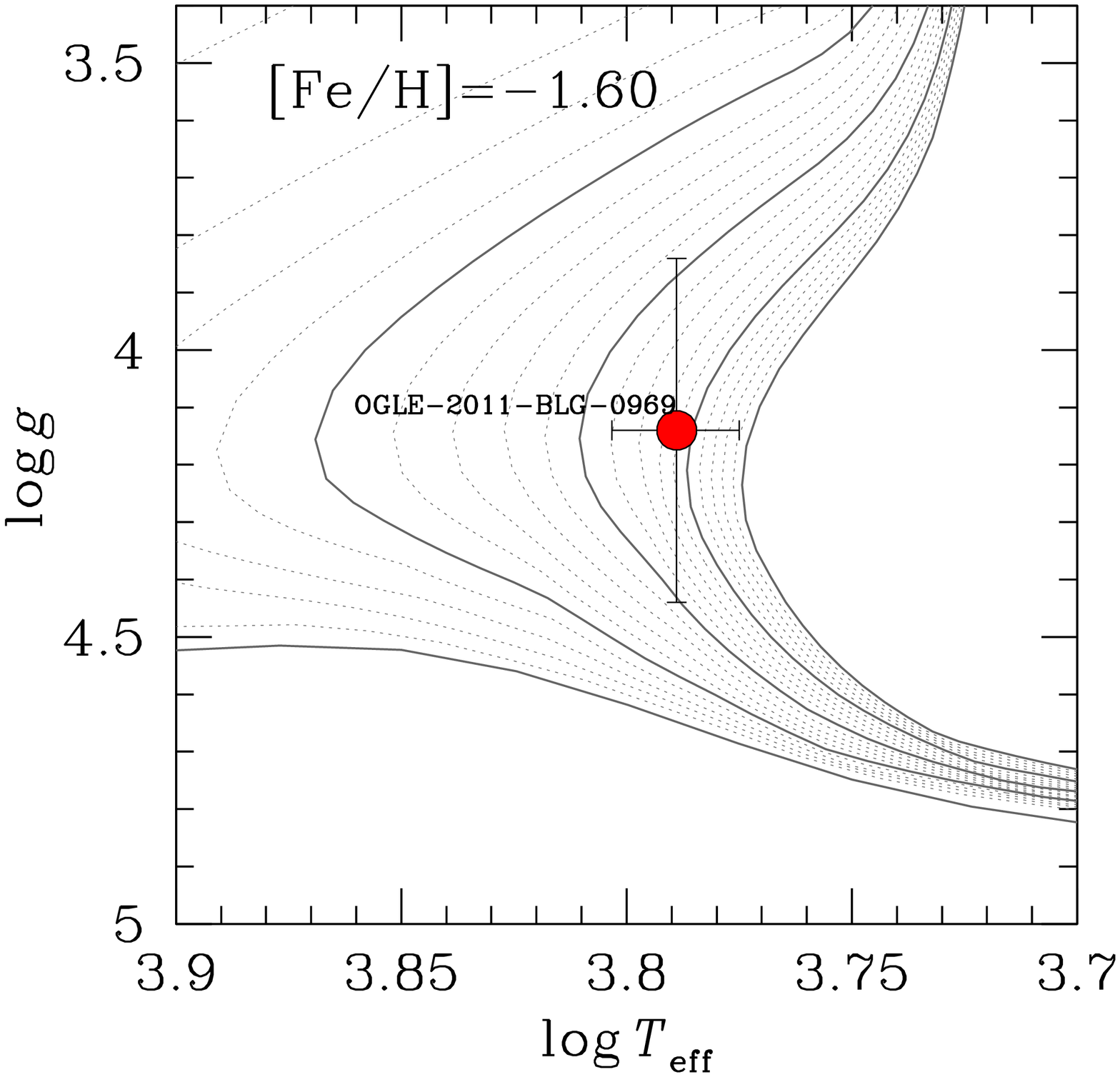}
\includegraphics[bb=47 178 580 718,clip]{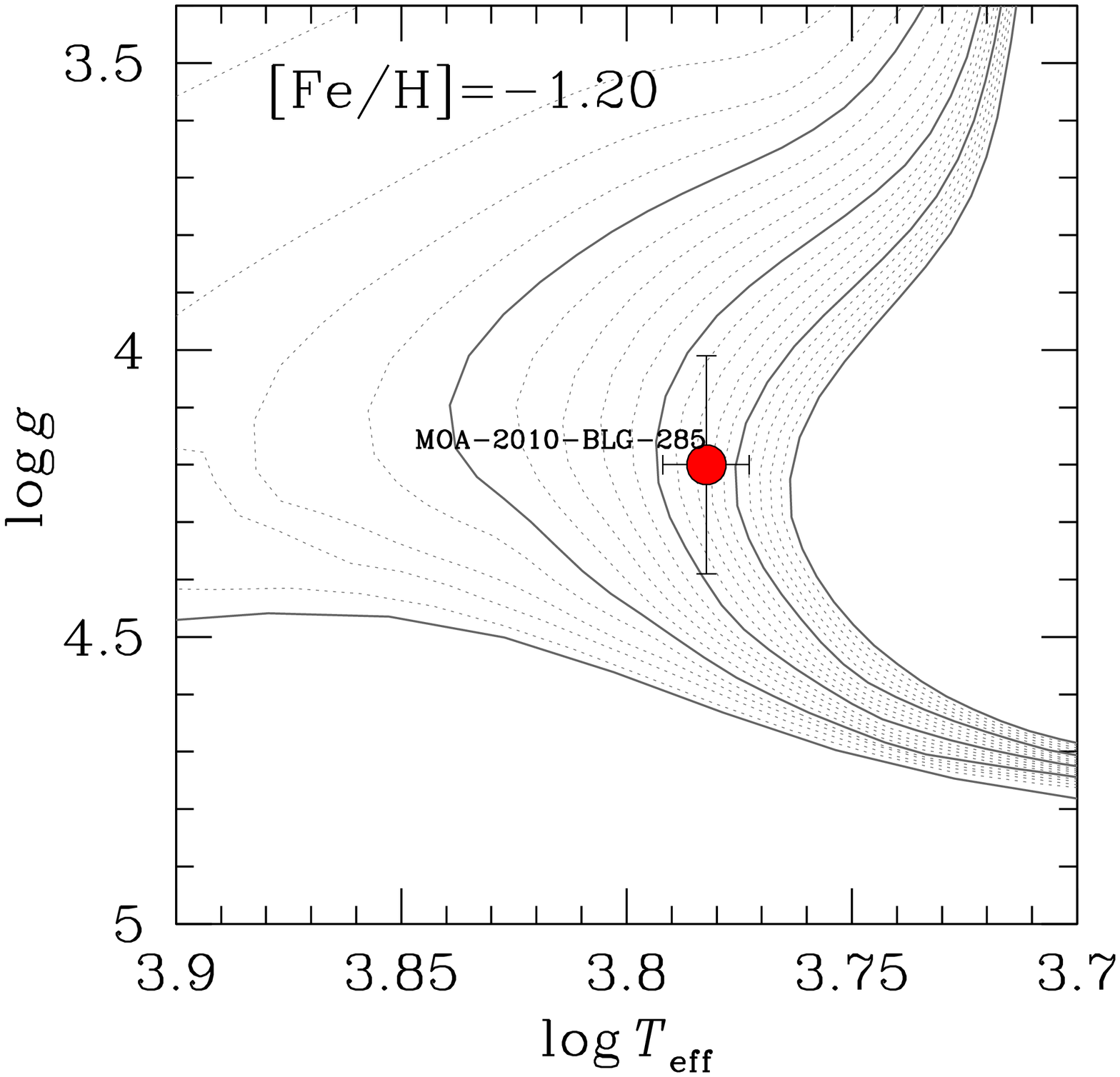}
\includegraphics[bb=47 178 592 718,clip]{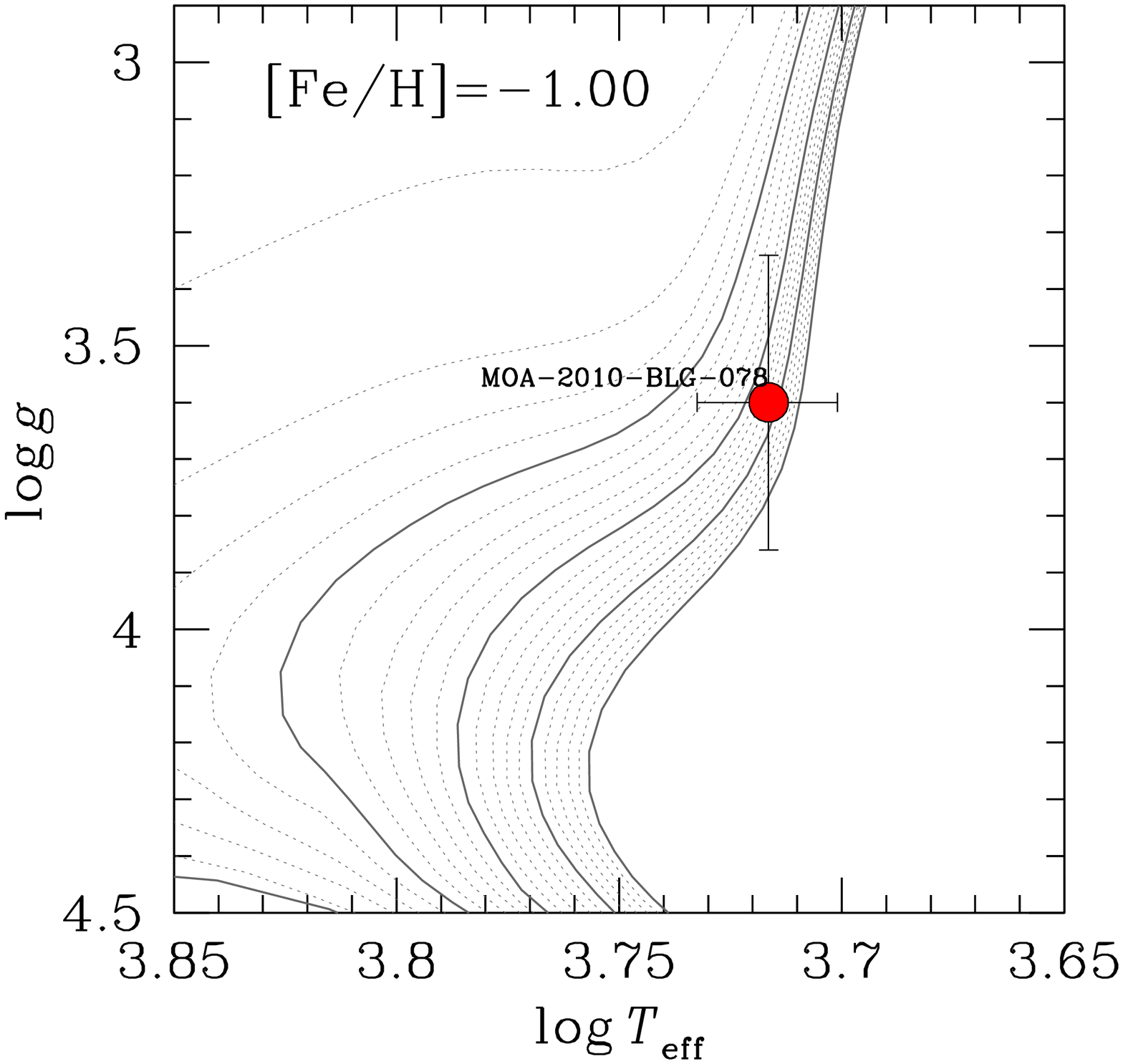}
}
\resizebox{\hsize}{!}{
\includegraphics[bb=18 178 580 708,clip]{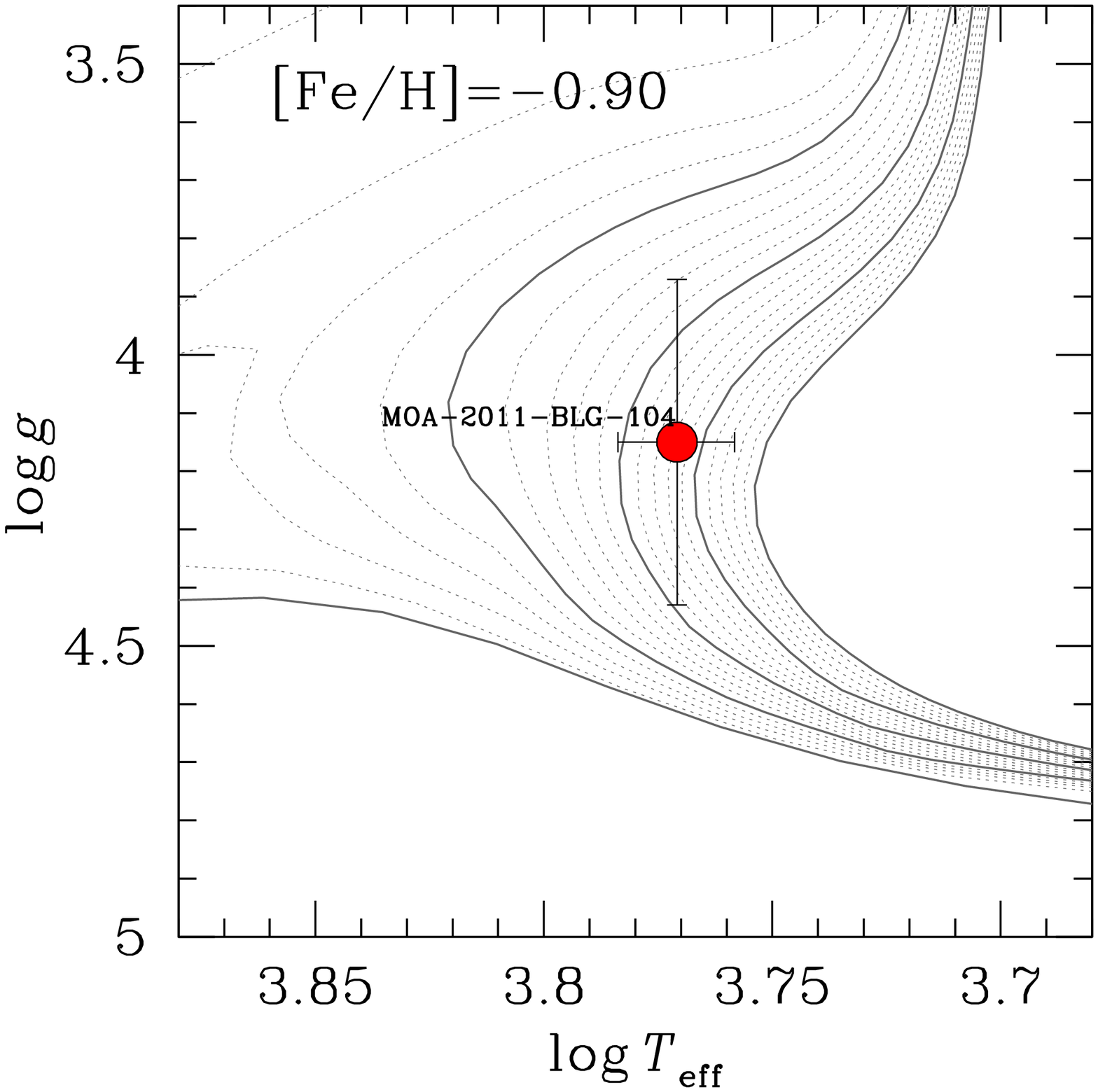}
\includegraphics[bb=47 178 580 708,clip]{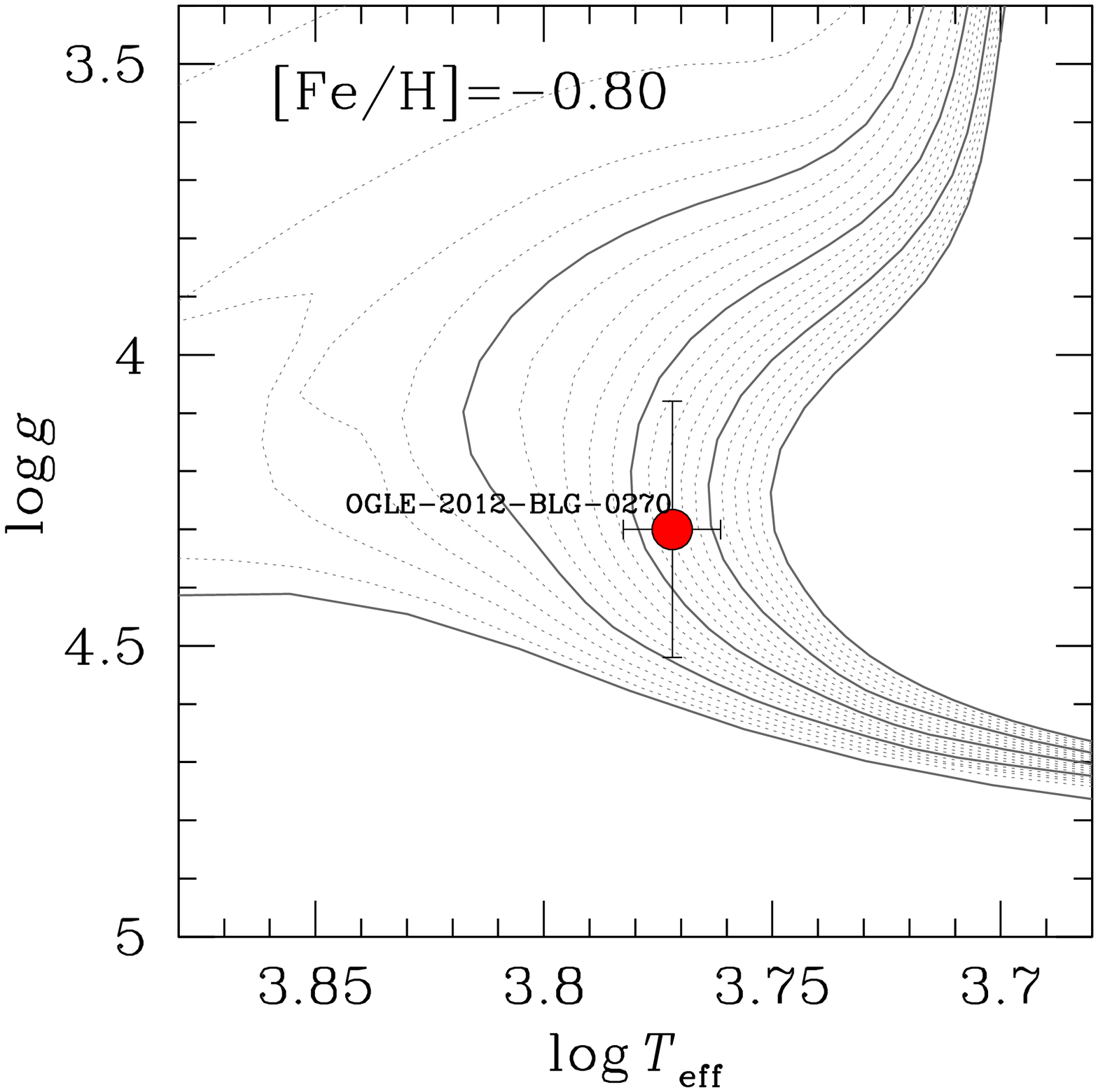}
\includegraphics[bb=47 178 580 708,clip]{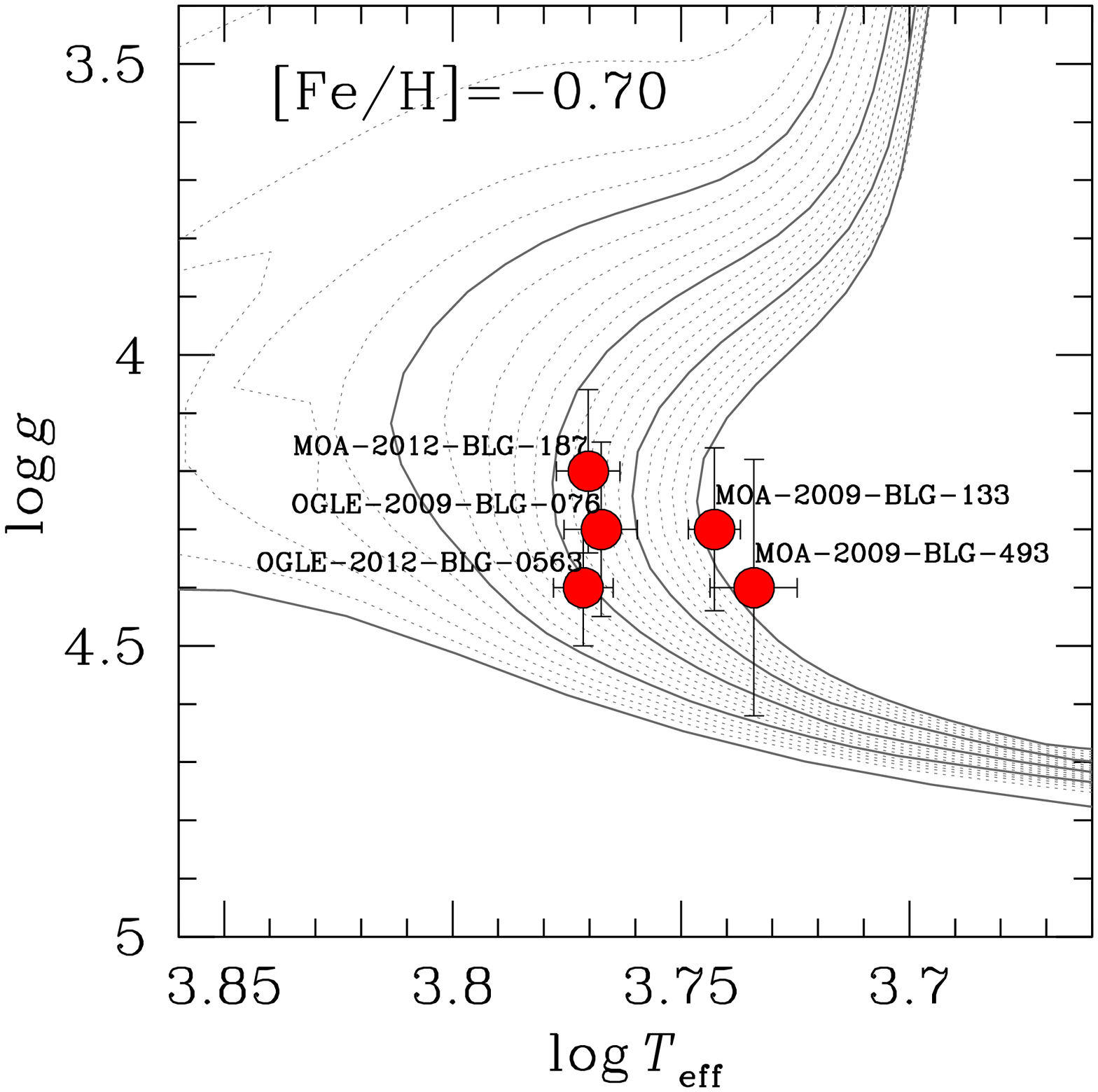}
\includegraphics[bb=47 178 592 708,clip]{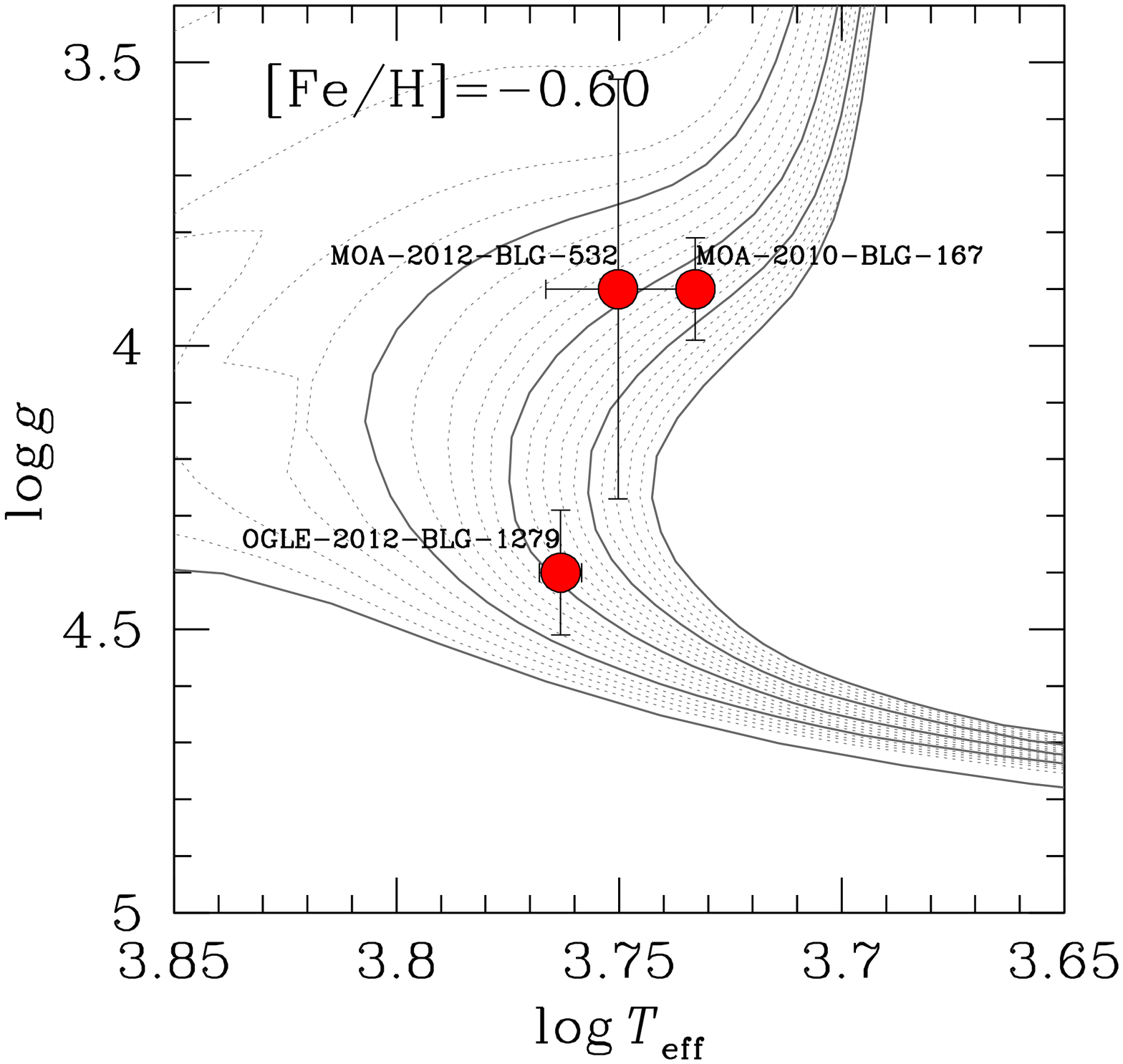}
}
\resizebox{\hsize}{!}{
\includegraphics[bb=18 178 580 708,clip]{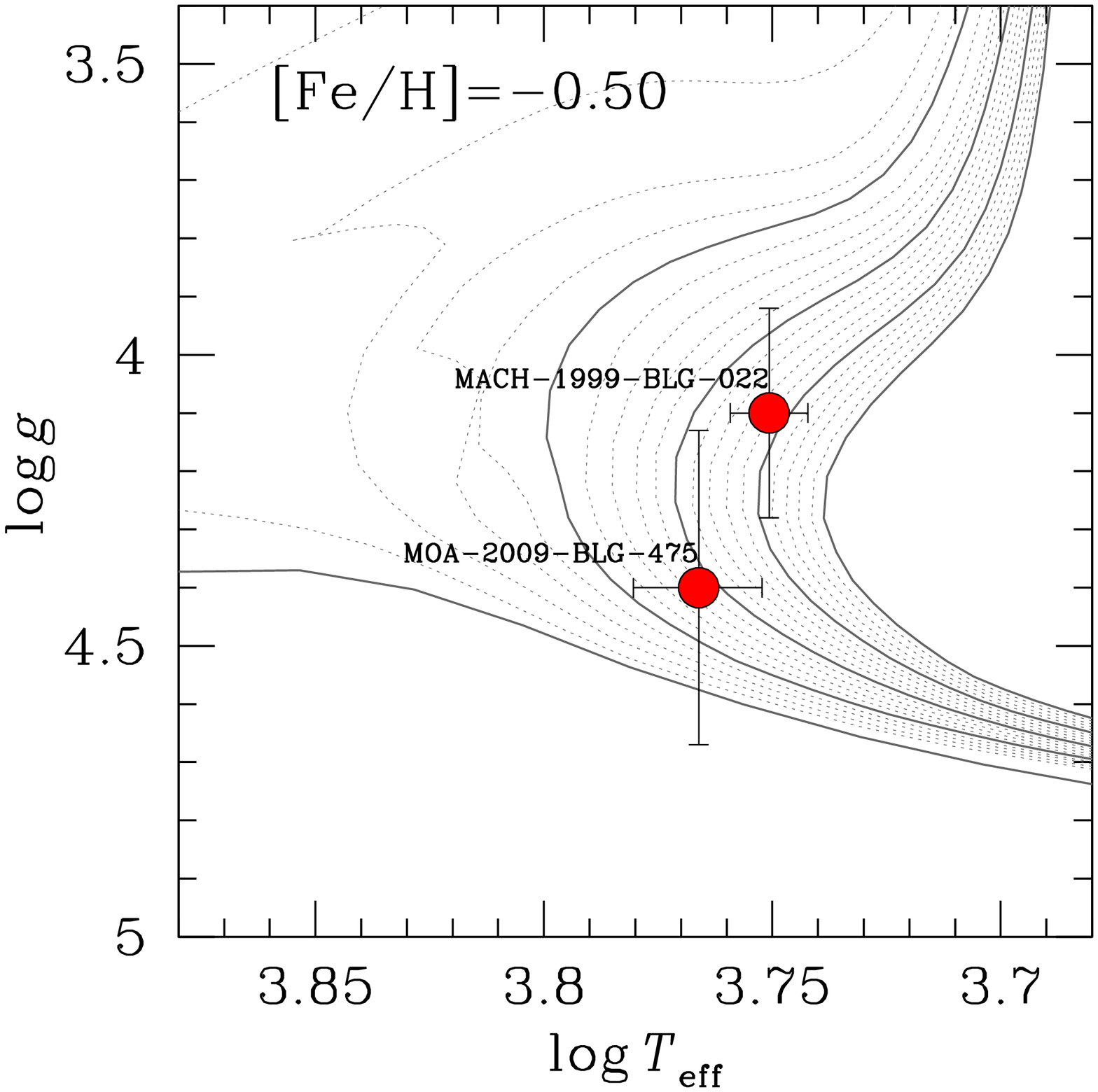}
\includegraphics[bb=47 178 580 708,clip]{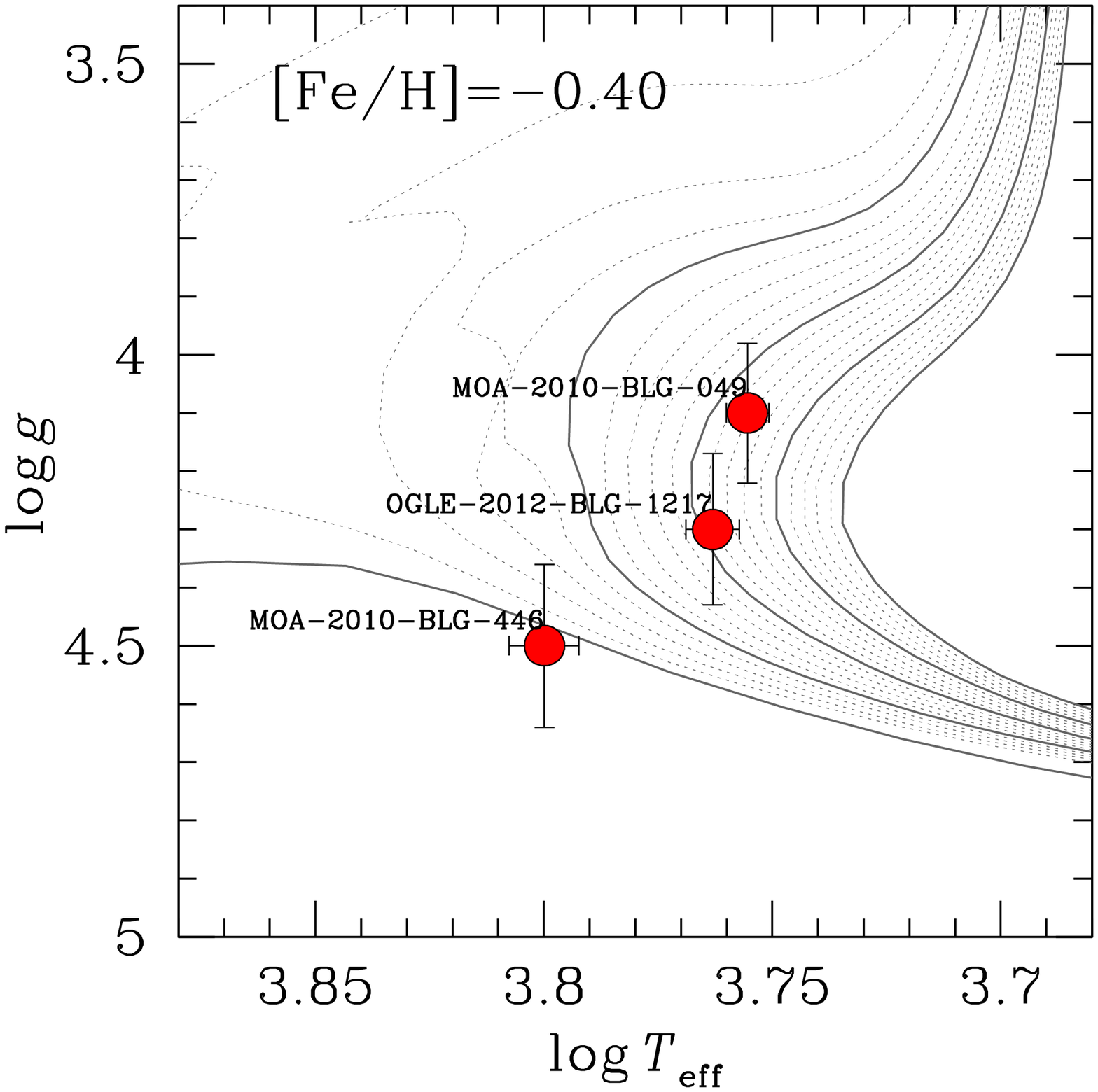}
\includegraphics[bb=47 178 580 708,clip]{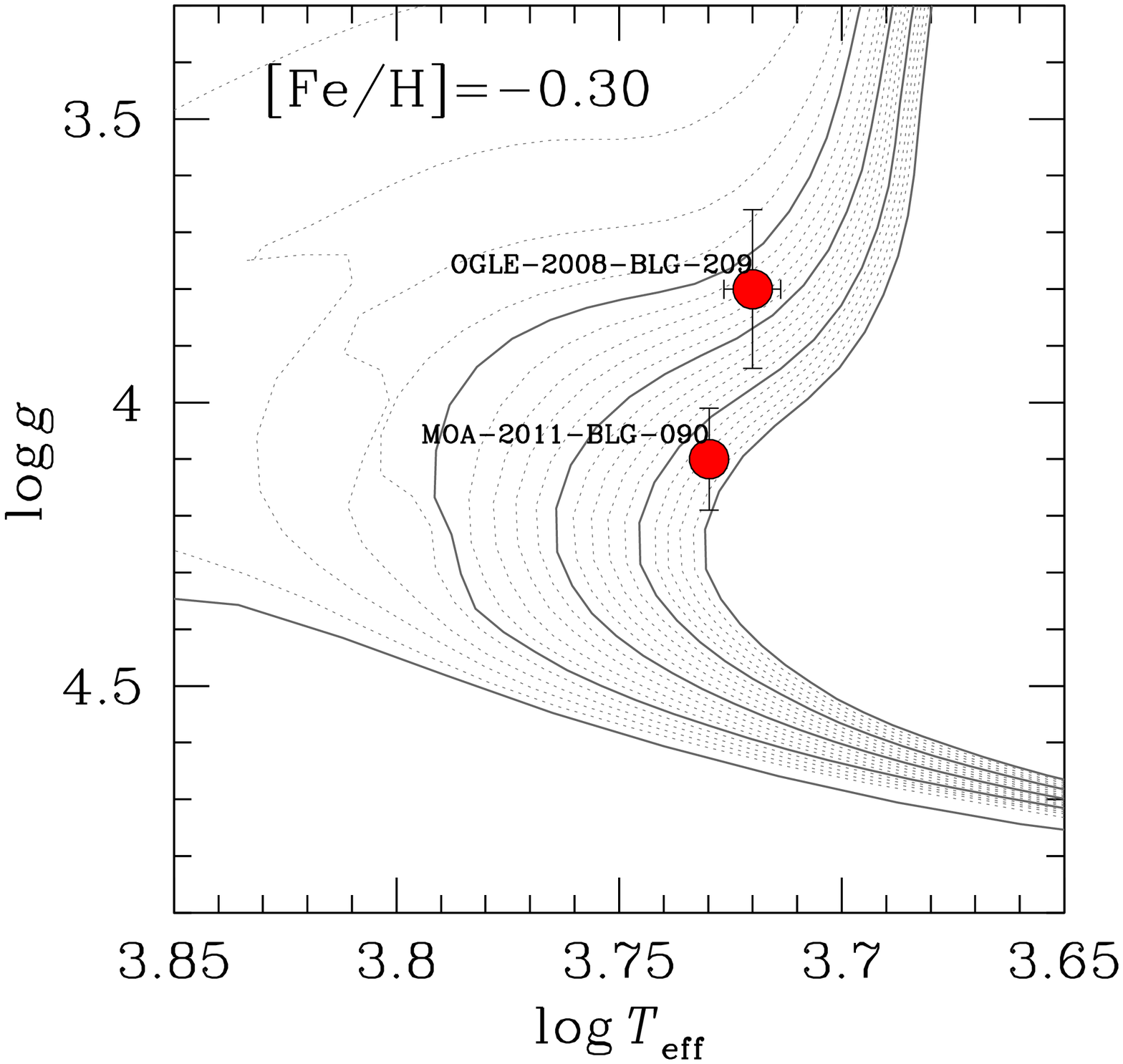}
\includegraphics[bb=47 178 592 708,clip]{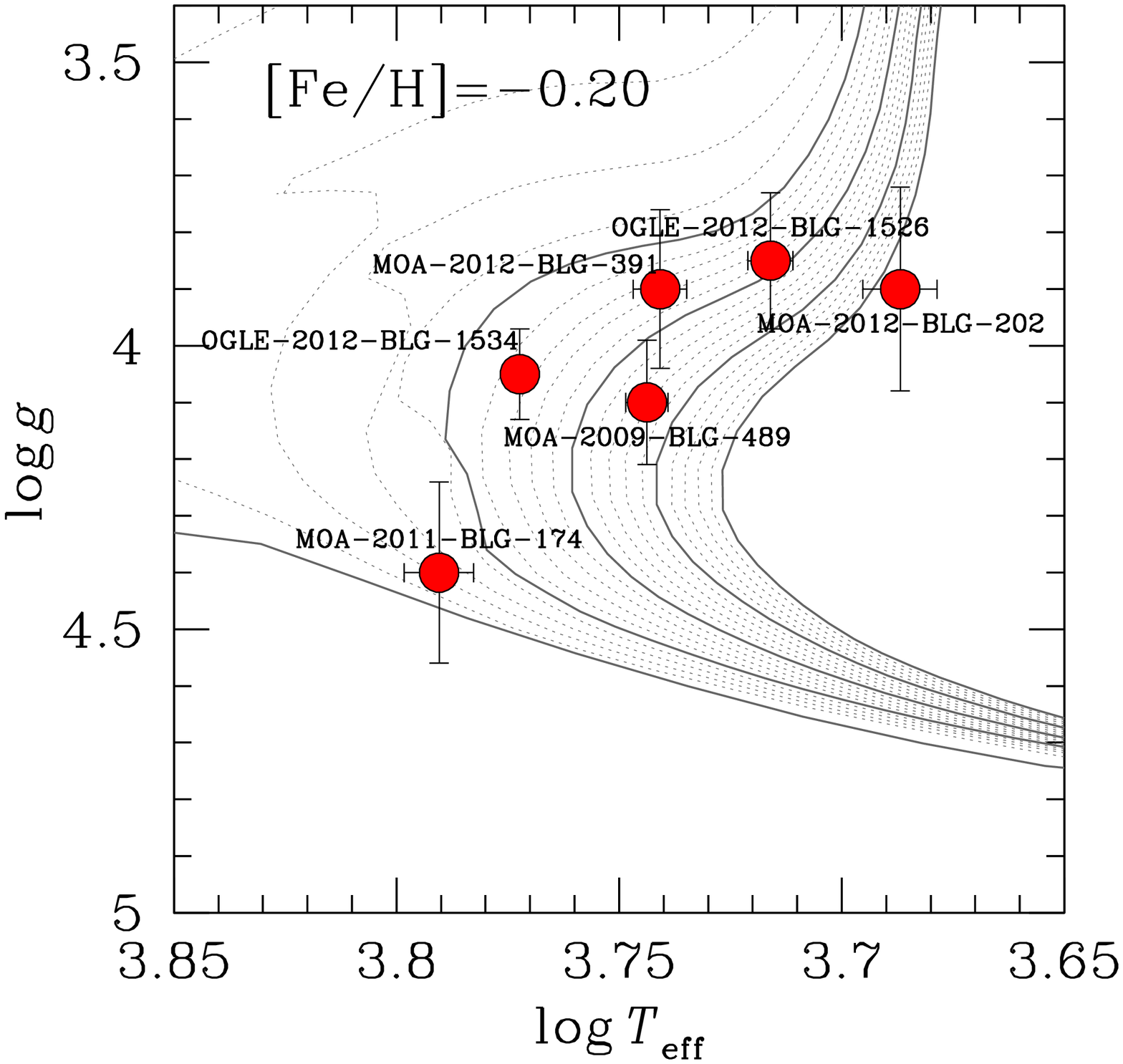}
}
\resizebox{\hsize}{!}{
\includegraphics[bb=18 178 580 708,clip]{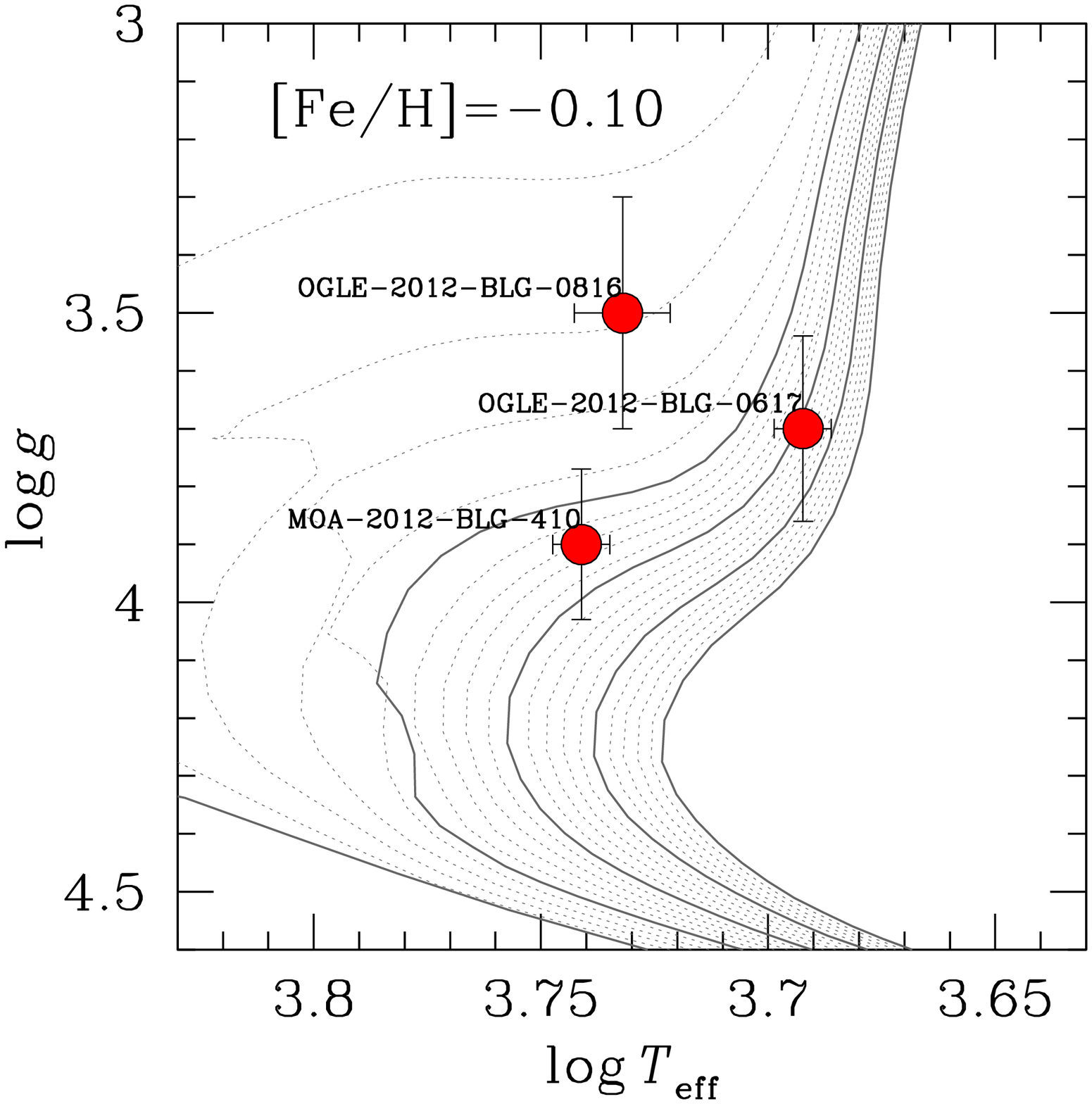}
\includegraphics[bb=47 178 580 708,clip]{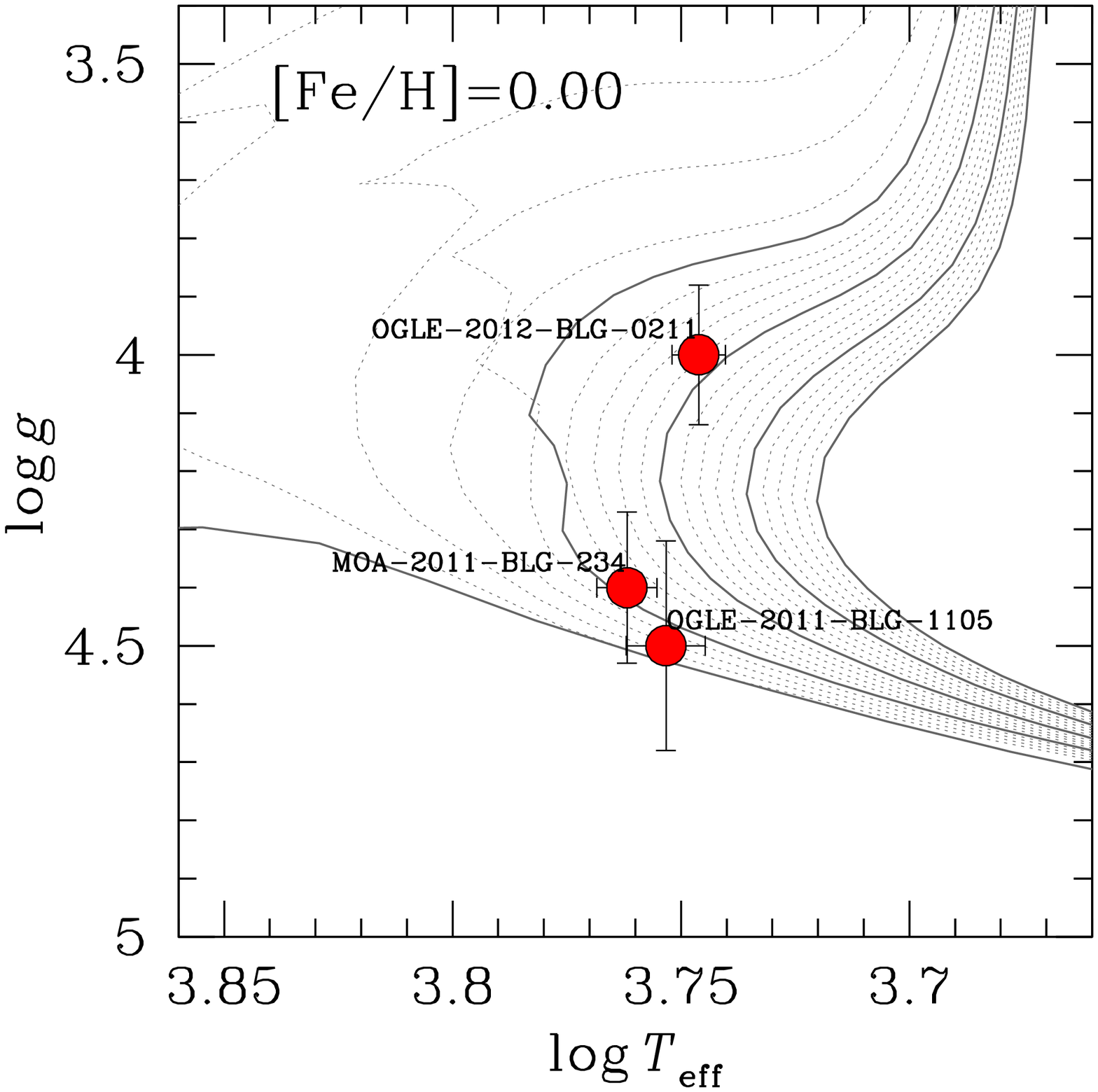}
\includegraphics[bb=47 178 580 708,clip]{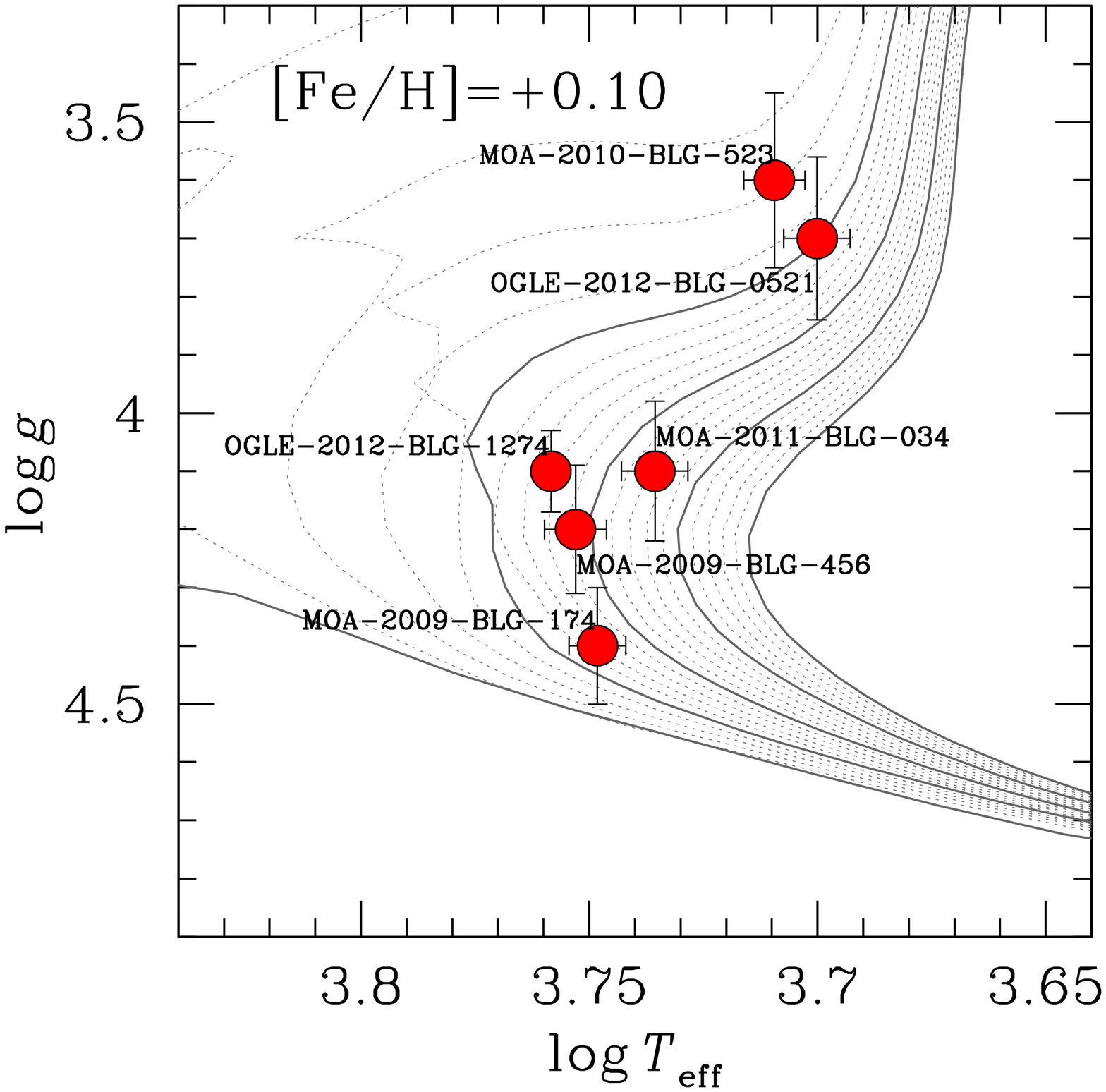}
\includegraphics[bb=47 178 592 708,clip]{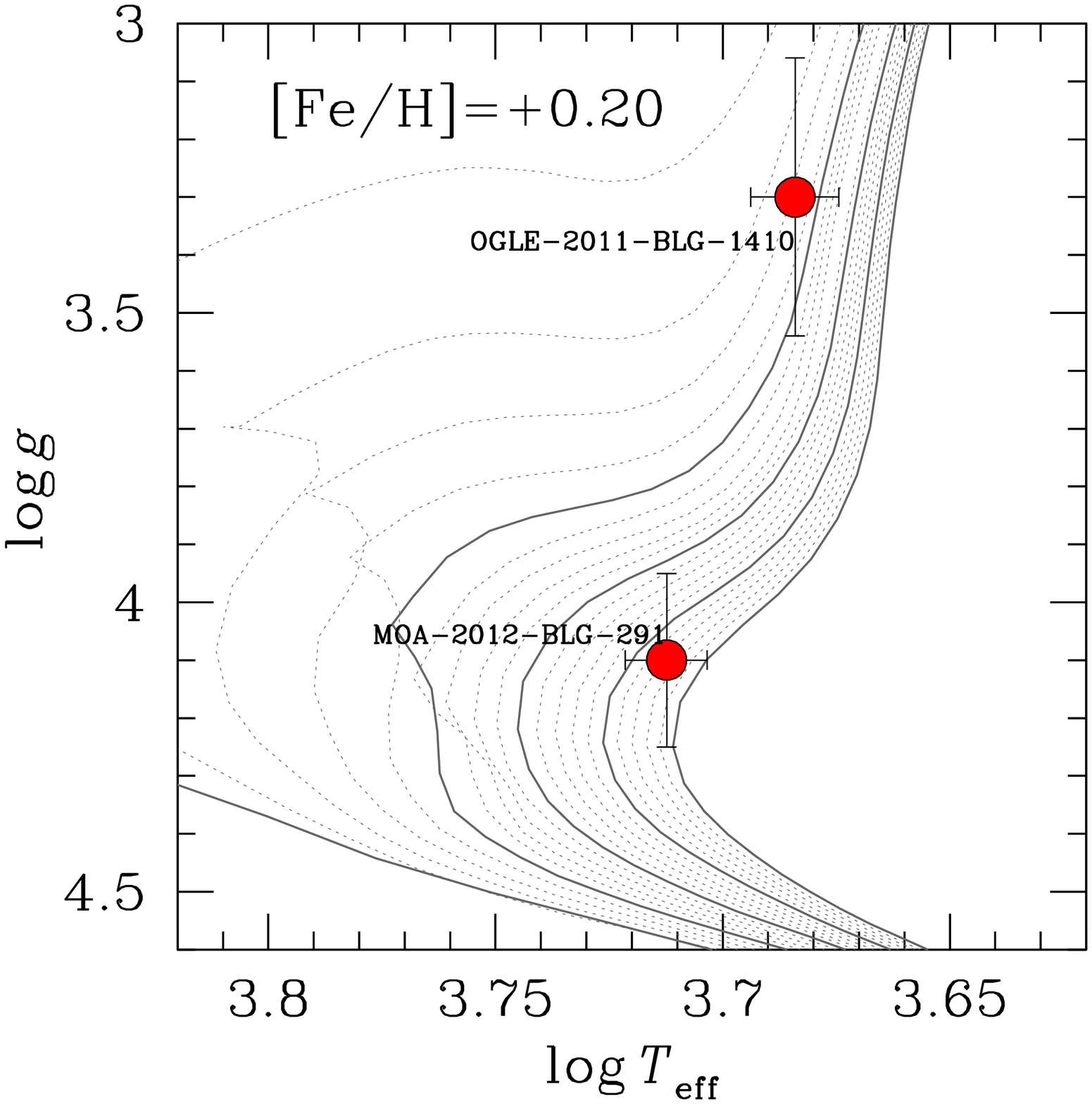}
}
\resizebox{\hsize}{!}{
\includegraphics[bb=18 144 580 708,clip]{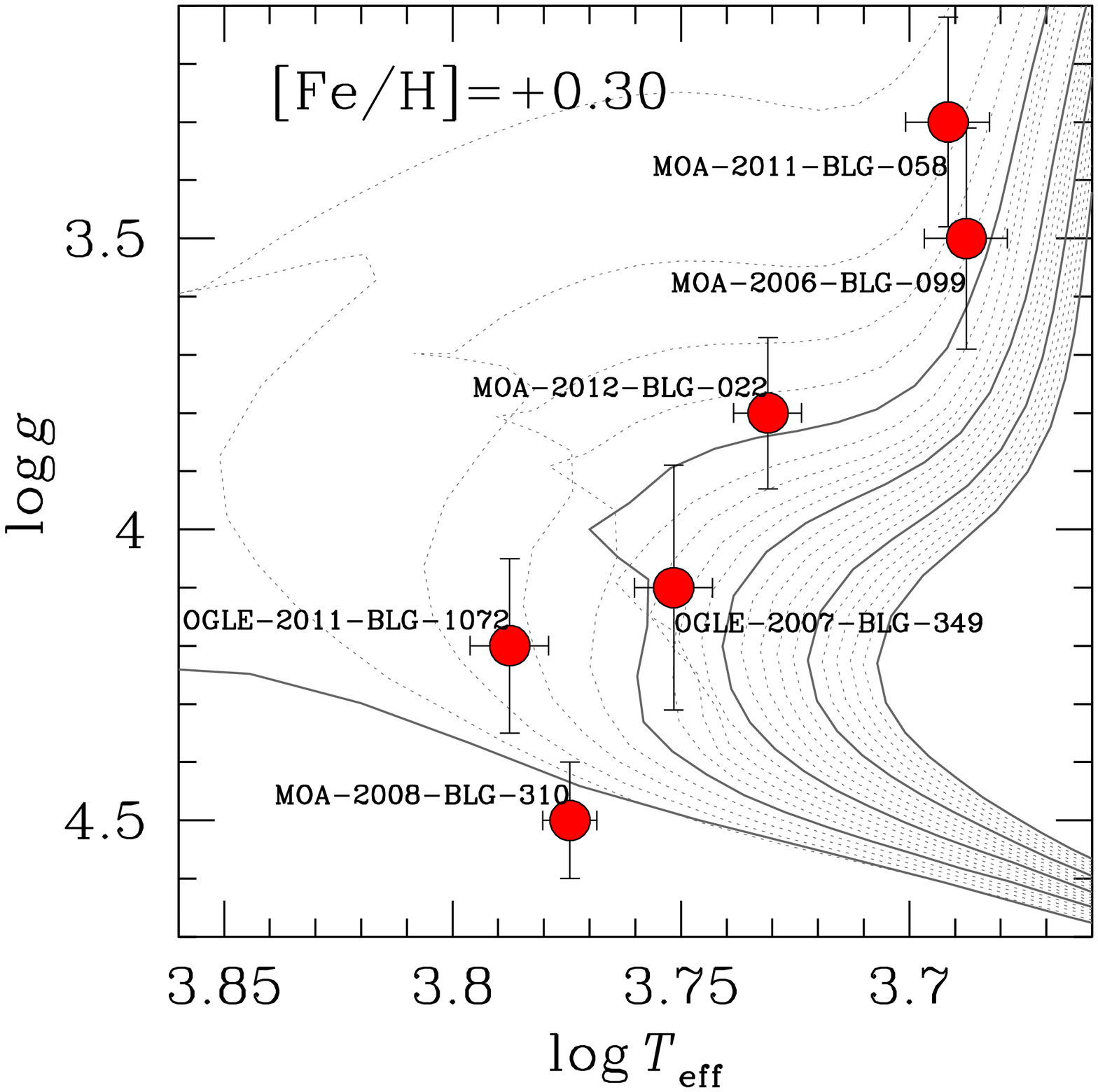}
\includegraphics[bb=47 144 580 708,clip]{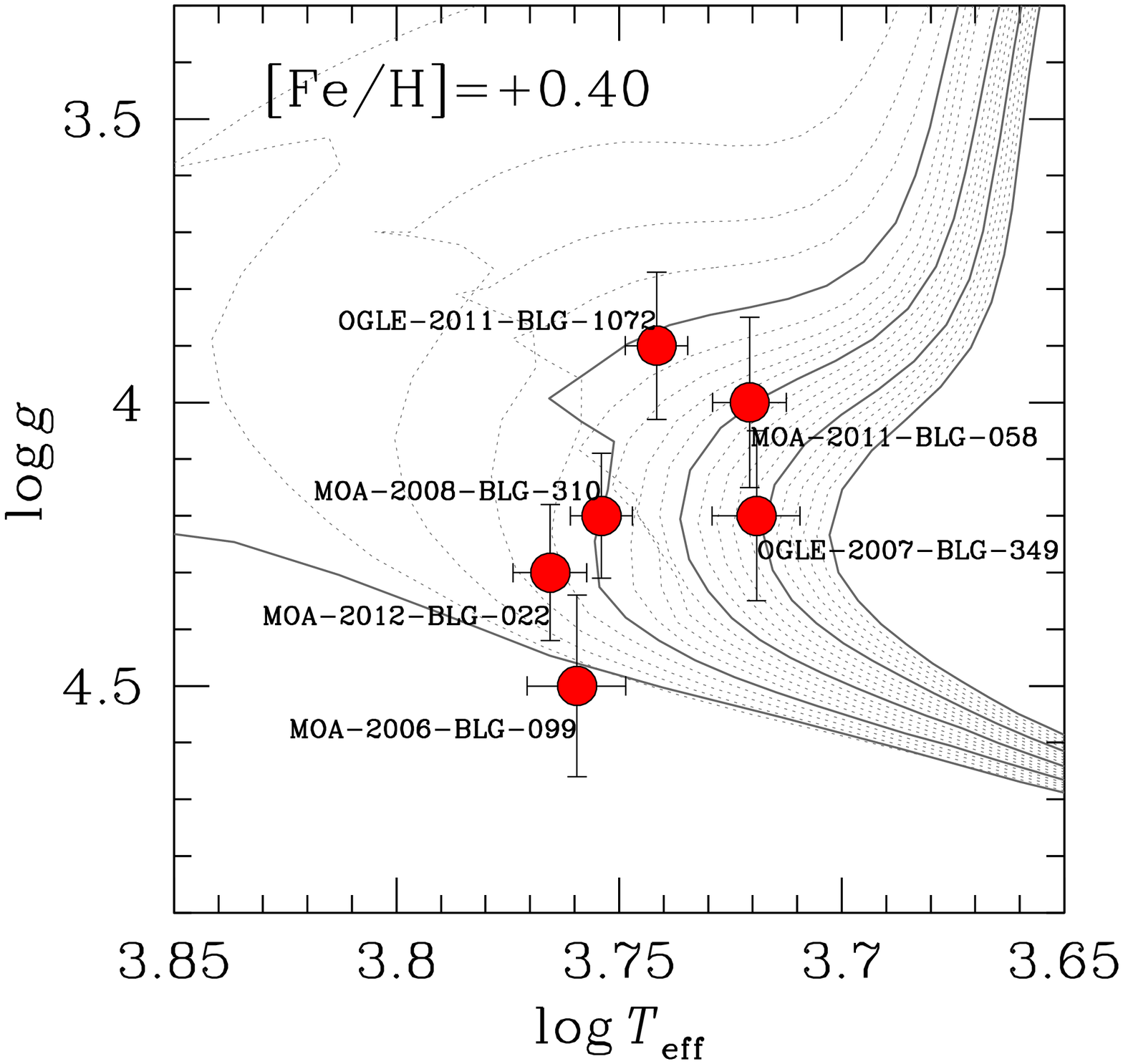}
\includegraphics[bb=47 144 580 708,clip]{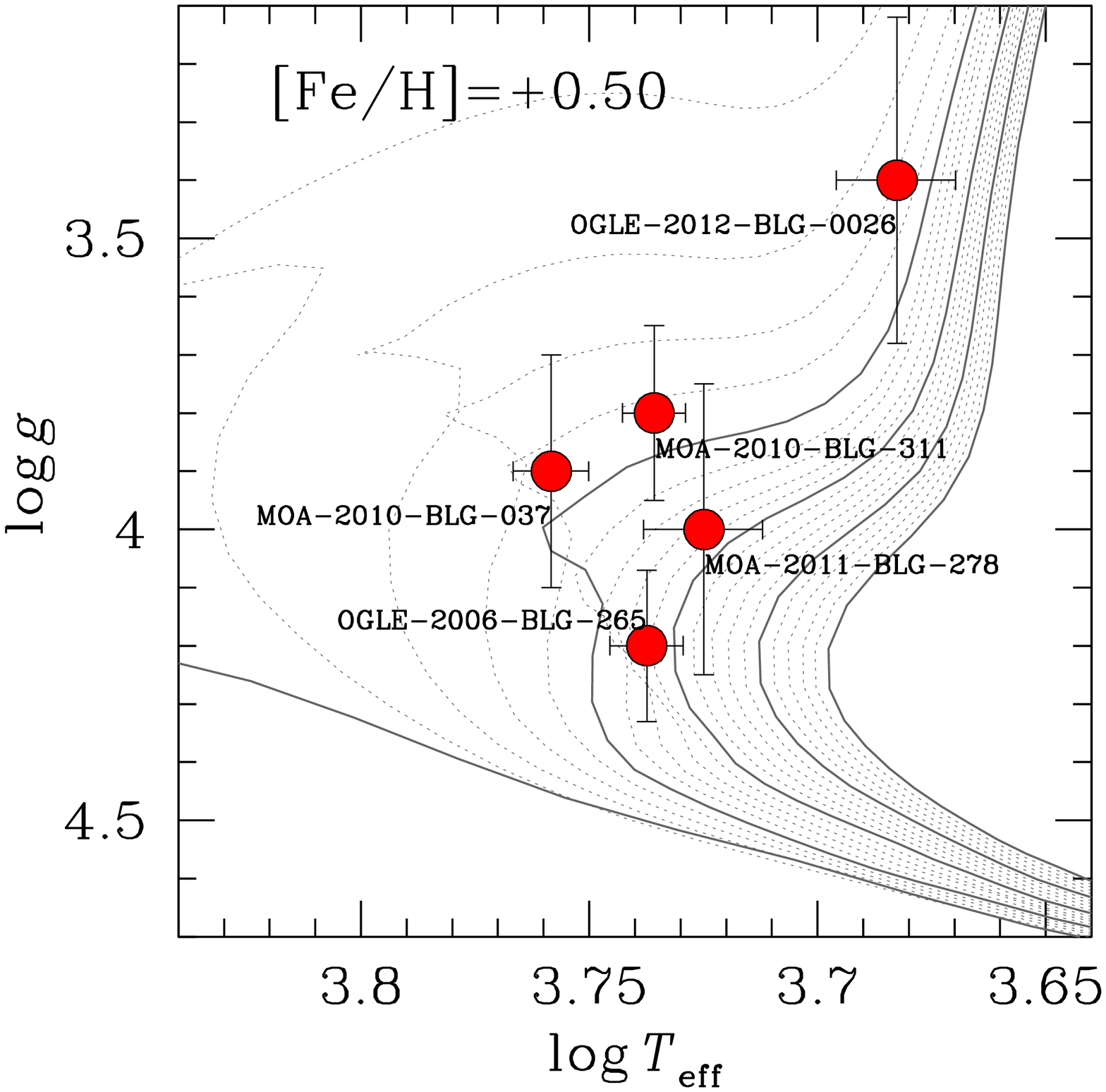}
\includegraphics[bb=47 144 592 145,clip]{hr_p050.eps}
}
\caption{
All 58 microlensed dwarf stars plotted on the $\alpha$-enhanced
isochrones from \cite{demarque2004}. The metallicity of the stars
in each panel is within $\pm 0.05$\,dex of the isochrone metallicity. 
In each plot the solid lines represent
isochrones with ages of 0.1, 5, 10, 15, and 20\,Gyr.
Dotted lines are isochrones in steps of 1\,Gyr. Error bars represent the uncertainties
in $\teff$ and $\log g$ as given in Table~\ref{tab:parameters}.
Note that the age determinations reported in Table~\ref{tab:ages} 
are based on probability distribution functions
as described in \cite{bensby2011}.
\label{fig:ages}
}
\end{figure*}
%-----------------------------------------------------------------------

NLTE corrections for \ion{Fe}{i} are now included in the analysis
compared to the analysis of the previous sample of 26 microlensed dwarf
stars from \cite{bensby2010,bensby2011}. 
As the inclusion of NLTE corrections for \ion{Fe}{i} lines
impacts the final values of the stellar parameters,
those 26 stars have been re-analysed, ensuring a homogeneous treatment
of the full sample.
The NLTE corrections are implemented on a 
line-by-line basis using an IDL script kindly provided by Karin Lind 
\citep{lind2012}. \ion{Fe}{ii} lines are on the other hand not sensitive
to NLTE effects \citep[e.g.,][]{thevenin1999,lind2012}, and need not be corrected. 
Apart from the implementation of \ion{Fe}{i} NLTE corrections, the 
methodology to find the stellar parameters is the same as before and is 
fully described in \cite{bensby2009,bensby2010,bensby2011}. 
In brief, the analysis is based on standard 1-D plane-parallel MARCS model 
stellar atmospheres \citep{gustafsson1975,edvardsson1993,asplund1997}, and 
abundances are calculated with the Uppsala EQWIDTH program using 
equivalent widths measured by hand using the IRAF task SPLOT. 
The effective temperatures were determined from excitation balance of 
NLTE corrected abundances from \ion{Fe}{i} lines, surface gravities from 
ionisation balance between NLTE corrected \ion{Fe}{i} abundances and 
abundances from \ion{Fe}{ii} lines, and the microturbulence parameters 
by requiring that the NLTE corrected  \ion{Fe}{i} abundances  are 
independent of line strength.

 As demonstrated by 
\cite{lind2012} the NLTE effects are very small, being smallest for 
the most metal-rich stars (generally below 0.01\,dex) and increasing 
for lower metallicities, and for stars with lower surface gravities. 
Figure~\ref{fig:nltecheck} shows comparisons of 
effective temperatures, metallicities, and 
microturbulence velocities for all 58 stars with and without the 
inclusion of \ion{Fe}{i} NLTE corrections, and as can be seen
the effects are truly minuscule for the type of stars analysed here
(mostly metal-rich F and G dwarfs).

Uncertainties in stellar parameters and elemental abundances were, as 
for the previous sample of 26 stars \citep{bensby2011},
calculated according to the method outlined in \cite{epstein2010}.
This method takes into account the uncertainties in the four observables 
used to find the stellar parameters, i.e. the uncertainty of 
the slope in the graph of \ion{Fe}{i} abundances versus lower excitation 
potential; the uncertainty of the slope in the graph of \ion{Fe}{i} 
abundances versus reduced line strength; the uncertainty in the difference between 
\ion{Fe}{i} and \ion{Fe}{ii} abundances; and the uncertainty in the 
difference between input and output metallicities. The method also accounts 
for abundance spreads (line-to-line scatter) as well as how the average 
abundances for each element reacts to changes in the stellar parameters 
(see also comments in \citealt{bensby2011}). 

Figure~\ref{fig:ages} shows the location in the $\log g - \teff$ plane,
for all 58 stars in the microlensed bulge dwarf sample
over-plotted on  the $\alpha$-enhanced
isochrones from \cite{demarque2004}. A majority of the
stars are located either on the main-sequence turn-off or the subgiant 
branch. Three or four stars have started to ascend the giant branch but not
far enough to have the chemical composition of their atmospheres altered. 
The fact that most of the stars are either turn-off
or subgiant stars, and that the uncertainties in the stellar parameters
are well-constrained with usually small uncertainties, 
means that it is possible to determine
relative ages accurately. Stellar ages, masses, luminosities, absolute 
$I$ magnitudes ($M_I$), and colours $(V-I)$ were estimated from Y$^2$ 
isochrones \citep{demarque2004} by maximising probability distribution 
functions as described in \cite{bensby2011}.

We have also double-checked the analysis of the previous 26 events  
and revised some equivalent width measurements. This lead to changes 
in the stellar parameters and abundance ratios for a few stars. 
Also, the inclusion of NLTE corrections for \ion{Fe}{i} lead
to some, although small, changes. Therefore, equivalent width 
measurements and abundances for individual lines are given 
in Table~\ref{tab:ews} for all 58 stars, and
stellar parameters, stellar ages, and elemental abundances in 
Tables~\ref{tab:abundances}, \ref{tab:parameters}, \ref{tab:ages}, 
for all 58 stars, i.e., all tables include the revised values for the 
26 stars in \cite{bensby2010,bensby2011}.
All abundances have been normalised to the Sun on a line-by-line 
basis using our own solar analysis as reference (see also \citealt{bensby2011}).

%-----------------------------------------------------------------------
\begin{figure}
\resizebox{\hsize}{!}{
\includegraphics[bb=18 430 592 720,clip]{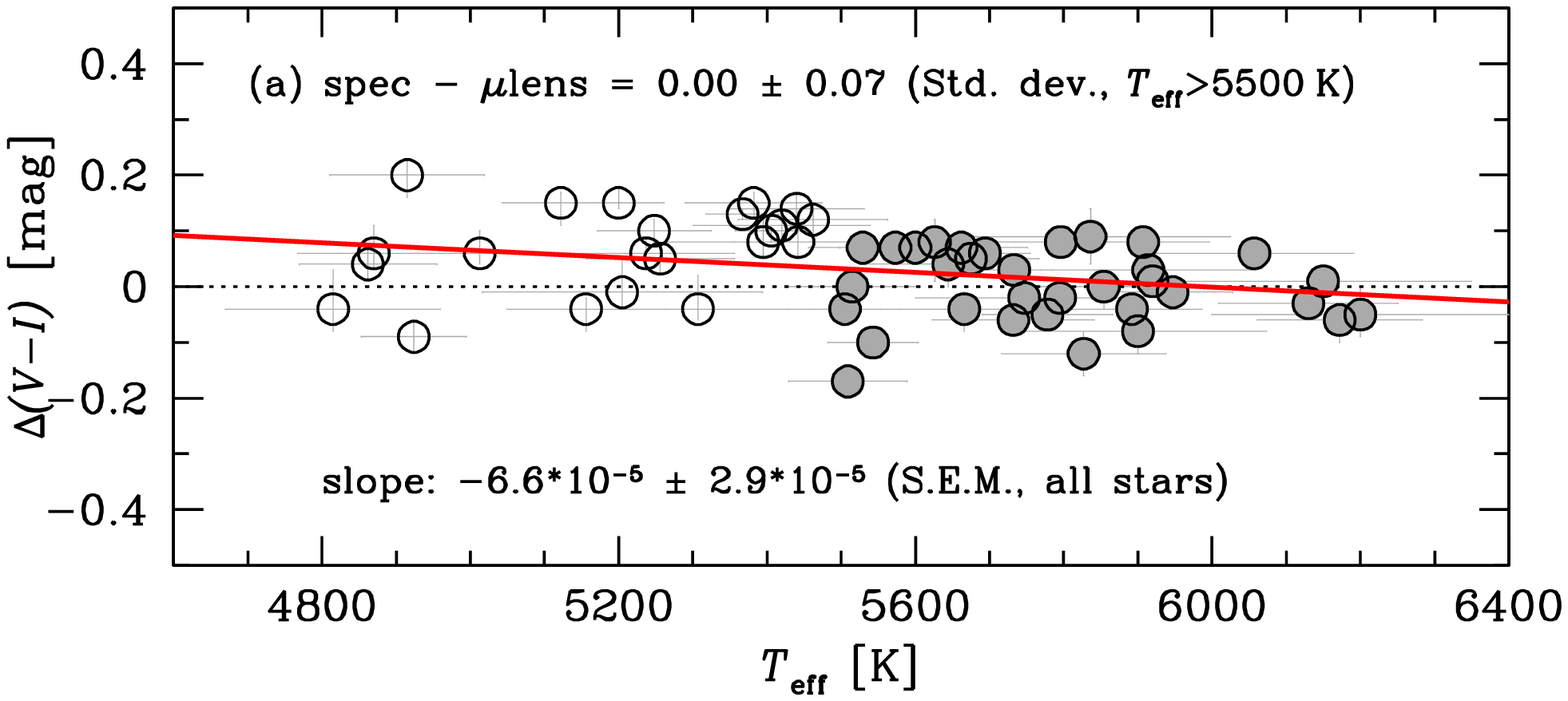}}
\resizebox{\hsize}{!}{
\includegraphics[bb=18 155 592 430,clip]{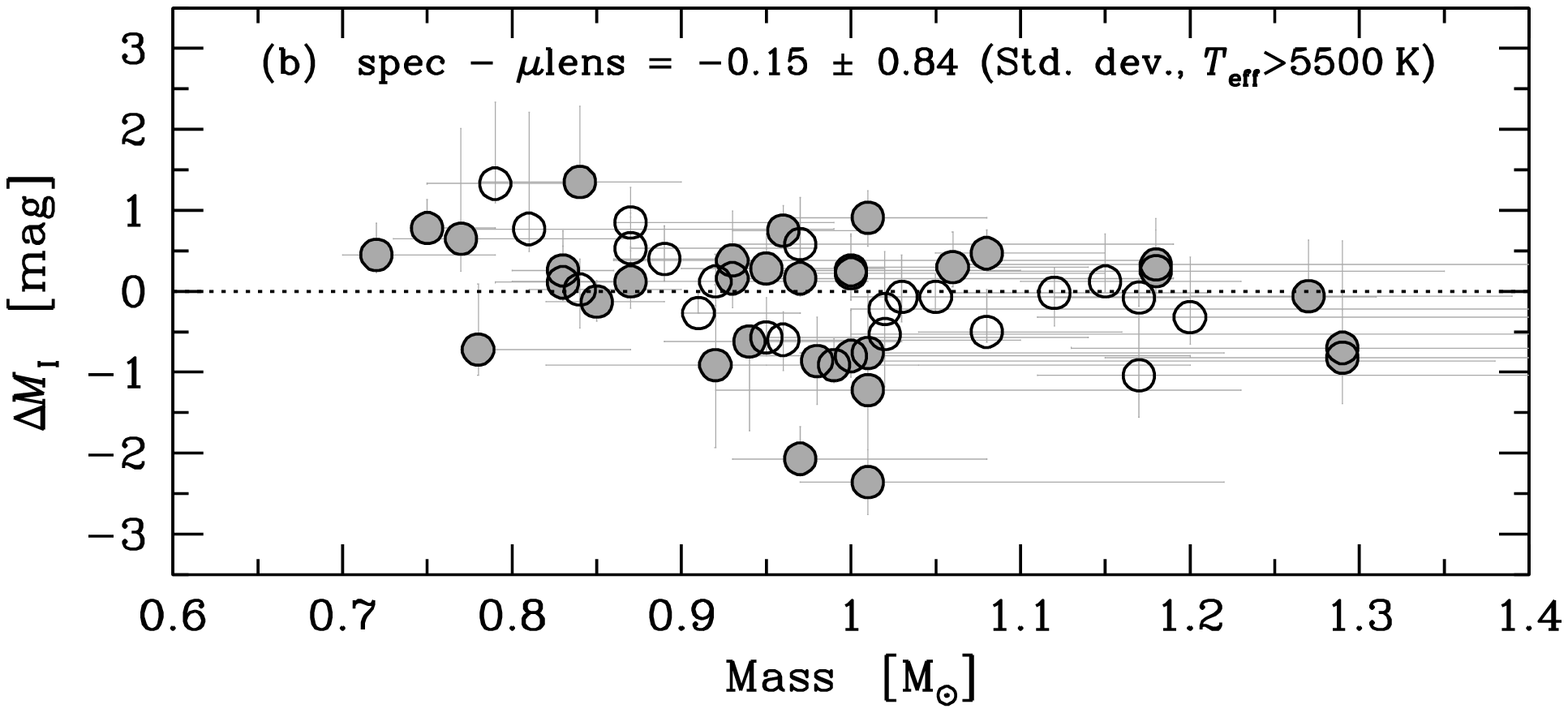}}
\caption{(a) Difference between microlensing colours and the 
spectroscopic colours versus spectroscopic $\teff$. 
(b) Difference between the absolute $I$ magnitudes from 
microlensing techniques and from spectroscopy versus stellar mass 
(derived from spectroscopy). Error bars represent the uncertainties in 
the spectroscopic values. The microlensing values used in the plots are 
based on the assumption that the bulge red clump has $(V-I)_0 = 1.06$ 
and $M_I=-0.12$.
Stars with $\teff>5500$\,K are marked by filled circles, otherwise
by open circles.
                    }
         \label{fig:delta_vi}
   \end{figure}
%-----------------------------------------------------------------------

%-$-$$-$$-$$-$$-$$-$$-$$-$$-$$-$$-$$-$$-$$-$$-$$-$$-$$-$$-$$-$$-$$-$$-$$-$$-$$-$$-$$-$$-$$-$$-$$-$$-$$-$$-$$-$$-$
\begin{table}[h]
\centering
\caption{
Measured equivalent widths and calculated elemental abundances for all 58 microlensed bulge dwarfs.$^\dag$
\label{tab:ews}
}
\tiny
\begin{tabular}{cccccccc}
\hline\hline
\noalign{\smallskip}
Element                           &  
$\lambda$                         &
$\chi_{\rm l}$                    &
\multicolumn{2}{c}{star 1}      &
 $\cdots$                         &
\multicolumn{2}{c}{star 12}       \\
\noalign{\smallskip}
                                  &
[{\AA}]                           &
[eV]                              &
$W_{\rm \lambda}$                 &
$\epsilon (X)$                    &
  $\cdots$                        &
$W_{\rm \lambda}$                 &
$\epsilon (X)$                    \\
\noalign{\smallskip}
\hline
\noalign{\smallskip}
\vdots &
\vdots &
\vdots &
\vdots &
\vdots &
$\cdots$ &
\vdots &
\vdots \\
\noalign{\smallskip}
\hline
\end{tabular}
\flushleft
{\tiny
$^{\dagger}$
For each line we give the $\log gf$ value, lower excitation energy ($\chi_{\rm l}$), equivalent width ($W_{\rm \lambda}$),
absolute abundance ($\log \epsilon (X)$).
The table is only available in electronic form at the CDS via anonymous ftp to 
{\tt cdsarc.u-strasbg.fr (130.79.128.5)} or via 
{\tt http://cdsweb.u-strasbg.fr/cgi-bin/qcat?J/A+A/XXX/AXX}.
}
\end{table}

%   blg      blgfe1     blgnafe     blgmgfe     blgalfe     blgsife     blgcafe     blgtife     blgcrfe     blgnife     blgznfe      blgyfe     blgbafe
%
%---------------------------------------------------------------------
\begin{table}[h]
\centering
\caption{
Elemental abundance ratios, errors in the abundance ratios, and number of lines used,  for all 58 microlensed dwarf stars$^\dag$.
\label{tab:abundances}
}
\setlength{\tabcolsep}{1.5mm}
\tiny
\begin{tabular}{ccccccc}
\hline\hline
\noalign{\smallskip}
Object              &  \ldots &  [Mg/Fe] &  $\rm \epsilon [Mg/Fe] $  & N(Mg) & $\rm \sigma (Mg)$ & \ldots \\
\noalign{\smallskip}
\hline
\noalign{\smallskip}
\vdots              &  \ldots & \vdots   & \vdots                & \vdots & \vdots    & \ldots  \\
\noalign{\smallskip}
\hline
\end{tabular}
\flushleft
{\tiny
$^\dagger$ 
For each star we give abundance ratios normalised to the Sun ([$X$/Fe]), number of lines
used for each element ($N(X)$), line-to-line scatter for each element ($\sigma(X)$),
and the calculated uncertainty for each abundance ratio ($\epsilon [X/X]$).
The table is available in electronic form at the CDS via anonymous ftp to
{\tt cdsarc.u-strasbg.fr (130.79.128.5)} or via
{\tt http://cdsweb.u-strasbg.fr/cgi-bin/qcat?J/A+A/XXX/AXX}.
}
\end{table}
%---------------------------------------------------------------------

%=======================================================================
\subsection{$(V-I)$ colours from spectroscopy and microlensing techniques}
\label{sec:mulens}

The $(V-I)_0$ colour and the $M_I$ magnitude can be estimated from 
microlensing techniques \citep{yoo2004} assuming that the reddening
towards the microlensed source is the same as towards the red clump in 
the same field, that $(V-I)_0$ and $M_I$ of the bulge red clump is 
known, and (for $M_I$) that the distance to the source and the red 
clump is the same. 
The de-reddened magnitude and colour of the source can then be 
derived from the offsets between the microlensing source and the red 
clump in the instrumental colour-magnitude diagram. 
The microlensing values for $M_I$ and $(V-I)$ given in 
Table~\ref{tab:ages} are based on the 
assumption that the bulge red clump has $(V-I)_0 = 1.06$
(as determined in \citealt{bensby2011}) and $M_I=-0.12$ \citep{nataf2012}.
For 55 of the 58 events, Fig.~\ref{fig:delta_vi} shows comparisons 
between the spectroscopic $M_{\rm I}$ magnitudes and $(V-I)$ colours 
(see Sect.~\ref{sec:analysis}) to those determined from microlensing 
techniques. Two events from 
\citealt{bensby2011} and one of the new events do not have any estimates of 
$(V-I)_0$ from microlensing techniques (see Table~\ref{tab:ages}).
The first thing to notice is that there appears to be a slope present 
between the colour difference ($\Delta (V-I)$) and the
spectroscopic $\teff$ 
(see red line in Fig.~\ref{fig:delta_vi}). The question is whether 
this trend is due to uncertainties in the spectroscopic analysis or due 
to the microlensing assumptions?

\begin{figure}
\resizebox{\hsize}{!}{
\includegraphics[bb=18 144 592 718,clip]{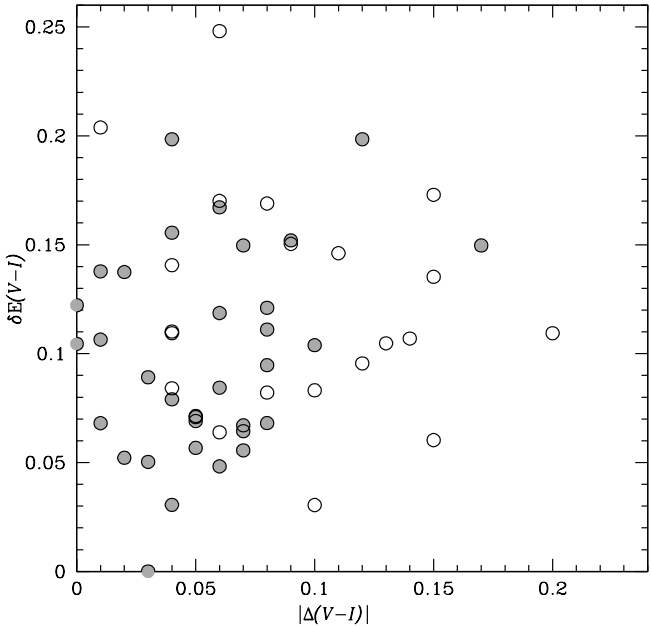}}
\caption{Differential reddening versus the absolute value of the
difference between microlensing colours and spectroscopic colours.
The differential reddening values for the individual 
sight lines are taken from \cite{nataf2013}.
Stars with $\teff>5500$\,K are marked by filled circles, otherwise
by open circles.
                    }
         \label{fig:n12red}
   \end{figure}

The bulge is known to have patchy and irregular reddening. 
\cite{nataf2013} measured reddening and differential reddening for 
more then 9\,000 sight lines towards the bulge.
Figure~\ref{fig:n12red} shows the differential 
reddening along the sight lines to the individual microlensing 
events in this study versus the absolute value of the $\Delta (V-I)$. 
While there is no 
one-to-one relationship between these two parameters it is evident that 
there are very few events with large colour discrepancies that at the 
same time have low differential reddening values (i.e., in the lower 
right corner of the Fig.~\ref{fig:n12red}). Note that the 
\cite{nataf2012} differential-reddening
dispersions are systemically larger than the offsets between
the colours determined from microlensing and spectroscopy.
For example, when Nataf estimates
$\delta E(V-I) \approx 0.15$, the microlensing offset averages about
0.06.  The reason for this apparent discrepancy is that to
estimate the microlensing $(V-I)_0$, we consider the minimum
possible area around the source needed to determine the colour of the
bulge clump, whereas \cite{nataf2012} use a robust algorithm
that chooses a larger area, in order to minimise failures
among their 9\,000 automated determinations.

In any case, Fig.~\ref{fig:n12red} gives a hint that
differential reddening could play a r\^ole
in explaining the discrepancies between spectroscopic and microlensing 
$(V-I)$ colours. However, differential reddening cannot
cause the (2$\sigma$ statistically significant) slope in the
$\Delta (V-I)-\teff$ diagram.
The only assumptions being made are that the extinction
is the same (on average) between the red clump and the source, and that
the bulge clump $(V-I)$ colour is known.  While either of these assumptions
might be violated, it is hard to seen why, e.g., the bulge red clump should
be redder in fields with reddish subgiants than turnoff stars, or why
there should be systematically more differential reddening (relative
to the mean of that  field) for fields containing reddish subgiants
compared to turnoff stars. 

%-----------------------------------------------------------------------
%-----------------------------------------------------------------------

%-----------------------------------------------------------------------
\begin{figure}
\resizebox{\hsize}{!}{
\includegraphics[bb=18 144 490 713,clip]{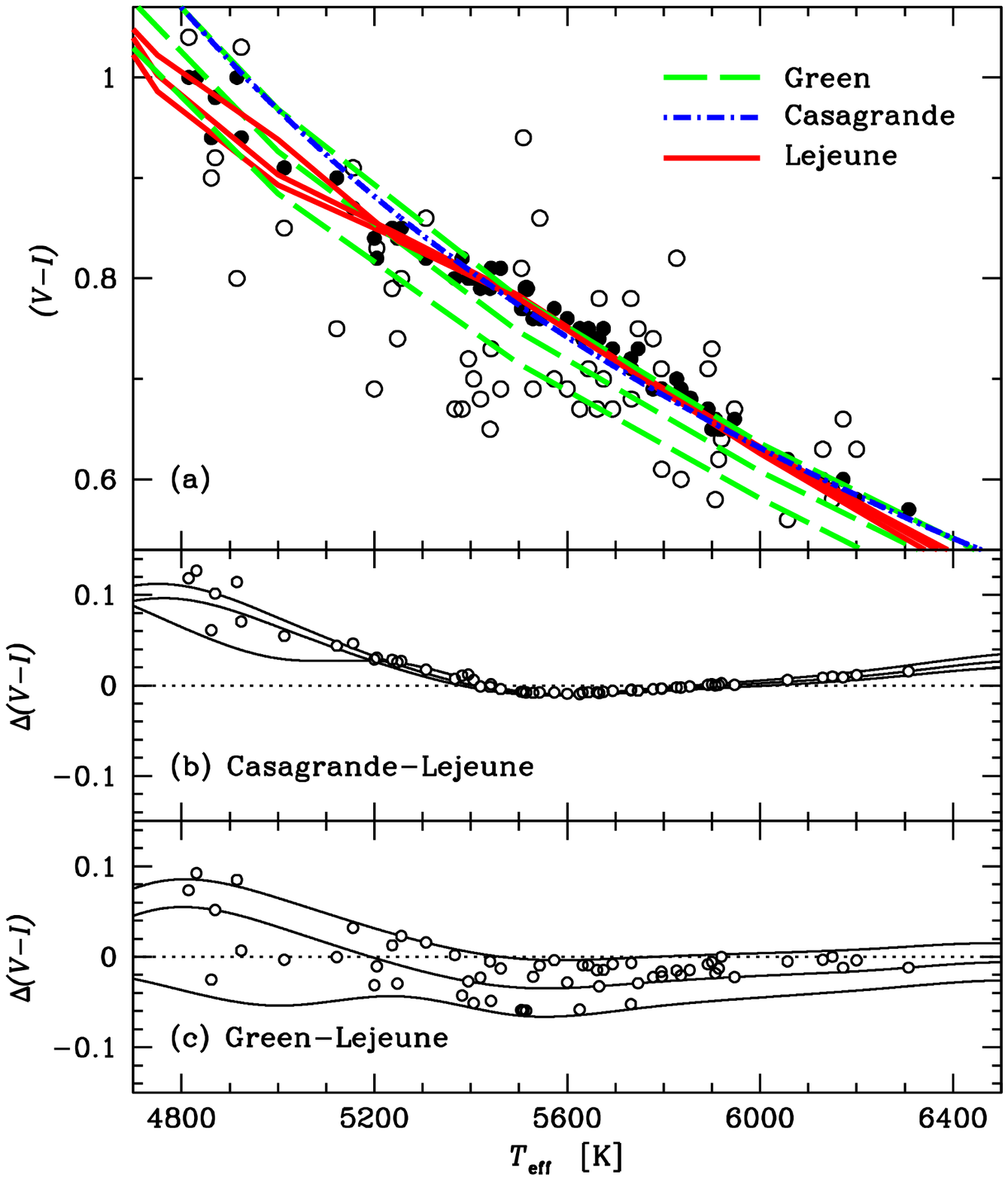}}
\caption{(a) Filled black circles mark the spectroscopic values, open
circles microlensing values. Dash-dotted blue line show the 
\cite{casagrande2010} $\teff$-colour transformation, dashed green 
lines the \cite{green1987} transformation (for three different $\log g$), 
and the solid red lines the \cite{lejeune1998} transformation 
(for three different $\log g$). (b) Differences between 
\cite{casagrande2010} and \cite{lejeune1998}. (c) Differences between 
\cite{green1987} and \cite{lejeune1998}. The markers in (b) and (c) 
show the differences that should
be applied to the microlensed dwarfs if the \cite{casagrande2010} 
and \cite{green1987} trandformations had been applied instead of the 
one from \cite{lejeune1998}.
                    }
         \label{fig:vi_comp}
   \end{figure}
%-----------------------------------------------------------------------

So, that turns the focus to the spectroscopic colours.
As explained by \cite{yi2001}, the Y$^2$ isochrones come with two different
$\teff$-to-colour transformations employed: either the one by
\cite{green1987} or the one by \cite{lejeune1998}. As also pointed out
by \cite{yi2001}, while the two sets of colour transformations 
agree fairly well for main sequence dwarf stars, there are large differences
for giant stars. The version of the Y$^2$ isochrones that we used
for the microlensed dwarf sample utilises the colour transformation 
from \cite{lejeune1998}. Figure~\ref{fig:vi_comp}a shows the spectroscopic
$(V-I)$ (black filled circles) and the microlensing ones (open circles).
Overplotted are three sets of $\teff$-colour transformations
for $\rm [Fe/H]=0$ and three surface gravities $\log g=3.5,\,4.0,\,4.5$:
\cite{lejeune1998}, \cite{green1987}, and \cite{casagrande2010}, which have 
no dependence on $\log g$. 
Figures~\ref{fig:vi_comp}b and c then show the differences
between the Casagrande-Lejeune and Green-Lejeune transformations,
respectively. What is interesting here is that the Lejeune and Casagrande
transformations are nearly identical for $\teff>5500$\,K,
and that there is an offset between Green and Lejeune that varies with 
$\log g$ but which is nearly constant for $\teff>5500$\,K.
However, for $\teff\lesssim$5500 the three sets of transformations
differ, and increasingly more for lower $\teff$. The points in 
Figs.~\ref{fig:vi_comp}b and c show the corrections that would have to be applied
to the microlensed dwarf stars if we were to change from the
Lejeune transformation to any of the other two. Generally, stars with
$\teff\gtrsim5500$ would be more or less untouched while the cooler
stars would get {\it higher} $(V-I)$ colours. This is a frustrating finding
as it means that the slope seen in Fig.~\ref{fig:delta_vi} would become
steeper with the Casagrande and Green $\teff$-colour transformations!

However, from Fig.~\ref{fig:vi_comp} it is apparent that the
agreement between the three calibrations is good for hotter stars.
Even though the \cite{green1987} transformations will show offsets
depending on $\log g$, it will be a constannt offset and no slope will 
be introduced. As further investigations
into the field of $\teff$-colour transformations is beyond the scope
of the current paper, we will, for now, restrict ourselves
to stars with
$\teff\gtrsim 5500$\,K when comparing spectroscopic and
microlensing colours. These stars are marked by filled circles in 
Fig.~\ref{fig:delta_vi}. The difference in the $(V-I)$ colour
is now $0.00\pm0.07$\,mag (while if we were to include all stars, 
i.e., also those with $\teff<5500$\,K, the difference would be 
$+0.03\pm0.08$\,mag).
Our conclusion is that the colour of the bulge red clump should 
remain at $(V-I)_0=1.06$.

%-----------------------------------------------------------------------
\begin{figure*}
\resizebox{\hsize}{!}{
\includegraphics[bb=18 200 450 718,clip]{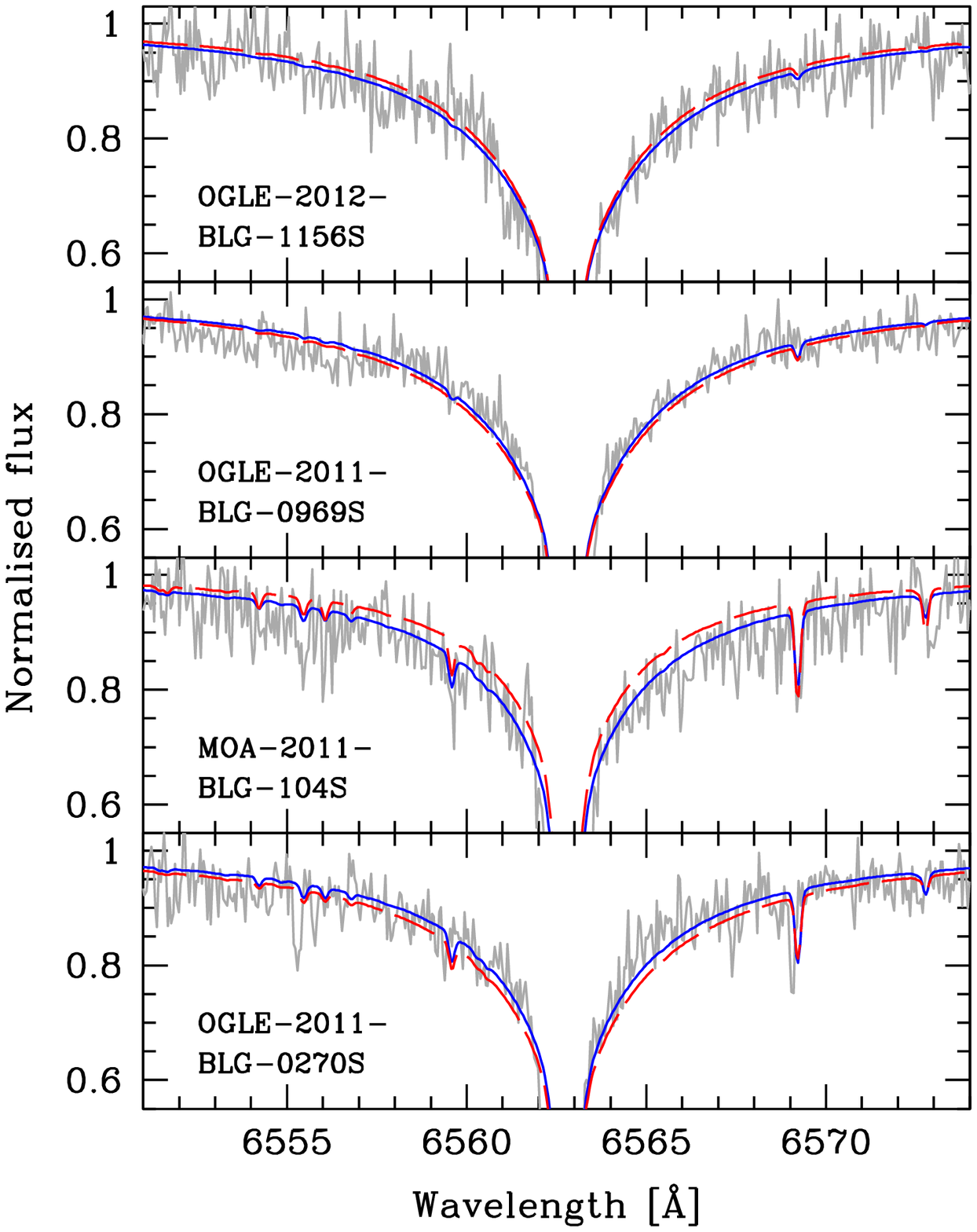}
\includegraphics[bb=75 200 450 718,clip]{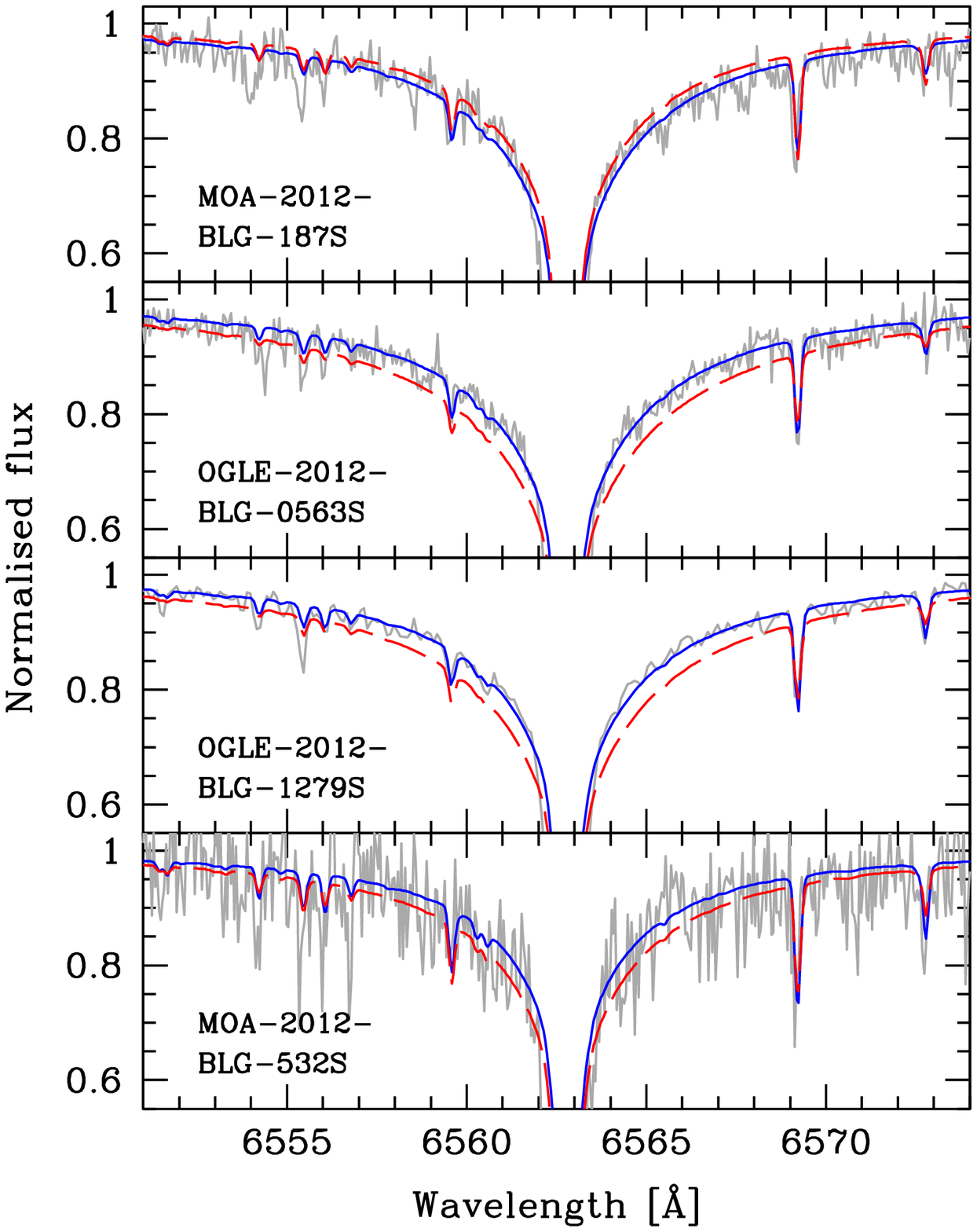}
\includegraphics[bb=75 200 470 718,clip]{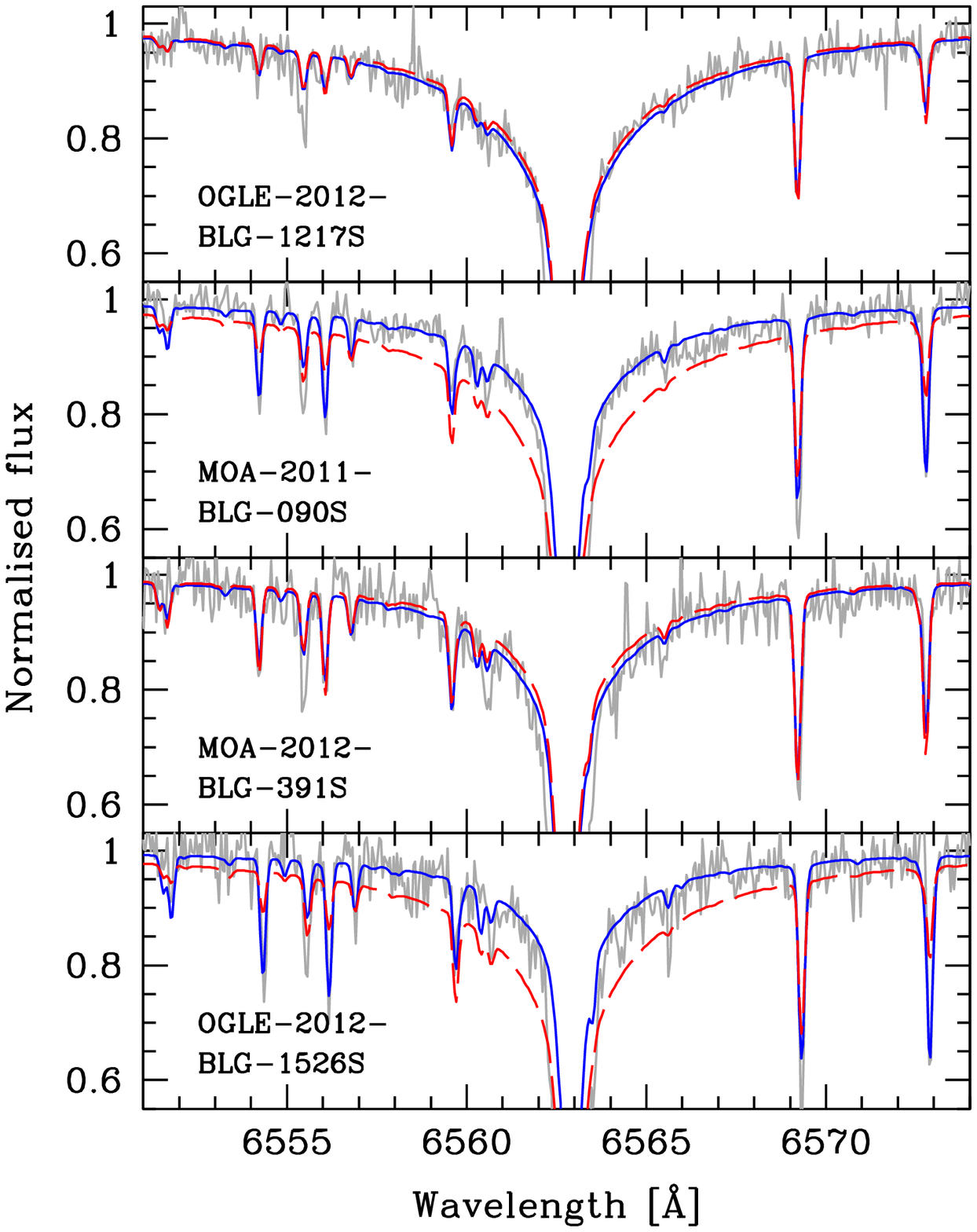}}
\resizebox{\hsize}{!}{
\includegraphics[bb=18 200 450 710,clip]{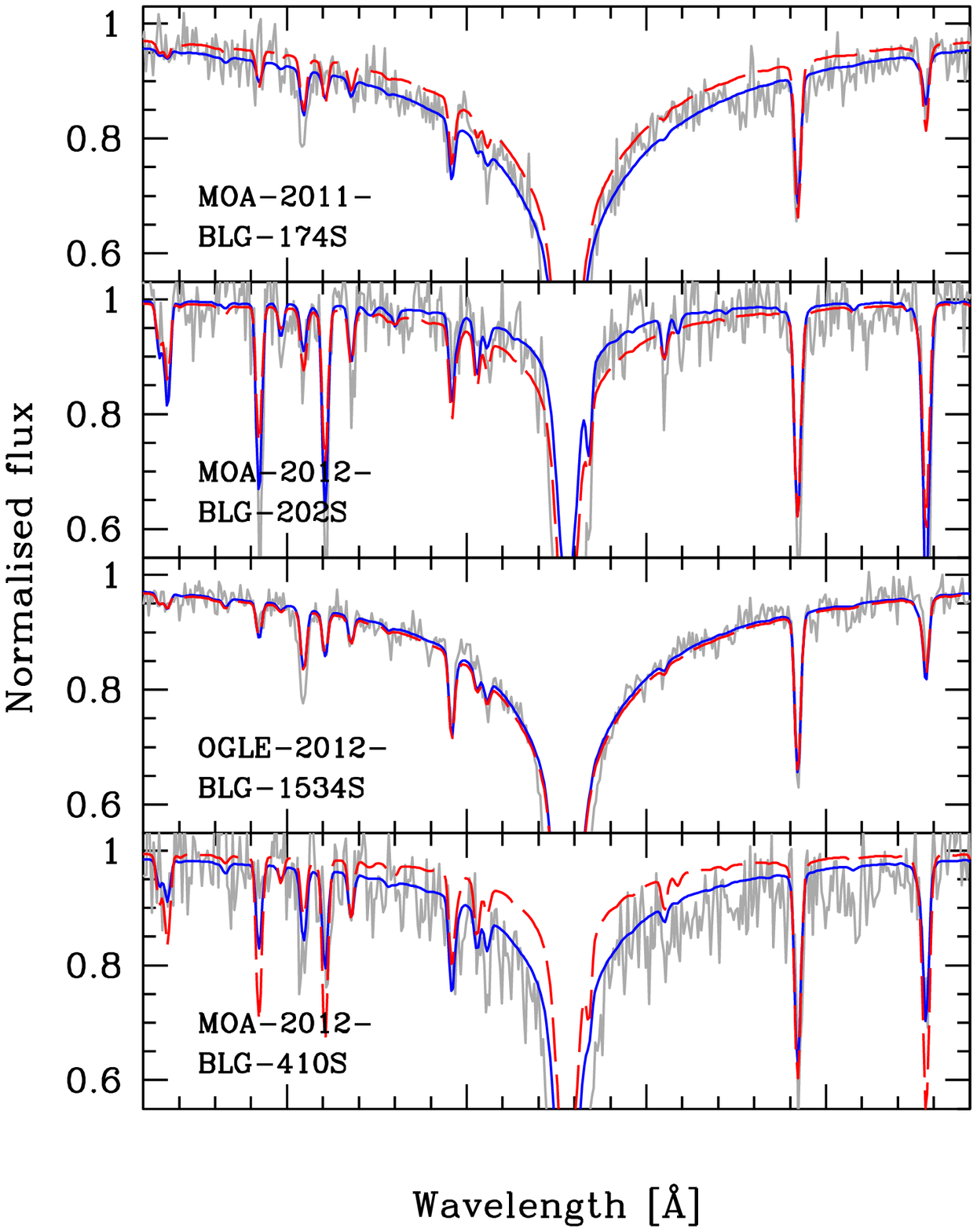}
\includegraphics[bb=75 200 450 710,clip]{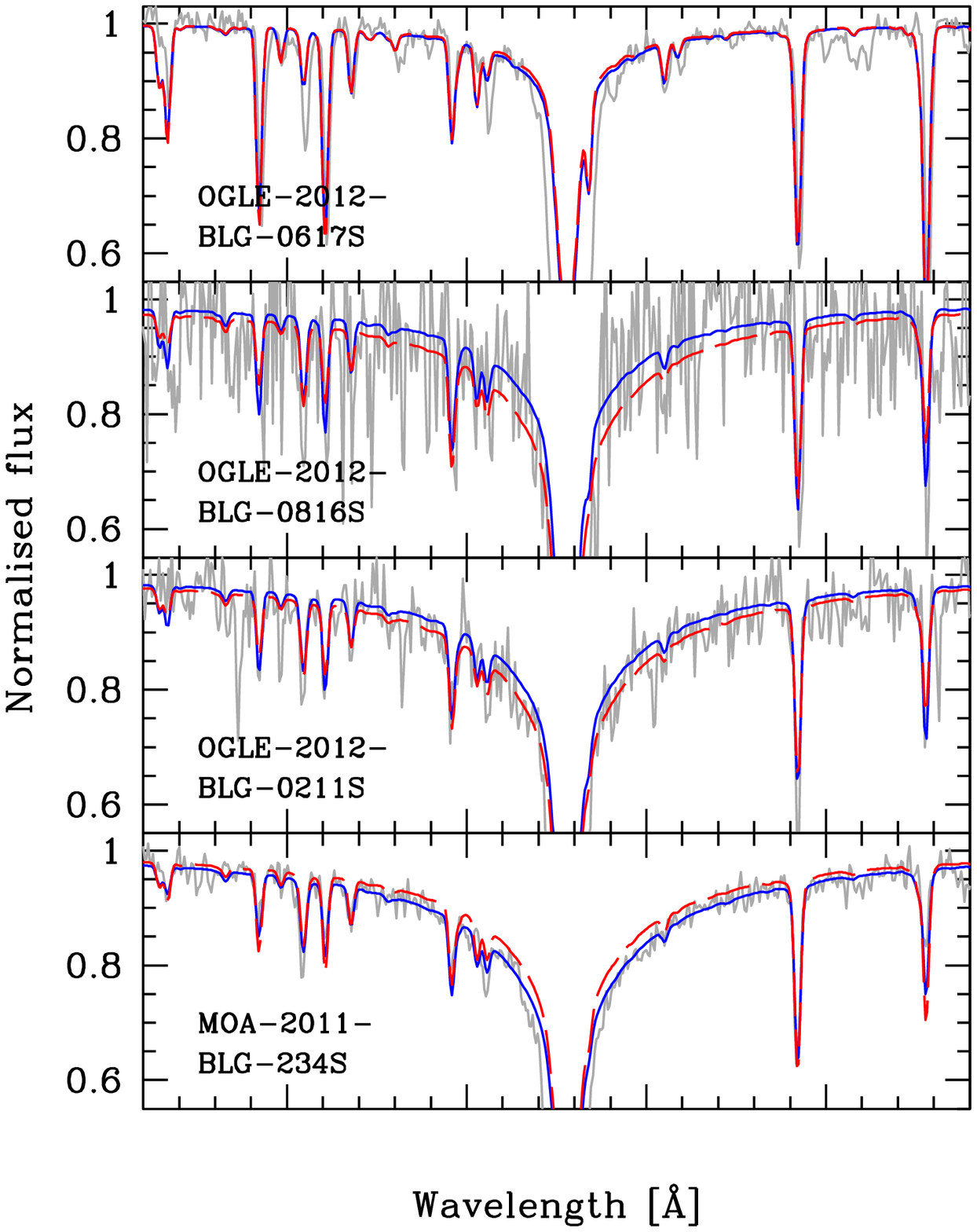}
\includegraphics[bb=75 200 470 710,clip]{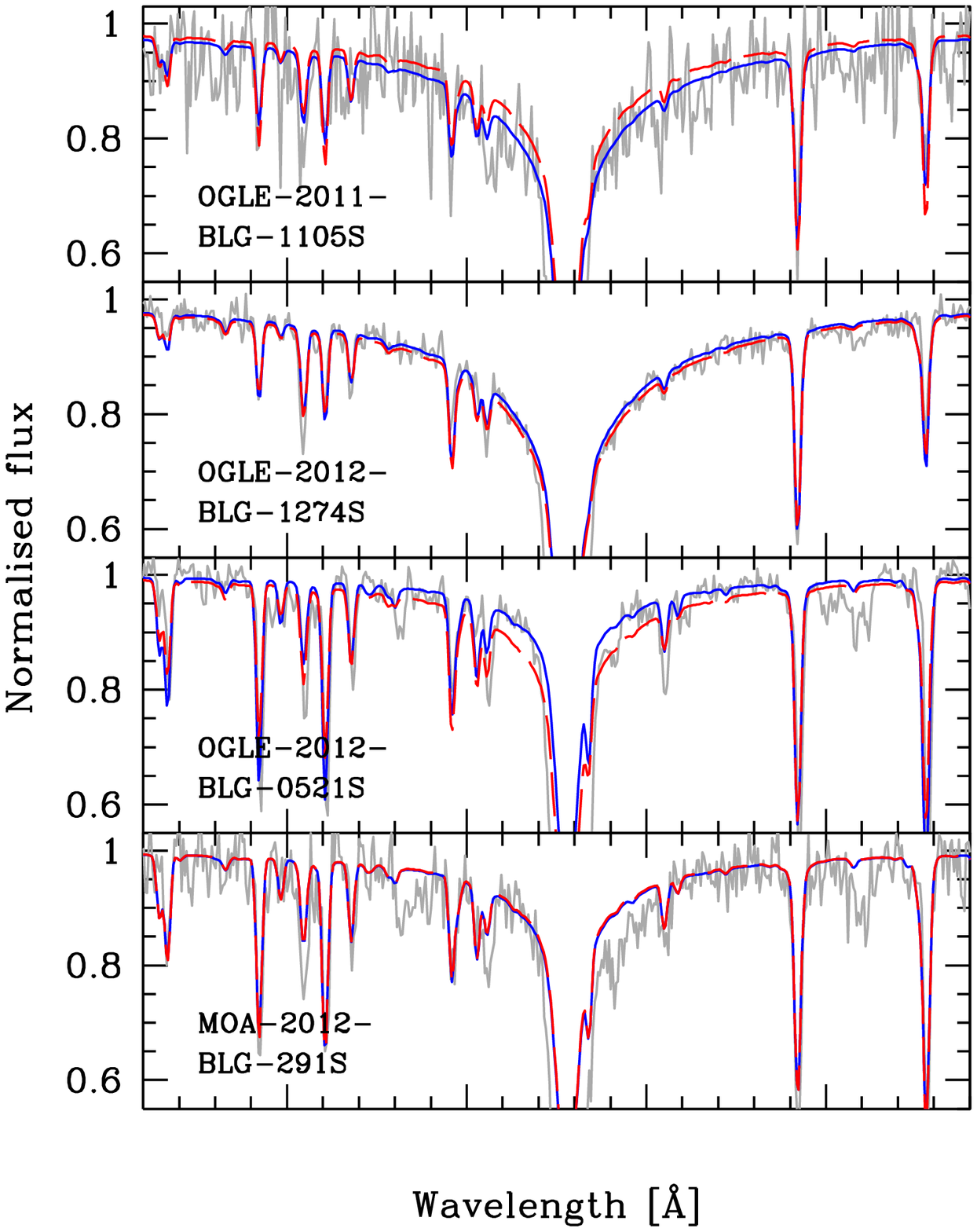}}
\resizebox{\hsize}{!}{
\includegraphics[bb=18 260 450 710,clip]{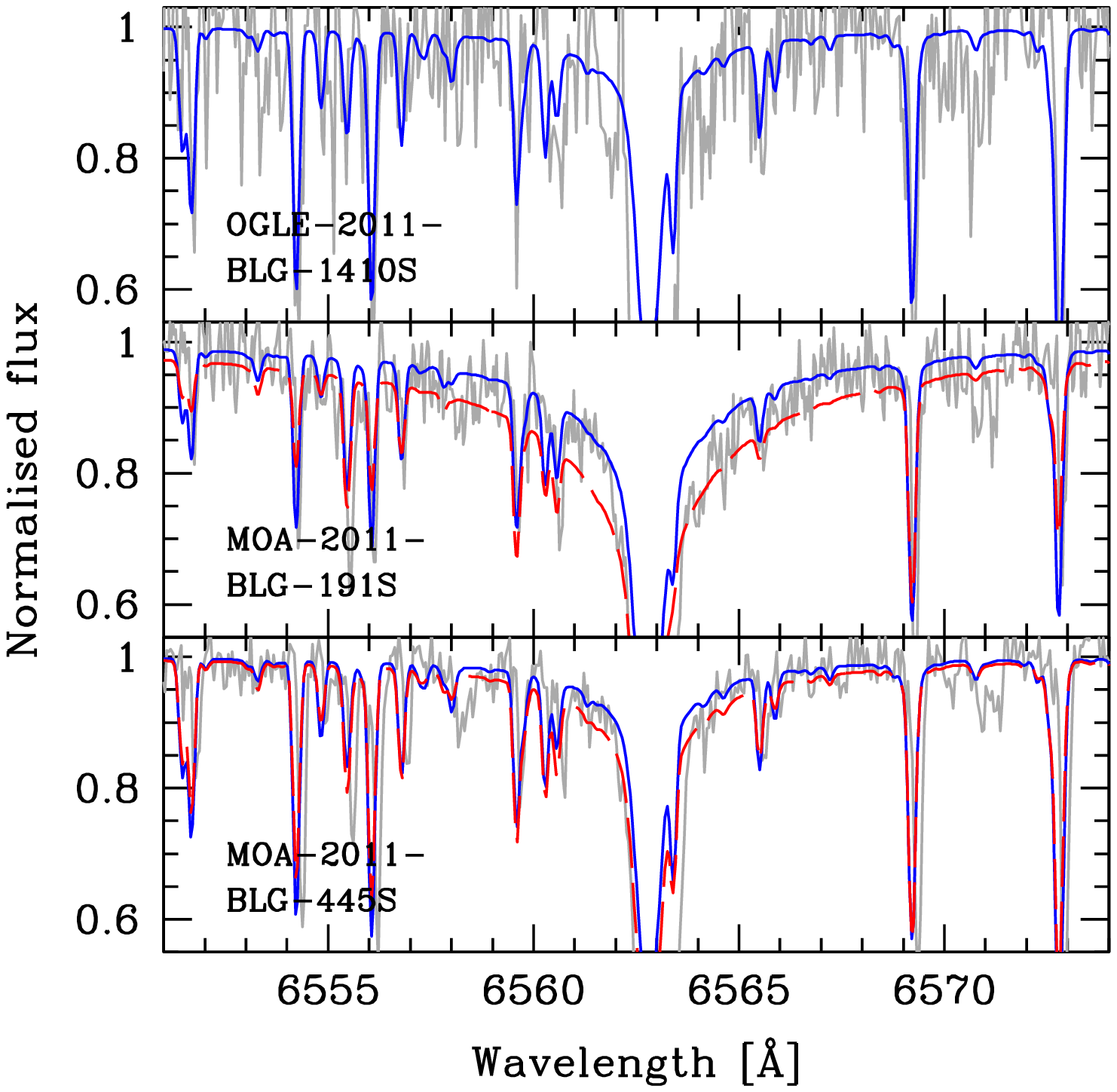}
\includegraphics[bb=75 260 450 710,clip]{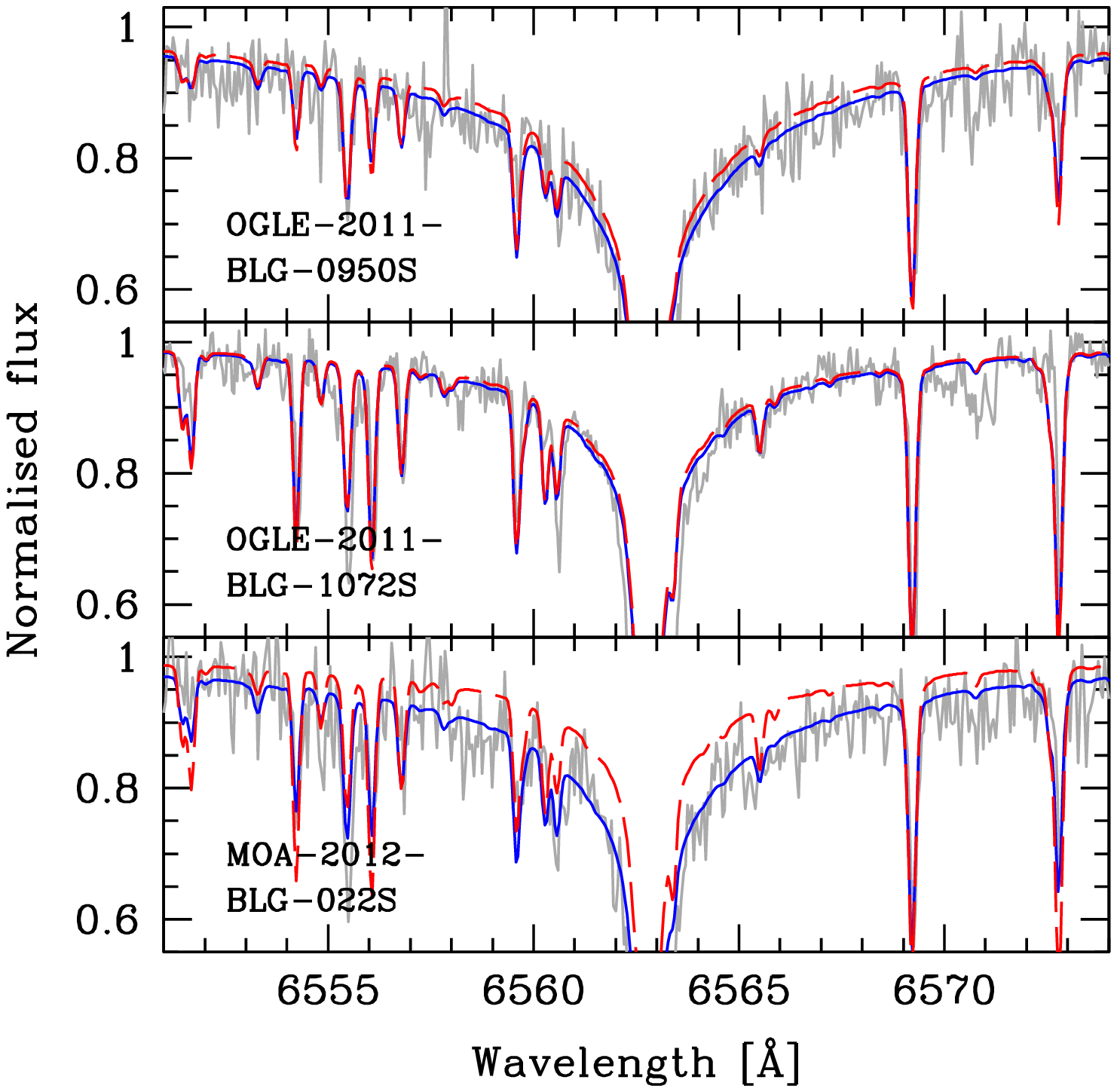}
\includegraphics[bb=75 260 470 710,clip]{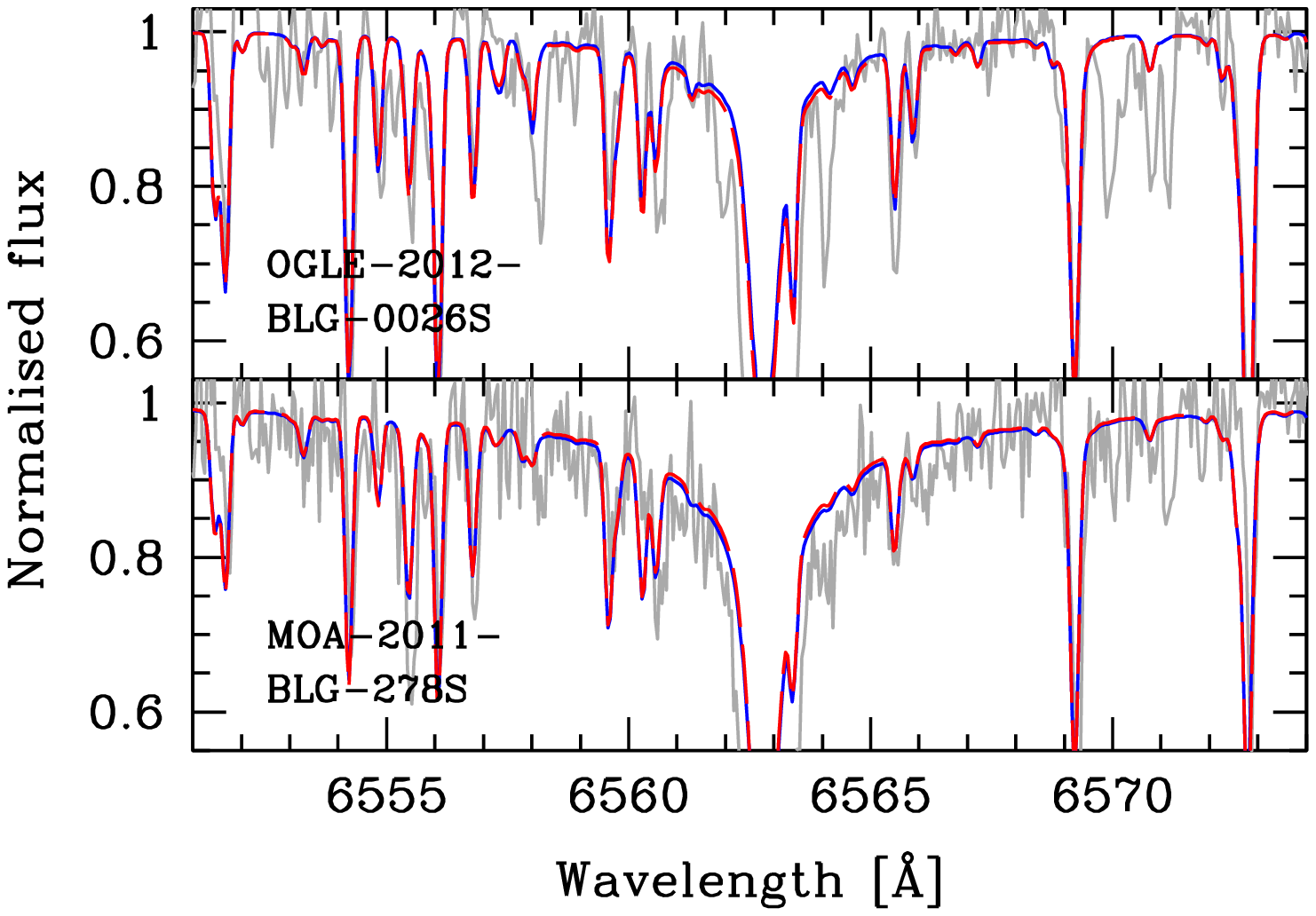}}
\caption{Illustration to see how well the spectroscopic temperatures 
(full blue lines) and temperatures from microlensing techniques (dashed 
red lines) reproduce the observed wing profiles of the H$\alpha$ line
at 6563\,{\AA} for the 32 new microlensed bulge dwarfs. Note that no 
estimate of the temperature from microlensing techniques is available 
for OGLE-2011-BLG-1410S. 
The stars are sorted by metallicity.                    }
         \label{fig:balmer}
   \end{figure*}
%-----------------------------------------------------------------------

%=======================================================================
\subsection{Comparisons to the Balmer H$\alpha$ line}
\label{sec:balmer}

To further check the effective temperatures of the microlensed dwarf stars
we have calculated synthetic spectra with the SME (Spectroscopy Made Easy, 
v.~2011-12-05, \citealt{valenti1996}) for the 32 new microlensing events.
Figure~\ref{fig:balmer} shows comparisons of the synthetic H$\alpha$ line 
profiles to the observed H$\alpha$ line profiles the spectroscopic 
temperatures (blue lines). There is generally very good agreement between 
synthetic and observed spectra for the spectroscopic effective temperatures.
Hence, we believe that the effective temperatures we have determined should be  
good. 

For the 55 stars that have $(V-I)_0$ colours from microlensing techniques
we calculate effective temperatures using the \cite{casagrande2010} 
$\teff$-colour calibration. These temperatures are on average 105\,K
higher than the spectroscopic ones. If we restrict ourselves to
stars with (spectroscopic) $\teff>5500$ the spectroscopic temperatures
are on average higher, but only by 10\,K. The dispersion in both cases 
is around 230\,K.

Figure~\ref{fig:balmer} also shows synthetic spectra based on temperatures
from the microlensing $(V-I)_0$ colours (red dashed lines). 
It is clear that the microlensing temperatures
produce spectra that do not match for a few cases 
(e.g., MOA-2011-BLG-090S and MOA-2012-BLG-410S).
The H$\alpha$ wing profile for the star with the largest colour discrepancy 
between spectroscopic and microlensing temperature, MOA-2009-BLG-259S,
is shown in Fig.~3 of \cite{bensby2011}. From that figure it is clear that 
the spectroscopic temperature is the better match. However, this star
has $\teff=4915$\,K and $\log g = 3.3$, which is in a region where
the $\teff$-colour calibrations appear to be very uncertain 
(see Fig.~\ref{fig:vi_comp}), meaning that the colour-$\teff$ transformation
could be substantially affected.

%=======================================================================
\subsection{A few special cases}
\label{sec:ogle2011blg0969}

\begin{figure}
\resizebox{\hsize}{!}{
\includegraphics[bb=18 144 592 718,clip]{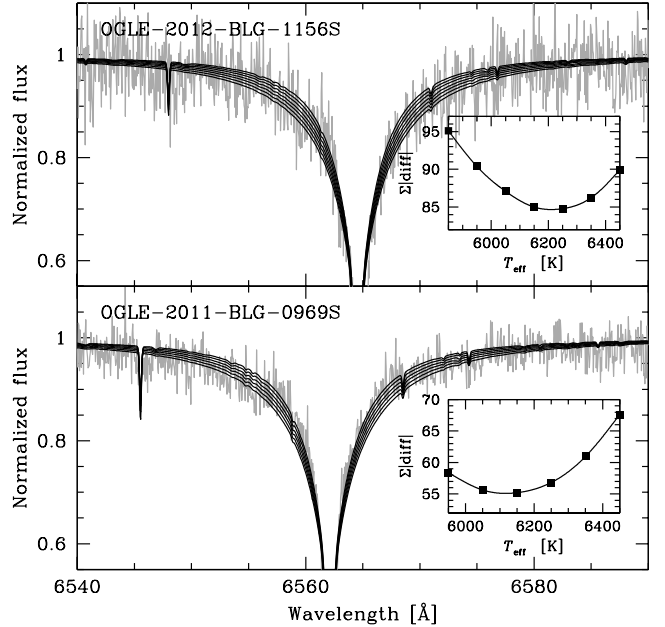}}
\caption{Fitting of the  H$\alpha$ wing profiles to find the
effective temperatures for OGLE-2011-BLG-0969S and OGLE-2012-BLG-1156S.
The insets show the summed absolute values of the residuals that was 
minimised to find the best fitting temperature.
               }
         \label{fig:hbeta}
   \end{figure}

The combination of low metallicity and low $S/N$ of the 
observed spectra of OGLE-2011-BLG-0969S and OGLE-2012-BLG-1156S meant
that very few equivalent widths could be measured for 
these stars. Hence, the effective temperatures and the microturbulence
parameters for these two stars could not be constrained in the standard 
way, through excitation and ionisation balance, as described above. 
As an alternative, the effective temperatures were constrained
by fitting the wing profiles of the Balmer H$\alpha$ line at 6563\,{\AA} 
for these two stars (as described in Sect.~\ref{sec:mulens}). 
This gives an effective temperature of around
6150\,K for OGLE-2011-BLG-0969S and around 6200\,K for OGLE-2012-BLG-1156S 
(see Fig.~\ref{fig:hbeta}). This in fair agreement with the
temperatures that can be estimated from microlensing techniques 
for these two stars, 6283\,K and 6012\,K, respectively 
(see Sect.~\ref{sec:mulens}).
The surface gravities were set based on microlensing techniques to 
$\log g=4.14$ for OGLE-2011-BLG-0969S and 
$\log g=4.25$ for OGLE-2012-BLG-1156S. 
First, $(V-I)_0$ and $I_0$ are determined from the offset to 
the bulge red clump. $(V-I)_0$ is then transferred to $(V-K)_0$ 
using the calibration by \cite{bessel1988}.
Combining $(V-K)_0$ with $V_0 = (V-I)_0 + I_0$ give 
$\theta_{star}$ using the calibration by
\cite{kervella2004}. We then assume that the microlensing source is at same 
distance as the red clump given by  \cite{nataf2013}. The radius of the 
star is then $r_{star} = D_{source} \cdot \theta_{star}$. Finally, 
we assume that
the stellar mass is $M_{star} = 1\,M_\odot$, so the gravity is given
by $g = GM_{star}/r_{star}^2$.
We estimate the uncertainty in the distance to be 15\% (0.3 mag),
and the mass uncertainty to be about 10\%.  
Hence, the dominant error is in the distance (because
it enters as the square), imply that the error in $\log g$ using this method
is about 0.1\,dex.
The microturbulence parameter was set to $\xi_{\rm t}=1.5\,\kms$ for 
both stars using the empirical calibration by 
M.~Bergemann~(private communication)\footnote{The empirical microturbulence relation by Maria Bergemann was derived for the Gaia-ESO 
public spectroscopic survey \citep{gilmore2012} and will be published in one of her
first science papers based on Gaia-ESO data.}.

%-----------------------------------------------------------------------
%-----------------------------------------------------------------------

%---------------------------------------------------------------------
\begin{table*}
\centering
\caption{
Stellar parameters and radial velocities for all 58 stars (sorted by metallicity).$^{\dagger}$
\label{tab:parameters}
}
\setlength{\tabcolsep}{4.9mm}
\tiny
\begin{tabular}{rllllcr}
\hline\hline
\noalign{\smallskip}
  \multicolumn{1}{c}{Object}                     &
  \multicolumn{1}{c}{$T_{\rm eff}$}              &
  \multicolumn{1}{c}{$\log g$}                   &
  \multicolumn{1}{c}{$\xi_{\rm t}$}              &
  \multicolumn{1}{c}{[Fe/H]}                     &
  $N_{\ion{Fe}{i},\,\ion{Fe}{ii}}$      &
  \multicolumn{1}{c}{$v_{\rm r}$}                \\
                                                 &
  \multicolumn{1}{c}{$\rm [K]$}                  &
  \multicolumn{1}{c}{$\rm [cgs]$}                &
  \multicolumn{1}{c}{$\rm [\kms]$}               &
                                                 &
                                                 &
  \multicolumn{1}{c}{[km/s]}                    \\
\noalign{\smallskip}
\hline
\noalign{\smallskip}
  OGLE-2012-BLG-1156S            &  $6200  \pm  200$  &  $4.25  \pm  0.30$  &  $1.50  \pm  0.30$  &  $-1.89  \pm  0.25$  &  $  3,\, 0$  &  $  72.6$  \\
  OGLE-2011-BLG-0969S            &  $6150  \pm  200$  &  $4.14  \pm  0.30$  &  $1.50  \pm  0.30$  &  $-1.57  \pm  0.25$  &  $ 11,\, 2$  &  $ -53.0$  \\
    MOA-2010-BLG-285S\phantom{0} &  $6057  \pm  134$  &  $4.20  \pm  0.19$  &  $1.88  \pm  0.42$  &  $-1.21  \pm  0.10$  &  $ 53,\, 9$  &  $  46.0$  \\
    MOA-2010-BLG-078S\phantom{0} &  $5205  \pm  189$  &  $3.60  \pm  0.26$  &  $1.31  \pm  0.32$  &  $-1.00  \pm  0.34$  &  $ 53,\, 4$  &  $  52.3$  \\
    MOA-2011-BLG-104S\phantom{0} &  $5900  \pm  200$  &  $4.15  \pm  0.30$  &  $1.30  \pm  0.30$  &  $-0.85  \pm  0.25$  &  $ 36,\, 6$  &  $ 197.3$  \\
  OGLE-2012-BLG-0270S            &  $5914  \pm  145$  &  $4.30  \pm  0.22$  &  $1.33  \pm  0.22$  &  $-0.84  \pm  0.13$  &  $ 45,\, 6$  &  $-128.7$  \\
    MOA-2009-BLG-493S\phantom{0} &  $5420  \pm  119$  &  $4.40  \pm  0.22$  &  $0.90  \pm  0.32$  &  $-0.74  \pm  0.15$  &  $ 80,\, 5$  &  $ -14.5$  \\
    MOA-2012-BLG-187S\phantom{0} &  $5895  \pm   94$  &  $4.20  \pm  0.14$  &  $1.60  \pm  0.18$  &  $-0.74  \pm  0.08$  &  $ 72,\, 8$  &  $ -40.3$  \\
   OGLE-2009-BLG-076S\phantom{0} &  $5854  \pm  108$  &  $4.30  \pm  0.15$  &  $1.63  \pm  0.22$  &  $-0.72  \pm  0.11$  &  $ 57,\, 7$  &  $ 128.7$  \\
    MOA-2009-BLG-133S\phantom{0} &  $5529  \pm   73$  &  $4.30  \pm  0.14$  &  $1.17  \pm  0.18$  &  $-0.69  \pm  0.07$  &  $ 66,\, 6$  &  $  91.6$  \\
  OGLE-2012-BLG-0563S            &  $5907  \pm   89$  &  $4.40  \pm  0.10$  &  $1.27  \pm  0.14$  &  $-0.66  \pm  0.07$  &  $ 73,\, 8$  &  $ -66.2$  \\
  OGLE-2012-BLG-1279S            &  $5796  \pm   63$  &  $4.40  \pm  0.11$  &  $0.99  \pm  0.12$  &  $-0.62  \pm  0.06$  &  $152,\,18$  &  $ 141.1$  \\
    MOA-2010-BLG-167S\phantom{0} &  $5406  \pm   49$  &  $3.90  \pm  0.09$  &  $1.15  \pm  0.09$  &  $-0.60  \pm  0.05$  &  $109,\,14$  &  $ -79.4$  \\
    MOA-2012-BLG-532S\phantom{0} &  $5626  \pm  207$  &  $3.90  \pm  0.37$  &  $0.88  \pm  0.25$  &  $-0.55  \pm  0.21$  &  $ 42,\, 2$  &  $  27.9$  \\
    MOA-2009-BLG-475S\phantom{0} &  $5836  \pm  189$  &  $4.40  \pm  0.27$  &  $1.35  \pm  0.37$  &  $-0.52  \pm  0.20$  &  $ 53,\, 4$  &  $ 137.8$  \\
   MACH-1999-BLG-022S\phantom{0} &  $5632  \pm  110$  &  $4.10  \pm  0.18$  &  $0.33  \pm  0.41$  &  $-0.49  \pm  0.17$  &  $ 97,\,10$  &  $  37.6$  \\
  OGLE-2012-BLG-1217S            &  $5795  \pm   73$  &  $4.30  \pm  0.13$  &  $1.28  \pm  0.15$  &  $-0.41  \pm  0.07$  &  $ 73,\, 9$  &  $ 124.4$  \\
    MOA-2010-BLG-446S\phantom{0} &  $6308  \pm  111$  &  $4.50  \pm  0.14$  &  $1.71  \pm  0.18$  &  $-0.40  \pm  0.08$  &  $ 66,\, 8$  &  $  56.5$  \\
    MOA-2010-BLG-049S\phantom{0} &  $5694  \pm   61$  &  $4.10  \pm  0.12$  &  $1.02  \pm  0.12$  &  $-0.40  \pm  0.07$  &  $ 96,\,14$  &  $-116.7$  \\
   OGLE-2008-BLG-209S\phantom{0} &  $5248  \pm   77$  &  $3.80  \pm  0.14$  &  $1.08  \pm  0.10$  &  $-0.31  \pm  0.09$  &  $147,\,18$  &  $-173.6$  \\
    MOA-2011-BLG-090S\phantom{0} &  $5367  \pm   49$  &  $4.10  \pm  0.09$  &  $0.87  \pm  0.09$  &  $-0.26  \pm  0.05$  &  $ 88,\, 8$  &  $  48.2$  \\
    MOA-2012-BLG-391S\phantom{0} &  $5505  \pm   76$  &  $3.90  \pm  0.14$  &  $1.12  \pm  0.16$  &  $-0.24  \pm  0.07$  &  $ 67,\, 7$  &  $ -65.0$  \\
  OGLE-2012-BLG-1526S            &  $5200  \pm   62$  &  $3.85  \pm  0.12$  &  $0.94  \pm  0.12$  &  $-0.24  \pm  0.06$  &  $ 76,\, 6$  &  $ -87.1$  \\
    MOA-2009-BLG-489S\phantom{0} &  $5543  \pm   61$  &  $4.10  \pm  0.11$  &  $0.62  \pm  0.10$  &  $-0.21  \pm  0.07$  &  $106,\,12$  &  $  96.5$  \\
    MOA-2011-BLG-174S\phantom{0} &  $6172  \pm  111$  &  $4.40  \pm  0.16$  &  $1.16  \pm  0.14$  &  $-0.18  \pm  0.09$  &  $ 57,\, 6$  &  $ -24.0$  \\
    MOA-2012-BLG-202S\phantom{0} &  $4862  \pm   93$  &  $3.90  \pm  0.18$  &  $0.90  \pm  0.17$  &  $-0.15  \pm  0.13$  &  $ 60,\, 4$  &  $  41.2$  \\
  OGLE-2012-BLG-1534S            &  $5920  \pm   52$  &  $4.05  \pm  0.08$  &  $1.37  \pm  0.10$  &  $-0.15  \pm  0.04$  &  $ 94,\,12$  &  $ 206.4$  \\
    MOA-2012-BLG-410S\phantom{0} &  $5509  \pm   80$  &  $3.90  \pm  0.13$  &  $1.09  \pm  0.14$  &  $-0.14  \pm  0.07$  &  $ 60,\, 7$  &  $  22.3$  \\
  OGLE-2012-BLG-0617S            &  $4924  \pm   71$  &  $3.70  \pm  0.16$  &  $1.06  \pm  0.11$  &  $-0.14  \pm  0.09$  &  $ 85,\, 8$  &  $ -68.0$  \\ 
  OGLE-2012-BLG-0816S            &  $5395  \pm  131$  &  $3.50  \pm  0.20$  &  $0.65  \pm  0.20$  &  $-0.10  \pm  0.14$  &  $ 37,\, 5$  &  $ -74.0$  \\
  OGLE-2012-BLG-0211S            &  $5573  \pm   75$  &  $4.00  \pm  0.12$  &  $0.67  \pm  0.13$  &  $-0.06  \pm  0.08$  &  $ 79,\,10$  &  $ -17.7$  \\
    MOA-2011-BLG-234S\phantom{0} &  $5778  \pm   88$  &  $4.40  \pm  0.13$  &  $0.93  \pm  0.08$  &  $-0.02  \pm  0.08$  &  $111,\,15$  &  $  44.0$  \\
  OGLE-2011-BLG-1105S            &  $5666  \pm  113$  &  $4.50  \pm  0.18$  &  $0.75  \pm  0.22$  &  $+0.00  \pm  0.17$  &  $ 37,\, 5$  &  $ 222.7$  \\
    MOA-2010-BLG-523S\phantom{0} &  $5122  \pm   79$  &  $3.60  \pm  0.15$  &  $1.68  \pm  0.20$  &  $+0.06  \pm  0.14$  &  $ 57,\,10$  &  $  97.3$  \\
  OGLE-2012-BLG-1274S            &  $5733  \pm   51$  &  $4.10  \pm  0.07$  &  $1.23  \pm  0.07$  &  $+0.07  \pm  0.04$  &  $ 94,\,12$  &  $ -25.0$  \\
  OGLE-2012-BLG-0521S            &  $5013  \pm   84$  &  $3.70  \pm  0.14$  &  $1.02  \pm  0.11$  &  $+0.09  \pm  0.15$  &  $ 75,\,10$  &  $ -68.8$  \\
    MOA-2009-BLG-174S\phantom{0} &  $5600  \pm   80$  &  $4.40  \pm  0.10$  &  $1.27  \pm  0.10$  &  $+0.11  \pm  0.08$  &  $117,\,15$  &  $ -21.4$  \\
    MOA-2011-BLG-034S\phantom{0} &  $5440  \pm   91$  &  $4.10  \pm  0.12$  &  $0.88  \pm  0.11$  &  $+0.11  \pm  0.15$  &  $ 77,\,10$  &  $ 127.0$  \\
    MOA-2009-BLG-456S\phantom{0} &  $5662  \pm   89$  &  $4.20  \pm  0.11$  &  $0.86  \pm  0.10$  &  $+0.13  \pm  0.10$  &  $ 88,\,12$  &  $-164.6$  \\
    MOA-2012-BLG-291S\phantom{0} &  $5156  \pm  107$  &  $4.10  \pm  0.15$  &  $0.79  \pm  0.13$  &  $+0.16  \pm  0.24$  &  $ 64,\, 9$  &  $  60.3$  \\
  OGLE-2011-BLG-1410S            &  $4831  \pm  108$  &  $3.30  \pm  0.24$  &  $0.62  \pm  0.15$  &  $+0.22  \pm  0.37$  &  $ 37,\, 4$  &  $ -75.2$  \\
    MOA-2011-BLG-191S\phantom{0} &  $5382  \pm   92$  &  $3.80  \pm  0.13$  &  $0.57  \pm  0.11$  &  $+0.26  \pm  0.14$  &  $ 61,\,11$  &  $ 134.0$  \\
    MOA-2011-BLG-445S\phantom{0} &  $4870  \pm  102$  &  $3.50  \pm  0.19$  &  $0.61  \pm  0.12$  &  $+0.26  \pm  0.31$  &  $ 57,\, 8$  &  $  72.6$  \\
   OGLE-2007-BLG-514S\phantom{0} &  $5644  \pm  111$  &  $4.10  \pm  0.21$  &  $1.47  \pm  0.21$  &  $+0.29  \pm  0.23$  &  $ 43,\, 7$  &  $ 158.8$  \\
  OGLE-2011-BLG-0950S            &  $6130  \pm  121$  &  $4.20  \pm  0.15$  &  $1.23  \pm  0.15$  &  $+0.33  \pm  0.10$  &  $ 58,\, 7$  &  $  91.5$  \\
    MOA-2009-BLG-259S\phantom{0} &  $4915  \pm  104$  &  $3.30  \pm  0.18$  &  $1.01  \pm  0.10$  &  $+0.34  \pm  0.19$  &  $ 64,\, 8$  &  $  81.5$  \\
    MOA-2008-BLG-311S\phantom{0} &  $5947  \pm   81$  &  $4.50  \pm  0.10$  &  $1.17  \pm  0.09$  &  $+0.35  \pm  0.08$  &  $111,\,14$  &  $ -34.1$  \\
  OGLE-2011-BLG-1072S            &  $5515  \pm   89$  &  $3.90  \pm  0.13$  &  $1.11  \pm  0.10$  &  $+0.36  \pm  0.12$  &  $ 69,\,13$  &  $ -62.2$  \\
    MOA-2011-BLG-058S\phantom{0} &  $5256  \pm  100$  &  $4.00  \pm  0.15$  &  $0.71  \pm  0.12$  &  $+0.37  \pm  0.25$  &  $ 62,\,10$  &  $-139.7$  \\
    MOA-2008-BLG-310S\phantom{0} &  $5675  \pm   91$  &  $4.20  \pm  0.11$  &  $1.03  \pm  0.08$  &  $+0.41  \pm  0.11$  &  $119,\,18$  &  $  77.5$  \\
    MOA-2012-BLG-022S\phantom{0} &  $5827  \pm  115$  &  $4.30  \pm  0.13$  &  $0.89  \pm  0.12$  &  $+0.42  \pm  0.10$  &  $ 53,\, 9$  &  $ -81.3$  \\
   OGLE-2007-BLG-349S\phantom{0} &  $5237  \pm  119$  &  $4.20  \pm  0.15$  &  $0.77  \pm  0.12$  &  $+0.42  \pm  0.26$  &  $103,\,17$  &  $ 113.0$  \\
    MOA-2006-BLG-099S\phantom{0} &  $5747  \pm  147$  &  $4.50  \pm  0.16$  &  $0.84  \pm  0.14$  &  $+0.44  \pm  0.21$  &  $117,\,17$  &  $  99.0$  \\
   OGLE-2006-BLG-265S\phantom{0} &  $5462  \pm  101$  &  $4.20  \pm  0.13$  &  $1.14  \pm  0.11$  &  $+0.46  \pm  0.19$  &  $ 92,\,14$  &  $-154.0$  \\
  OGLE-2012-BLG-0026S            &  $4815  \pm  145$  &  $3.40  \pm  0.28$  &  $0.62  \pm  0.14$  &  $+0.50  \pm  0.44$  &  $ 42,\, 7$  &  $ 132.2$  \\
    MOA-2010-BLG-311S\phantom{0} &  $5442  \pm   86$  &  $3.80  \pm  0.15$  &  $1.16  \pm  0.11$  &  $+0.51  \pm  0.19$  &  $ 52,\,11$  &  $  44.4$  \\
    MOA-2011-BLG-278S\phantom{0} &  $5307  \pm  159$  &  $4.00  \pm  0.25$  &  $0.54  \pm  0.18$  &  $+0.52  \pm  0.39$  &  $ 48,\, 7$  &  $ 229.6$  \\
    MOA-2010-BLG-037S\phantom{0} &  $5732  \pm  109$  &  $3.90  \pm  0.20$  &  $1.52  \pm  0.19$  &  $+0.55  \pm  0.20$  &  $ 56,\,12$  &  $  -8.4$  \\
\noalign{\smallskip}
\hline
\end{tabular}
\flushleft
{\tiny
{\bf Notes.}
$^{\dagger}$
Given for each star is: effective temperature ($\teff$); surface gravity ($\log g$); microturbulence
parameter ($\xi_{\rm t}$); [Fe/H]; number of \ion{Fe}{i} and \ion{Fe}{ii} lines used in the analysis;
stellar age; radial velicity ($v_{\rm r}$).\\
}
\end{table*}
%---------------------------------------------------------------------

 %---------------------------------------------------------------------
\begin{table*}
\centering
\caption{
Stellar ages, masses, and colours as determined
from spectroscopy and microlensing techniques for all 58 stars.$^\dag$
%\red{microlensing data for ob121534 and ob121526 missing}.
\label{tab:ages}
}
\setlength{\tabcolsep}{2.1mm}
\tiny
\begin{tabular}{rccrrrrccccc|ccc}
\hline\hline
\noalign{\smallskip}
  \multicolumn{1}{c}{Object}                     &
  \multicolumn{1}{c}{$T_{\rm eff}$}              &
  \multicolumn{1}{c}{$\mathcal{M}$}              &
  \multicolumn{1}{c}{$\log L$}              &
  \multicolumn{1}{c}{Age}                        &
  \multicolumn{1}{c}{$-$1$\sigma$}                        &
  \multicolumn{1}{c}{$+$1$\sigma$}                        &
  \multicolumn{1}{c}{$(V$--$I)_{0}$}             &
  \multicolumn{1}{c}{$M_{\rm V}$}                      &
  \multicolumn{1}{c}{$-$1$\sigma$}                        &
  \multicolumn{1}{c}{$+$1$\sigma$}                        &
  \multicolumn{1}{c|}{$M_{\rm I}$}                      &
  \multicolumn{1}{c}{$M_{\rm I,\,\mu}$}                  &
  \multicolumn{1}{c}{$(V$--$I)_{\rm 0,\,\mu}$}       &
  \multicolumn{1}{c}{$\teff^{\mu}$}              \\
                                                 &
  \multicolumn{1}{c}{$\rm [K]$}                  &
  \multicolumn{1}{c}{$M_{\odot}$}                &
  \multicolumn{1}{c}{$L_{\odot}$}                &
  \multicolumn{1}{c}{[Gyr]}                      &
  \multicolumn{1}{c}{[Gyr]}                      &
  \multicolumn{1}{c}{[Gyr]}                      &
  \multicolumn{1}{c}{[mag]}                      &
  \multicolumn{1}{c}{[mag]}                      &
  \multicolumn{1}{c}{[mag]}                      &
  \multicolumn{1}{c}{[mag]}                      &
  \multicolumn{1}{c|}{[mag]}                     &
  \multicolumn{1}{c}{[mag]}                      &
  \multicolumn{1}{c}{[mag]}                      &
  \multicolumn{1}{c}{[K]}                        \\
\noalign{\smallskip}
\hline
\noalign{\smallskip}
 OGLE-2012-BLG-1156 & 6200 & 0.77 & $ 0.05$ & 13.1 &  6.4 & 13.4 & 0.58 & 4.79 & 3.43 & 5.19 & 4.21 & 3.56 & 0.63 & 6058 \\
 OGLE-2011-BLG-0969 & 6150 & 0.79 & $ 0.02$ & 11.6 &  5.9 & 13.2 & 0.59 & 4.86 & 3.85 & 5.10 & 4.27 & 2.94 & 0.58 & 6296 \\
 MOA-2010-BLG-285   & 6057 & 0.78 & $ 0.09$ & 11.6 &  8.2 & 13.4 & 0.62 & 4.67 & 3.86 & 4.99 & 4.05 & 4.77 & 0.56 & 6384 \\
 MOA-2010-BLG-078   & 5205 & 0.93 & $ 0.77$ & 14.6 &  4.9 & 13.1 & 0.82 & 2.99 & 2.38 & 3.57 & 2.17 & 1.79 & 0.83 & 5317 \\
 MOA-2011-BLG-104   & 5900 & 0.81 & $ 0.00$ & 10.2 &  5.5 & 12.9 & 0.65 & 4.92 & 3.48 & 5.20 & 4.27 & 3.50 & 0.73 & 5644 \\
 OGLE-2012-BLG-0270 & 5914 & 0.84 & $-0.07$ & 11.6 &  6.1 & 13.0 & 0.65 & 5.07 & 4.13 & 5.25 & 4.42 & 3.07 & 0.62 & 6081 \\
 MOA-2009-BLG-493   & 5420 & 0.72 & $-0.40$ & 13.1 &  4.7 & 13.3 & 0.79 & 5.98 & 5.59 & 6.20 & 5.19 & 4.74 & 0.68 & 5829 \\
 MOA-2012-BLG-187   & 5895 & 0.84 & $ 0.19$ & 11.6 &  9.2 & 13.4 & 0.67 & 4.40 & 4.02 & 4.87 & 3.73 & 3.71 & 0.71 & 5715 \\
 OGLE-2009-BLG-076  & 5854 & 0.83 & $-0.03$ & 11.6 &  7.6 & 13.4 & 0.68 & 4.96 & 4.32 & 5.18 & 4.28 & 4.17 & 0.68 & 5828 \\
 MOA-2009-BLG-133   & 5529 & 0.75 & $-0.31$ & 14.6 &  7.4 & 13.9 & 0.76 & 5.71 & 5.36 & 5.81 & 4.95 & 4.17 & 0.69 & 5790 \\
 OGLE-2012-BLG-0563 & 5907 & 0.85 & $-0.04$ & 10.2 &  5.7 & 12.0 & 0.66 & 4.89 & 4.63 & 5.13 & 4.23 & 4.36 & 0.58 & 6257 \\
 OGLE-2012-BLG-1279 & 5796 & 0.83 & $-0.07$ & 11.6 &  6.6 & 13.1 & 0.69 & 5.06 & 4.76 & 5.26 & 4.37 & 4.11 & 0.61 & 6117 \\
 MOA-2010-BLG-167   & 5406 & 0.87 & $ 0.37$ & 13.5 &  9.0 & 13.8 & 0.80 & 4.04 & 3.65 & 4.07 & 3.24 & 2.71 & 0.70 & 5751 \\
 MOA-2012-BLG-532   & 5626 & 0.92 & $ 0.37$ & 11.6 &  4.2 & 12.7 & 0.75 & 3.99 & 3.00 & 5.01 & 3.24 & 4.15 & 0.67 & 5864 \\
 MOA-2009-BLG-475   & 5836 & 0.87 & $-0.05$ &  8.7 &  4.4 & 12.3 & 0.69 & 5.04 & 4.19 & 5.37 & 4.35 & 4.23 & 0.60 & 6157 \\
 MACH-1999-BLG-022  & 5632 & 0.85 & $ 0.29$ & 13.1 &  8.0 & 13.6 & 0.74 & 4.17 & 3.66 & 4.85 & 3.43 &  --  &  --  &  --  \\
 OGLE-2012-BLG-1217 & 5795 & 0.89 & $ 0.00$ & 11.6 &  7.2 & 12.4 & 0.69 & 4.76 & 4.35 & 5.03 & 4.07 & 3.67 & 0.71 & 5712 \\
 MOA-2010-BLG-049   & 5694 & 0.87 & $ 0.25$ & 12.4 &  9.4 & 13.6 & 0.73 & 4.25 & 3.82 & 4.55 & 3.52 & 2.67 & 0.67 & 5861 \\
 MOA-2010-BLG-446   & 6308 & 1.06 & $ 0.20$ &  2.6 &  1.3 &  4.4 & 0.57 & 4.32 & 4.00 & 4.51 & 3.75 &  --  &  --  &  --  \\
 OGLE-2008-BLG-209  & 5248 & 0.97 & $ 0.40$ &  8.6 &  5.3 & 12.5 & 0.84 & 3.92 & 3.34 & 4.09 & 3.08 & 2.50 & 0.74 & 5608 \\
 MOA-2011-BLG-090   & 5367 & 0.91 & $ 0.26$ & 14.3 & 11.1 & 14.4 & 0.80 & 4.30 & 4.02 & 4.30 & 3.50 & 3.77 & 0.67 & 5857 \\
 MOA-2012-BLG-391   & 5505 & 1.00 & $ 0.40$ &  8.1 &  5.3 & 12.2 & 0.77 & 3.89 & 3.33 & 4.15 & 3.12 & 3.92 & 0.81 & 5390 \\
OGLE-2012-BLG-1526  & 5200 & 0.93 & $ 0.37$ &  9.6 &  6.4 & 13.0 & 0.84 & 3.99 & 3.54 & 4.17 & 3.15 & 2.99 & 0.69 & 5782 \\
 MOA-2009-BLG-489   & 5543 & 0.92 & $ 0.26$ & 14.2 &  9.8 & 14.0 & 0.76 & 4.25 & 3.91 & 4.55 & 3.49 & 3.37 & 0.86 & 5251 \\
 MOA-2011-BLG-174   & 6172 & 1.08 & $ 0.18$ &  3.1 &  1.7 &  5.2 & 0.60 & 4.37 & 3.85 & 4.50 & 3.77 & 4.27 & 0.66 & 5893 \\
 MOA-2012-BLG-202   & 4862 & 0.97 & $ 0.41$ & 14.7 &  8.6 & 13.9 & 0.94 & 3.96 & 3.56 & 4.17 & 3.02 & 5.09 & 0.90 & 5151 \\
 OGLE-2012-BLG-1534 & 5920 & 1.08 & $ 0.44$ &  6.8 &  5.8 &  7.7 & 0.65 & 3.73 & 3.44 & 3.94 & 3.08 & 2.61 & 0.64 & 5971 \\
 MOA-2012-BLG-410   & 5509 & 1.01 & $ 0.46$ &  6.8 &  5.2 & 11.7 & 0.77 & 3.73 & 3.33 & 4.13 & 2.96 & 5.32 & 0.94 & 5056 \\
 OGLE-2012-BLG-0617 & 4924 & 0.95 & $ 0.48$ & 13.3 &  7.0 & 13.4 & 0.94 & 3.79 & 3.30 & 4.02 & 2.85 & 3.42 & 1.03 & 4867 \\
 OGLE-2012-BLG-0816 & 5395 & 1.29 & $ 0.81$ &  1.6 &  1.7 &  7.0 & 0.80 & 3.30 & 1.98 & 3.63 & 2.50 & 3.20 & 0.72 & 5673 \\
 OGLE-2012-BLG-0211 & 5573 & 1.00 & $ 0.35$ &  8.6 &  6.5 & 11.9 & 0.77 & 3.98 & 3.52 & 4.31 & 3.21 & 2.99 & 0.70 & 5742 \\
 MOA-2011-BLG-234   & 5778 & 0.99 & $ 0.00$ &  6.4 &  3.3 &  8.8 & 0.69 & 4.76 & 4.44 & 4.98 & 4.07 & 4.98 & 0.74 & 5606 \\
 OGLE-2011-BLG-1105 & 5666 & 0.95 & $-0.07$ &  6.4 &  3.1 & 10.4 & 0.74 & 5.05 & 4.62 & 5.26 & 4.31 & 4.03 & 0.78 & 5480 \\
 MOA-2010-BLG-523   & 5122 & 1.14 & $ 0.58$ &  3.3 &  2.8 &  8.8 & 0.90 & 3.53 & 2.80 & 3.90 & 2.63 &  --  & 0.75 & 5573 \\
 OGLE-2012-BLG-1274 & 5733 & 1.05 & $ 0.32$ &  8.3 &  7.2 &  9.2 & 0.71 & 4.03 & 3.75 & 4.21 & 3.32 & 3.39 & 0.68 & 5811 \\
 OGLE-2012-BLG-0521 & 5013 & 1.03 & $ 0.47$ &  5.7 &  4.6 & 11.8 & 0.91 & 3.82 & 3.30 & 4.13 & 2.91 & 2.98 & 0.85 & 5283 \\
 MOA-2009-BLG-174   & 5600 & 0.96 & $-0.05$ &  8.7 &  4.2 & 10.4 & 0.76 & 5.01 & 4.70 & 5.18 & 4.25 & 3.50 & 0.69 & 5774 \\
 MOA-2011-BLG-034   & 5440 & 0.94 & $ 0.23$ & 12.4 &  8.7 & 13.6 & 0.79 & 4.31 & 3.92 & 4.67 & 3.52 &  --  & 0.65 & 5922 \\
 MOA-2009-BLG-456   & 5662 & 1.01 & $ 0.14$ &  8.7 &  6.7 & 10.9 & 0.74 & 4.39 & 4.06 & 4.74 & 3.65 & 2.74 & 0.67 & 5846 \\
 MOA-2012-BLG-291   & 5156 & 0.96 & $ 0.27$ & 14.6 &  8.2 & 13.8 & 0.87 & 4.29 & 3.94 & 4.67 & 3.42 & 4.02 & 0.91 & 5135 \\
 OGLE-2011-BLG-1410 & 4831 & 1.02 & $ 0.86$ &  6.5 &  3.4 & 12.1 & 1.00 & 2.91 & 2.20 & 3.51 & 1.91 & 2.13 &  --  &  --  \\
 MOA-2011-BLG-191   & 5382 & 1.18 & $ 0.50$ &  3.9 &  3.7 &  9.2 & 0.82 & 3.65 & 3.08 & 4.00 & 2.83 & 2.50 & 0.67 & 5841 \\
 MOA-2011-BLG-445   & 4870 & 0.98 & $ 0.68$ &  8.2 &  4.1 & 12.1 & 0.98 & 3.32 & 2.78 & 3.86 & 2.34 & 3.20 & 0.92 & 5114 \\
 OGLE-2007-BLG-514  & 5644 & 1.01 & $ 0.22$ &  6.4 &  4.6 & 10.3 & 0.75 & 4.33 & 3.58 & 4.79 & 3.58 & 4.34 & 0.71 & 5700 \\
 OGLE-2011-BLG-0950 & 6130 & 1.27 & $ 0.34$ &  2.9 &  1.6 &  3.6 & 0.60 & 3.95 & 3.26 & 4.14 & 3.35 & 3.41 & 0.63 & 5990 \\
 MOA-2009-BLG-259   & 4915 & 1.17 & $ 0.88$ &  3.0 &  2.0 &  9.0 & 1.00 & 2.84 & 2.06 & 3.36 & 1.84 & 2.88 & 0.80 & 5421 \\
 MOA-2008-BLG-311   & 5947 & 1.17 & $ 0.13$ &  1.7 &  0.9 &  3.0 & 0.66 & 4.49 & 4.25 & 4.60 & 3.83 & 3.91 & 0.67 & 5838 \\
 OGLE-2011-BLG-1072 & 5515 & 1.18 & $ 0.47$ &  5.5 &  4.4 &  8.8 & 0.79 & 3.70 & 3.19 & 4.08 & 2.91 & 2.66 & 0.79 & 5450 \\
 MOA-2011-BLG-058   & 5256 & 0.97 & $ 0.26$ & 10.8 &  6.5 & 13.1 & 0.85 & 4.30 & 3.74 & 4.52 & 3.45 & 3.29 & 0.80 & 5421 \\
 MOA-2008-BLG-310   & 5675 & 1.12 & $ 0.27$ &  4.9 &  4.4 &  8.0 & 0.75 & 4.17 & 3.86 & 4.58 & 3.42 & 3.44 & 0.70 & 5731 \\
 MOA-2012-BLG-022   & 5827 & 1.15 & $ 0.17$ &  3.5 &  2.1 &  5.5 & 0.70 & 4.52 & 3.93 & 4.65 & 3.82 & 3.70 & 0.82 & 5366 \\
 OGLE-2007-BLG-349  & 5237 & 0.94 & $-0.13$ & 13.1 &  8.1 & 13.6 & 0.85 & 4.37 & 4.28 & 5.47 & 3.52 & 4.14 & 0.79 & 5450 \\
 MOA-2006-BLG-099   & 5747 & 1.06 & $ 0.02$ &  3.5 &  1.8 &  7.1 & 0.73 & 4.82 & 4.39 & 5.06 & 4.09 & 3.79 & 0.75 & 5568 \\
 OGLE-2006-BLG-265  & 5462 & 1.00 & $ 0.10$ &  8.7 &  6.3 & 11.8 & 0.81 & 4.64 & 4.19 & 5.00 & 3.83 & 3.57 & 0.69 & 5764 \\
 OGLE-2012-BLG-0026 & 4815 & 1.02 & $ 0.74$ &  6.5 &  3.7 & 12.4 & 1.00 & 3.15 & 2.36 & 3.80 & 2.15 & 2.68 & 1.04 & 4878 \\
 MOA-2010-BLG-311   & 5442 & 1.20 & $ 0.50$ &  4.4 &  3.3 &  8.0 & 0.81 & 3.63 & 2.89 & 3.96 & 2.82 & 3.14 & 0.73 & 5630 \\
 MOA-2011-BLG-278   & 5307 & 1.01 & $ 0.34$ & 10.2 &  5.2 & 12.7 & 0.82 & 4.09 & 3.59 & 5.06 & 3.27 & 4.49 & 0.86 & 5263 \\
 MOA-2010-BLG-037   & 5732 & 1.29 & $ 0.56$ &  4.2 &  3.1 &  6.3 & 0.72 & 3.44 & 2.85 & 4.01 & 2.72 & 3.54 & 0.78 & 5478 \\
\noalign{\smallskip}
\hline
\end{tabular}
\flushleft
{\tiny
{\bf Notes.}
$^{\dagger}$
Column 2 effective temperature;
col.~3 stellar mass;
col.~4 luminosity,
col.~5 stellar age;
cols.~6 and 7 the 1-sigma lower and upper age limits;
col.~8 ``spectroscopic" colour,
col.~9 ``spectroscopic" $M_{\rm V}$ magnitude,
cols.~10 and 11 the 1-sigma lower and upper limits on $M_{\rm V}$;
col.~12 ``spectroscopic" $M_{\rm I}$ magnitude, 
col.~13 $M_{\rm I,\,\mu}$ from microlensing techniques (with the bulge red clump at $-0.12$);
col.~14 $(V-I)_{\rm \,\,0,\,\mu}$ from microlensing techniques (with the bulge red clump at 1.06);
col.~15 inferred effective temperatures from microlensing $(V-I)_{\,\,0}$ colour using the \cite{casagrande2010}
$(V-I)-\teff$ calibration.
}
\end{table*}
%---------------------------------------------------------------------

Following the standard procedure to determine the stellar parameters
(see above) for MOA-2011-BLG-104S we get a microturbulence value of 
$\xi_{\rm t} = 2.7$\,$\kms$.  This value seems too high for these
types of stars. According to the empirical calibration 
by M.~Bergemann, stars with similar stellar parameters ($\teff=6106$\,K, 
$\log g=4.4$, and $\rm [Fe/H]=-0.88$)  have microturbulence
values around 1.5\,$\kms$. Also, the discrepancy between the spectroscopic 
temperature and the temperature deduced from microlensing techniques is 
unusually large for this star ($\sim 500\,K$). Hence, we believe that due 
to the quite low metallicity of this star, combined with a relatively low 
signal-to-noise ratio of the observed spectrum ($S/N\approx 25$), the stellar  
parameters have been poorly constrained from the \ion{Fe}{i} balance plots.

The discrepancies indicate that it is likely that the temperature
should be lower than the 6100\,K derived from excitation balance. However, 
based on the Balmer H$\alpha$ line profile it should probably not be as 
low as the temperature deduced from the microlensing techniques (5644\,K, 
see Table~\ref{tab:ages} and Fig.~\ref{fig:balmer}). Hence, for
MOA-2011-BLG-104S, we set the temperature to 5900\,K. Following the
empirical calibration by M.~Bergemann we set the microturbulence to 1.3\,$\kms$.
To get ionisation balance between \ion{Fe}{i} and \ion{Fe}{ii}
abundances the surface gravity needs to be set to $\log g = 4.15$.
The effect on the metallicity for MOA-2011-BLG-104S, when changing 
the temperature from 6100\,K to 5900\,K is only $-0.03$\,dex
(down from $\rm [Fe/H]= -0.88$ to $-0.85$). The age of the star 
increased from 8.7\,Gyr to 10.2\,Gyr (which is more in line with 
ages seen for the other metal-poor stars).

%=======================================================================
\section{Results}

%=======================================================================
\subsection{Positions on the sky}

%-----------------------------------------------------------------------
\begin{figure}
\resizebox{\hsize}{!}{
\includegraphics[bb=18 144 592 718,clip]{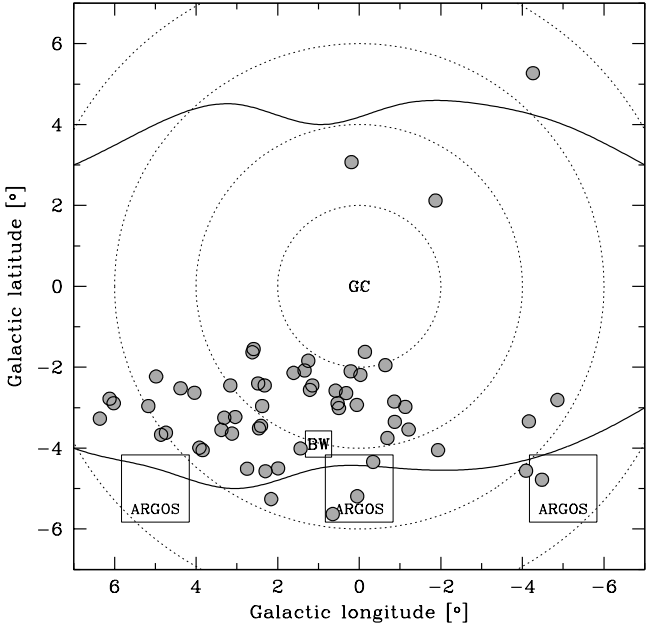}}
\caption{Positions on the sky for the microlensed dwarf
sample. The bulge contour 
lines based on observations with the COBE satellite are shown as solid
lines \citep{weiland1994}. The positions of Baade's window (BW), the
Galactic centre (GC), and three of the ARGOS survey fields 
\citep[see][]{ness2012} 
are delineated. The dotted lines are concentric circles in steps of 
2{\degr}.
                    }
         \label{fig:glonglat}
   \end{figure}
%-----------------------------------------------------------------------

Figure~\ref{fig:glonglat} shows the positions on the sky for 
the 58 microlensed dwarf stars. 
Except for the first three detections at positive latitudes,
essentially all stars are located between 2{\degr} to 5{\degr} below the 
the Galactic plane and mainly at positive longitudes. This distribution 
is a reflection of the locations of the fields surveyed by MOA and OGLE.
At the distance of the bulge, 8.2\,kpc \citep[e.g.,][]{nataf2013},
2{\degr} to 5{\degr} corresponds to a vertical distance of 150 to 750\,pc,
respectively, below the plane.

%=======================================================================
\subsection{Radial velocities - bulge membership}

Radial velocities can help constraining dynamical models of the bulge. 
The BRAVA survey
\citep[see,][for latest release]{kunder2012} presents radial velocities
for $\sim10\,000$ red giant stars covering $-10<l<10$ and $-8<b<-4$.
They find that the rotation curves at $l=-4,-6$ and $-8$ almost perfectly
follow the cylindrical rotation models by \cite{shen2010}. 
The velocity dispersion
varies from around $100\,\kms$ at $(l,\,b)=(0,\,-4)$ to around $80\,\kms$
at $(l,\,b)=(\pm10,\,-4)$.

\begin{figure}
\resizebox{\hsize}{!}{
\includegraphics[bb=18 144 580 718,clip]{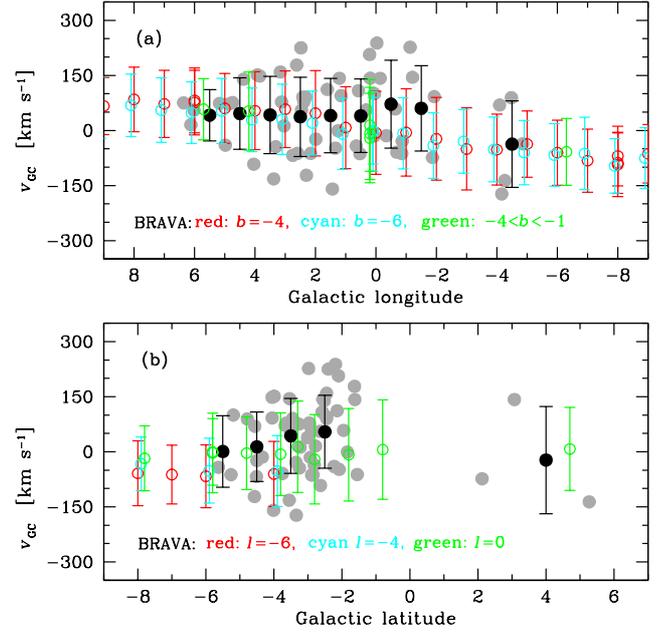}}
\caption{Galactocentric velocity versus Galactic longitude
and latitude for the microlensed dwarf sample (grey circles).
The black circles represent the mean radial velocity and the 1$\sigma$
rms dispersion for the microlensed bulge dwarfs, binned into
3{\degr} wide bins.
The results from the BRAVA survey \citep{rich2007_brava,kunder2012} have been
over-plotted, and the different colours (red, cyan, and green) represent
the values at different $l$ and $b$ as given in the plots.
                    }
         \label{fig:rvel}
   \end{figure}

We correct our measured heliocentric radial velocities for the microlensed 
bulge dwarfs to the Galactic centre
using the relation\footnote{Obtained from {\tt  http://leda.univ-lyon1.fr}.}
$v_{GC} = v_{r,HC} + 232\sin(l)\cos(b) + 9\cos(l)\cos(b) + 7\sin(b)$
which takes the motion of the relative to the local standard of rest (LSR), 
and the motion of the LSR relative to the Galactic centre, into account.
Figure~\ref{fig:rvel} then shows the galactocentric radial velocities versus
$l$ and $b$ for the microlensed dwarf sample (grey circles) as well
as the results from the BRAVA survey in different $l$ and $b$ bins
(as indicated in the figure; red, cyan, and green). The black circles
mark the mean velocity and the rms dispersion around the mean the 
microlensed dwarf stars, binned into 3{\degr} wide bins. The average velocities
as well as the dispersions for the microlensed dwarf stars follow 
nicely the results from the BRAVA survey, especially versus $l$.
Given that the BRAVA sample is representative of the bulge population,
this shows that the radial velocities and the velocity dispersion
seen in the microlensed dwarf sample are at the expected levels if they
belong to a bulge population.

%-----------------------------------------------------------------------
%-----------------------------------------------------------------------

%=======================================================================
\subsection{Metallicity distribution}

%-----------------------------------------------------------------------
\begin{figure}
\resizebox{\hsize}{!}{
\includegraphics[bb=30 160 470 710,clip]{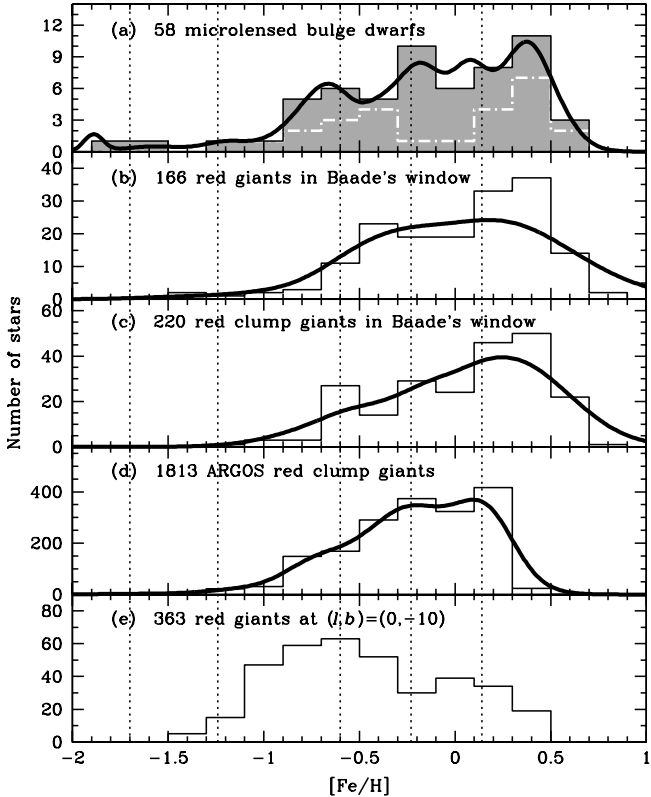}}
\caption{(a) Metallicity distribution for the microlensed dwarf sample 
(white dashed line shows the distribution of the 26 microlensed dwarf 
stars from \citealt{bensby2011});
(b) 166 red giant stars in Baade's window from \cite{hill2011};
(c) 220 red clump stars in Baade's window from \cite{hill2011};
(d) 1813 red giant stars with from the ARGOS survey fields at 
$(l,b) = (0,-5),\,(5,-5),\,(-5,-5)$ from Ness et al.~(submitted).
(e) 363 red giants at $(l,b)=(0,-10)$ from \cite{uttenthaler2012};
The curved lines in (a)-(d) represent generalised histograms.
Dotted vertical lines mark the peaks claimed by Ness et al. in (d).
        }
\label{fig:mdf}
\end{figure}
%-----------------------------------------------------------------------
%-----------------------------------------------------------------------
\begin{figure}
\resizebox{\hsize}{!}{
\includegraphics[bb=18 160 592 718,clip]{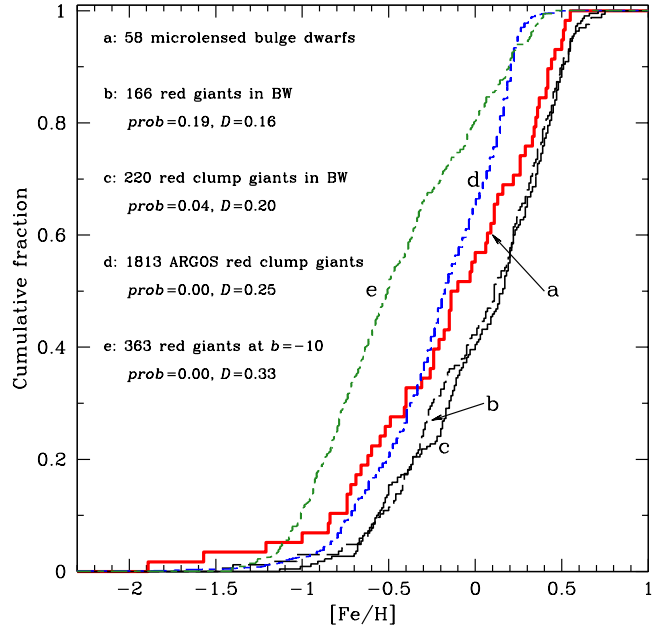}}
\caption{Cumulative histograms and the Kolmogorov-Smirnow 
tests between the microlensed dwarf sample and the giant 
star samples shown in Fig.~\ref{fig:mdf}. The solid thicker red line
shows the microlensed bulge dwarfs, (b) the RGB sample from
\cite{hill2011}, (c) the red clump sample from \cite{hill2011},
(d) the red clump sample from ARGOS (Ness et al.),
and (e) the RGB sample from \cite{uttenthaler2012}.
        }
\label{fig:mdfcum}
\end{figure}
%-----------------------------------------------------------------------
%-----------------------------------------------------------------------
\begin{figure}
\resizebox{\hsize}{!}{
\includegraphics[bb=18 160 592 550,clip]{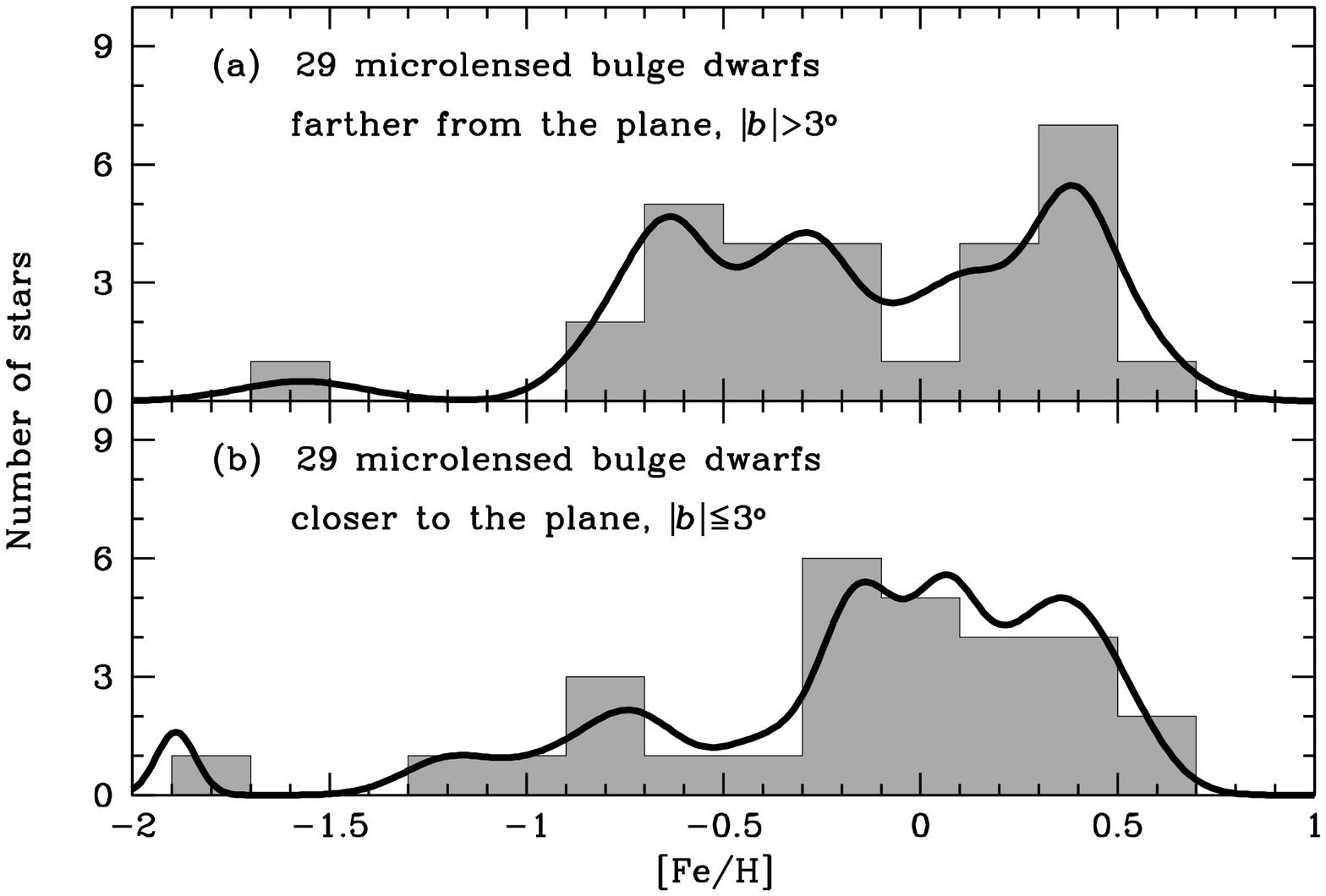}}
\resizebox{\hsize}{!}{
\includegraphics[bb=18 160 592 550,clip]{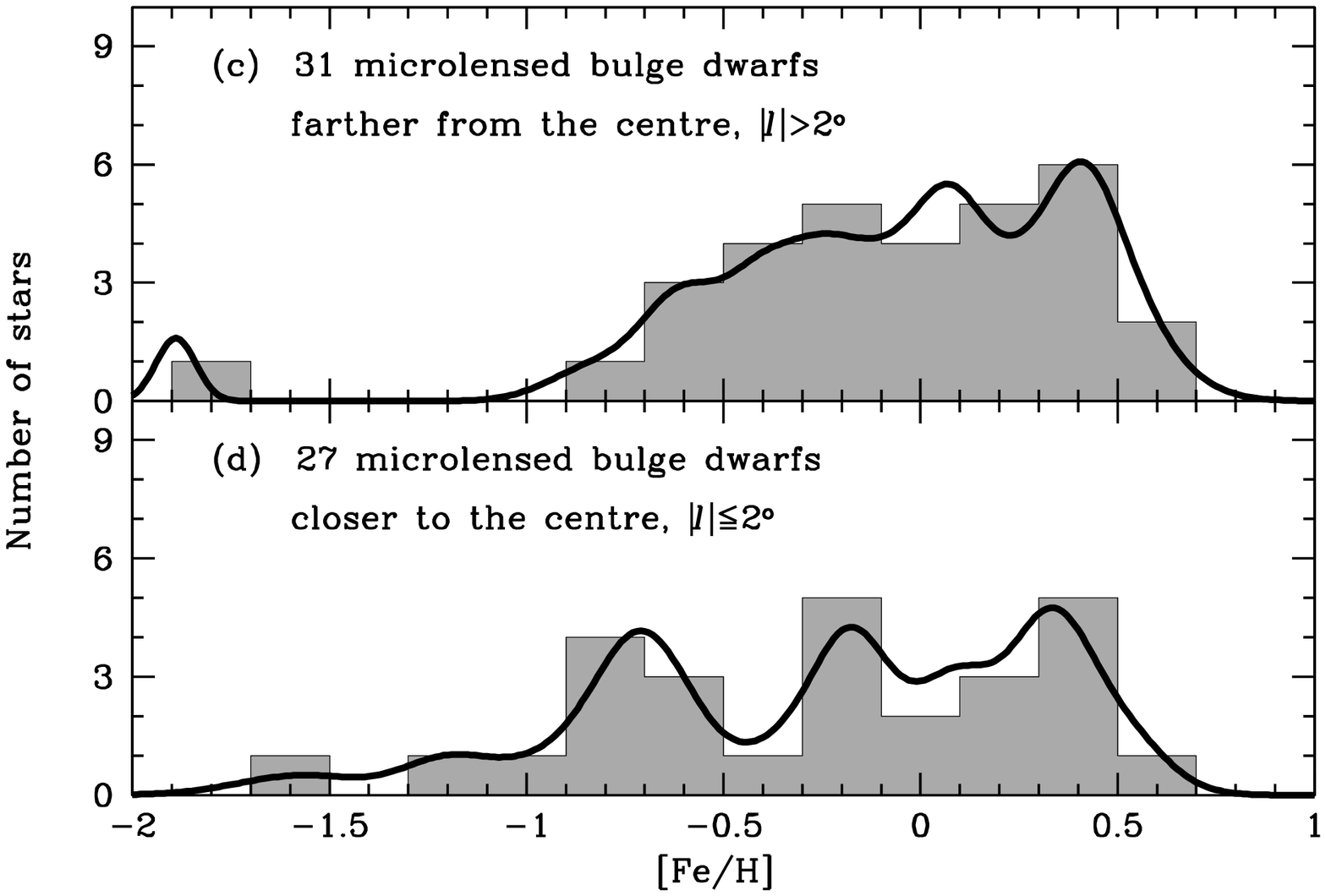}}
\caption{
(a) and (b) show the MDFs when the microlensed dwarf 
sample is split into stars farther from the plane ($|b|\gtrsim 3$), and 
closer to the plane ($|b|<3$), respectively.
(c) - (d) the MDFs when sample is split into stars farther from the centre 
($|l|>2$), and closer to the centre ($|l|\lesssim2$), respectively. 
The curved lines represent generalised histograms.                    
}
         \label{fig:mdf_lb}
   \end{figure}
%-----------------------------------------------------------------------

Figure~\ref{fig:mdf}a shows the metallicity distribution for the
microlensed dwarf sample. The finding
from \cite{bensby2011}, that the MDF is bi-modal, with a paucity of stars
around solar metallicity, is weakened as we now have several stars with 
metallicities within $\pm 0.1$\,dex of the Sun. One of those stars, 
MOA-2010-BLG-523S, was already published in \cite{bensby2011} but has 
now been revised 0.08\,dex lower metallicity following the current analysis
(see Sect.~\ref{sec:analysis}).

A major issue in \cite{bensby2011} was the apparent discrepancy between 
the microlensed dwarf star MDF, which was clearly bi-modal, and the red 
giant MDF in Baade's window by \cite{zoccali2008}, which then did not 
appear bi-modal. Since then, the \cite{zoccali2008} RGB sample has been 
re-analysed by \cite{hill2011} with the result that the RGB sample
shrank from 213 stars to 166 stars and that the MDF now appears 
bi-modal (see Fig.~\ref{fig:mdf}b). With the current microlensed 
dwarf sample, and the re-analysed RGB sample, the two distributions are 
quite similar. A two-sided KS test yields a $p$-value of 0.17
(see Fig.~\ref{fig:mdfcum}), i.e., the  hypothesis that the two 
samples have been drawn from the same underlying population cannot be
rejected. 

\cite{hill2011} also presented a sample of 220 red clump giants in 
Baade's window. The MDF of these red clump stars (Fig.~\ref{fig:mdf}c) 
is similar to the RGB sample, and a two-sided KS test between the red 
clump sample and the microlensed dwarf sample
yields a $p$-value of 0.06 (see Fig.~\ref{fig:mdfcum}). Hence, the 
conclusions are the same as for the RGB sample.

Furthermore, ARGOS (Abundances and Radial velocity Galactic Origins Survey) 
observed 28,000 stars, mostly red clump giants, in 28 fields 
of the Galactic bulge. In Fig.~\ref{fig:mdf}d we show the MDF for 
the 1813 red clump giants that are located in the three ARGOS fields 
at $(l,b) = (0,-5),\,(5,-5),\,(-5,-5)$ (the fields are delineated 
out in Fig.~\ref{fig:glonglat}). The internal random 
uncertainty in [Fe/H] for individual stars in ARGOS is 0.13\,dex 
(Ness et al.~submitted). By fitting Gaussians to the MDF, 
Ness et al. claim detection of three components at 
$\rm [Fe/H]=+0.14,\,-0.23,\,-0.60$, and possibly another two at 
$\rm [Fe/H]=-1.24,\,-1.7$. The peaks are indicated by the vertical dotted lines in Fig.~\ref{fig:mdf}. It is 
interesting to see that the three metal-rich ARGOS peaks more or less
coincide with the ``bumps'' in the generalised dwarf star MDF in 
Fig.~\ref{fig:mdf}a. There might be a slight offset, but the three 
peaks are clearly within 0.1\,dex of the ``bumps''. It is also interesting 
to note that the ARGOS MDF does not show the relatively large
fraction of metal-rich stars ($\rm [Fe/H]>+0.3$)
seen in the microlensed dwarf sample and the RGB and RC samples in Baade's 
window. Since the ARGOS fields are slightly farther from the Galactic plane
than the other samples, this trend could be explained if there were a
metal-rich population that drops off very rapidly with distance from the
plane. The MDF of 363 RGB stars at
$(l,b) = (0,-10)$ by \cite{uttenthaler2012} in Fig.~\ref{fig:mdf}e
shows a more dominant metal-poor peak, possibly confirming the
drop-off with $|b|$ for the metal-rich peak. 
However, contrary to the ARGOS MDF, the $b=-10\degr$ MDF 
by \cite{uttenthaler2012}
contains a larger fraction of metal-rich stars ($\rm [Fe/H]>+0.3$).
Since it is located even farther from the plane than the 
ARGOS fields it could mean that the ARGOS study suffers from
a systematic shift in [Fe/H] (i.e., it should be moved to higher [Fe/H]).

Regarding the regions closer to the plane, both \cite{ramirez2000} and
\cite{rich2012} used infrared spectroscopy to study M giants in several 
fields in the innermost 600\,pc between Baade's window and $\sim 150$\,pc 
(i.e., 4{\degr} to 1{\degr}) below the plane. Neither of these two studies
find a major vertical abundance or composition gradient. 
Hence, if there is a gradient, or gradual change of stellar populations,
it appears to disappear close to the plane. What is intriguing about the 
\cite{rich2012} study is that the M giant MDFs peak at $\rm [Fe/H]\approx -0.2$
at all latitudes, and that there are essentially no M giants with 
$\rm [Fe/H]>0$. This is in stark contrast to the findings from both 
the microlensed bulge dwarf sample as well as the RGB and RC samples
in Baade's window from \cite{hill2011}, which all see a large fraction 
of stars at super-solar metallicities. 
The reason might be that very metal-rich M stars
actually never reach the RGB phase due to strong stellar 
winds \citep[see discussion in, e.g.,][]{cohen2008}, or that
the metallicities of M giants may be systematically 
underestimated. In any case, the metal-rich stars are 
lacking from studies that probe the MDF using M giants

To further investigate the variation of the MDF we split the microlensed
dwarf sample 
in latitude and then in longitude. Figures~\ref{fig:mdf_lb}a and b
show the MDFs when the microlensed dwarf sample is divided
into two samples: one with stars closer than 3{\degr} to the plane, and 
one with stars farther than 3{\degr} from the plane. 
A two-sided KS-test ($prob =0.32$, and $D=0.24$) shows that 
the hypothesis that the two subsamples have been drawn from the same underlying
distribution cannot be rejected. However, to the eye,
two distributions appear to have distinctly different features,
and not just a systematic shift which would as in the case of
a vertical metallicity gradient.
The outer sample appears bi-modal with a paucity of stars around solar 
metallicities. The inner sample on the other hand contain many stars
around solar values. Both MDFs span the same metallicity range,
but with different peaks, indicating
a gradual change with $b$ of the fraction of the different stellar population
components \citep[see also discussion in][]{babusiaux2010}. 
When going towards the plane, the fraction of 
metal-poor and metal-rich stars drops off while the fraction of solar 
metallicity stars increases. The fraction of metal-poor stars appears to drop
faster than the fraction of metal-rich stars. 
A tentative interpretation could be that there are three populations with 
different scale heights; could it be the thin disk, the thick disk, and a bar 
population?
This interpretation would be in line with the results by Ness et al.~(in prep.) 
of their multicomponent MDF (see Fig.~\ref{fig:mdf}d). They interpret their 
$\rm [Fe/H]=+0.14$ peak as belonging 
to the thin disk surrounding the bulge, the $\rm [Fe/H]=-0.23$ peak 
as the true boxy/peanut bulge, and the $\rm [Fe/H]=-0.60$ peak as 
the old thick disk, maybe being part of the bulge. 

Figure~\ref{fig:mdf_lb}e-f shows the MDFs for the microlensed dwarf 
stars split in two subsamples: one with $|l|>2$ and one with $|l|\lesssim2$, 
respectively. The two-sided $KS$-test yields a 
similar result ($prob =0.32$, and $D=0.24$) as when splitting the sample in 
latitude, i.e. it is not possible by statistics to reject the claim that 
the two subsamples have been drawn from the same underlying distribution. 

%-----------------------------------------------------------------------
\begin{figure}
\resizebox{\hsize}{!}{
\includegraphics[bb=18 160 592 718,clip]{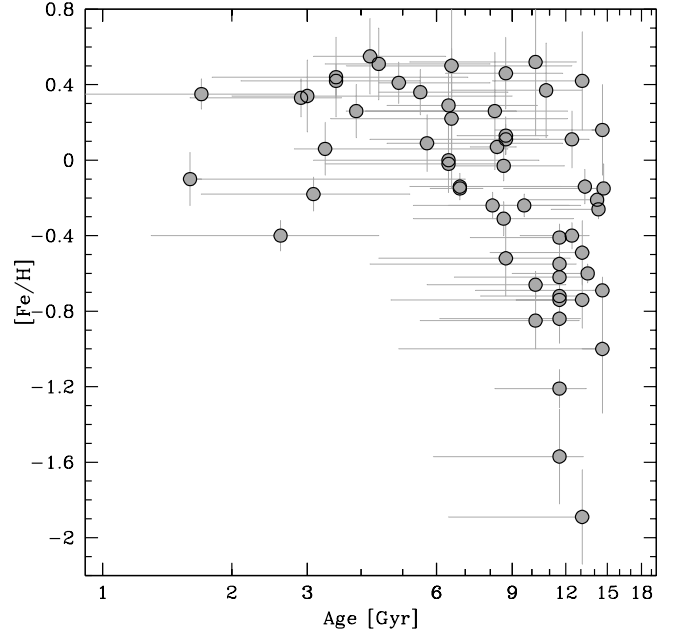}}
\caption{Age versus [Fe/H] for the microlensed dwarf sample. 
\label{fig:agefe}}
\end{figure}
%-----------------------------------------------------------------------

%=======================================================================
\subsection{Stellar ages}
\label{sec:ages}

Figure~\ref{fig:agefe} shows the age-metallicity diagram for the microlensed
bulge dwarfs. The metal-poor stars below $\rm [Fe/H]\approx -0.4$ are 
consistently old with ages around 10 to 12\,Gyr. The metal-rich stars 
on the other hand show a large dispersion, with ages ranging from a few
billion years up to $\sim 13$\,Gyr.

\begin{figure}
\resizebox{\hsize}{!}{
\includegraphics[bb=18 160 592 718,clip]{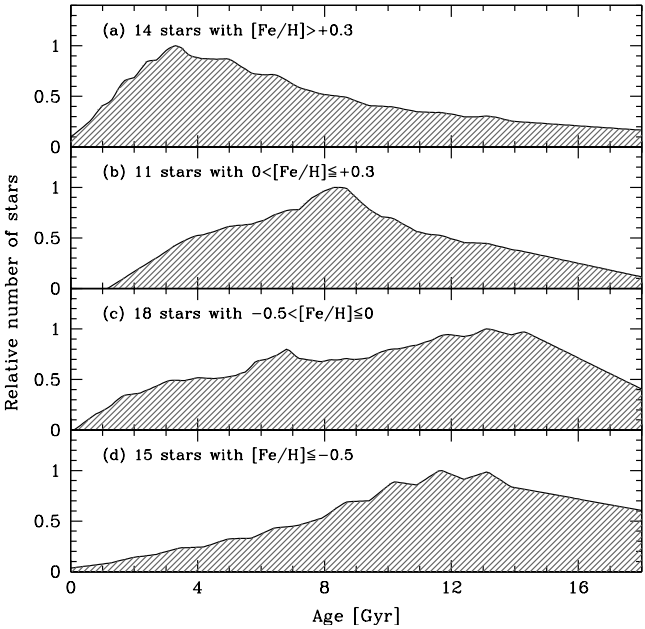}}
\caption{Sums of individual age probability distribution functions 
for the microlensed dwarf sample for four metallicity bins 
(as indicated).
\label{fig:agepdf}}
\end{figure}

The varying age distribution is further highlighted in 
Fig.~\ref{fig:agepdf} which shows the summed age distribution 
functions (ADF) from the individual stars in different metallicity 
bins (an ADF for an individual star include all possible ages from the isochrones
that are encompassed by the uncertainties of the stellar parameters). 
The summed ADF of the most metal-rich population 
($\rm [Fe/H]>0.3$) is dominated by young stars and peaks around 
3-4\,Gyr with a long tail towards higher ages. The ADF
of the most metal-poor stars ($\rm [Fe/H]<-0.5$) is on the other hand dominated 
by stars older that 10-12\,Gyr.  The ADF for the stars with $\rm -0.5<[Fe/H]<0$
shows a mixed age structure with a broad peak around 5-9\,Gyr. The ADF for
the stars with $\rm -0.5<[Fe/H]<0$ divides into two peaks, one 
coinciding with the metal-rich ADF ($\sim 4$\,Gyr) and one coinciding with the
metal-poor ADF (10-12\,Gyr). 

%-----------------------------------------------------------------------
%-----------------------------------------------------------------------

%=======================================================================
\subsection{On the presence of young stars in the bulge}

%-----------------------------------------------------------------------
\begin{figure*}
\resizebox{\hsize}{!}{
\includegraphics[bb=60 185 570 330,clip]{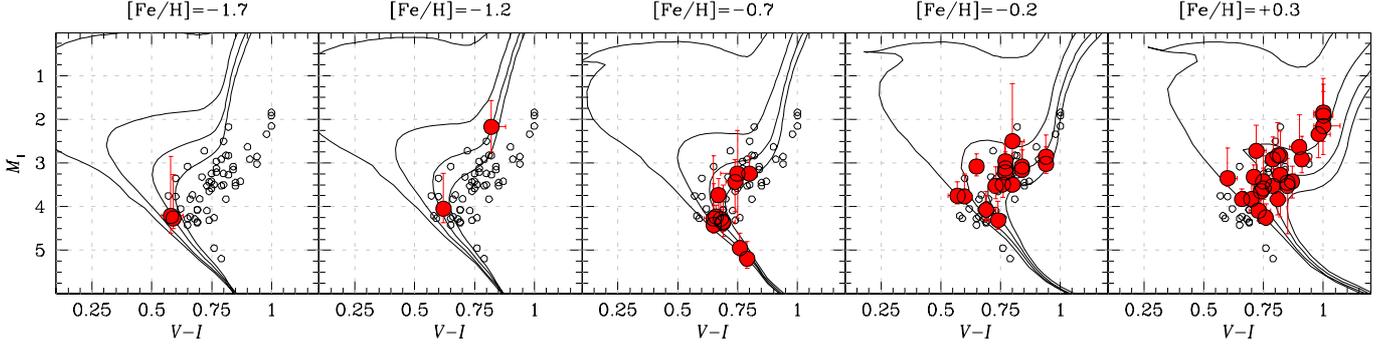}}
\caption{Y$^2$ isochrone plots for different metallicities. Four different 
ages are plotted(1, 5, 10, and 15\,Gyr). In each plot, all 58 stars
in microlensed dwarf sample are included, and those with [Fe/H] 
within $\pm 0.25$\,dex of the isochrone metallicity have been marked by
bigger red circles. 
\label{fig:bias}}
\end{figure*}
%-----------------------------------------------------------------------

The microlensed dwarf sample signals the existence of
a significant fraction of low- and intermediate age stars in the bulge. 
As deep colour magnitude diagrams of the Galactic bulge
\citep[e.g.,][]{holtzman1993,ortolani1995,feltzing2000b,zoccali2003,clarkson2011,brown2010}
show a faint red turn-off, indicative of an exclusively old and
metal-rich population, this is a surprising result. 
For instance, the \cite{clarkson2011} study of blue stragglers concluded
that no more than 3\,\% of the bulge population could
be younger than 5\,Gyr. In our sample of microlensed dwarf
stars 13 out of 58 stars (23\,\%) have ages lower than 5\,Gyr.
However, if the claimed age uncertainties are taken into account,
we only have 3 stars out of 58 ($\sim 5$\,\%) that within 
1\,$\sigma$ have an age lower than 5\,Gyr, which is more or less
in agreement with the estimate from \cite{clarkson2011}.
It should be noted that our sample does indeed contain 13 
stars out of 58 (23\,\%) that have an {\it upper} age limit 
below 9\,Gyr, i.e. taking the age uncertainties into account these stars
are for certain younger than 9\,Gyr, pointing to a significant 
intermediate-age population of stars in the bulge.
While it is feasible to hide a small number of metal-rich 
and young stars ``inside'' a red and old turn-off, but this cannot 
be large.

In Fig.~\ref{fig:bias} we show the microlensed dwarf stars plotted on top
of isochrones with five different metallicities (as indicated on the top
of the figure). Each plot contains all 58 microlensed dwarf stars, 
with the stars that have [Fe/H] within $\pm 0.25$\,dex of the isochrone
metallicity highlighted in red. It is evident that a single set of isochrones
cannot be used for all stars. Especially at low metallicity,
many of the stars that have higher metallicities fall outside the
isochrones. If we were 
to assign a single set of isochrones to the sample, say the $\rm [Fe/H]=-0.2$ set 
of isochrones, it would be difficult to claim a significant 
intermediate-age population in the bulge. As we have no stars in the upper left 
corner of the CMD, i.e., within the 1 to 5\,Gyr isochrones, one would be tempted 
to claim that the bulge turn-off is around 10\,Gyr and that the bulge population
is all old. This is not the case, as can be seen in panel with the $\rm [Fe/H]=+0.3$,
many stars fall on young isochrones, stars that would be classified as old
on a set of more metal-poor isochrones.
This illustrates the importance of metallicity 
as a key ingredient in the determination of the age of the bulge. In many cases, 
such as in photometric studies, the metallicities of the stars are lacking, and 
a metallicity has to be assumed. The microlensed stars show that even though it is 
very difficult to claim stars younger than say 3-4\,Gyr, it is clear that there 
is a substantial fraction of intermediate age (5 to 8\,Gyr).

\cite{nataf2011} found evidence from studying the red giant branch bump
that the bulge should be enriched in helium. This led to  
an attempt by \citet{nataf2012} to reconcile the age situation by 
suggesting that the reason for the apparent miss-match in age between
spectroscopic and photometric results can be related to an enhanced
He/metals content, $Y/Z$, in the bulge as opposed to that in the Sun.
As the stellar isochrones used to infer the ages are calculated using
the standard He abundance the resulting spectroscopic ages might be 
erroneously too low. They discuss two systematic effects that 
arise as a consequence of an age offset that should be a function of
the evolutionary state of the star. The first prediction was that 
this age offset, when applying standard $\log g - \teff$ isochrones,
should be maximised on the subgiant branch and minimised around
the main-sequence turn-off. The second prediction was that the
difference between the true absolute magnitude and the 
spectroscopically inferred absolute magnitude should correlate 
positively with the spectroscopically inferred stellar mass. 
Using the, at the time, 26 available microlensed bulge dwarfs in 
\cite{bensby2010,bensby2011} they found a positive correlation
between the difference in absolute magnitude and stellar mass for 
the stars with $\rm [Fe/H]>0$. This led \cite{nataf2012} to
tentatively claim that the $Y$ abundance should be higher in the 
bulge than in the Sun. 

\begin{figure}
\resizebox{\hsize}{!}{
\includegraphics[bb=18 160 592 718,clip]{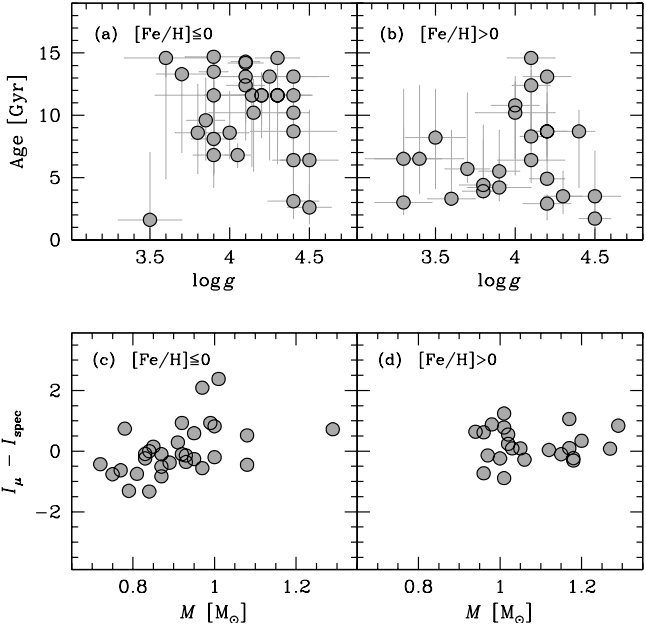}}
\caption{(a) and (b) show the spectroscopic ages as a function of
spectroscopic surface gravity for the metal-poor and metal-rich 
bulge dwarfs, respectively. 
(c) and (d) show the difference between absolute magnitude inferred from 
microlensing techniques and from spectroscopy as a function of inferred 
spectroscopic mass for the metal-poor and metal-rich bulge dwarfs, respectively.
(Compare Figs.~2 and 3 in \citealt{nataf2012}),
\label{fig:I_mass}}
\end{figure}

\begin{figure}
\resizebox{\hsize}{!}{
\includegraphics[bb=18 160 592 395,clip]{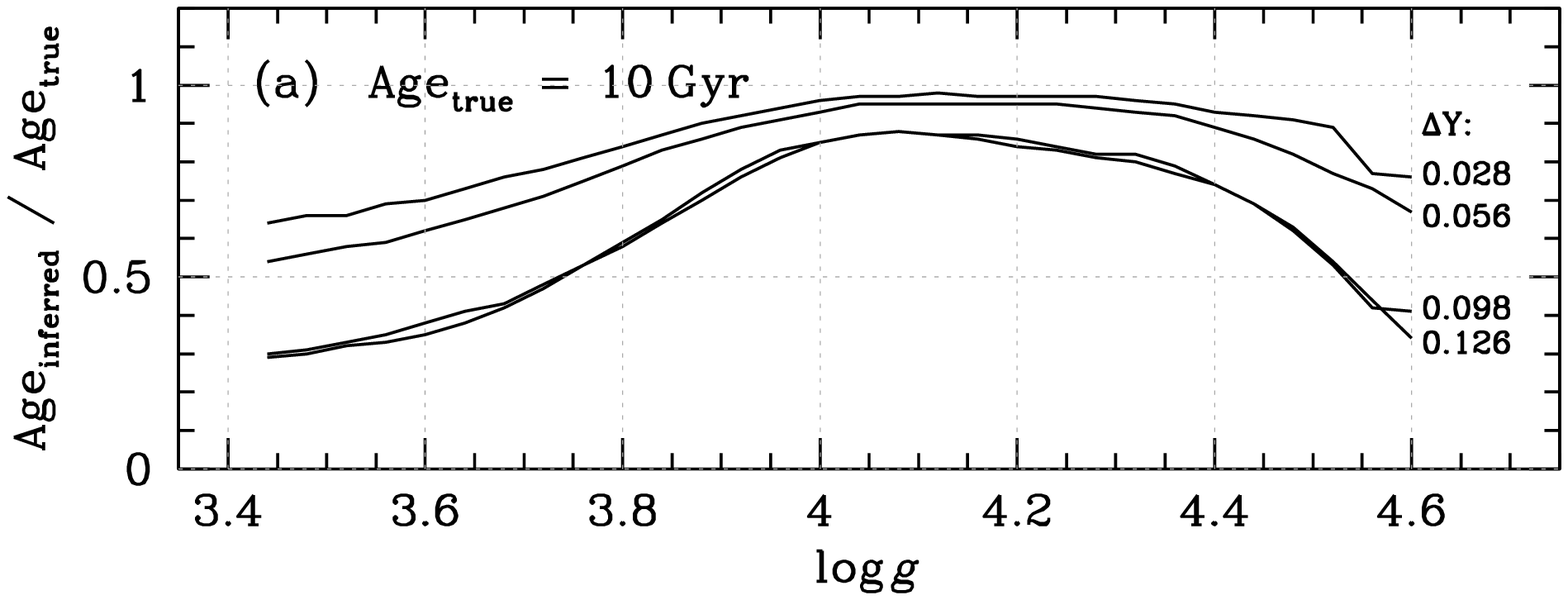}}
\resizebox{\hsize}{!}{
\includegraphics[bb=18 140 592 718,clip]{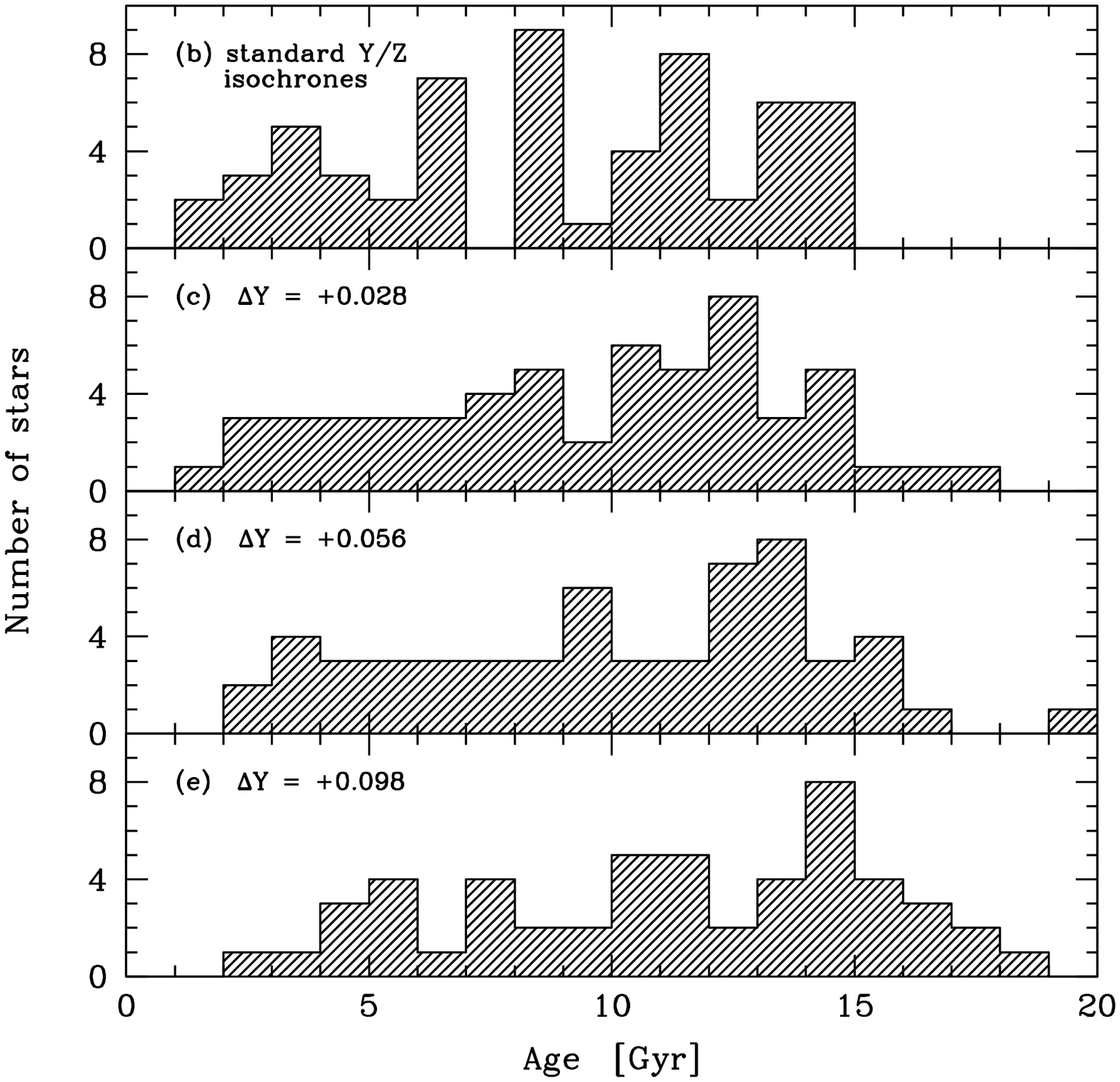}}
\caption{
(a) shows the predicted ratios (taken from Table~1 in 
\cite{nataf2012} for a true age of 10\,Gyr) between the inferred
spectroscopic age and the true age versus surface gravity for four 
different increases of the He abundance ($\Delta$Y, as indicated on 
the right-hand side). 
(b) shows the age distribution for the microlensed dwarf sample. 
(c)-(e) show how the age distribution changes with $\Delta Y$
if the ages are corrected following the relationships in (a).
\label{fig:nataf2012}}
\end{figure}

The plots in Fig.~\ref{fig:nataf2012} are 
reproductions of the bottom plots in Figs.~2 and 3 of 
\cite{nataf2012} but with the now much expanded sample of 58
microlensed bulge dwarfs. 
Fig.~\ref{fig:I_mass}a and b show the stellar age versus 
$\log g$ for the metal-poor and the metal-rich bulge dwarfs, 
respectively. Fig.~\ref{fig:I_mass}c and d show the difference between 
the absolute $I$ magnitude inferred from microlensing techniques and
spectroscopy versus the spectroscopically inferred stellar mass for the
metal-poor and metal-rich bulge dwarfs, respectively.
It is clear that the trend with stellar mass
that was found for the metal-rich bulge dwarfs is
no longer present. Also, for the stars with sub-solar [Fe/H], where
they found a negative trend with stellar mass, there is now a 
positive trend. If anything, this shows that a statistically significant
sample of microlensed bulge dwarf stars is needed to firmly establish
whether there are any trends or not. 

%-----------------------------------------------------------------------
%-----------------------------------------------------------------------
%-----------------------------------------------------------------------
%-----------------------------------------------------------------------

Figure~\ref{fig:nataf2012}a shows the \cite{nataf2012}
prediction (their Table~1) of the ratio between the spectroscopically
inferred age to a ``true" age versus surface gravity in the case of
a true age of 10\,Gyr. 
As can be seen it would be possible to derive a too low 
He abundance in the models responsible for some of the young stars we see in 
the bulge. For instance, old ages could be recovered
for some of the stars with ages less the 5\,Gyr that have
$\log g\lesssim 3.8$ or $\log g\gtrsim4.5$ if the He abundance were
increased by 0.1 or more. However, this is a dramatic change, and
there would still be stars with ages around, or less than, 5\,Gyr
between $4\lesssim \log g\lesssim 4.3$, where an increase in the He 
abundance will not have an equally large impact on the spectroscopically
inferred stellar ages. In Fig.~\ref{fig:nataf2012}b we show the age
distribution for the 58 micro lensed dwarf stars and 
Figs.~\ref{fig:nataf2012}c-e then show how this age distribution 
changes when the effects predicted in Fig.~\ref{fig:nataf2012}a are applied 
to the stars. Even though there are stars that become substantially older, 
it is not enough, even in the extreme case of $\Delta Y = +0.098$, to 
to contradict the conclusion that the age distribution has 
a substantial fraction of young 
and intermediate age stars.

There is other evidence, from AGB stars, of an intermediate age
population in the bulge \citep[e.g.,][]{vanloon2003,cole2002,uttenthaler2007}.
\cite{uttenthaler2007} find evidence for Tc in a sub-sample of
their C-stars, indicative of third dredge up and a minimum stellar
mass of 1.5\,M$_{\odot}$ which implies an upper age limit of 3\,Gyr.
\cite{vanloon2003} used near infrared CMDs from DENIS and ISOGAL
to simultaneously derive the extinction, metallicity and age of
individual stars. They found that the inner region (the inner
10{\degr}) are dominated by an old population ($>7$\,Gyr), but that
an intermediate age population is also present. This is consistent
with their finding of a few hundred AGB stars with heavy mass loss.
Using data from 2MASS, \cite{cole2002} were able to identify
a population of carbon stars in the Galaxy that traced the
bar. Their analysis finds that these stars very likely are of
intermediate age. \cite{groenewegen2005} also found Mira variables
with ages of only 1-3\,Gyr at latitudes between 
$-$1.2{\degr} to $-$5.8{\degr}. That stars have formed recently ``around'' 
the bar is not unexpected \citep[e.g.,][]{sanchezblazquez2011}. 
\cite{cole2002} discuss the possibility that the carbon stars have 
wandered in to the bulge/bar region, but are unable to draw firm conclusions.

%=======================================================================
\subsection{The $A_{max}$ puzzle}
\label{sec:amax}

%-----------------------------------------------------------------------
\begin{figure}
\resizebox{\hsize}{!}{
\includegraphics[bb=18 160 460 710,clip]{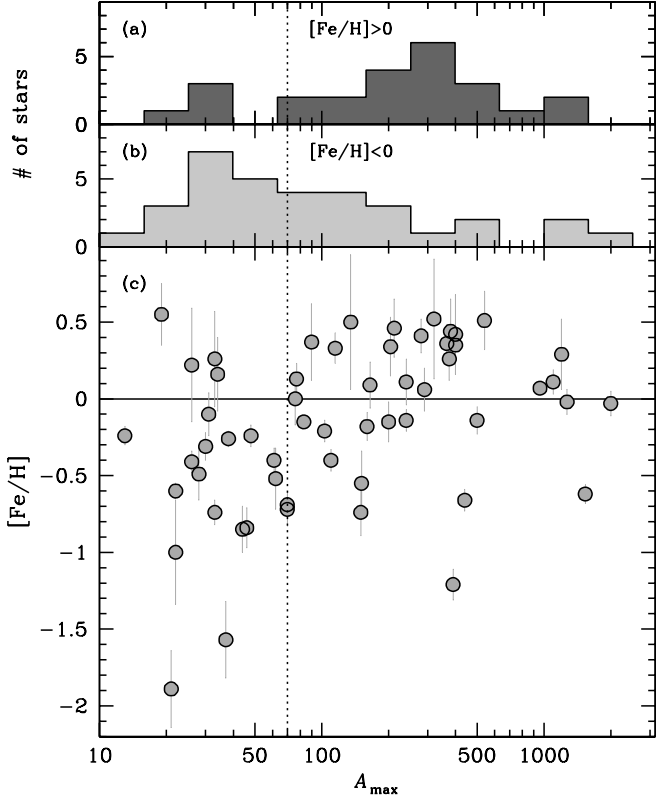}}
\caption{[Fe/H] versus $A_{\rm max}$ for the microlensed dwarf sample.
\label{fig:fehamax}}
\end{figure}
%-----------------------------------------------------------------------

Using at the time 16 available microlensing events
\cite{cohen2010puzzle} reported an unexpected correlation between
the metallicity of the microlensed star and the maximum magnification
of the microlensing event in the sense that metal-rich stars had higher 
maximum magnifications than metal-poor ones. Several possibilities 
were investigated but none could satisfactory resolve the situation.

Figure~\ref{fig:fehamax}c shows [Fe/H] versus $A_{\rm max}$ for the now much
larger sample of microlensed dwarf stars. Compared to when the analysis
by \cite{cohen2010puzzle} was done, the sample now contains high-magnification
events at low metallicities and low-magnification events at 
high metallicities. However, looking at the histograms in Fig.~\ref{fig:fehamax}a 
and b, it is clear that the metal-rich part of the sample contains 
a higher fraction of high-magnification events than the
metal-poor part. Classifying a microlensing event with $A_{\rm max}\geq70$
as a high-magnification event, and otherwise a low-magnification event
(vertical dotted line in Fig.~\ref{fig:fehamax}),
the metal-rich sample contain 84\,\% high-mag events, compared to 51\,\%
for the metal-poor stars. As it is highly unlikely that the maximum 
magnification should have any dependence on the chemical composition 
of the background star, this correlation
must be a result of some selection bias or being due to that the
environment of the metal-rich stars is different from the 
environment of the metal-poor stars. 
In the following section we will investigate
how big such a bias might be and whether the differences in the $A_{\rm max}$
distributions between the metal-rich and metal-poor samples can be accounted 
for.

%=======================================================================
\subsection{Sampling bias?}

The selection of candidate microlensed bulge dwarfs for spectroscopy 
is made from the OGLE and MOA surveys, which image through $I$ and $R/I$ 
filters, respectively.
The cutoff in brightness for effective high dispersion spectroscopy even with
8-10\,m telescopes, given that target of opportunity observations cannot
exceed 1 or 2 hours,  is $I$ brighter than about 15.0\,mag.

We attempt to model our sampling process to look for selection
effects.  We assume a distance to the bulge of 8.2\,kpc and
a mean extinction of $A_{\rm I}\approx 1.0$. %  I(obs) - M_I  = 15.57 MAG
We adopt version 2 of the Yale-Yonsei evolutionary tracks and 
isochrones \citep{yi2003,demarque2004} and select five representative 
choices of [Fe/H], $-0.90$, $-0.43$, +0.05, +0.39, and +0.60~dex.
Stars with subsolar metallicities are assumed to have
$\rm [\alpha/Fe]= +0.3$, while higher metallicity stars have no
$\alpha$-enhancement.  We adopt the Salpeter IMF.

These isochrones are normalized to have 1000 stars in their 
initial mass function with masses between 0.5 and 1.0$M_{\odot}$.  
The summed stellar mass for each our five metallicity choices for 
isochrones with age 4\,Gyr was compared; there is a total range of 2\,\%,
which we can ignore for present purposes.

If we assume that all the bulge stars have the same age,
irrespective of metallicity, then at ages older than
$\sim5$\,Gyr, the main sequence turnoff is $\sim$0.4\,mag fainter
for bulge dwarfs with [Fe/H] super-solar than it is for
dwarfs with $\rm -0.9\lesssim[Fe/H]\lesssim-0.5$.  This would suggest
that in the microlensed sample, the intrinsically fainter
high-metallicity dwarfs would be under-represented.

\begin{figure}
\resizebox{\hsize}{!}{
\includegraphics[bb=18 140 592 713,clip]{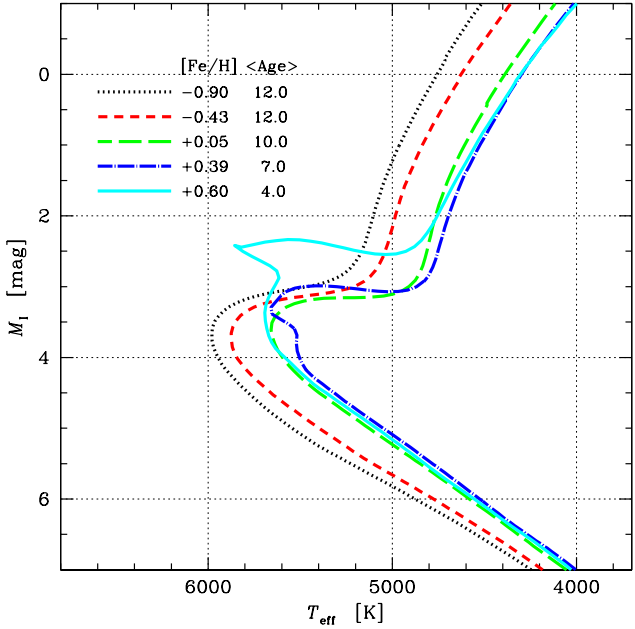}}
\caption{Y$^2$ isochrones for five different metallicities
with different mean ages that decrease with increasing metallicity.
}
\label{fig:iso_mixed}
\end{figure}

\begin{figure}
\resizebox{\hsize}{!}{
\includegraphics[bb=18 140 592 550,clip]{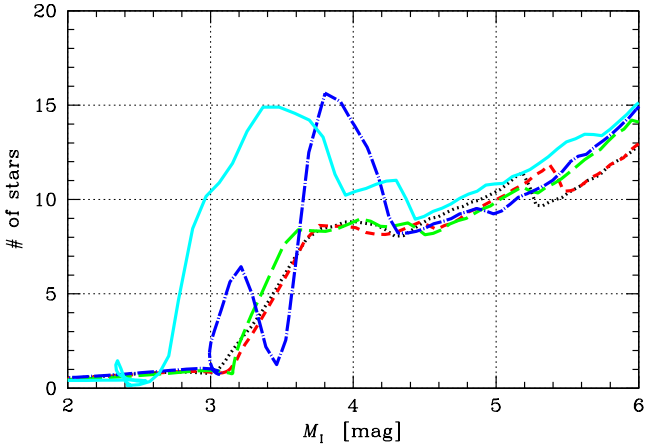}}
\caption{Number of stars along the isochrones shown in 
Fig.~\ref{fig:iso_mixed}.
}
\label{fig:iso_mixed_counts}
\end{figure}

However, a key result of this paper is the apparent dependence
of mean age of the bulge dwarfs with metallicity (see Fig.~\ref{fig:agefe}).  
If we adopt this as true, and allow the stellar mean age to decrease as the 
metallicity increases, with the most metal-poor stars having an age of 
12\,Gyr and the most metal-rich ones 4\,Gyr,  a different result emerges; 
the younger population of super-solar metal-rich bulge dwarfs then has a 
main sequence turnoff that is essentially identical to that of the more 
metal-poor, older bulge populations, with a substantial blue hook as is 
shown in Fig.\ref{fig:iso_mixed}. This is manifested in the star counts 
in the regime $2.2 < M_{\rm I} < 3.8$, which are dominated by the youngest, 
most metal-rich bulge turnoff region stars.  In this regime, the star 
counts from this young population are a factor of $\sim2$ or more higher 
than those from populations with ages of 10 to 12\,Gyr 
(see Fig.~\ref{fig:iso_mixed_counts}).

%-----------------------------------------------------------------------
%-----------------------------------------------------------------------
%-----------------------------------------------------------------------
\begin{figure}
\resizebox{\hsize}{!}{
\includegraphics[bb=18 140 592 550,clip]{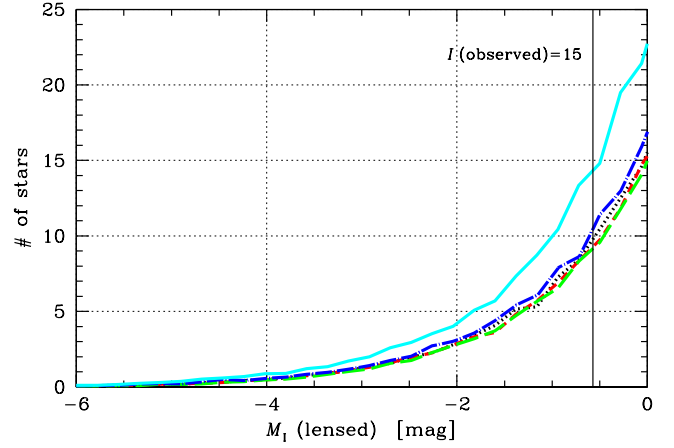}}
\caption{Number of stars along the isochrones shown in 
Fig.~\ref{fig:iso_mixed}, but now with microlensing of the
sample included.
}
\label{fig:lens_mixed}
\end{figure}
%-----------------------------------------------------------------------

When we simulate lensing of this sample, this difference of a factor
of two is smoothed out somewhat, but is still present
(see Fig.~\ref{fig:lens_mixed}).
At $M_{\rm I} {\rm (lensed)} = -0.6$, corresponding to the limiting magnitude of
our spectroscopic sample, the youngest, most metal-rich population
has star counts which are about 50\,\% higher
than those of any of the older, more metal-poor populations  with the same 
total stellar mass.  Hence we would
overestimate the fractional contribution to the total bulge population
of young, metal-rich stars. For a system
of uniform age 10\,Gyr, the super-solar stars would be under-represented
in the lensed star counts by about a factor of 30\,\%, and would then
be more difficult to detect in our microlensed bulge dwarf sample.

Thus, assuming the adopted metallicity -- age relation shown
in Fig.~\ref{fig:agefe} is valid, the fraction of the bulge stellar 
population which is metal-rich and young is over-estimated in our current 
microlensed bulge dwarf sample of 58 stars by $\sim$50\,\%.  Note that 
a turnoff star from the lower metallicity populations with 
$M_{\rm I}\approx +3.8$ has to be lensed by a factor of 55 to be bright 
enough to be included in the spectroscopic microlensed bulge stellar sample.

%-----------------------------------------------------------------------
\begin{figure}
\centering
\resizebox{\hsize}{!}{
\includegraphics[bb=18 140 592 550,clip]{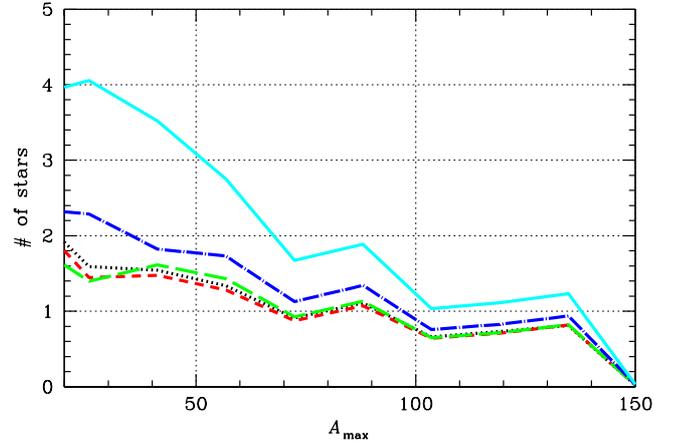}}
\caption{Simulated magnification distribution for the isochrones shown in 
Fig.~\ref{fig:iso_mixed}.
}
\label{fig:amp_mixed}
\end{figure}
%-----------------------------------------------------------------------

Next, we simulate the distribution of $A_{\rm max}$ in an attempt to
determine whether selection effects arising from the
age-metallicity relation of Fig.~\ref{fig:agefe} could be producing the
apparent deficiency of low metallicity, high $A_{\rm max}$ stars
seen in Fig.~\ref{fig:fehamax}.

Our simulation for $A_{\rm max}$ is shown in Fig.~Ê\ref{fig:amp_mixed}.
We find that there is a strong trend for the young, metal-rich
population to dominate over the more metal-poor and older
populations which is strongest at the lowest magnifications
and decreases towards higher magnification.  Since there
are more intrinsically luminous stars in the former, they
will dominate at the lower magnification levels.  At high magnification, 
the main sequence turnoff can be boosted
into the observed sample for the entire range of metallicities
probed, and hence the bias favoring the highest metallicity young
population is reduced.
We thus deduce that our sample of microlensed bulge
dwarfs is biased in favor of high-metallicity young stars,
and hence our derived MDF is also biased.
However, it does not appear possible to reproduce
the apparent relative absence of low-metallicity high magnification
events with our simulation.

%=======================================================================
\subsection{Abundance trends}

%-----------------------------------------------------------------------
\begin{figure*}
\resizebox{\hsize}{!}{
\includegraphics[bb=18 220 580 460,clip]{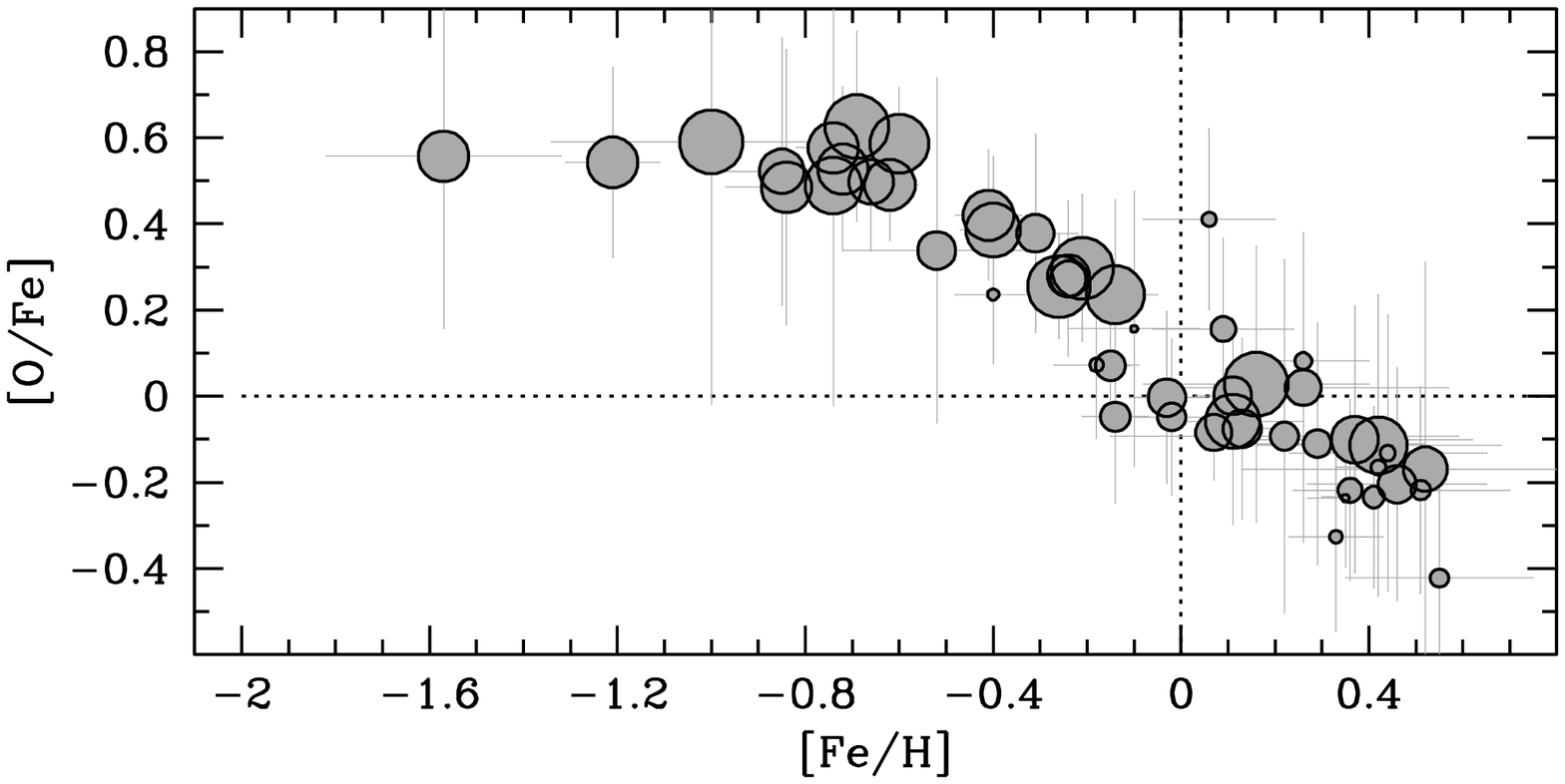}
\includegraphics[bb=30 220 592 460,clip]{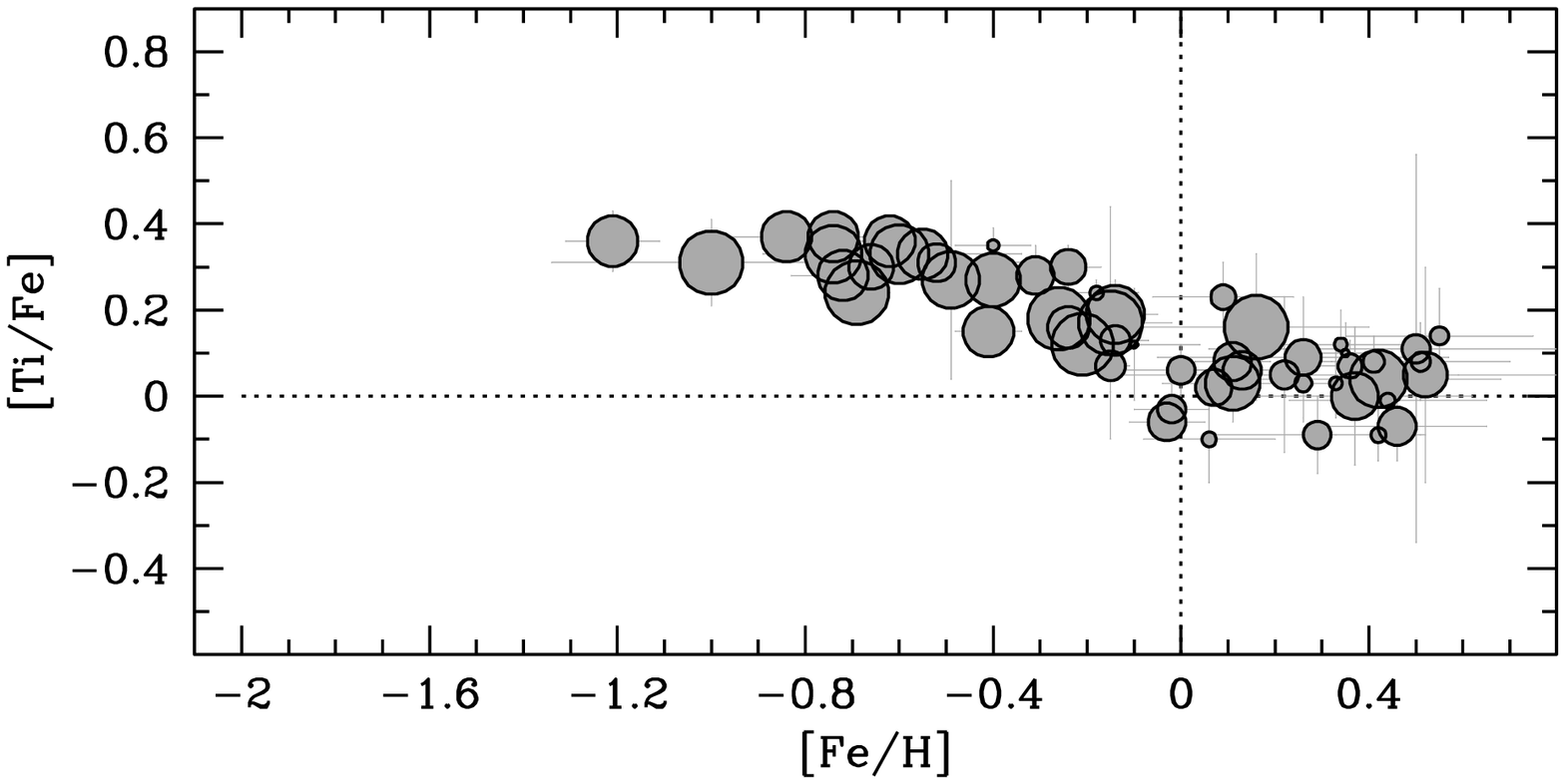}}
\resizebox{\hsize}{!}{
\includegraphics[bb=18 220 580 455,clip]{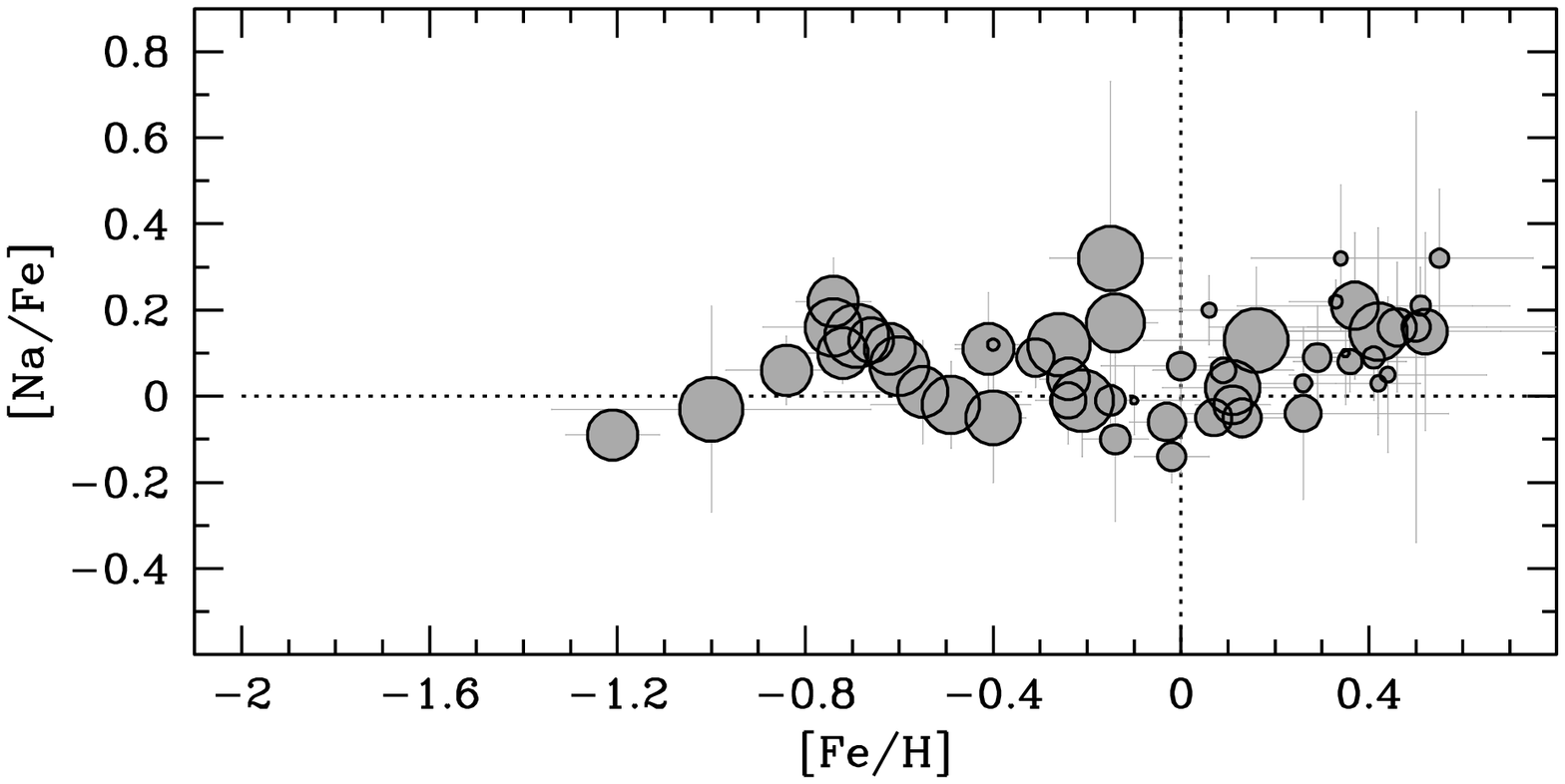}
\includegraphics[bb=30 220 592 455,clip]{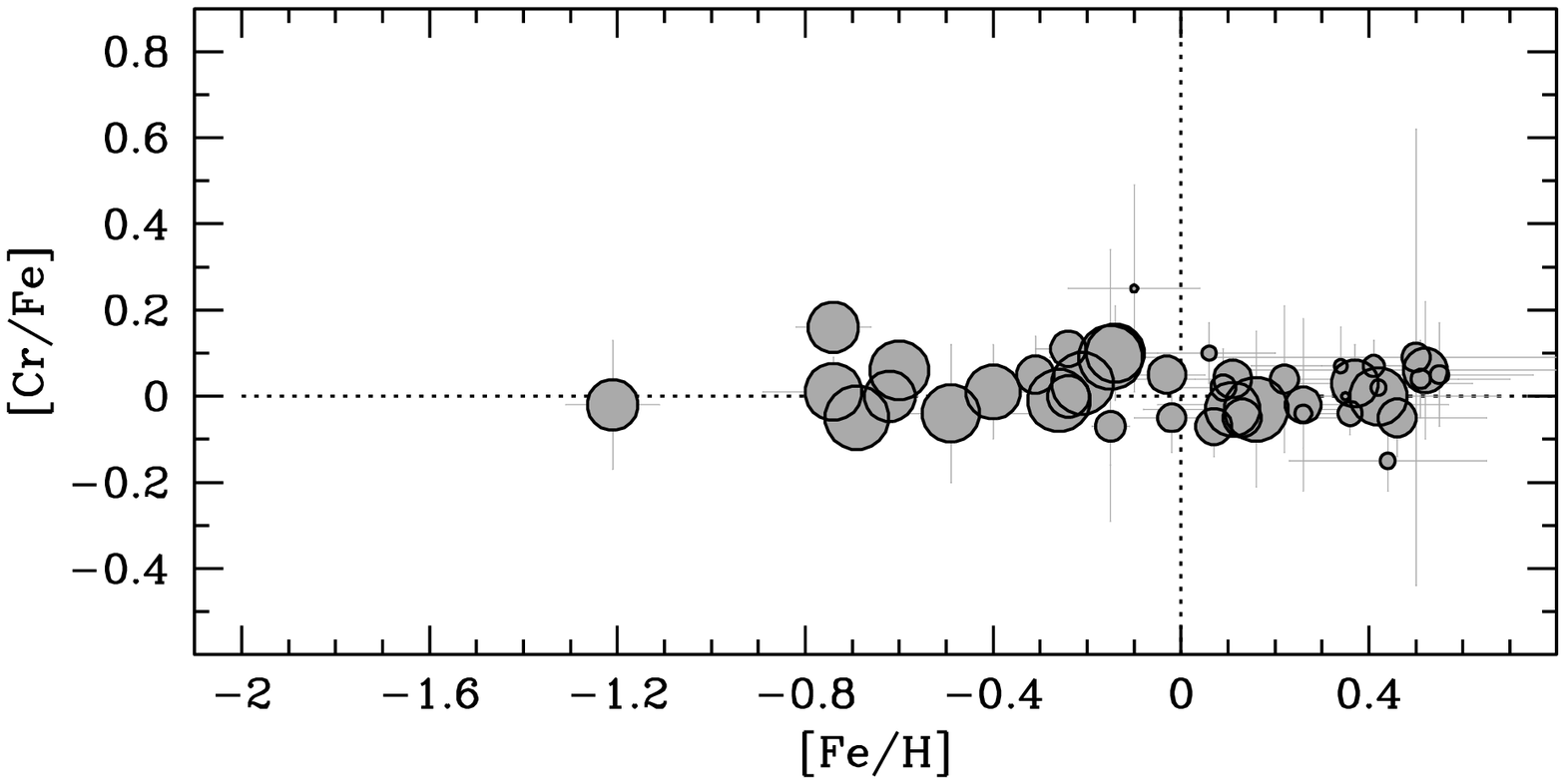}}
\resizebox{\hsize}{!}{
\includegraphics[bb=18 220 580 455,clip]{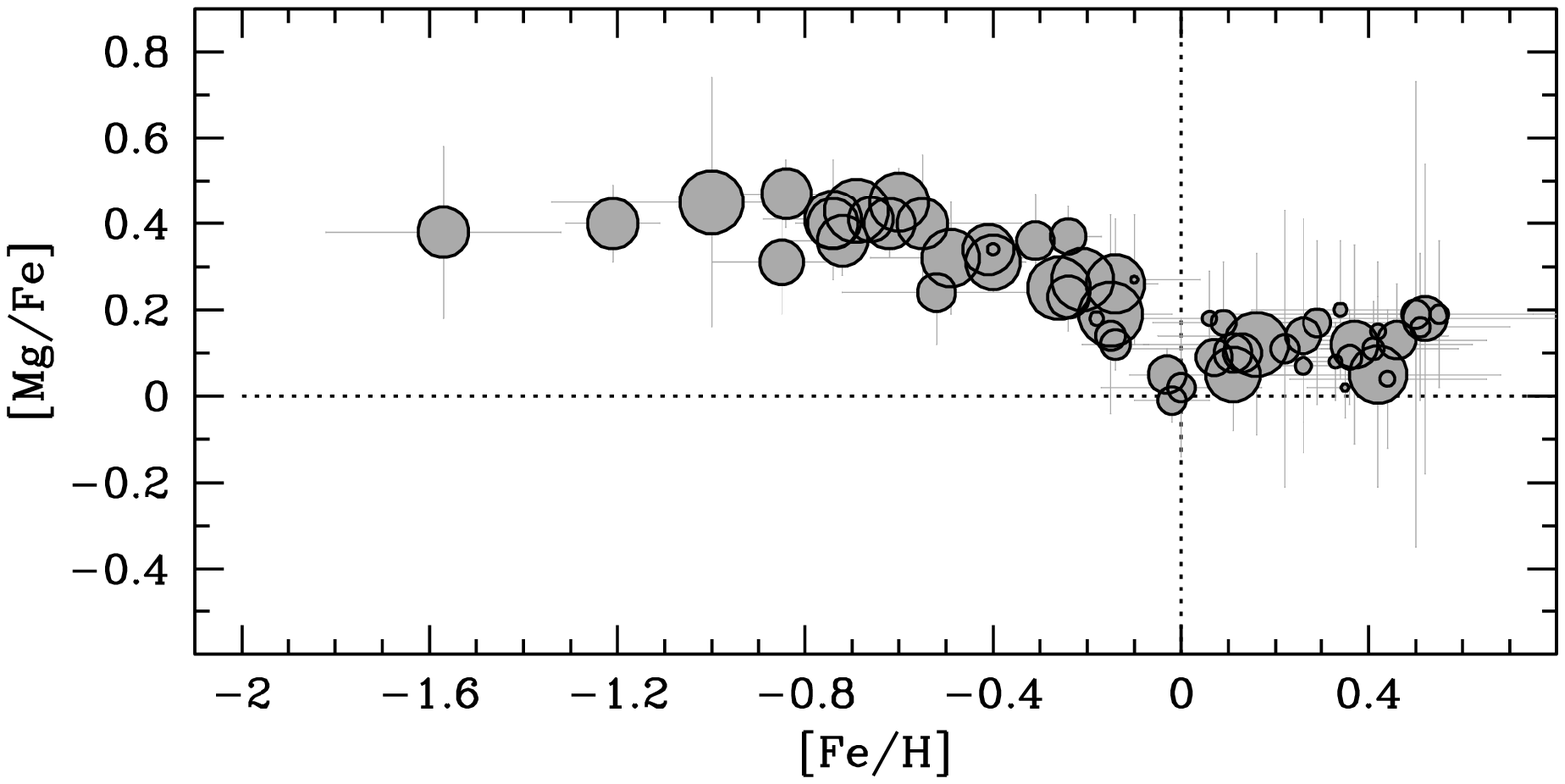}
\includegraphics[bb=30 220 592 455,clip]{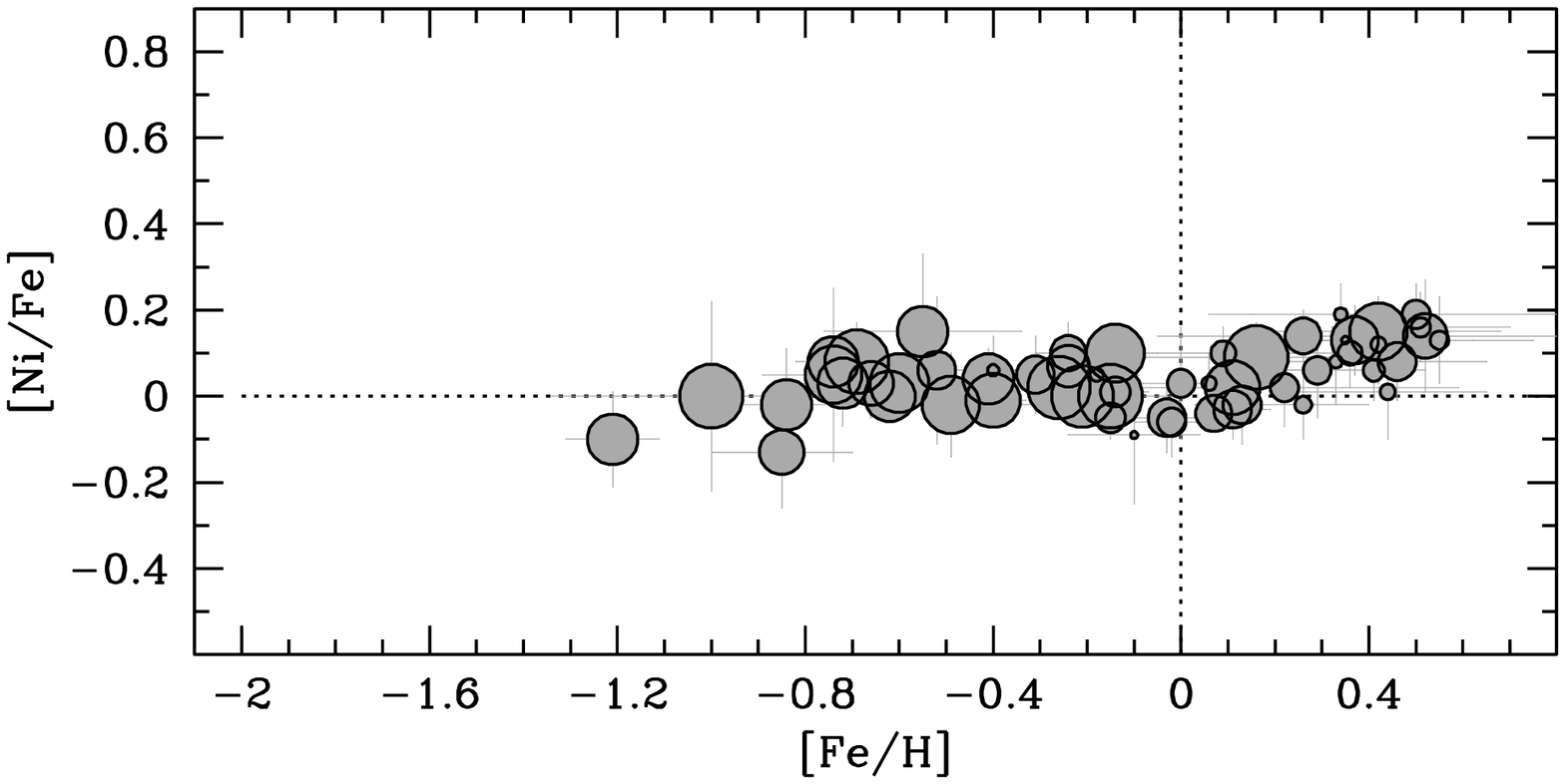}}
\resizebox{\hsize}{!}{
\includegraphics[bb=18 220 580 455,clip]{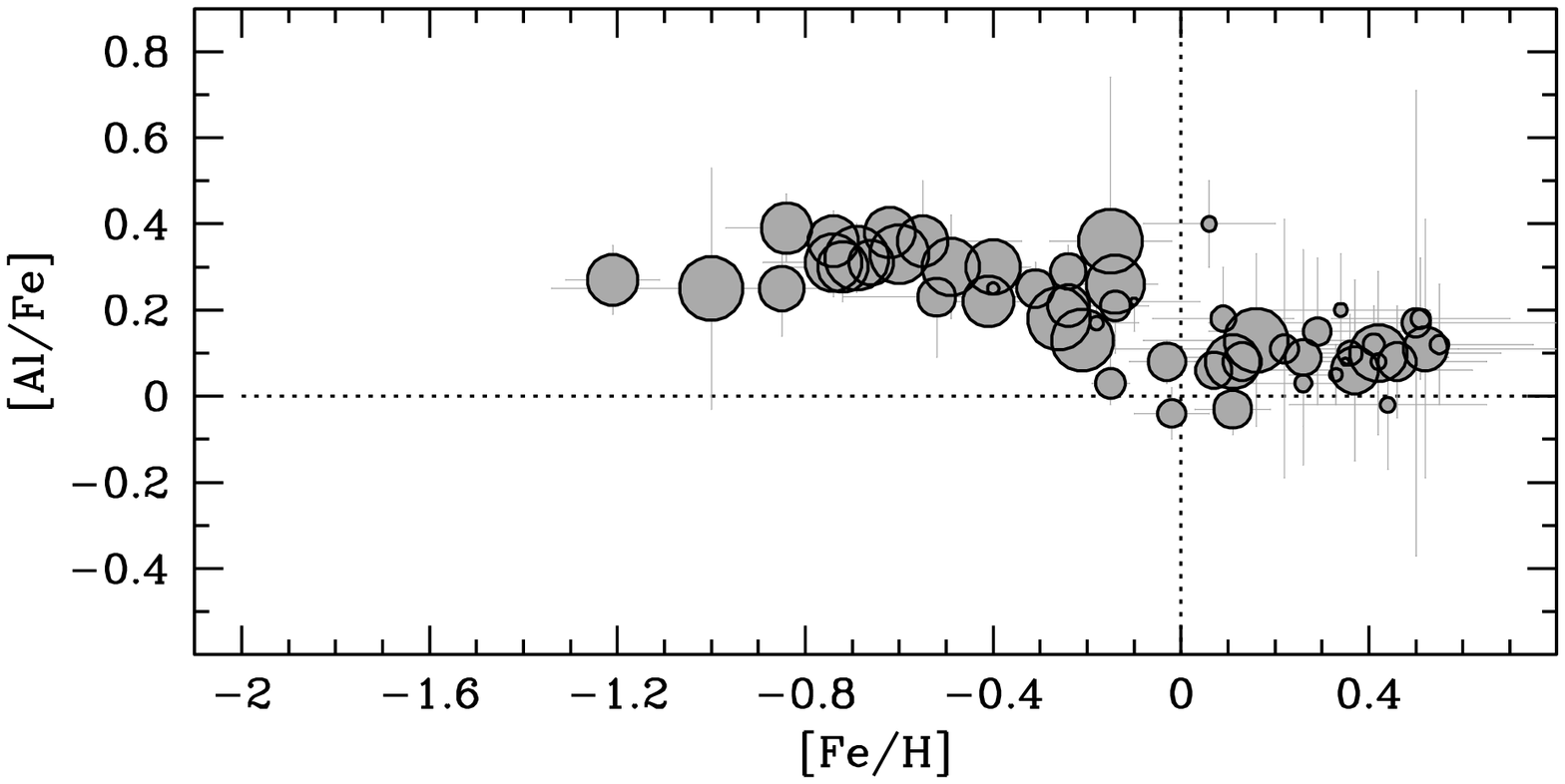}
\includegraphics[bb=30 220 592 455,clip]{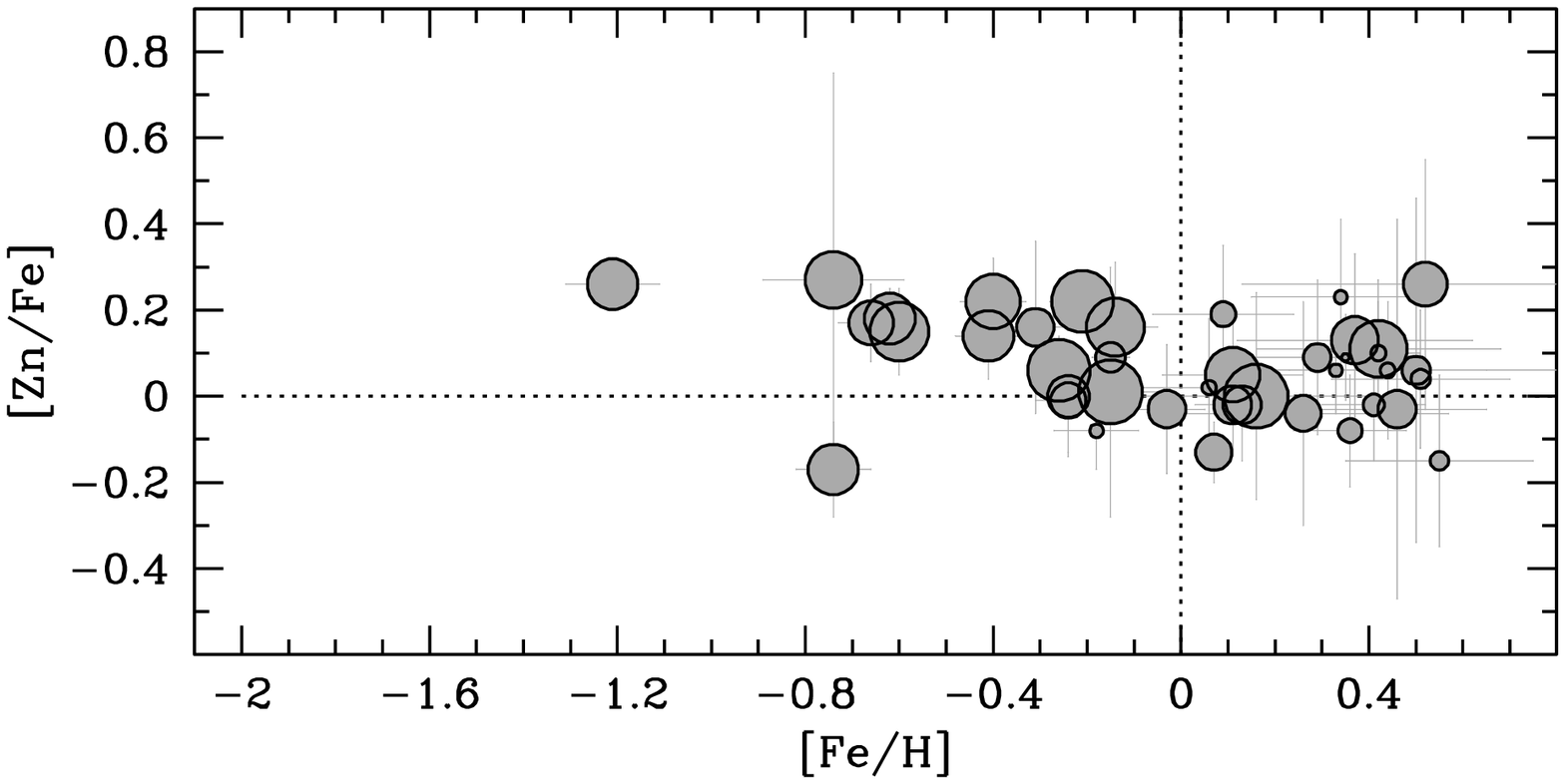}}
\resizebox{\hsize}{!}{
\includegraphics[bb=18 220 580 455,clip]{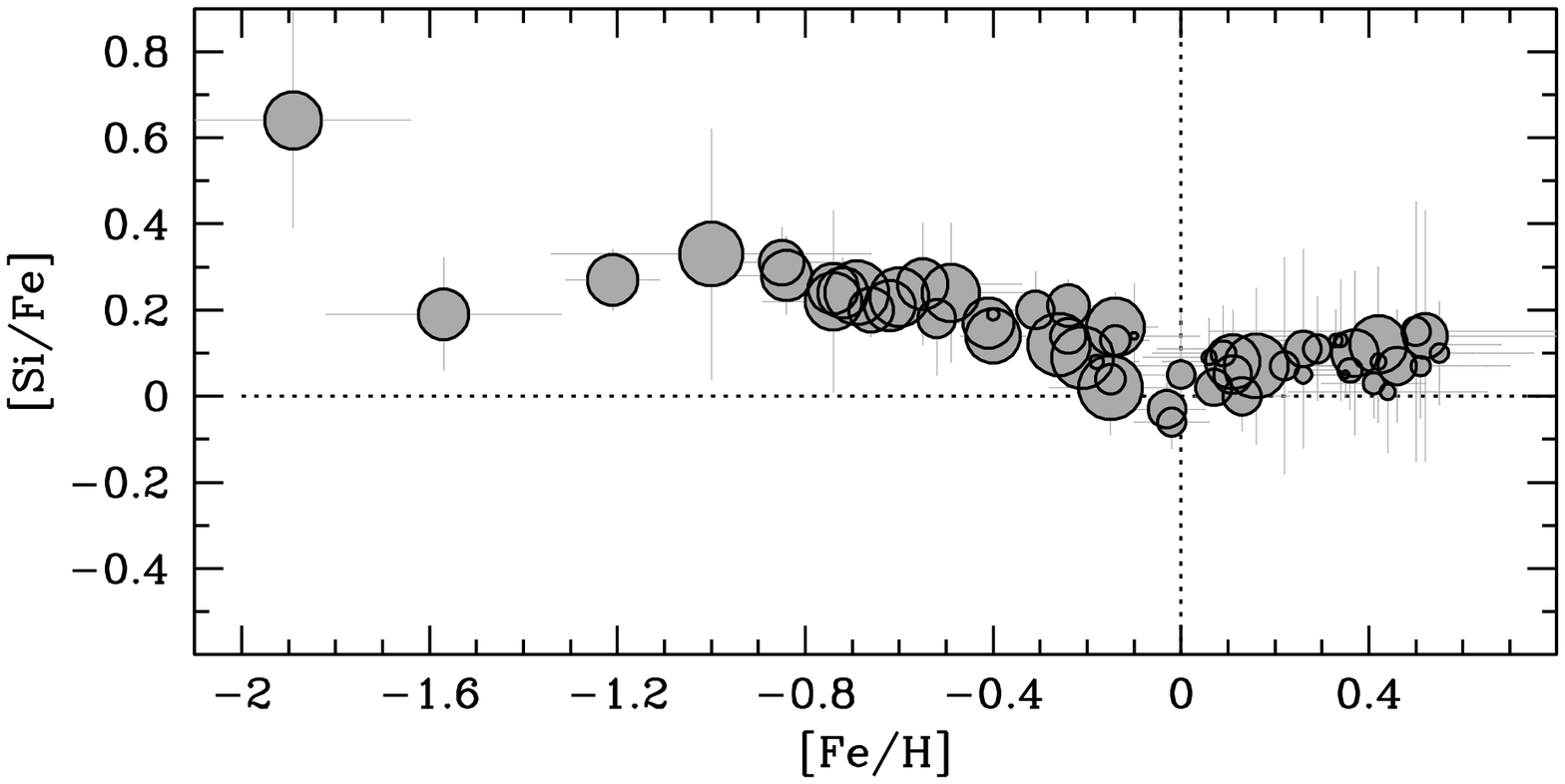}
\includegraphics[bb=30 220 592 455,clip]{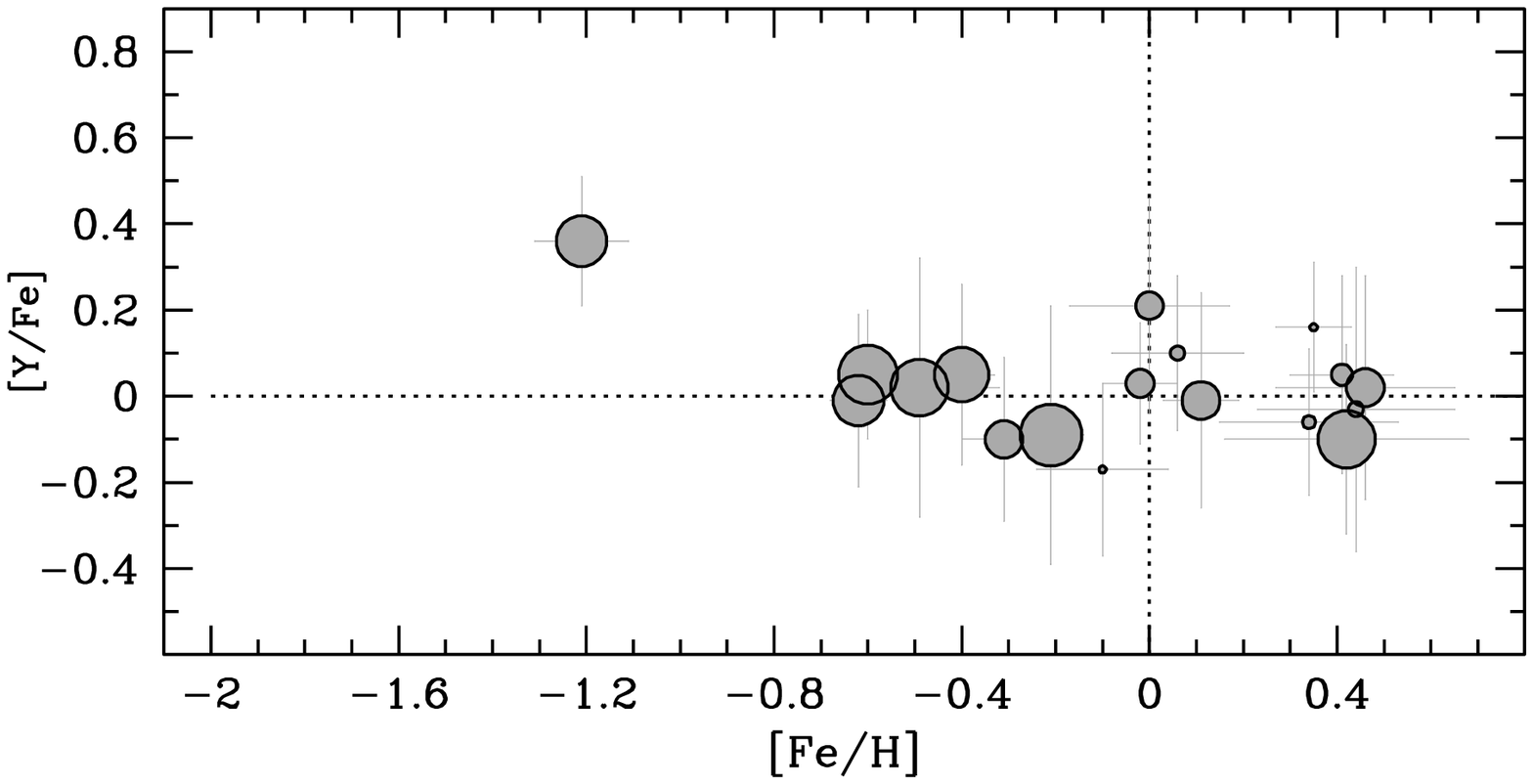}}
\resizebox{\hsize}{!}{
\includegraphics[bb=18 180 580 455,clip]{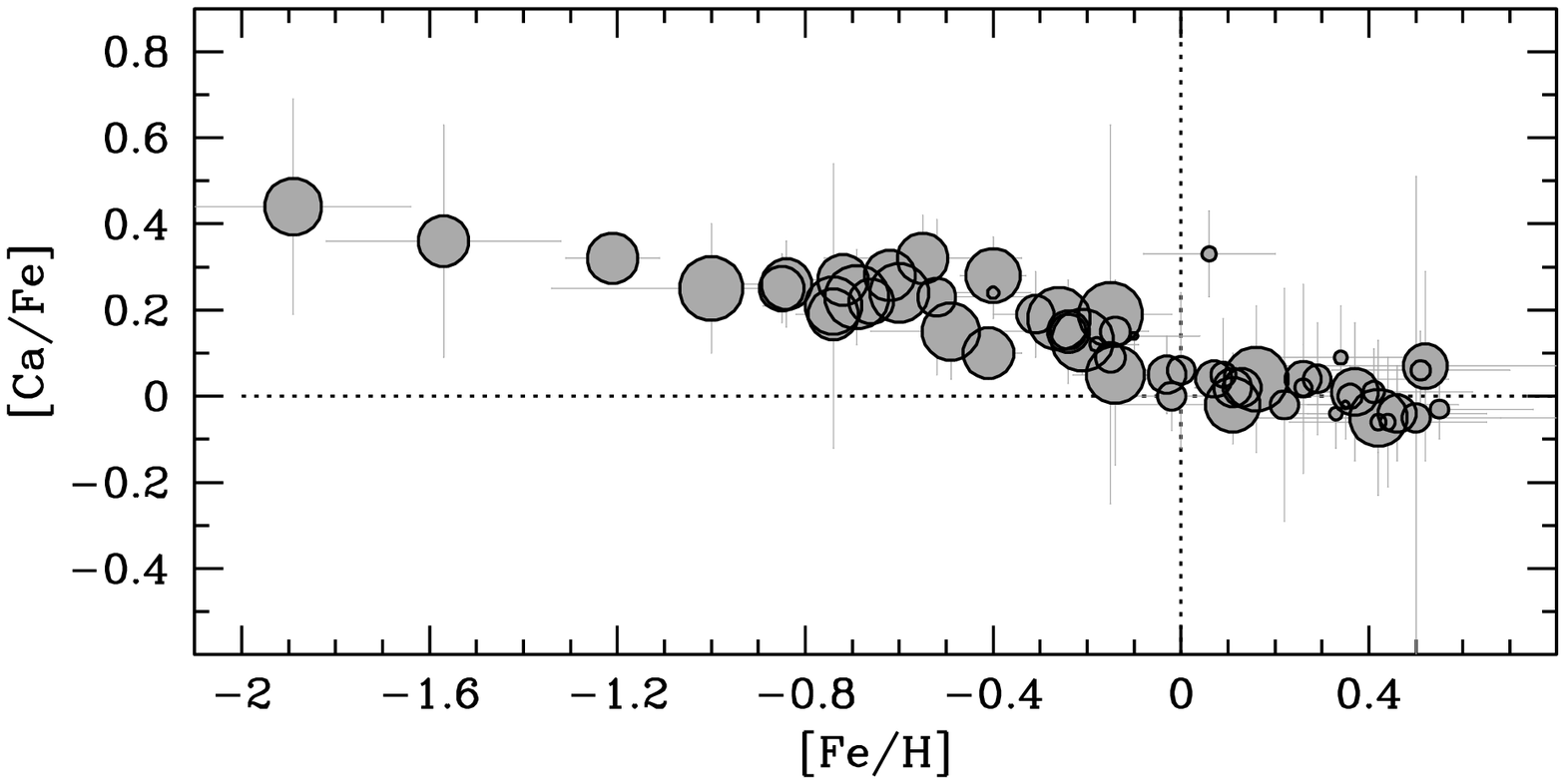}
\includegraphics[bb=30 180 592 455,clip]{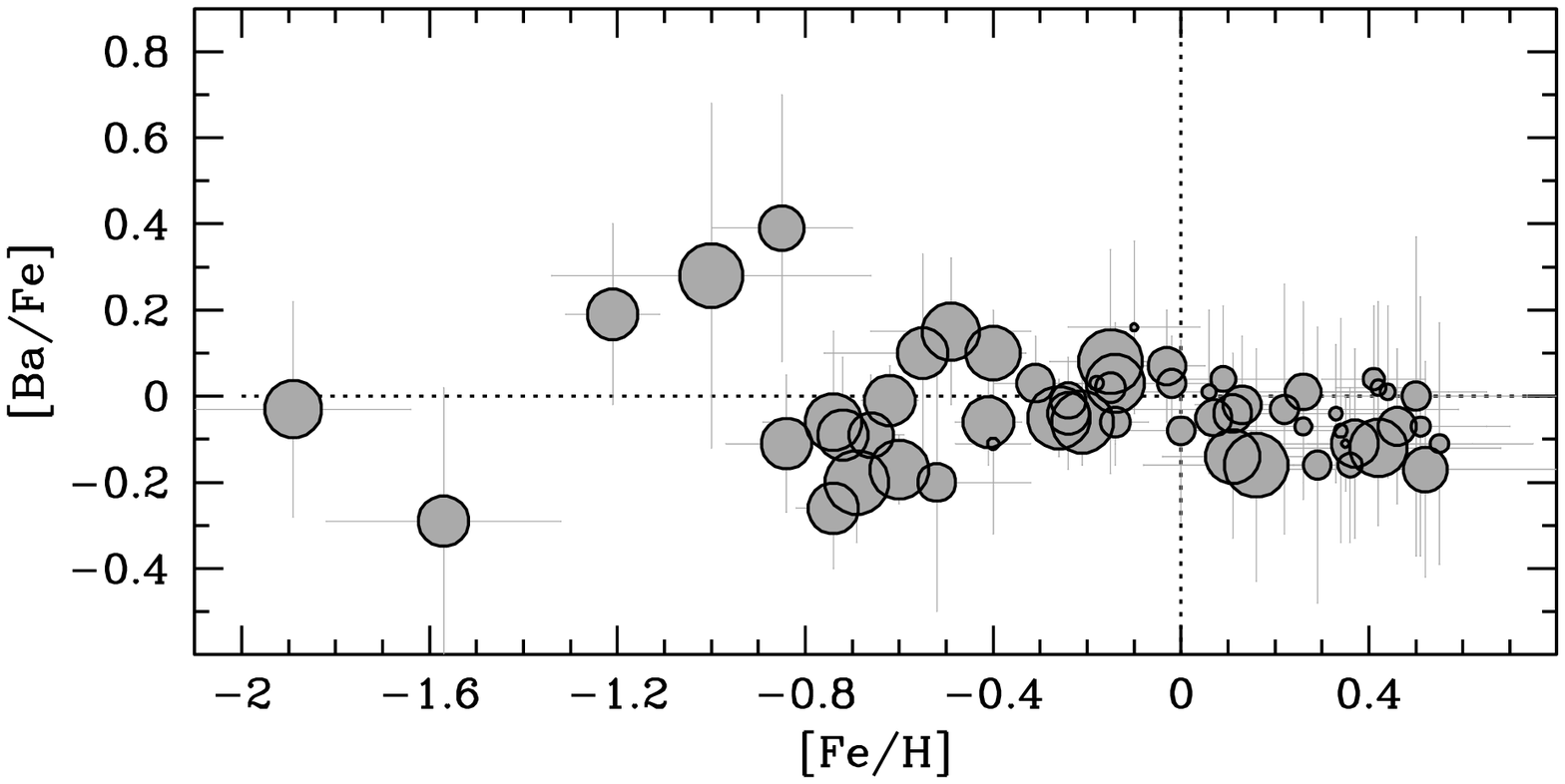}}
\caption{Abundance trends with Fe as reference element for the microlensed 
dwarf sample. Circles have been scaled with the ages of the stars.
\label{fig:abundances}}
\end{figure*}
%-----------------------------------------------------------------------
%-----------------------------------------------------------------------
\begin{figure*}
\resizebox{\hsize}{!}{
\includegraphics[bb=18 220 580 460,clip]{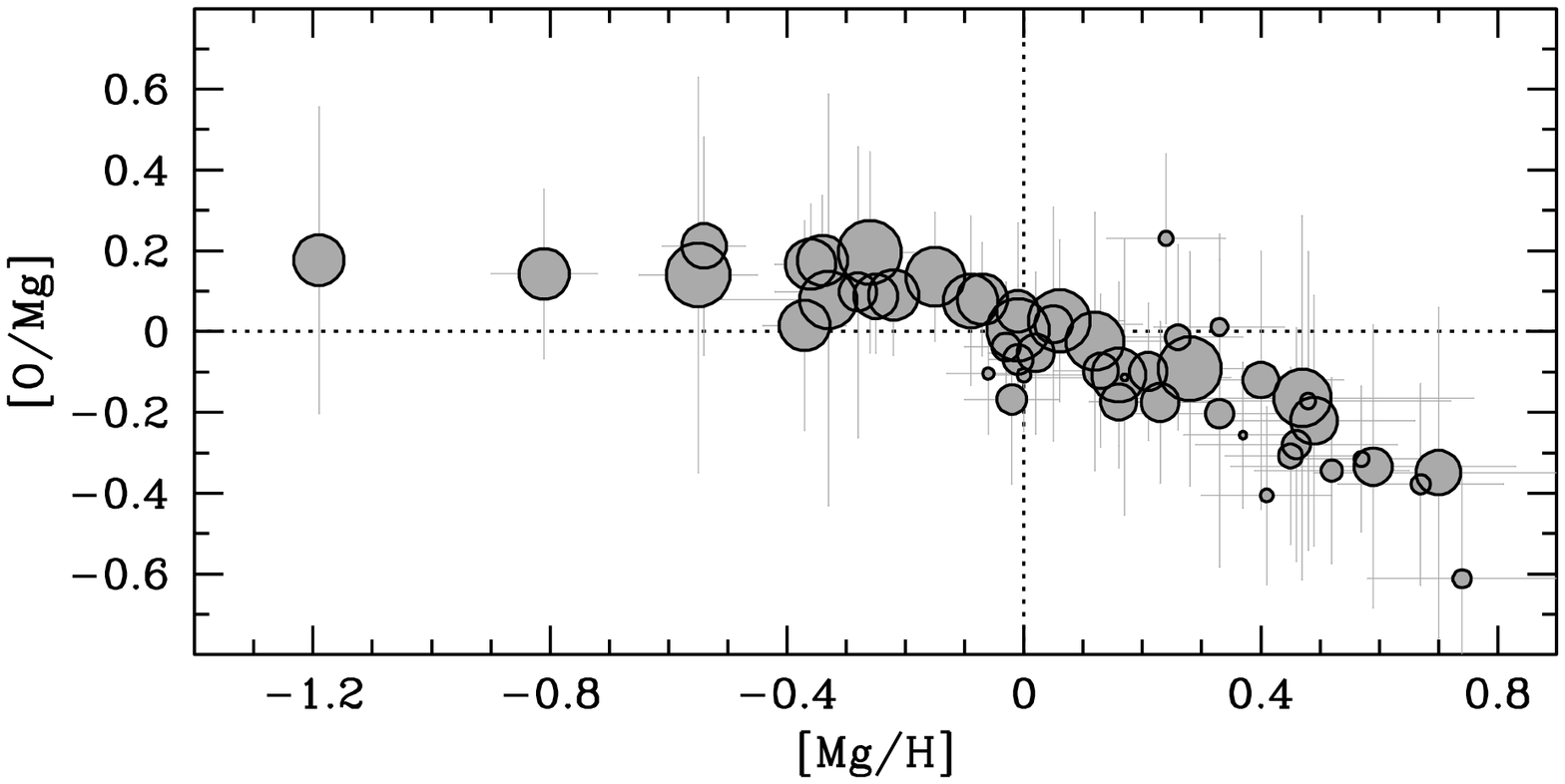}
\includegraphics[bb=30 220 592 460,clip]{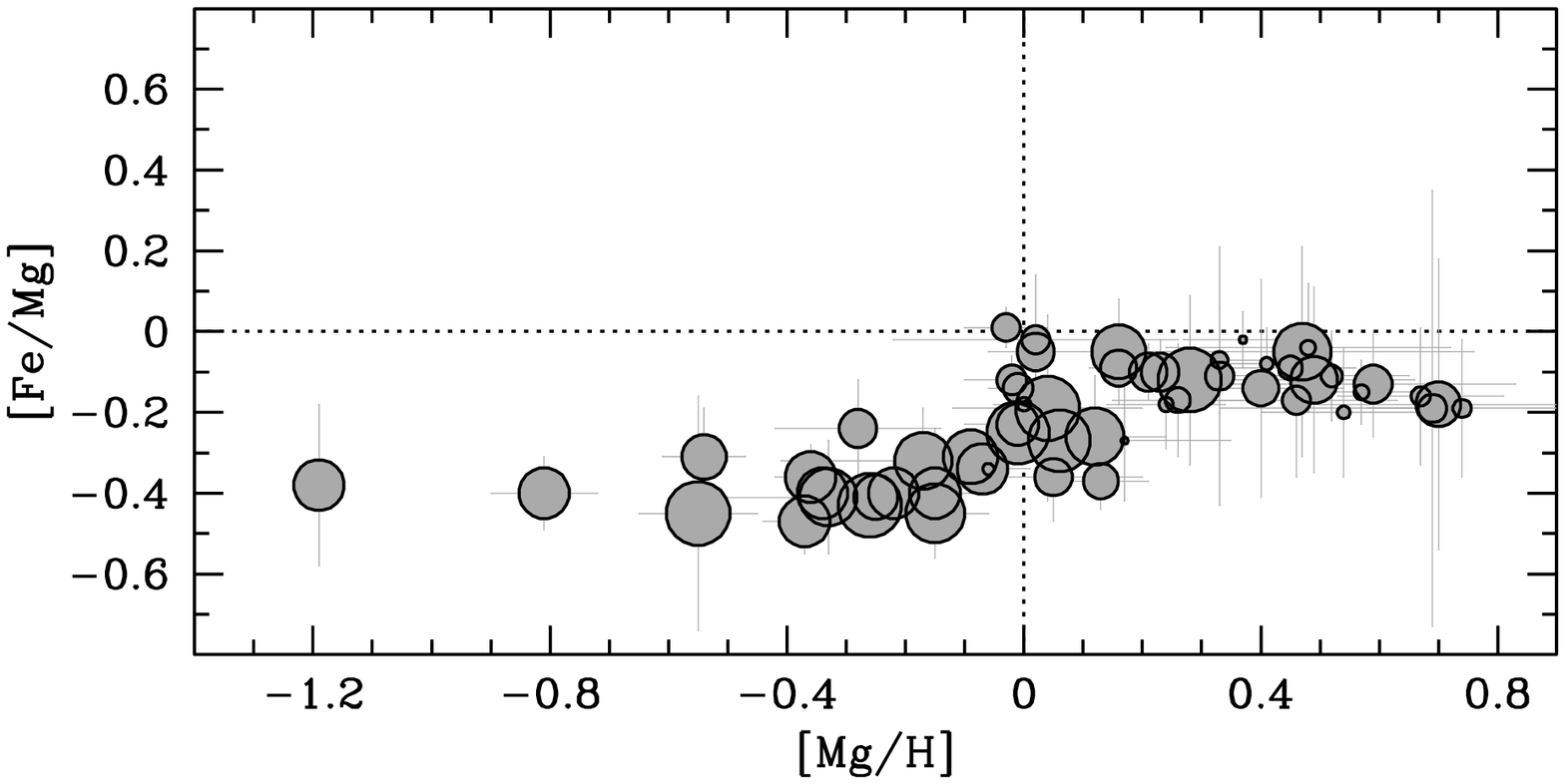}}
\resizebox{\hsize}{!}{
\includegraphics[bb=18 220 580 455,clip]{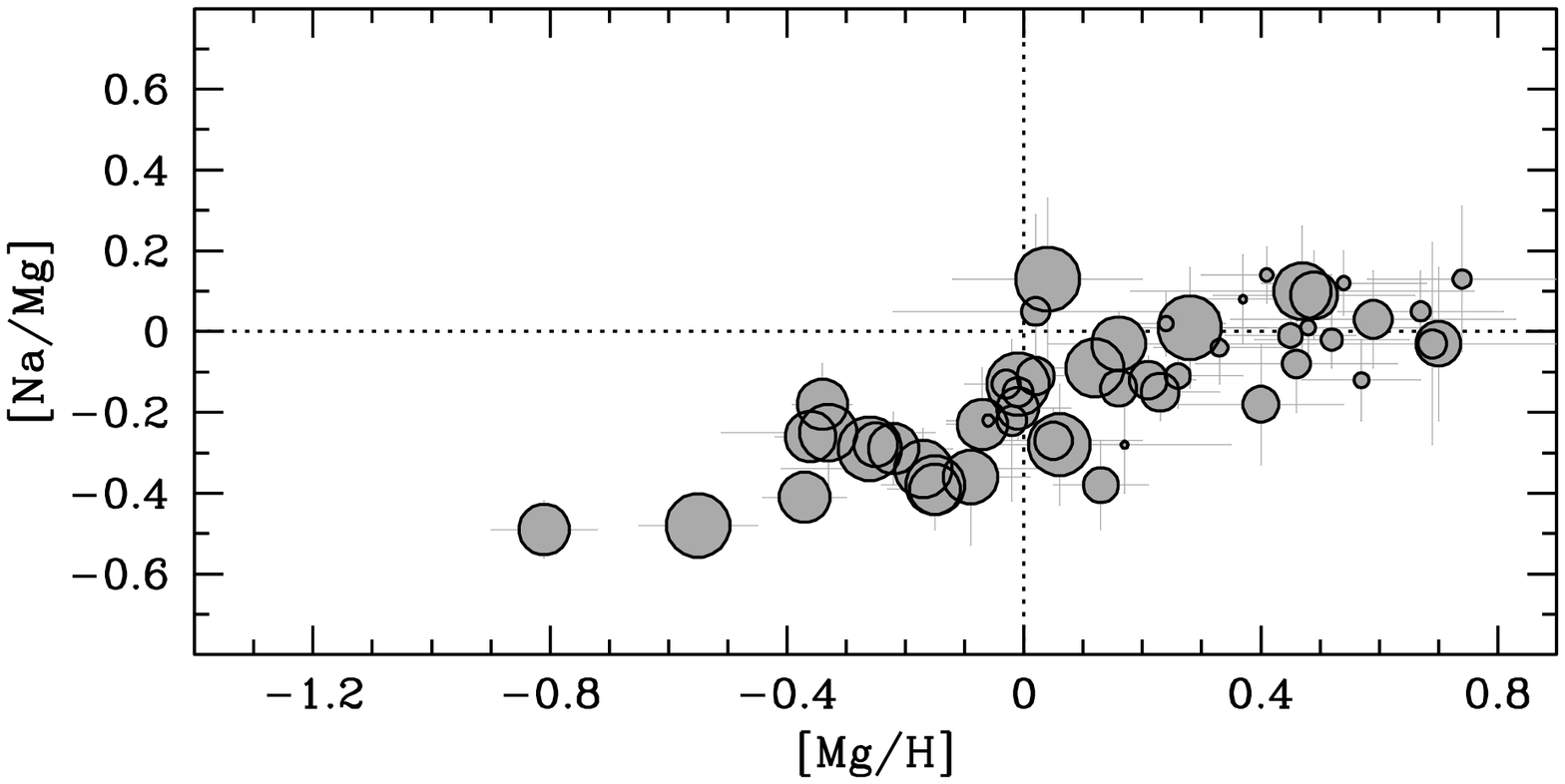}
\includegraphics[bb=30 220 592 455,clip]{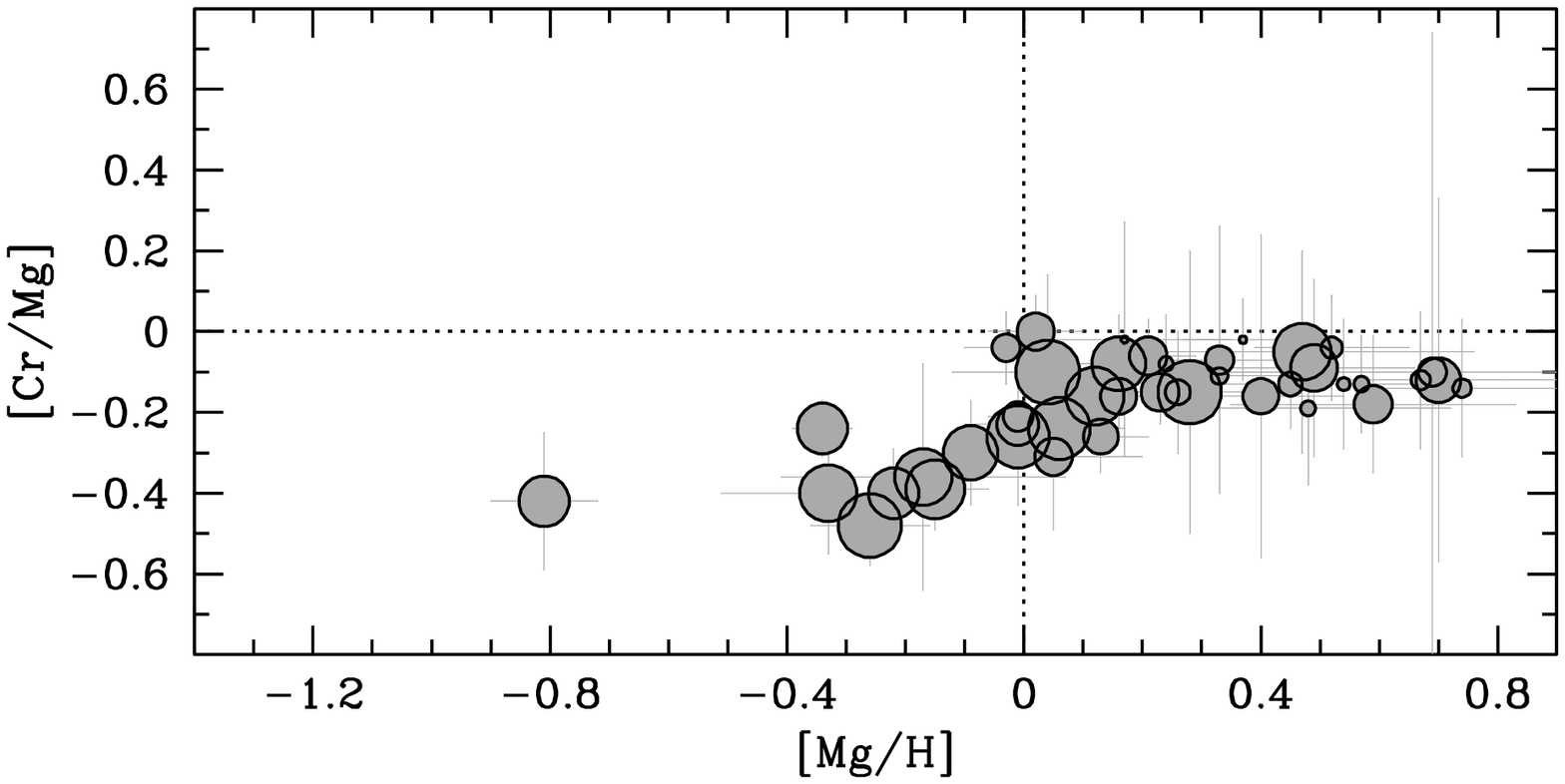}}
\resizebox{\hsize}{!}{
\includegraphics[bb=18 220 580 455,clip]{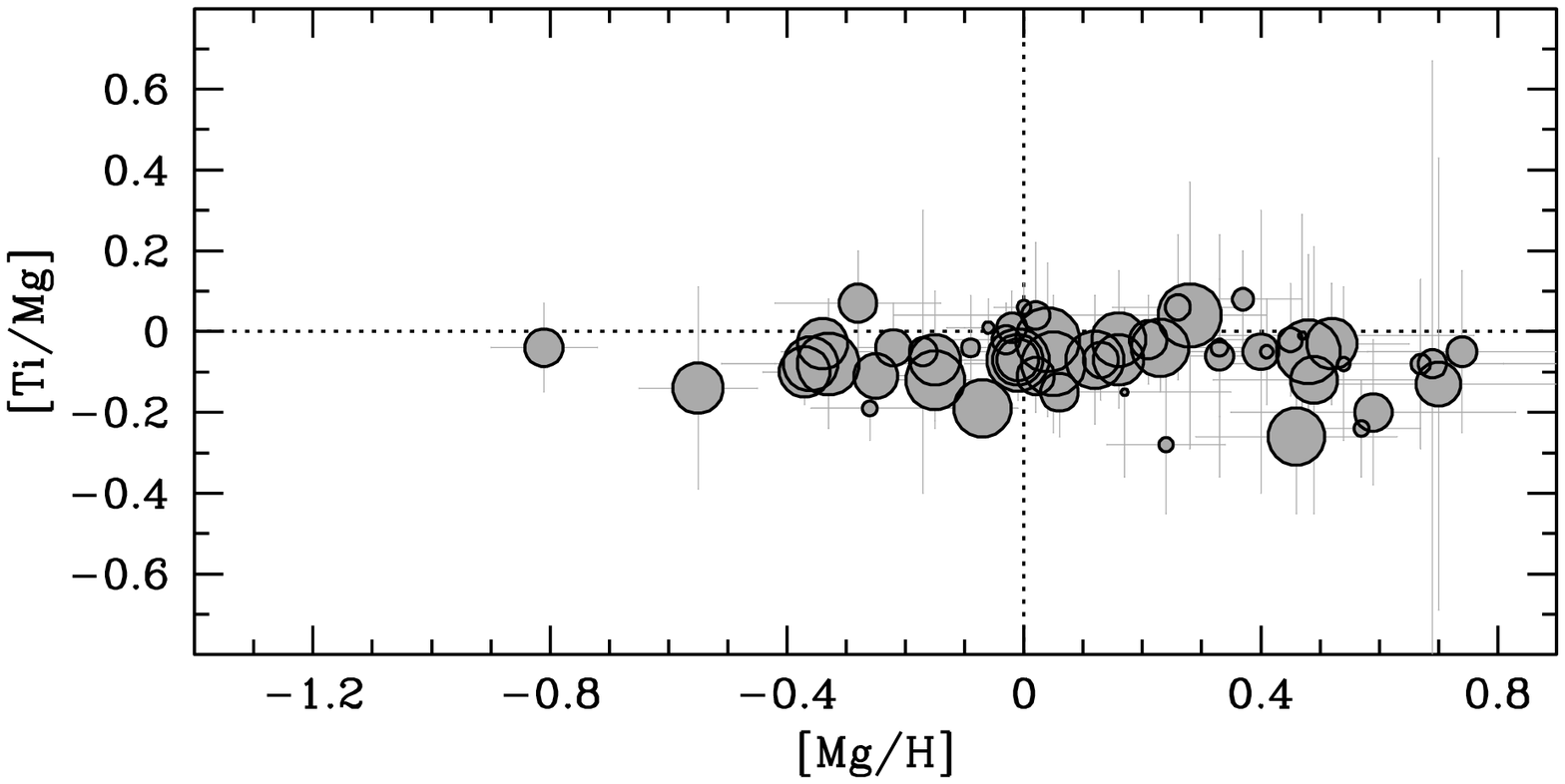}
\includegraphics[bb=30 220 592 455,clip]{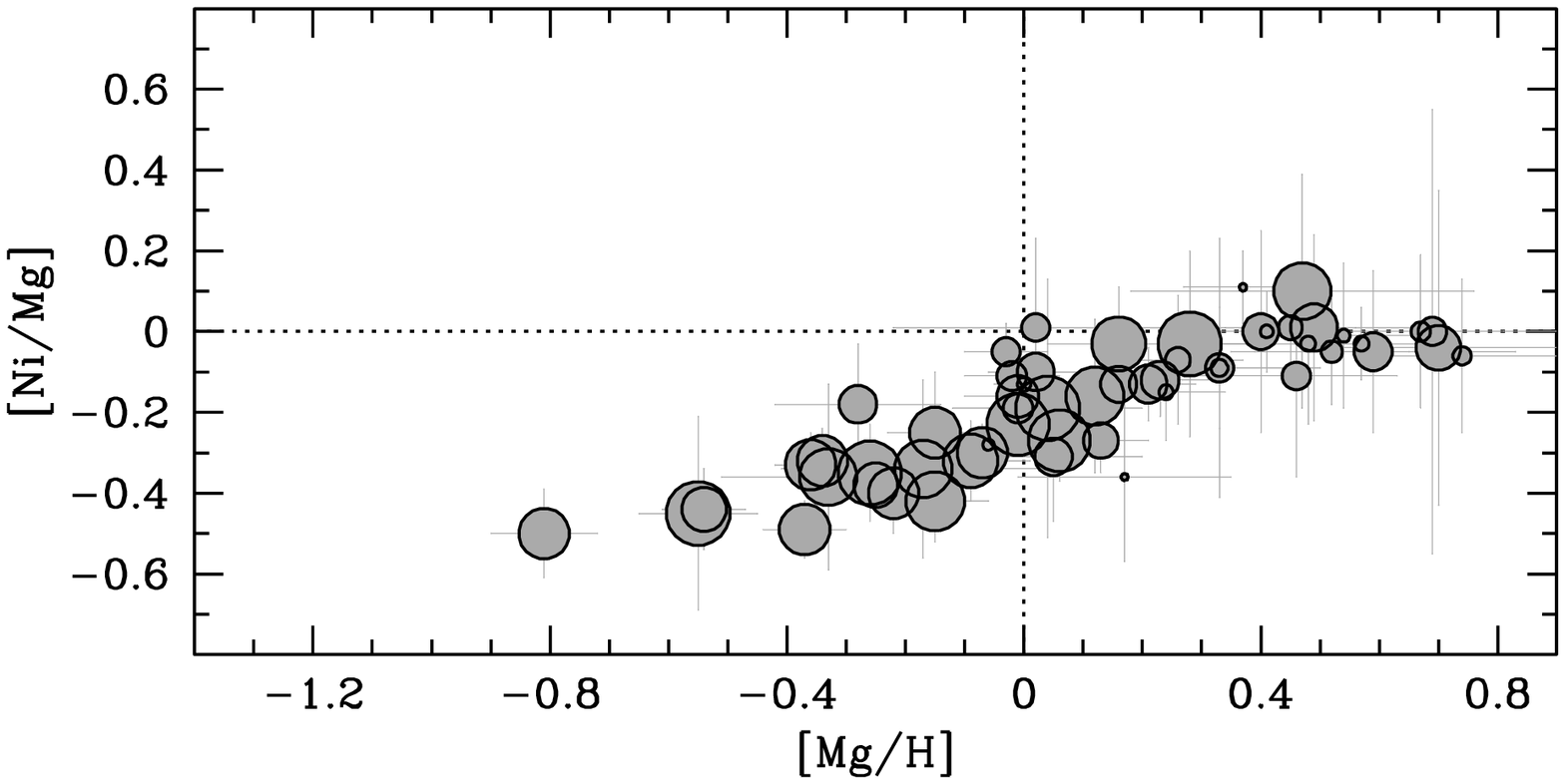}}
\resizebox{\hsize}{!}{
\includegraphics[bb=18 220 580 455,clip]{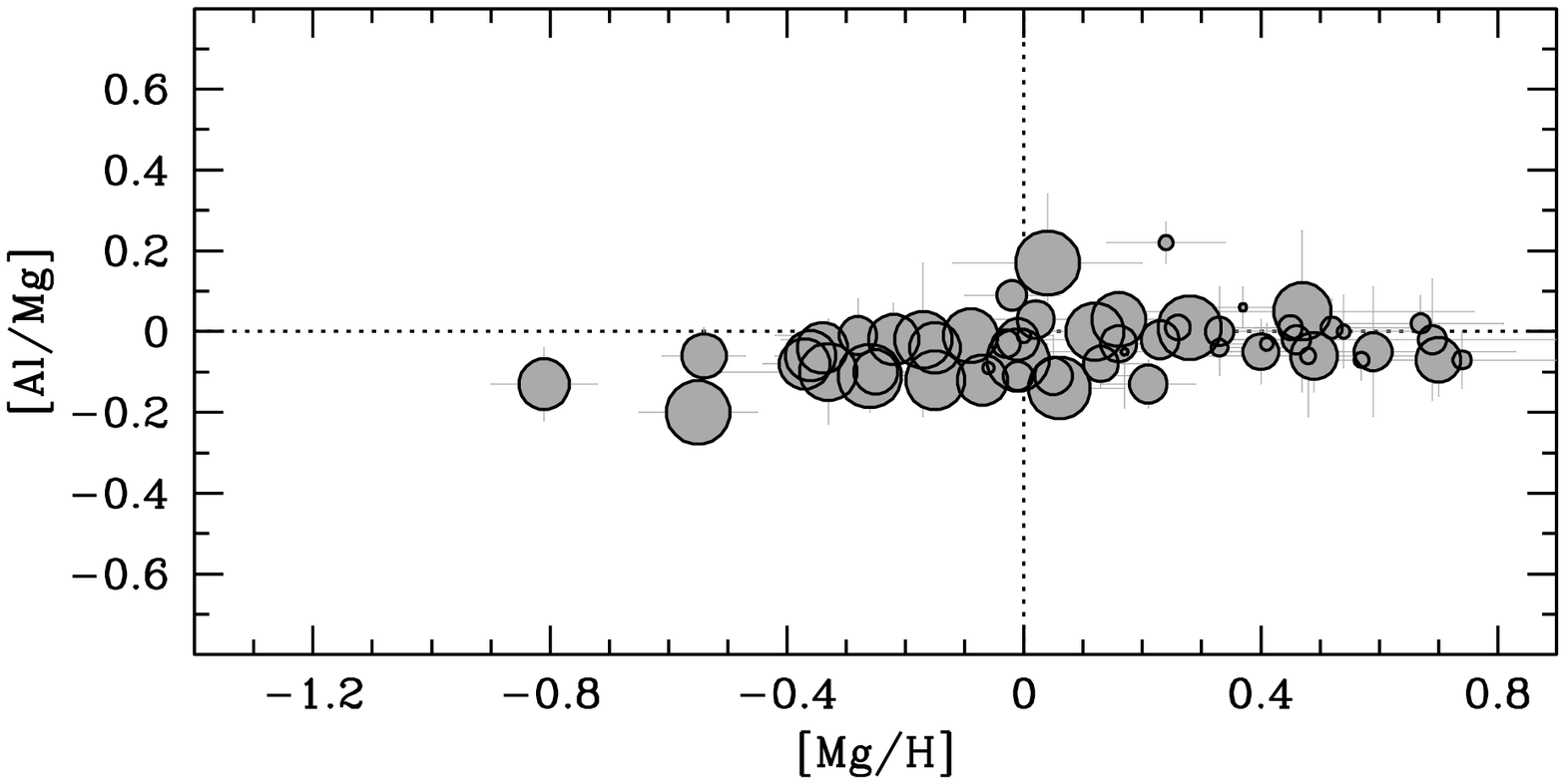}
\includegraphics[bb=30 220 592 455,clip]{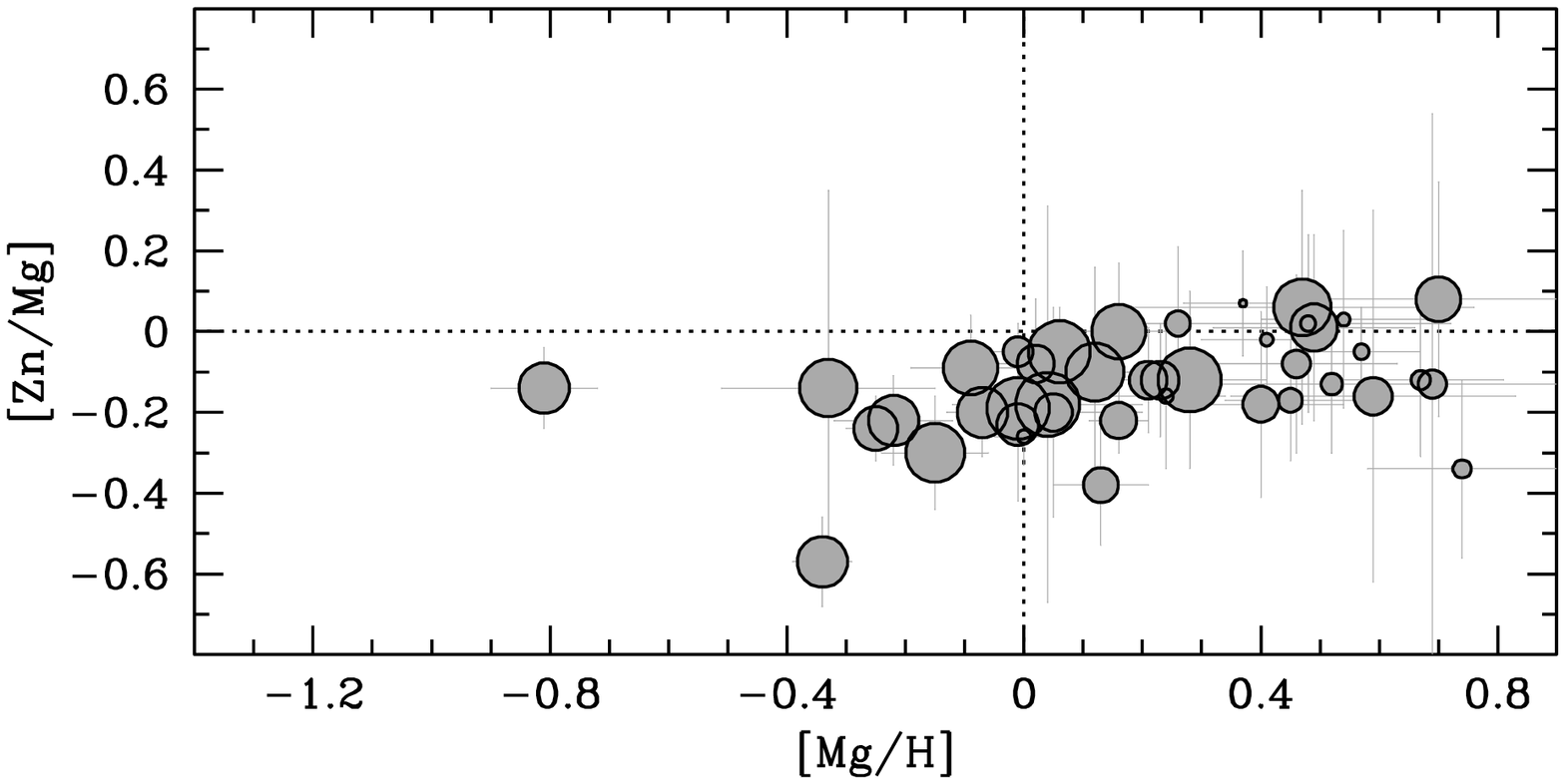}}
\resizebox{\hsize}{!}{
\includegraphics[bb=18 220 580 455,clip]{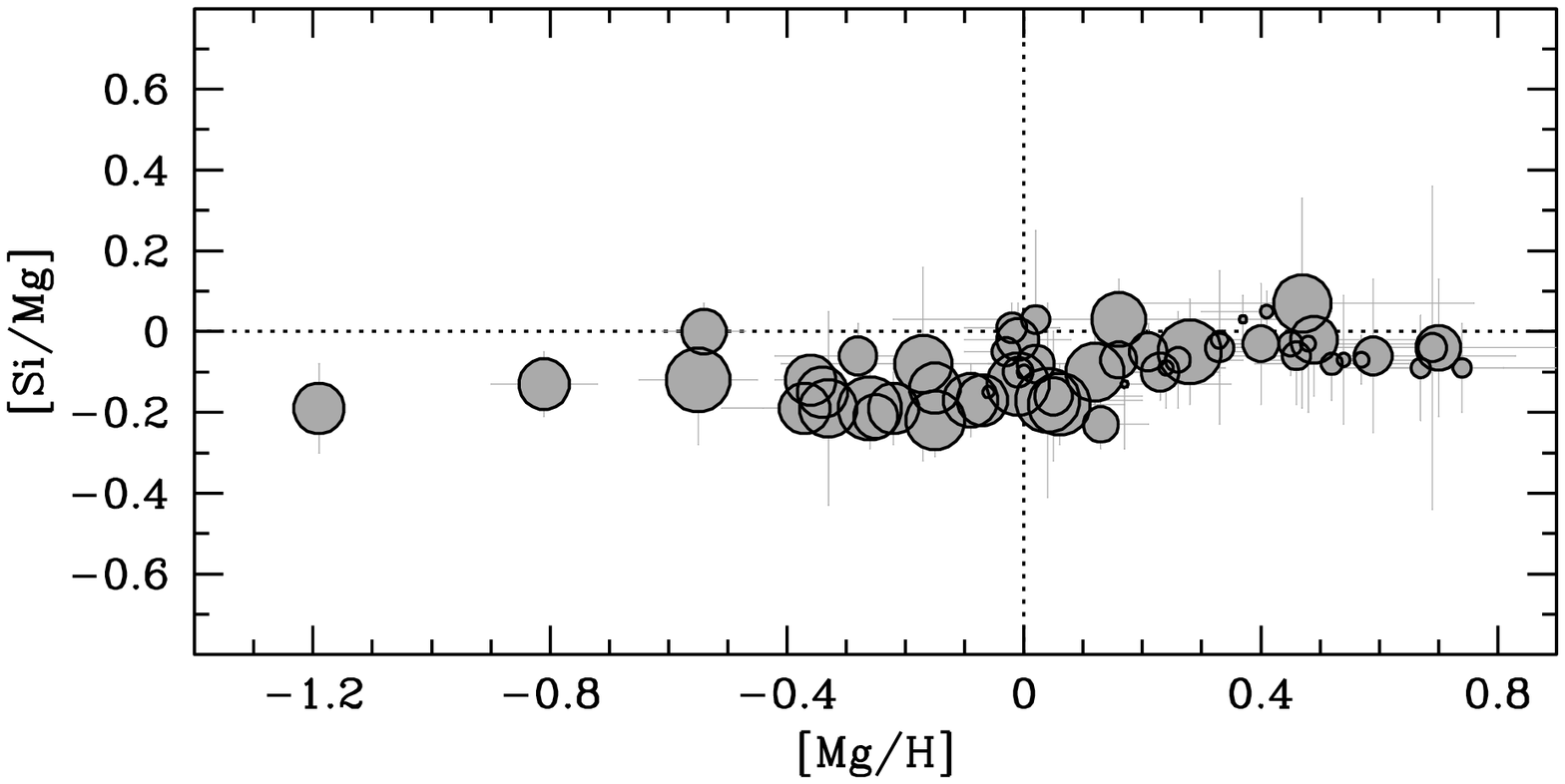}
\includegraphics[bb=30 220 592 455,clip]{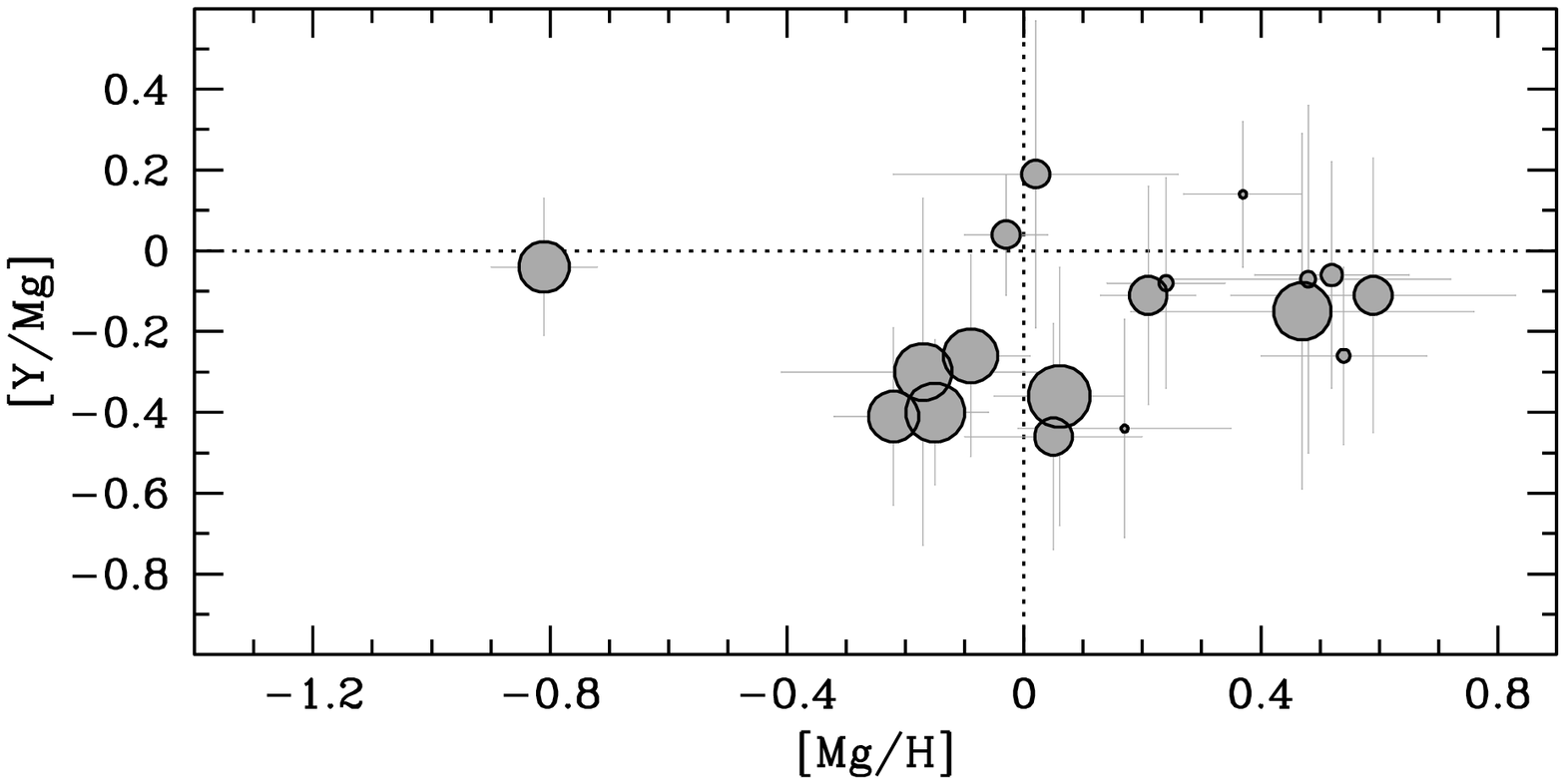}}
\resizebox{\hsize}{!}{
\includegraphics[bb=18 180 580 455,clip]{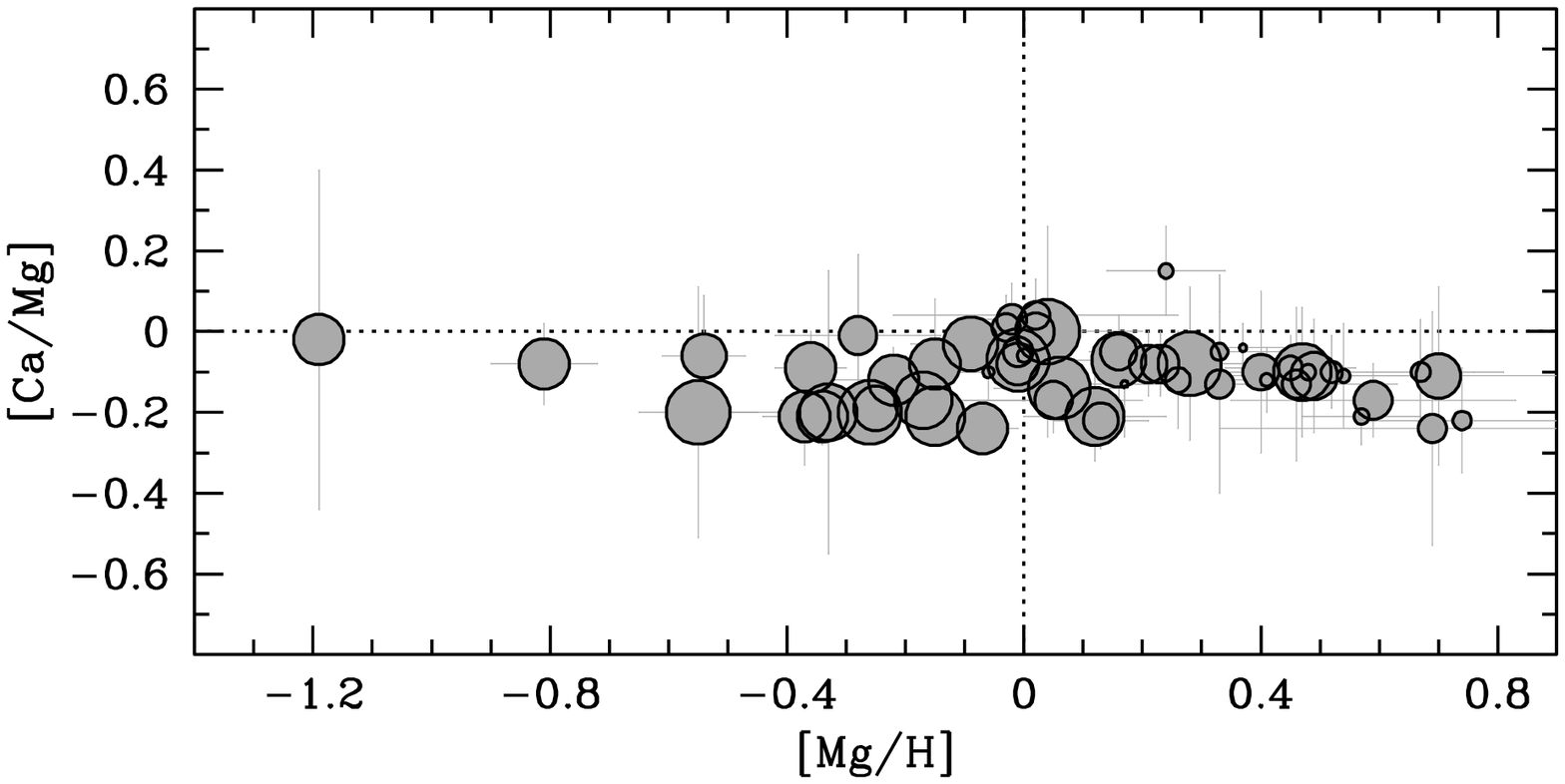}
\includegraphics[bb=30 180 592 455,clip]{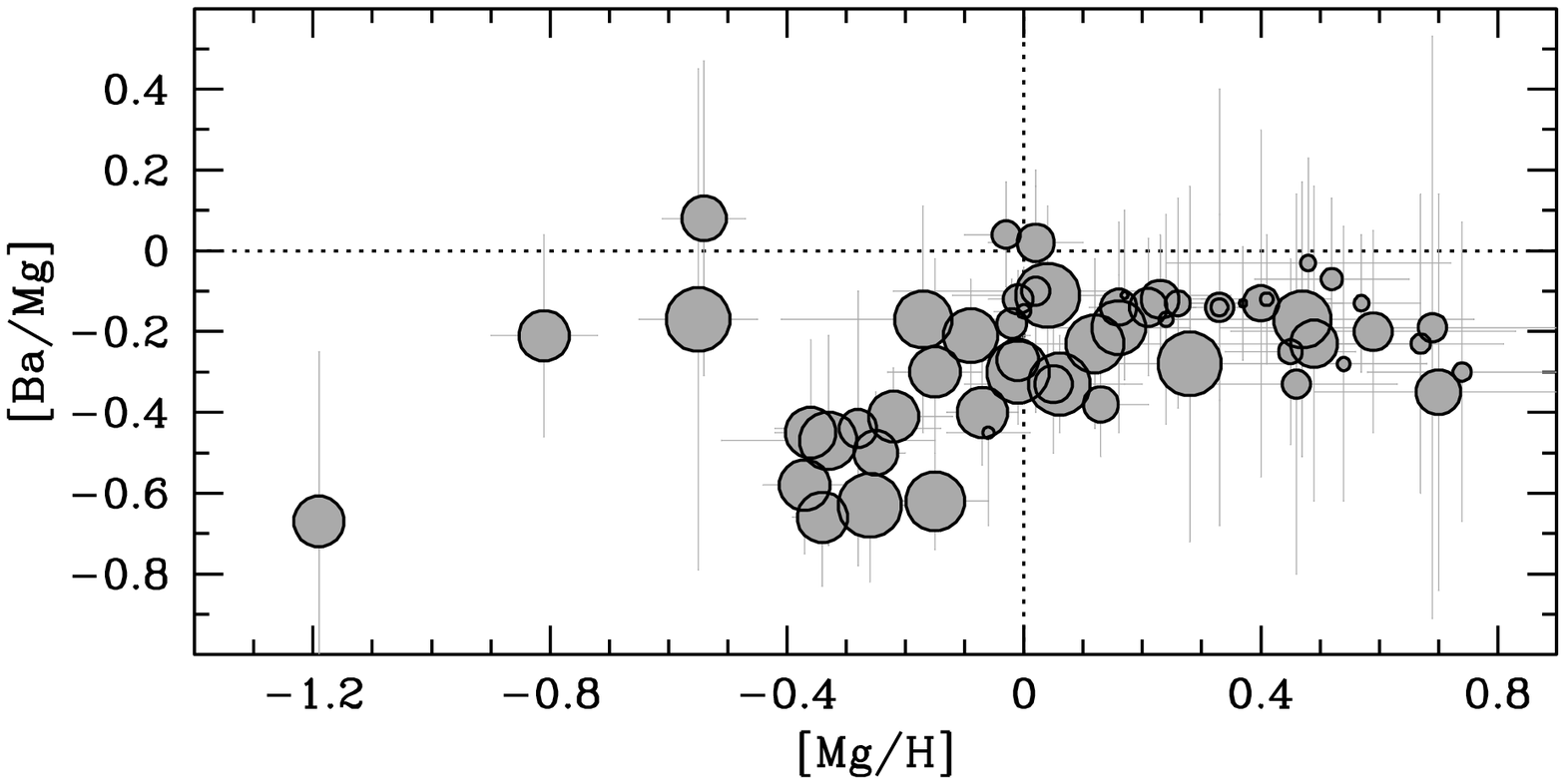}}
\caption{Abundance trends with Mg as reference element for the microlensed
dwarf sample. Circles have been scaled with the ages of the stars.
\label{fig:abundances2}}
\end{figure*}
%-----------------------------------------------------------------------
%-----------------------------------------------------------------------
\begin{figure*}
\resizebox{\hsize}{!}{
\includegraphics[bb=18 445 592 718,clip]{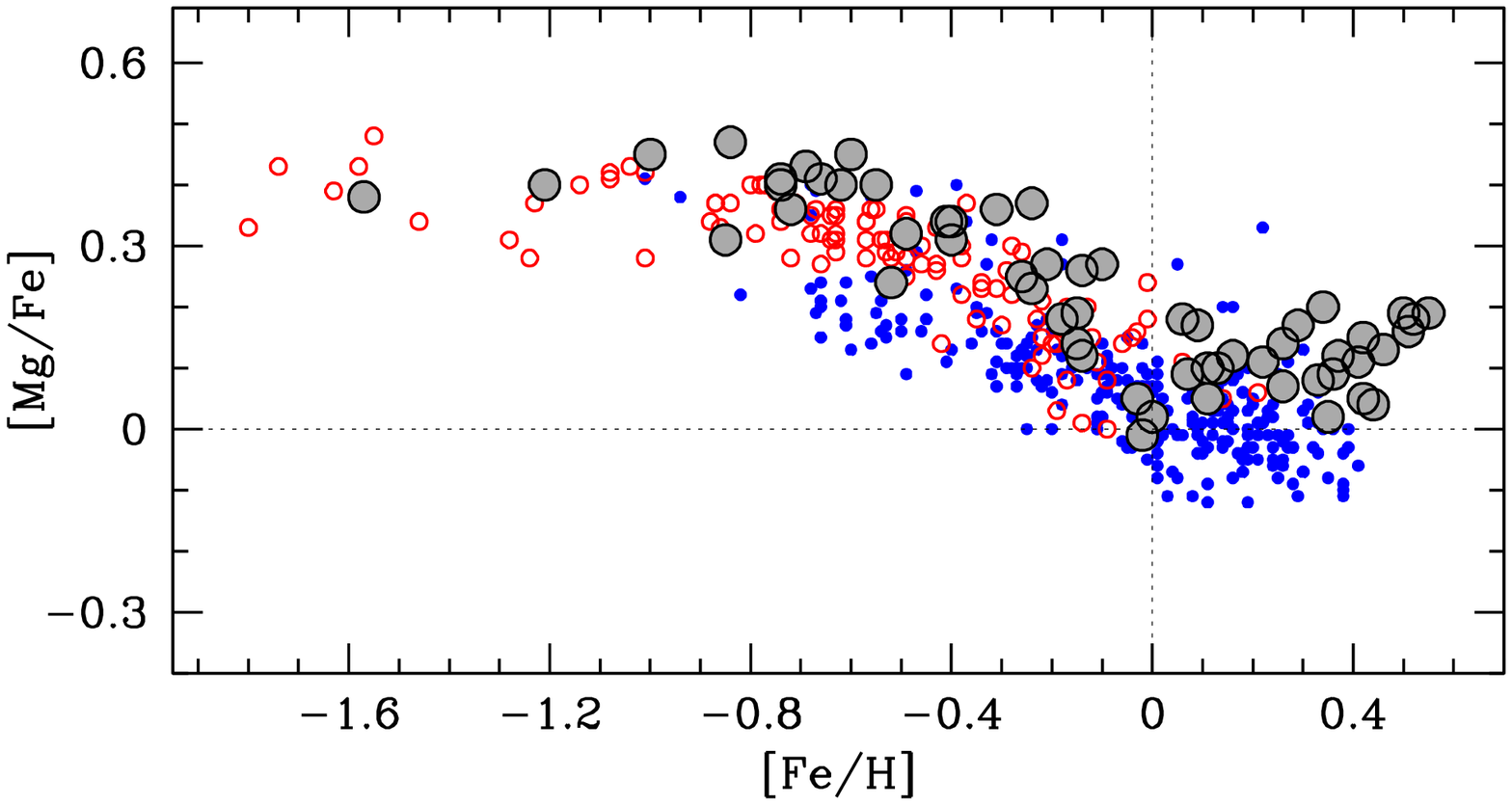}
\includegraphics[bb=18 445 592 718,clip]{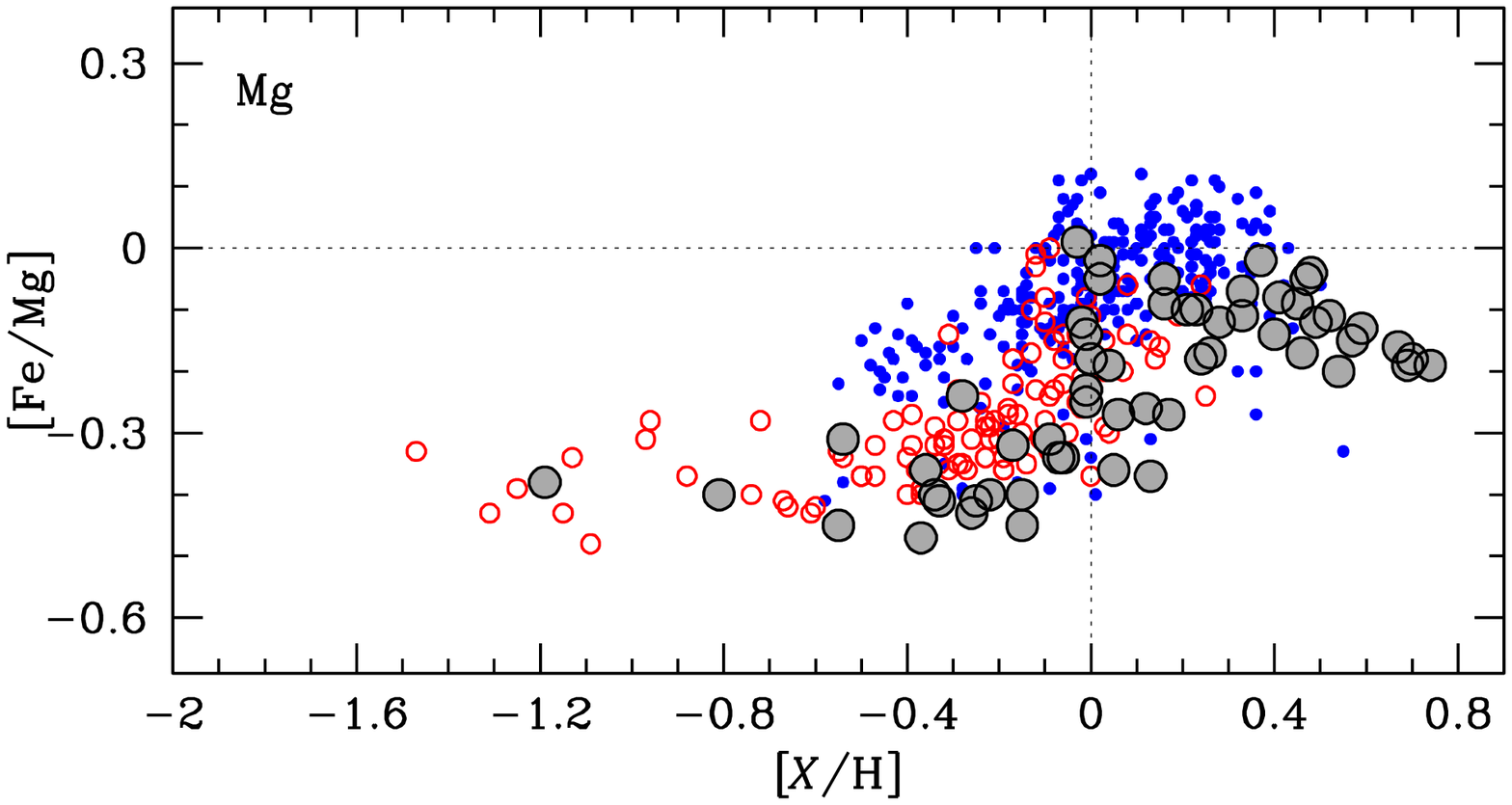}
}
\resizebox{\hsize}{!}{
\includegraphics[bb=18 445 592 700,clip]{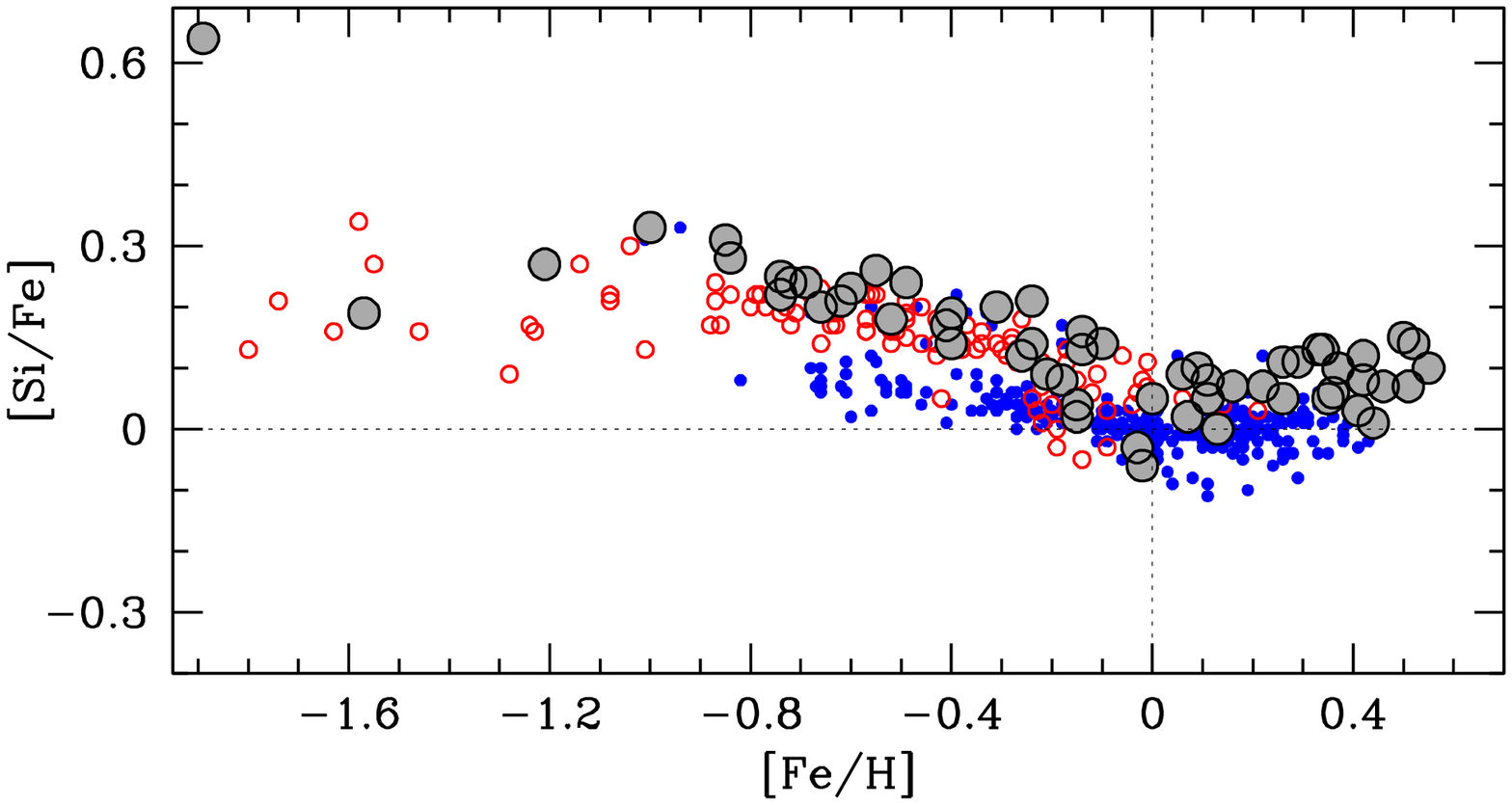}
\includegraphics[bb=18 445 592 700,clip]{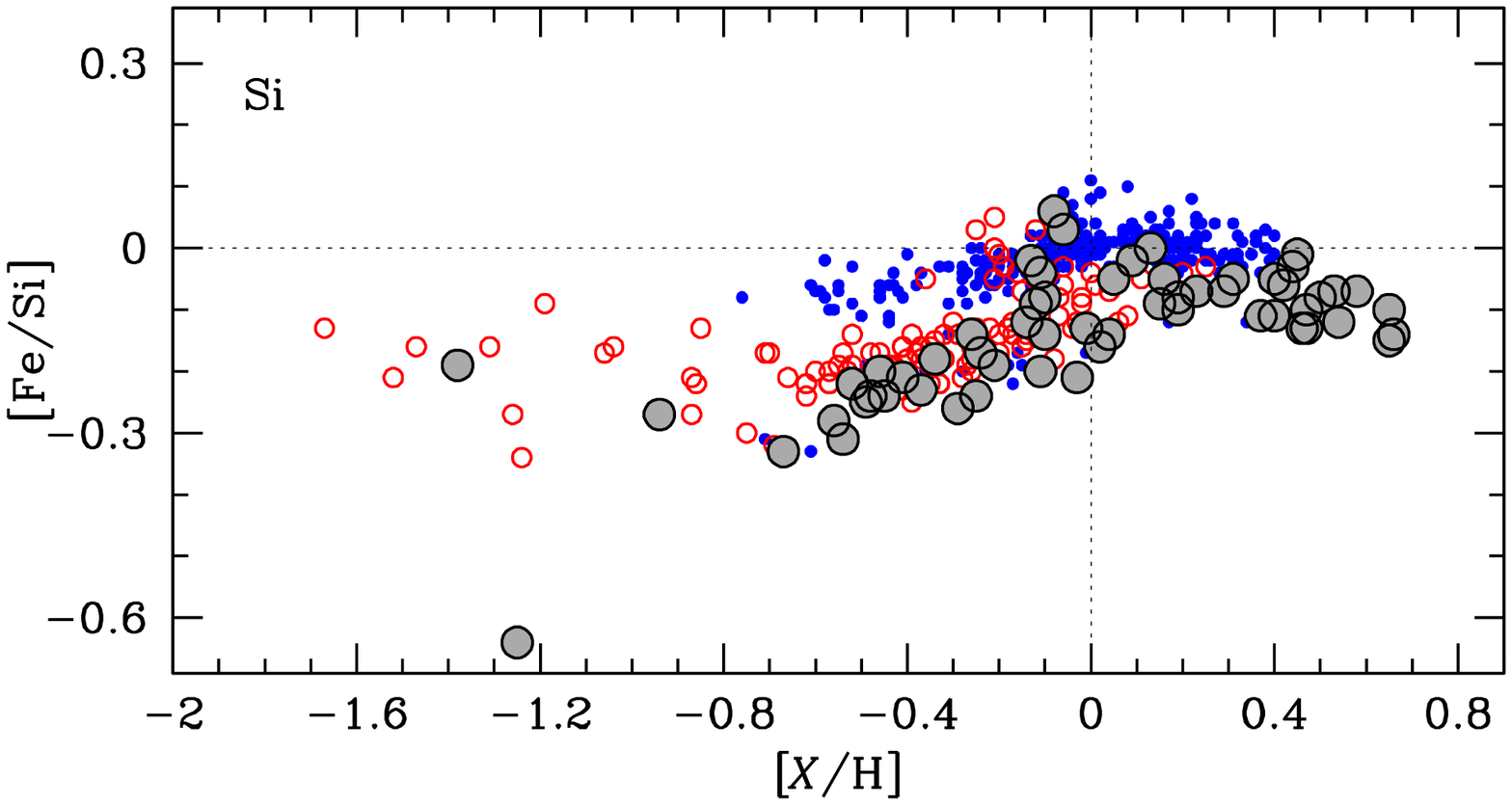}
}
\resizebox{\hsize}{!}{
\includegraphics[bb=18 445 592 700,clip]{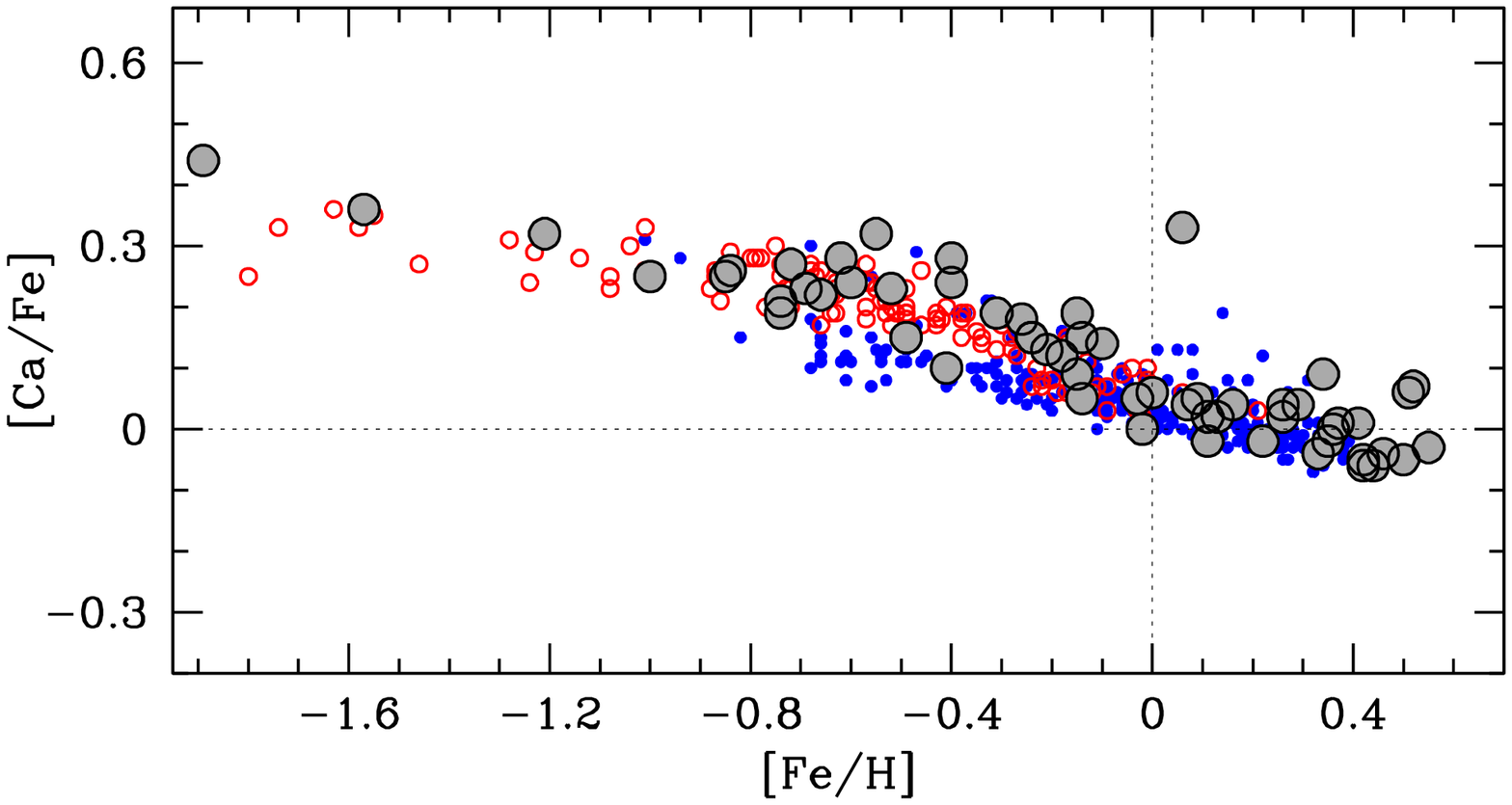}
\includegraphics[bb=18 445 592 700,clip]{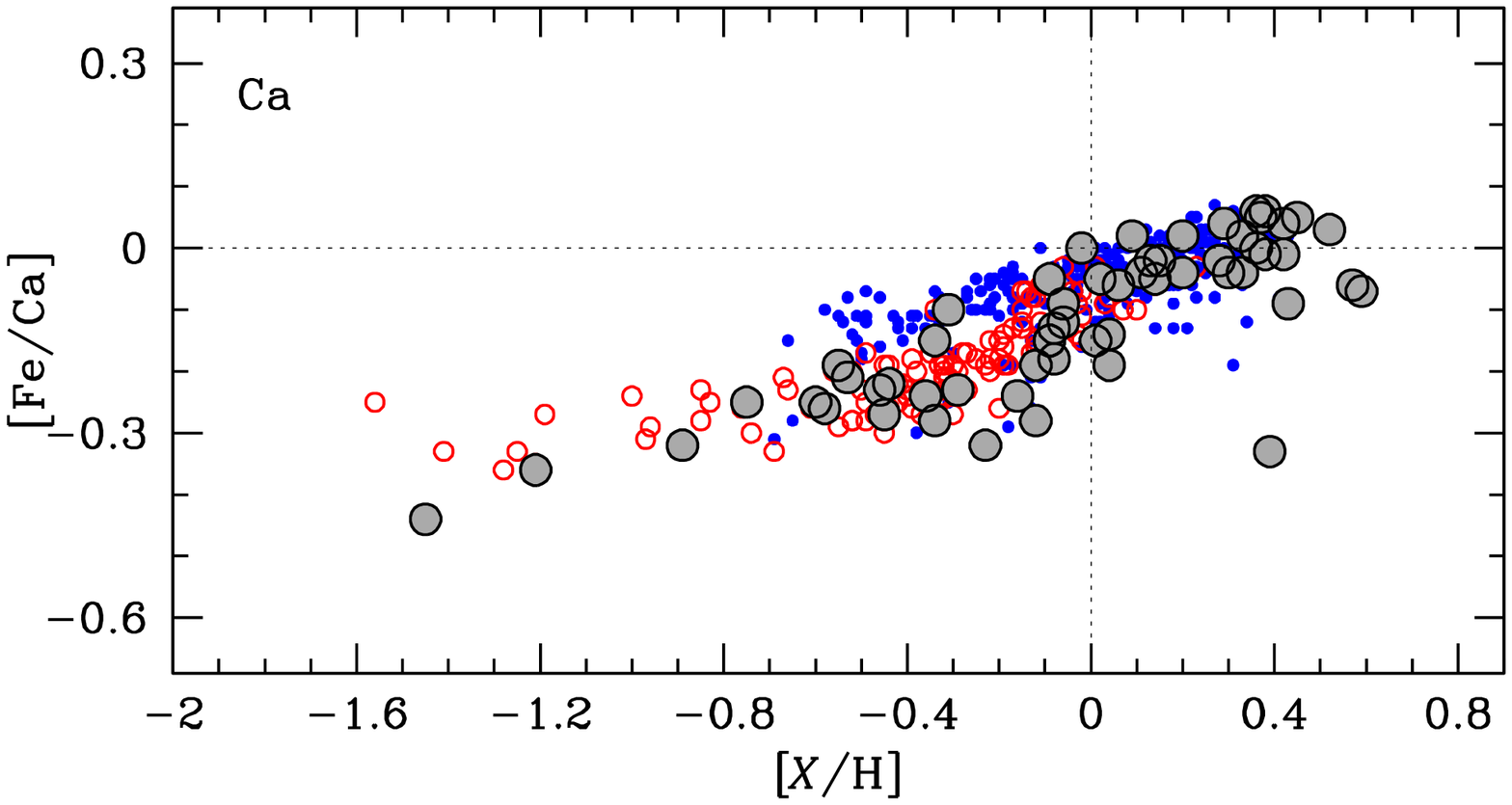}
}
\resizebox{\hsize}{!}{
\includegraphics[bb=18 400 592 700,clip]{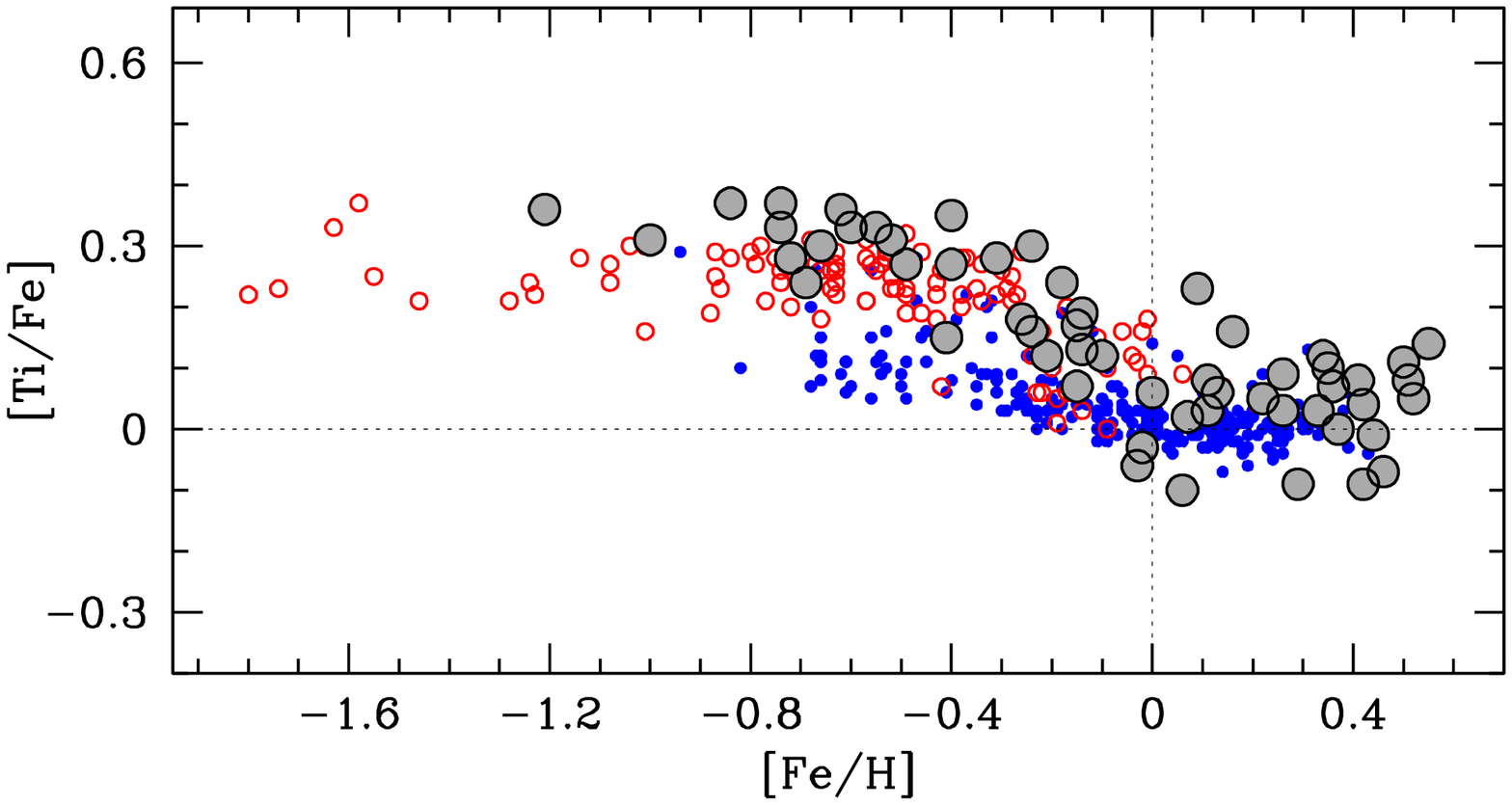}
\includegraphics[bb=18 400 592 700,clip]{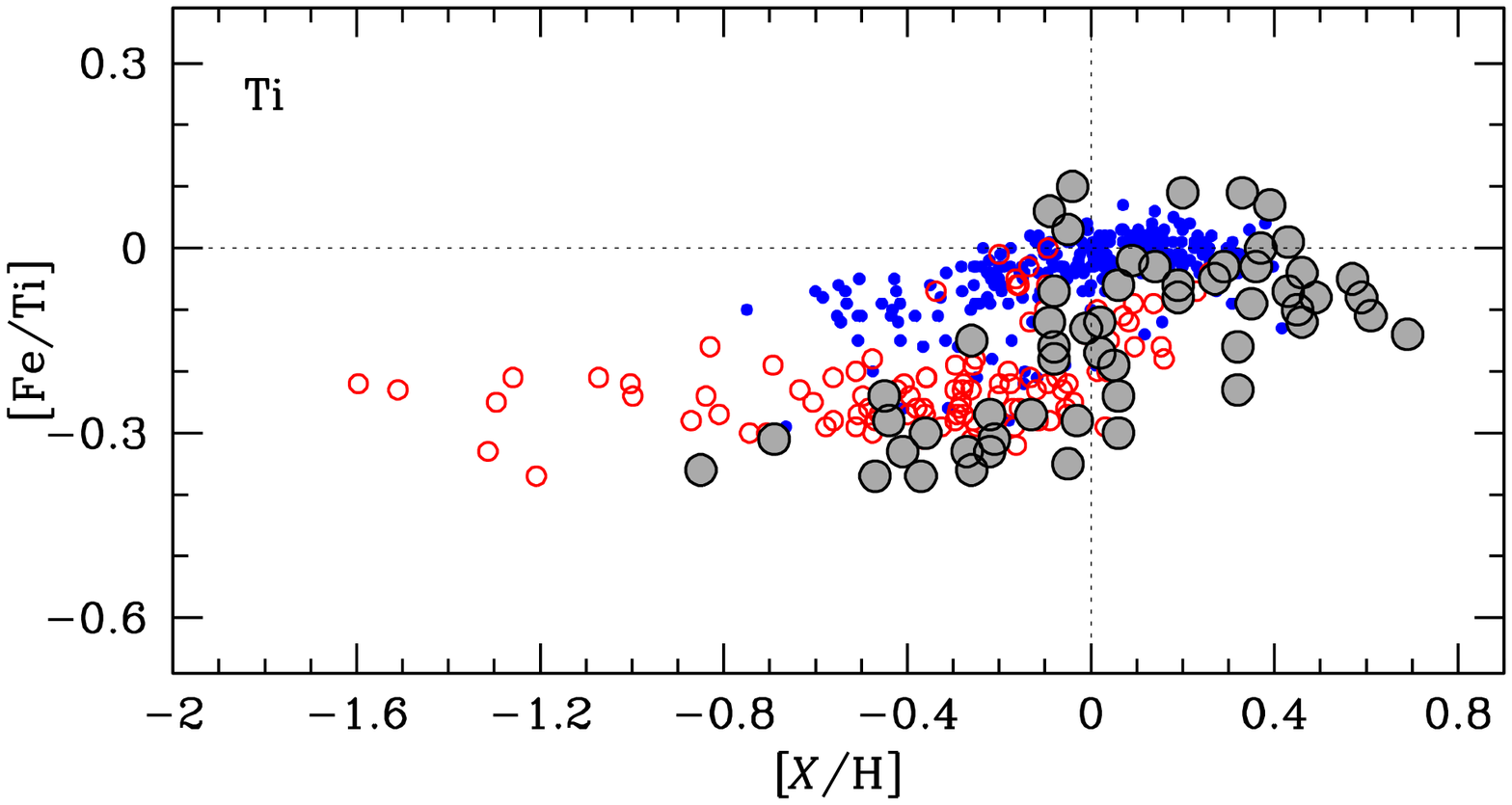}
}
\caption{Abundance trends for the $\alpha$-elements Mg, Si, Ca, and Ti.
Left-hand-side panels show [$X$/Fe] versus [Fe/H], and 
right-hand panels show [Fe/$X$] versus [$X$/H].
Filled grey circles mark the microlensed bulge dwarfs and the
Red and blue circles are nearby thick and thin disk dwarf stars,
respectively, from \cite[][and in prep.]{bensby2003,bensby2005}.
\label{fig:alpha}}
\end{figure*}
%-----------------------------------------------------------------------

%=======================================================================
\subsubsection{General appearance for 12 elements}
\label{sec:abundancetrends}

The plots in Figs~\ref{fig:abundances} and \ref{fig:abundances2} show the
abundance trends for O, Na, Mg,Al, Si, Ca, Ti, Cr, Ni, Fe, Zn, Y, and Ba
for the 58 microlensed bulge dwarfs. In Fig.~\ref{fig:abundances} Fe is the
reference element and in Fig.~\ref{fig:abundances2} Mg is the reference element. 
In all plots the sizes of the circles have been scaled with the ages of 
the stars (larger circles equal higher ages). The $\alpha$-element plots
in Fig.~\ref{fig:abundances} (i.e., [O, Mg, Si, Ca, Ti] versus [Fe/H])
show a plateau of elevated $\rm [\alpha/Fe]$ abundance ratios
at metallicities below $\rm [Fe/H]\approx -0.4$, signalling fast 
chemical enrichment by massive stars. These stars are all old.
The $\alpha$-trends then decline towards solar values, and even though most of
the stars are old, there are one or two that start to show up with lower ages.
At super-solar metallicities the trends generally level out, meaning that there
is some kind of equilibrium between SN\,II and SN\,Ia enrichment. 
The age structure is also very complex with a mix of stars at all ages, 2 to 15\,Gyr. 
What is notable is that the abundance trends are very well-defined and generally
have very low scatter. 

By using an $\alpha$-element that mainly comes from a single source 
(in this case core-collapse supernovae), one might get a clearer picture 
on the chemical history of a stellar population. The ``$x$-axis" could 
then be interpreted as a proxy for time. Hence, 
Fig.~\ref{fig:abundances2} shows the abundance trends again, but now 
with Mg as the reference element. Especially interesting plots
are now the Fe, Cr, and Ni plots where we can see flat, under-abundant, 
[(Fe, Ni, Cr)/Mg] trends for low [Mg/H] values. This is consistent with 
rapid enrichment by massive stars.
The [(Fe, Ni, Cr)/Mg] ratios start to increase around 
$\rm [Mg/H]\approx-0.2$ to $-0.1$, signalling that low-mass stars are 
becoming the dominant source for chemical enrichment of the interstellar 
medium.

%=======================================================================
\subsubsection{Fast enrichment - revising the position of the knee}
\label{sec:alpha}

In \cite{bensby2010} we noted that even though the metal-poor bulge dwarf 
stars generally follow the abundance trends outlined by the thick disk it was also 
apparent that they were slightly more $\alpha$-enhanced, 
mainly being located on the upper rim of the thick disk abundance trends. 
Figure~\ref{fig:alpha} shows the abundance trends for the 
$\alpha$-elements Mg, Si, Ca, and Ti
with the nearby thick disk stars from \cite[][and in prep.]{bensby2003,bensby2005}
included for comparison purposes. As found by \cite{bensby2012_lgb},
discriminating between thin and thick disk stars based on stellar age
appears to produce ``cleaner'' results when applied to local samples, 
and we have therefore
marked stars with ages greater than 9\,Gyr (and age uncertainties
less than 4\,Gyr) as thick disk stars (red circles), and stars with ages
less than 7\,Gyr (and age uncertainties less than 4\,Gyr) as thin disk
stars (blue dots).

At low [Fe/H], the metal-poor bulge dwarfs show an
$\alpha$-to-iron ratio that is similar to what is seen the nearby thick disk
(which also is similar to what is observed in the stellar halo, see, e.g.,
\citealt{nissen2010}). However, it is possible
that the declining [$\alpha$/Fe] ratios (or rising [Fe/$\alpha$] ratios) occur 
at slightly higher [Fe/H] (or [$\alpha$/H]) than for the local thick disk. 
As we now have many more microlensed dwarf stars around and slightly below solar 
[Fe/H], where the thick disk $\rm [\alpha/Fe]$ trends decline, 
we are more confident to claim that this slight shift might be real.
This slight shift in the position of the ``knee'' is clearly evident
in Mg and Ti, slightly less in Si, and barely in Ca. If it is real it 
can be interpreted as being due to the bulge having experienced a faster 
enrichment than the local thick disk, leading to the
onset of enrichment from low-mass stars occur at higher metallicities.

Does this mean that the possible connection between the bulge and the
thick disk becomes weaker? As of now we are only
able to compare the bulge with the thick disk in the solar neighbourhood
while it would be desirable to compare with the thick disk in the
inner regions of the Galaxy. The only such study to date is the one
by \cite{bensby2010letter} but their sample of 44 red giants
is unfortunately too small to whether or not the inner thick disk is different 
from the nearby thick disk. If there is a radial metallicity gradient in the 
thick disk it would mean that the inner thick disk regions  have 
experienced a more rapid evolution, and hence pushed the ``knee" to 
higher metallicities. What we see in the bulge is a ``knee" that starts 
to decline perhaps 0.1\,dex higher in [Fe/H] than seen in the nearby 
thick disk.

%=======================================================================
\section{Discussion}

The first microlensed dwarf stars that were studied provided a real
surprise, an MDF heavily skewed to super-solar metallicities and
significantly different from what had been found from giants
\citep{johnson2007,cohen2008}.  
The difference was somewhat elevated by the very first of our
VLT ToO observations from which we found many of the microlensed dwarfs
to have super-solar metallicities. But as the number of microlensing events grew
we started to also get metal-poor dwarfs in larger numbers, and a 
clearly bi-modal MDF emerged \citep{bensby2010,bensby2011}. That MDF 
was still very different from the best MDF from high-resolution spectra 
of a representative sample of bulge giants \citep{zoccali2008}. 
A recent re-analysis of the \cite{zoccali2008} data by \cite{hill2011} 
and further increased number statistics of the microlensed dwarf stars 
has changed this, and there is now good agreement between the overall
appearance of the MDFs traced by microlensed dwarfs and the giant stars in 
the Galactic bulge (see Figs.~\ref{fig:mdf} and \ref{fig:mdfcum}).

The MDF is an important constraint to models of galactic chemical
evolution \citep[e.g.,][]{matteucci1989}. The MDF of the Galactic bulge
is best probed through the detailed elemental abundances of stars.
Most studies use the intrinsically bright red giant stars for this.
The first studies showed a metal-rich population with a somewhat
broadened distribution of [Fe/H]. Coupled with the high $\alpha$
abundances this pointed to a short formation time-scale
\citep{rich1988,mcwilliam1994,fulbright2007}. However, the mean [Fe/H]
changed between these studies leaving substantial room for
interpretation. Recently, with the arrival of the multi-fibre
spectrograph FLAMES \citep{pasquini2002} on the VLT, it has become 
feasible to probe further down the red giant branch and reach 
warmer stars that are less prone to evolutionary effects (compare 
lack of metal-rich M giants in \citealt{rich2012}) and also to the 
red clump which should give a reliable tracer of the MDF. 

\begin{figure}
\resizebox{\hsize}{!}{
\includegraphics[bb=18 140 580 520,clip]{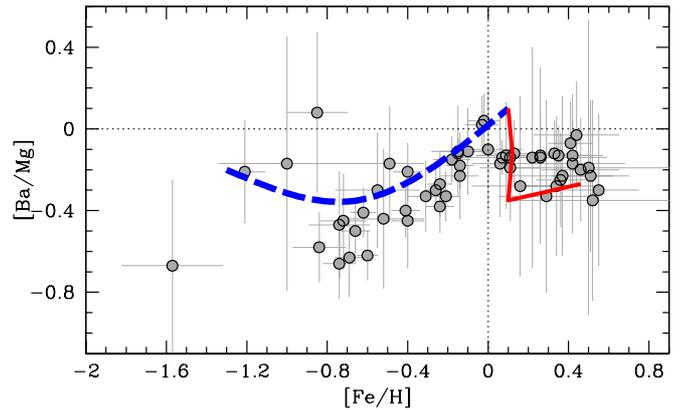}}
\caption{[Ba/Mg] versus [Fe/H]  for the microlensed 
dwarf sample. The two-component bulge model by \cite{tsujimoto2012} 
is shown (blue dashed line is the first component that forms quickly, 
and red solid line is the second  component that forms from pre-enriched gas).
\label{fig:bamgfe}}
\end{figure}

Chemical evolution models have been created to attempt to explain
the broad spread in [Fe/H] and to explain the fact that the bulge appears
to have two populations: one with a low mean [Fe/H] and one with
super-solar [Fe/H]. For example, the two-component
chemical evolution model by \cite{tsujimoto2012}
is able to explain the [Ba/Mg] versus [Fe/H]
pattern seen in the dwarf stars (Fig.~\ref{fig:bamgfe}) as well 
as a bi-modal MDF. They also find that a top-heavy IMF is essential
to explain the high metallicity of the metal-rich component of the MDF.

%-----------------------------------------------------------------------
%-----------------------------------------------------------------------

\citet{grieco2012} also explores a multi-component model of the
chemical evolution of the Galactic bulge. They model a two component
bulge with one representing the classical bulge and the other the
pseudo-bulge (bar population?). One prediction from their model is
that there could be an abundance gradient within the classical, spherical
bulge but that also the differences between the chemical evolution
of the spherical component and the bar component could be (erroneously)
interpreted as a gradient. 

In simulations of the formation and evolution of the Milky Way in a 
cosmological context, the star formation of the bulge lasts for longer 
than a single starburst 
\citep{rahimi2010,kobayashi2011,scannapieco2011,domenechmoral2012}.
For example, while most of the bulge stars in \cite{kobayashi2011}
are old (80\%$>10$\,Gyr, and 90\%$>8$\,Gyr), ~10\% of stars form later, 
at super-solar metallicities. While \cite{kobayashi2011}
simulated a single history for a MW-like galaxy, 
\cite{scannapieco2011} and \cite{domenechmoral2012}
each modelled several halo histories. The galaxies showed different star 
formation histories for all components, and it is therefore likely that 
star formation histories are not universal in each galaxy type. However, 
several of their simulationsshow extended  star formation histories for 
the bulge, although the interpretation of \cite{scannapieco2011} is 
complicated by the fact that their "inner spheroid" extends $\sim10$\,kpc. 
While it is not yet clear how well these simulations model the behaviour 
of baryons through their sub-grid physics, that some fraction of the bulge 
consists of stars younger than 8 Gyr is not surprising within the 
framework of current theories.

\paragraph{So what is the bulge?}
Where do all the observational evidence point to? 
One thing is clear, there is overwhelming evidence for more than
one population in the bulge. Whether it is two, or even more, it is
a fact that the inner region of the Galaxy is where the major Galactic 
stellar populations (if they all exist in this region) should meet and overlap.
For instance, the interpretation by Ness et al.~(in prep.) of their
multicomponent MDF is that the $\rm [Fe/H]=+0.14$ peak belongs 
to the thin disk surrounding the bulge, the $\rm [Fe/H]=-0.23$ peak 
is the true boxy/peanut bulge, and the $\rm [Fe/H]=-0.60$ peak is 
the old thick disk, which may or may not be part of the bulge. 
Adding the Galactic bar (possibly formed from disk stars), and 
the influence that it has on its surroundings, the situation is 
truly intricate. However, at the same time the metal-poor bulge
shows abundance trends that are very well-defined with low scatter,
which could argue against a bulge
composed of more than two populations (as proposed by 
Ness et al. in prep., see Fig.~\ref{fig:mdf}d).

As noted in Sect.~\ref{sec:alpha}, the $\alpha$-element trends 
for the metal-poor bulge dwarfs now appear to be slightly different 
from those of the nearby thick disk, possibly implying that the metal-poor 
bulge population experienced a somewhat more intense star formation than 
the local thick disk. This does not necessarily weaken a possible connection
between the metal-poor bulge component and the thick disk. It might just 
instead be a reflection of the more intense star formation in the inner 
regions of the Galaxy than in and around the solar circle. 

The abundance trends seen for the metal-rich bulge is partly consistent 
with what is seen for metal-rich nearby thin disk dwarf stars 
\citep[e.g.,][]{bensby2003,bensby2005,reddy2006,bensby2007letter2}, 
although disk stars with metallicities higher than 
$\rm [Fe/H]\gtrsim +0.4$ are rarely seen. Given the wide range in 
ages, the metal-rich bulge could be the most complex region in the 
Galaxy, possibly hosting stars from several populations.

A bimodal age distribution in the bulges of external spirals galaxies 
with bars has been recently found through a stellar population 
synthesis analysis by \cite{coelho2011}, 
with mean ages of 4.7 and 10.4 Gyr. Interestingly, unbarred galaxies
do not show this bimodal distribution, perhaps implying that bars
have the effect of rejuvenating bulges. The above results are in line 
with our findings based on Galactic bulge dwarf stars, and a larger 
sample will allow us to obtain firmer results for comparison with 
external bulges.

%=======================================================================
\section{Summary}

In this study we have presented a detailed elemental abundance 
analysis of 32 microlensed dwarf and subgiant stars in the bulge.
Together with the previous sample of 26 microlensed bulge dwarfs
from \cite{bensby2010,bensby2011} the sample now contains 58 stars.
In summary, the findings and main results from this data set are:
\begin{itemize}
\item   The metallicities span the full range between $\rm [Fe/H] = -2$ to $+0.6$.
The MDF that appeared clearly bi-modal, even with a paucity of stars around
$\rm [Fe/H]=0$, when based on the 26 stars in \cite{bensby2011}, is now
less so. The overall shape of the MDF is now similar to the MDF in 
Baade's window found by \cite{hill2011}, as traced by both red giants and red 
clump giants. Furthermore, signatures of more than two components
is starting to emerge in the microlensed dwarf star MDF. 
\item   The microlensed dwarf stars more metal-poor than 
$\rm [Fe/H]\lesssim-0.4$ are all old with ages around or greater than 10\,Gyr. 
The more metal-rich stars, on the other hand, have a wide age distribution 
with a significant fraction of stars having ages lower than 5 to 6\,Gyr,
and quite a large number of stars with intermediate ages around 5-7\,Gyr. 
These are essentially indistinguishable from an old turn-off when the metallicities 
are unknown.  We have investigated in some detail 
whether an He enriched bulge, as proposed by \cite{nataf2012}, could 
explain the large fraction of young stars in the microlensed dwarf sample, 
but the effects are not large enough. 
We have also investigated sampling biases and find that our microlensed dwarf 
sample could be biased in the sense that the number of young metal-rich stars 
is over-estimated (perhaps by as much as 50\,\%). In any case, neither sampling bias,
nor a He enriched bulge, can fully account for the large fraction of young
and intermediate age stars in the microlensed bulge dwarf star sample.
\item With the now much expanded sample of 58 stars we see that the metal-poor 
bulge population have abundance trends slightly different from what is observed 
in the nearby Galactic thick disk. The ``knee'' in the $\rm [\alpha/Fe]-[Fe/H]$
trends occurs at a slightly higher metallicity in the bulge,  
indicating that the metal-poor bulge component has experienced a somewhat 
faster star formation rate. 
\item The microlensed bulge dwarfs with solar and super-solar [Fe/H] show
abundance trends similar to what is seen in the nearby
metal-rich thin and thick disks. If the bar population is formed from
disk material, this is what could be expected.
\end{itemize}
Even though it is difficult to pinpoint the exact origin of the bulge,
observational evidence now clearly indicates that the bulge has a complex
structure with wide age and metallicity distributions. Combined
with the cylindrical rotation of the bulge, 
the bulge appears to be a conglomerate of Galactic stellar populations
under the influence of the bar. 

%=======================================================================
\begin{acknowledgement}

 We are grateful to and thank Patrick Baumann who obtained the MIKE
 spectrum for MOA-2011-BLG-234S, Julio Chaname who obtained the MIKE
 spectrum for MOA-2011-BLG-278S, G.W.~Marcy who obtained the HIRES 
 spectra for MOA-2011-BLG-191S and OGLE-2012-BLG-0816S, and Elizabeth Wylie
 and Mario Mateo who obtained the MIKE spectrum for OGLE-2012-BLG-1279S.
 We would also like to thank Bengt Gustafsson, Bengt 
 Edvardsson, and Kjell Eriksson for usage of the MARCS model atmosphere 
 program and their suite of stellar abundance programs. 
 T.B. was funded by grant No. 621-2009-3911 from The Swedish 
 Research Council.
 S.F. was partly funded by grant No. 2008-4095 from The Swedish Research Council.
 Work by J.C.Y. was supported by an SNSF Graduate Research Fellowship under
 Grant No. 2009068160. A.G. and J.C.Y. acknowledge support from NSF AST-1103471.
 M.A. gratefully acknowledges funding from
 the Australian Research Council (FL110100012).
 J.L.C. is grateful to NSF award AST-0908139 for partial support.
 S.L. reasearch is partially supported by the INAF PRIN grant 
 "Multiple populations in Globular Clusters: their role in the Galaxy assembly" 
 J.M. thanks support from FAPESP (2010/50930-6), 
 USP (Novos Docentes) and CNPq (Bolsa de produtividade).
 A.G.-Y. acknowledges support by the Lord Sieff of Brimpton Fund.
 The OGLE project has received funding from the European Research Council
 under the European Community's Seventh Framework Programme
 (FP7/2007-2013) / ERC grant agreement no. 246678 to AU.
 Work by C.H. was supported by the Creative Research
 Initiative Program (2009-0081561) of National Research
 Foundation of Korea. 
 T.B, S.F., and J.A.J. are grateful to the Aspen Center for Physics and 
 the NSF Grant 1066293 for hospitality during the ``Galactic bulge and bar'' 
 workshop in August 2011, where many inspiring discussions were held.
\end{acknowledgement}
%=======================================================================
\bibliographystyle{aa}
\bibliography{referenser}

\end{document}